\documentclass[prx,superscriptaddress,twocolumn,nopreprintnumbers,floatfix,nofootinbib,]{aastex62}

\usepackage{amssymb,amsmath,verbatim,mathtools,needspace,enumitem,etoolbox,graphicx,physics,microtype,ulem,units,}
\usepackage[T1]{fontenc} 
\usepackage{xcolor} 
\usepackage{colortbl} 
\usepackage{cleveref}
\usepackage[caption=false]{subfig}
\usepackage{multirow}
\usepackage{wrapfig} 
\usepackage{xstring} 

\newcommand{\PrimaryMassMayOnePercentile}{\ensuremath{67}}

\newcommand{\PrimaryMassMayNinetyNinePercentile}{\ensuremath{83}}

\newcommand{\peakNoAugNoEvolutionMTwoMinPercentBelow}{\ensuremath{0.02}}

\newcommand{\peakNoAugNoEvolutionMassRatioMinPercentBelow}{\ensuremath{0.02}}

\newcommand{\peakNoAugNoEvolutionObsPrimaryMassGrEightyPercent}{\ensuremath{32}}

\newcommand{\peakNoAugNoEvolutionMassRatioTenPercentile}{\ensuremath{0.26^{+0.14}_{-0.08}}}
 
\newcommand{\MassRatioAprNinetyNinePercentile}{\ensuremath{0.53}}

\newcommand{\peakNoAugNoEvolutionRateMassBandFourToTen}{\ensuremath{8.91^{+11.28}_{-5.25}}}
 
\newcommand{\BPLNoAugNoEvolutionRateMassBandFourToTen}{\ensuremath{7.86^{+11.17}_{-4.81}}}
 
\newcommand{\multipeakNoAugNoEvolutionRateMassBandFourToTen}{\ensuremath{9.54^{+11.48}_{-5.78}}}
 
\newcommand{\peakNoAugNoEvolutionRateMassBandTenToTwenty}{\ensuremath{7.07^{+4.58}_{-2.98}}}
 
\newcommand{\BPLNoAugNoEvolutionRateMassBandTenToTwenty}{\ensuremath{8.89^{+5.29}_{-3.57}}}
 
\newcommand{\multipeakNoAugNoEvolutionRateMassBandTenToTwenty}{\ensuremath{7.02^{+4.54}_{-3.06}}}
 
\newcommand{\peakNoAugNoEvolutionRateMassBandTwentyToThirty}{\ensuremath{2.75^{+1.94}_{-1.14}}}
 
\newcommand{\BPLNoAugNoEvolutionRateMassBandTwentyToThirty}{\ensuremath{3.71^{+1.58}_{-1.1}}}
 
\newcommand{\multipeakNoAugNoEvolutionRateMassBandTwentyToThirty}{\ensuremath{2.6^{+1.88}_{-1.3}}}
 
\newcommand{\peakNoAugNoEvolutionRateMassBandThirtyToForty}{\ensuremath{3.05^{+1.86}_{-1.13}}}
 
\newcommand{\BPLNoAugNoEvolutionRateMassBandThirtyToForty}{\ensuremath{2.04^{+1.16}_{-0.78}}}
 
\newcommand{\multipeakNoAugNoEvolutionRateMassBandThirtyToForty}{\ensuremath{3.19^{+2.02}_{-1.22}}}
 
\newcommand{\truncatedNoAugNoEvolutionMOnePercentile}{\ensuremath{6.0^{+0.8}_{-2.1}}}
 
\newcommand{\truncatedNoAugNoEvolutionMNinetyNinePercentile}{\ensuremath{65.5^{+10.4}_{-9.6}}}
 
\newcommand{\peakNoAugNoEvolutionMOnePercentile}{\ensuremath{6.0^{+1.0}_{-1.6}}}
 
\newcommand{\peakNoAugNoEvolutionMNinetyNinePercentile}{\ensuremath{59.7^{+13.9}_{-12.8}}}
 
\newcommand{\BPLNoAugNoEvolutionMOnePercentile}{\ensuremath{5.6^{+1.3}_{-1.6}}}
 
\newcommand{\BPLNoAugNoEvolutionMNinetyNinePercentile}{\ensuremath{57.8^{+12.5}_{-8.7}}}
 
\newcommand{\multipeakNoAugNoEvolutionMOnePercentile}{\ensuremath{6.0^{+1.0}_{-1.6}}}
 
\newcommand{\multipeakNoAugNoEvolutionMNinetyNinePercentile}{\ensuremath{66.0^{+12.1}_{-16.4}}}
 
\newcommand{\truncatedNoMayNoEvolutionMOnePercentile}{\ensuremath{5.8^{+0.9}_{-2.8}}}
 
\newcommand{\truncatedNoMayNoEvolutionMNinetyNinePercentile}{\ensuremath{52.3^{+8.9}_{-5.2}}}
 
\newcommand{\peakNoMayNoEvolutionMOnePercentile}{\ensuremath{5.9^{+1.0}_{-1.5}}}
 
\newcommand{\peakNoMayNoEvolutionMNinetyNinePercentile}{\ensuremath{52.4^{+13.2}_{-7.0}}}
 
\newcommand{\BPLNoMayNoEvolutionMOnePercentile}{\ensuremath{5.6^{+1.3}_{-1.8}}}
 
\newcommand{\BPLNoMayNoEvolutionMNinetyNinePercentile}{\ensuremath{52.3^{+9.5}_{-5.9}}}
 
\newcommand{\multipeakNoMayNoEvolutionMOnePercentile}{\ensuremath{6.2^{+0.8}_{-1.5}}}
 
\newcommand{\multipeakNoMayNoEvolutionMNinetyNinePercentile}{\ensuremath{58.9^{+11.0}_{-11.7}}}
 
\newcommand{\truncatedAllNoEvolutionMOnePercentile}{\ensuremath{2.4^{+0.2}_{-0.3}}}
 
\newcommand{\truncatedAllNoEvolutionMNinetyNinePercentile}{\ensuremath{60.3^{+10.1}_{-13.2}}}
 
\newcommand{\peakAllNoEvolutionMOnePercentile}{\ensuremath{2.5^{+0.3}_{-0.3}}}
 
\newcommand{\peakAllNoEvolutionMNinetyNinePercentile}{\ensuremath{57.8^{+15.3}_{-14.5}}}
 
\newcommand{\BPLAllNoEvolutionMOnePercentile}{\ensuremath{2.6^{+0.5}_{-0.3}}}
 
\newcommand{\BPLAllNoEvolutionMNinetyNinePercentile}{\ensuremath{56.1^{+12.5}_{-9.0}}}
 
\newcommand{\multipeakAllNoEvolutionMOnePercentile}{\ensuremath{2.6^{+0.4}_{-0.3}}}
 
\newcommand{\multipeakAllNoEvolutionMNinetyNinePercentile}{\ensuremath{63.8^{+11.2}_{-19.1}}}

\newcommand{\truncatedNoAugNoEvolutionBF}{\ensuremath{0.01}}
 
\newcommand{\truncatedNoAugNoEvolutionLogBF}{\ensuremath{-1.91}}
 
\newcommand{\peakNoAugNoEvolutionBF}{\ensuremath{1.0}}
 
\newcommand{\peakNoAugNoEvolutionLogBF}{\ensuremath{0.0}}
 
\newcommand{\BPLNoAugNoEvolutionBF}{\ensuremath{0.12}}
 
\newcommand{\BPLNoAugNoEvolutionLogBF}{\ensuremath{-0.92}}
 
\newcommand{\multipeakNoAugNoEvolutionBF}{\ensuremath{0.5}}
 
\newcommand{\multipeakNoAugNoEvolutionLogBF}{\ensuremath{-0.3}}
 
\newcommand{\BPLPeakNoAugNoEvolutionBF}{\ensuremath{0.74}}
 
\newcommand{\BPLPeakNoAugNoEvolutionLogBF}{\ensuremath{-0.13}}
 
\newcommand{\peakNoAugNoEvolutionNoPeakBF}{\ensuremath{0.05}}
 
\newcommand{\peakNoAugNoEvolutionNoPeakLogBF}{\ensuremath{-1.34}}
 
\newcommand{\peakNoAugNoEvolutionNoSmoothingBF}{\ensuremath{0.87}}
 
\newcommand{\peakNoAugNoEvolutionNoSmoothingLogBF}{\ensuremath{-0.06}}
 
\newcommand{\BPLNoAugNoEvolutionNoSmoothingBF}{\ensuremath{0.35}}
 
\newcommand{\BPLNoAugNoEvolutionNoSmoothingLogBF}{\ensuremath{-0.46}}

\newcommand{\peakNoAprNoEvolutionbeta}{\ensuremath{4.0^{+6.4}_{-3.2}}}

\newcommand{\BPLNoAprNoEvolutionbeta}{\ensuremath{4.5^{+5.9}_{-3.5}}}

\newcommand{\truncatedNoAugNoEvolutionMminUpperNinety}{\ensuremath{6.6}}
 
\newcommand{\truncatedNoAugNoEvolutionmmax}{\ensuremath{78.5^{+14.1}_{-9.4}}}

\newcommand{\truncatedNoAugNoEvolutionrate}{\ensuremath{33^{+22}_{-12}}}
 
\newcommand{\peakNoAugNoEvolutionbetaqPGrZero}{\ensuremath{92}}
 
\newcommand{\peakNoAugNoEvolutionbetaqUpperNinety}{\ensuremath{2.9}}
 
\newcommand{\peakNoAugNoEvolutionMminUpperNinety}{\ensuremath{5.7}}
 
\newcommand{\peakNoAugNoEvolutionbeta}{\ensuremath{1.3^{+2.4}_{-1.5}}}
 
\newcommand{\peakNoAugNoEvolutionmmax}{\ensuremath{86^{+12}_{-13}}}
 
\newcommand{\peakNoAugNoEvolutionmpp}{\ensuremath{33.1^{+4.0}_{-5.6}}}

\newcommand{\peakNoAugNoEvolutionrate}{\ensuremath{23.9^{+14.3}_{-8.6}}}
 
\newcommand{\BPLNoAugNoEvolutionPAlphaOneLessAlphaTwoPercent}{\ensuremath{98}}
 
\newcommand{\BPLNoAugNoEvolutionmbreak}{\ensuremath{39.7^{+20.3}_{-9.1}}}
 
\newcommand{\BPLNoAugNoEvolutionbetaqPGrZero}{\ensuremath{94}}
 
\newcommand{\BPLNoAugNoEvolutionbetaqUpperNinety}{\ensuremath{3.1}}
 
\newcommand{\BPLNoAugNoEvolutionalphaOne}{\ensuremath{1.58^{+0.82}_{-0.86}}}
 
\newcommand{\BPLNoAugNoEvolutionalphaTwo}{\ensuremath{5.6^{+4.1}_{-2.5}}}
 
\newcommand{\BPLNoAugNoEvolutionbeta}{\ensuremath{1.4^{+2.5}_{-1.5}}}

\newcommand{\multipeakNoAugNoEvolutionalpha}{\ensuremath{2.9^{+1.9}_{-1.4}}}
 
\newcommand{\multipeakNoAugNoEvolutionbeta}{\ensuremath{0.9^{+1.9}_{-1.3}}}
 
\newcommand{\multipeakNoAugNoEvolutionmmax}{\ensuremath{65^{+31}_{-30}}}
 
\newcommand{\multipeakNoAugNoEvolutionmmin}{\ensuremath{4.6^{+1.3}_{-1.8}}}
 
\newcommand{\multipeakNoAugNoEvolutionmppOne}{\ensuremath{33.4^{+4.4}_{-4.9}}}
 
\newcommand{\multipeakNoAugNoEvolutionmppTwo}{\ensuremath{68^{+18}_{-14}}}

\newcommand{\truncatedNoMayNoEvolutionmmax}{\ensuremath{57.0^{+11.9}_{-6.6}}}

\newcommand{\peakAllNoEvolutionmmin}{\ensuremath{2.18^{+0.27}_{-0.16}}}

\newcommand{\peakAllNoEvolutionrate}{\ensuremath{52^{+52}_{-26}}}

\newcommand{\peakNoAugEvolutionRateAtRedshiftOneOverZero}{\ensuremath{2.5^{+8.0}_{-1.9}}}
 
\newcommand{\peakNoAugEvolutionMDPercentileInLambda}{\ensuremath{86}}
 
\newcommand{\peakNoAugEvolutionLambdaPGrZero}{\ensuremath{85}}

\newcommand{\peakNoAugEvolutionlamb}{\ensuremath{1.3^{+2.1}_{-2.1}}}
 
\newcommand{\peakNoAugEvolutionrate}{\ensuremath{19.3^{+15.1}_{-9.0}}}
 
\newcommand{\BPLNoAugEvolutionMDPercentileInLambda}{\ensuremath{77}}
 
\newcommand{\BPLNoAugEvolutionLambdaPGrZero}{\ensuremath{91}}

\newcommand{\BPLNoAugEvolutionlamb}{\ensuremath{1.8^{+2.1}_{-2.2}}}

\newcommand{\peakNoAugNoEvolutionRateAbovePrimaryFortyFive}{\ensuremath{0.70^{+0.65}_{-0.35}}}

\newcommand{\peakNoAugNoEvolutionFractionAbovePrimaryFortyFive}{\ensuremath{2.9^{+3.5}_{-1.7}}}

\newcommand{\peakNoAugNoEvolutionFractionBelowPrimaryFortyFive}{\ensuremath{97.1^{+1.7}_{-3.5}}}

\newcommand{\peakNoAugNoEvolutionTurnoverMass}{\ensuremath{7.8^{+1.8}_{-2.0}}}

\newcommand{\peakNoAugNoEvolutionTurnoverMassPGrThree}{\ensuremath{99.9}}

\newcommand{\BPLNoAugNoEvolutionTurnoverMass}{\ensuremath{6.02^{+0.78}_{-1.96}}}

\newcommand{\BPLNoAugNoEvolutionTurnoverMassPGrThree}{\ensuremath{98.5}}

\newcommand{\BPLAllNoEvolutionTurnoverMass}{\ensuremath{2.59^{+0.78}_{-0.39}}}

\newcommand{\MMinWithAugPercentileInNoAugpeak}{\ensuremath{0.78^{+2.22}_{-0.69}}}
 
\newcommand{\MMinWithAugPercentileInNoAugtruncated}{\ensuremath{0.25^{+0.44}_{-0.20}}}
 
\newcommand{\massonesourceminusA}[1]{\IfEqCase{#1}{{GW170817}{0.2}{GW150914}{3.1}{GW151012}{5.6}{GW151226}{3.2}{GW170104}{5.8}{GW170608}{1.7}{GW170729}{10.8}{GW170809}{6.0}{GW170814}{3.0}{GW170818}{4.8}{GW170823}{7.1}}}
\newcommand{\massonesourcemedA}[1]{\IfEqCase{#1}{{GW170817}{1.6}{GW150914}{35.7}{GW151012}{23.4}{GW151226}{13.7}{GW170104}{31.2}{GW170608}{10.9}{GW170729}{51.9}{GW170809}{35.2}{GW170814}{30.6}{GW170818}{35.5}{GW170823}{40.0}}}
\newcommand{\massonesourceplusA}[1]{\IfEqCase{#1}{{GW170817}{0.3}{GW150914}{4.7}{GW151012}{15.1}{GW151226}{8.7}{GW170104}{7.3}{GW170608}{5.5}{GW170729}{16.5}{GW170809}{8.5}{GW170814}{5.6}{GW170818}{7.4}{GW170823}{11.5}}}
\newcommand{\masstwosourceminusA}[1]{\IfEqCase{#1}{{GW170817}{0.2}{GW150914}{4.4}{GW151012}{4.9}{GW151226}{2.6}{GW170104}{4.7}{GW170608}{2.2}{GW170729}{10.4}{GW170809}{5.2}{GW170814}{4.1}{GW170818}{5.3}{GW170823}{8.4}}}
\newcommand{\masstwosourcemedA}[1]{\IfEqCase{#1}{{GW170817}{1.2}{GW150914}{30.6}{GW151012}{13.6}{GW151226}{7.7}{GW170104}{20.0}{GW170608}{7.6}{GW170729}{33.9}{GW170809}{23.8}{GW170814}{25.3}{GW170818}{26.8}{GW170823}{29.1}}}
\newcommand{\masstwosourceplusA}[1]{\IfEqCase{#1}{{GW170817}{0.2}{GW150914}{3.0}{GW151012}{4.1}{GW151226}{2.2}{GW170104}{4.9}{GW170608}{1.4}{GW170729}{10.2}{GW170809}{5.3}{GW170814}{2.8}{GW170818}{4.4}{GW170823}{7.2}}}

\newcommand{\TIME}[1]{\IfEqCase{#1}{{GW190413A}{05:29:54}{GW190719A}{21:55:14}{GW190620A}{03:04:21}{GW190514A}{06:54:16}{GW190731A}{14:09:36}{GW190503A}{18:54:04}{GW190602A}{17:59:27}{GW190929A}{01:21:49}{GW190517A}{05:51:01}{GW190915A}{23:57:02}{GW190425A}{08:18:05}{GW190512A}{18:07:14}{GW190630A}{18:52:05}{GW190521A}{03:02:29}{GW190413B}{13:43:08}{GW190924A}{02:18:46}{GW190930A}{13:35:41}{GW190706A}{22:26:41}{GW190408A}{18:18:02}{GW190909A}{11:41:49}{GW190728A}{06:45:10}{GW190426A}{15:21:55}{GW190412A}{05:30:44}{GW190720A}{00:08:36}{GW190521B}{07:43:59}{GW190910A}{11:28:07}{GW190803A}{02:27:01}{GW190519A}{15:35:44}{GW190708A}{23:24:57}{GW190527A}{09:20:55}{GW190513A}{20:54:28}{GW190424A}{18:06:48}{GW190727A}{06:03:33}{GW190814A}{21:10:39}{GW190707A}{09:33:26}{GW190828A}{06:34:05}{GW190828B}{06:55:09}{GW190701A}{20:33:06}{GW190421A}{21:38:56}}}

\newcommand{\DATE}[1]{\IfEqCase{#1}{{GW190413A}{2019-04-13 05:29:54}{GW190719A}{2019-07-19 21:55:14}{GW190620A}{2019-06-20 03:04:21}{GW190514A}{2019-05-14 06:54:16}{GW190731A}{2019-07-31 14:09:36}{GW190503A}{2019-05-03 18:54:04}{GW190602A}{2019-06-02 17:59:27}{GW190929A}{2019-09-29 01:21:49}{GW190517A}{2019-05-17 05:51:01}{GW190915A}{2019-09-15 23:57:02}{GW190425A}{2019-04-25 08:18:05}{GW190512A}{2019-05-12 18:07:14}{GW190630A}{2019-06-30 18:52:05}{GW190521A}{2019-05-21 03:02:29}{GW190413B}{2019-04-13 13:43:08}{GW190924A}{2019-09-24 02:18:46}{GW190930A}{2019-09-30 13:35:41}{GW190706A}{2019-07-06 22:26:41}{GW190408A}{2019-04-08 18:18:02}{GW190909A}{2019-09-09 11:41:49}{GW190728A}{2019-07-28 06:45:10}{GW190426A}{2019-04-26 15:21:55}{GW190412A}{2019-04-12 05:30:44}{GW190720A}{2019-07-20 00:08:36}{GW190521B}{2019-05-21 07:43:59}{GW190910A}{2019-09-10 11:28:07}{GW190803A}{2019-08-03 02:27:01}{GW190519A}{2019-05-19 15:35:44}{GW190708A}{2019-07-08 23:24:57}{GW190527A}{2019-05-27 09:20:55}{GW190513A}{2019-05-13 20:54:28}{GW190424A}{2019-04-24 18:06:48}{GW190727A}{2019-07-27 06:03:33}{GW190814A}{2019-08-14 21:10:39}{GW190707A}{2019-07-07 09:33:26}{GW190828A}{2019-08-28 06:34:05}{GW190828B}{2019-08-28 06:55:09}{GW190701A}{2019-07-01 20:33:06}{GW190421A}{2019-04-21 21:38:56}}}

\newcommand{\NAME}[1]{\IfEqCase{#1}{{GW190413A}{GW190413\_052954}{GW190719A}{GW190719\_215514}{GW190620A}{GW190620\_030421}{GW190514A}{GW190514\_065416}{GW190731A}{GW190731\_140936}{GW190503A}{GW190503\_185404}{GW190602A}{GW190602\_175927}{GW190929A}{GW190929\_012149}{GW190517A}{GW190517\_055101}{GW190915A}{GW190915\_235702}{GW190425A}{GW190425}{GW190512A}{GW190512\_180714}{GW190630A}{GW190630\_185205}{GW190521A}{GW190521}{GW190413B}{GW190413\_134308}{GW190924A}{GW190924\_021846}{GW190930A}{GW190930\_133541}{GW190706A}{GW190706\_222641}{GW190408A}{GW190408\_181802}{GW190909A}{GW190909\_114149}{GW190728A}{GW190728\_064510}{GW190426A}{GW190426\_152155}{GW190412A}{GW190412}{GW190720A}{GW190720\_000836}{GW190521B}{GW190521\_074359}{GW190910A}{GW190910\_112807}{GW190803A}{GW190803\_022701}{GW190519A}{GW190519\_153544}{GW190708A}{GW190708\_232457}{GW190527A}{GW190527\_092055}{GW190513A}{GW190513\_205428}{GW190424A}{GW190424\_180648}{GW190727A}{GW190727\_060333}{GW190814A}{GW190814}{GW190707A}{GW190707\_093326}{GW190828A}{GW190828\_063405}{GW190828B}{GW190828\_065509}{GW190701A}{GW190701\_203306}{GW190421A}{GW190421\_213856}}}

\newcommand{\SID}[1]{\IfEqCase{#1}{{GW190413A}{S190413i}{GW190719A}{S190719an}{GW190620A}{S190620e}{GW190514A}{S190514n}{GW190731A}{S190731aa}{GW190503A}{S190503bf}{GW190602A}{S190602aq}{GW190929A}{S190929d}{GW190517A}{S190517h}{GW190915A}{S190915ak}{GW190425A}{S190425z}{GW190512A}{S190512at}{GW190630A}{S190630ag}{GW190521A}{S190521g}{GW190413B}{S190413ac}{GW190924A}{S190924h}{GW190930A}{S190930s}{GW190706A}{S190706ai}{GW190408A}{S190408an}{GW190909A}{S190909w}{GW190728A}{S190728q}{GW190426A}{S190426c}{GW190412A}{S190412m}{GW190720A}{S190720a}{GW190521B}{S190521r}{GW190910A}{S190910s}{GW190803A}{S190803e}{GW190519A}{S190519bj}{GW190708A}{S190708ap}{GW190527A}{S190527w}{GW190513A}{S190513bm}{GW190424A}{S190424ao}{GW190727A}{S190727h}{GW190814A}{S190814bv}{GW190707A}{S190707q}{GW190828A}{S190828j}{GW190828B}{S190828l}{GW190701A}{S190701ah}{GW190421A}{S190421ar}}}

\newcommand{\PUBLIC}[1]{\IfEqCase{#1}{{GW190413A}{\bf}{GW190719A}{\bf}{GW190620A}{\bf}{GW190514A}{\bf}{GW190731A}{\bf}{GW190503A}{}{GW190602A}{}{GW190929A}{\bf}{GW190517A}{}{GW190915A}{}{GW190425A}{}{GW190512A}{}{GW190630A}{}{GW190521A}{}{GW190413B}{\bf}{GW190924A}{}{GW190930A}{}{GW190706A}{}{GW190408A}{}{GW190909A}{\bf}{GW190728A}{}{GW190426A}{}{GW190412A}{}{GW190720A}{}{GW190521B}{}{GW190910A}{\bf}{GW190803A}{\bf}{GW190519A}{}{GW190708A}{\bf}{GW190527A}{\bf}{GW190513A}{}{GW190424A}{\bf}{GW190727A}{}{GW190814A}{}{GW190707A}{}{GW190828A}{}{GW190828B}{}{GW190701A}{}{GW190421A}{}}}

\newcommand{\INSTRUMENTS}[1]{\IfEqCase{#1}{{GW190413A}{HL}{GW190719A}{HL}{GW190620A}{LV}{GW190514A}{HL}{GW190731A}{HL}{GW190503A}{HLV}{GW190602A}{HLV}{GW190929A}{HLV}{GW190517A}{HLV}{GW190915A}{HLV}{GW190425A}{LV}{GW190512A}{HLV}{GW190630A}{LV}{GW190521A}{HLV}{GW190413B}{HLV}{GW190924A}{HLV}{GW190930A}{HL}{GW190706A}{HLV}{GW190408A}{HLV}{GW190909A}{HL}{GW190728A}{HLV}{GW190426A}{HLV}{GW190412A}{HLV}{GW190720A}{HLV}{GW190521B}{HL}{GW190910A}{LV}{GW190803A}{HLV}{GW190519A}{HLV}{GW190708A}{LV}{GW190527A}{HL}{GW190513A}{HLV}{GW190424A}{L}{GW190727A}{HLV}{GW190814A}{LV}{GW190707A}{HL}{GW190828A}{HLV}{GW190828B}{HLV}{GW190701A}{HLV}{GW190421A}{HL}}}

\newcommand{\PARTINSTRUMENTS}[1]{\IfEqCase{#1}{{GW190413A}{HL}{GW190719A}{HL}{GW190620A}{L}{GW190514A}{HL}{GW190731A}{HL}{GW190503A}{HL}{GW190602A}{HL}{GW190929A}{HL}{GW190517A}{HL}{GW190915A}{HL}{GW190425A}{L}{GW190512A}{HL}{GW190630A}{LV}{GW190521A}{HL}{GW190413B}{HL}{GW190924A}{HL}{GW190930A}{HL}{GW190706A}{HL}{GW190408A}{HL}{GW190909A}{HL}{GW190728A}{HL}{GW190426A}{HL}{GW190412A}{HL}{GW190720A}{HLV}{GW190521B}{HL}{GW190910A}{L}{GW190803A}{HL}{GW190519A}{HL}{GW190708A}{L}{GW190527A}{HL}{GW190513A}{HL}{GW190424A}{L}{GW190727A}{HL}{GW190814A}{LV}{GW190707A}{HL}{GW190828A}{HL}{GW190828B}{HL}{GW190701A}{HLV}{GW190421A}{HL}}}

\newcommand{\CWBALLSKYPTERRES}[1]{\IfEqCase{#1}{{GW190413A}{--}{GW190719A}{--}{GW190620A}{--}{GW190514A}{--}{GW190731A}{--}{GW190503A}{--}{GW190602A}{--}{GW190929A}{--}{GW190517A}{--}{GW190915A}{--}{GW190425A}{--}{GW190512A}{--}{GW190630A}{--}{GW190521A}{--}{GW190413B}{--}{GW190924A}{--}{GW190930A}{--}{GW190706A}{--}{GW190408A}{--}{GW190909A}{--}{GW190728A}{--}{GW190426A}{--}{GW190412A}{--}{GW190720A}{--}{GW190521B}{--}{GW190910A}{--}{GW190803A}{--}{GW190519A}{--}{GW190708A}{--}{GW190527A}{--}{GW190513A}{--}{GW190424A}{--}{GW190727A}{--}{GW190814A}{--}{GW190707A}{--}{GW190828A}{--}{GW190828B}{--}{GW190701A}{--}{GW190421A}{--}}}

\newcommand{\CWBALLSKYPASTRO}[1]{\IfEqCase{#1}{{GW190413A}{--}{GW190719A}{--}{GW190620A}{~}{GW190514A}{--}{GW190731A}{--}{GW190503A}{--}{GW190602A}{--}{GW190929A}{--}{GW190517A}{--}{GW190915A}{--}{GW190425A}{~}{GW190512A}{--}{GW190630A}{~}{GW190521A}{--}{GW190413B}{--}{GW190924A}{--}{GW190930A}{--}{GW190706A}{--}{GW190408A}{--}{GW190909A}{--}{GW190728A}{--}{GW190426A}{--}{GW190412A}{--}{GW190720A}{--}{GW190521B}{--}{GW190910A}{~}{GW190803A}{--}{GW190519A}{--}{GW190708A}{~}{GW190527A}{--}{GW190513A}{--}{GW190424A}{~}{GW190727A}{--}{GW190814A}{~}{GW190707A}{--}{GW190828A}{--}{GW190828B}{--}{GW190701A}{--}{GW190421A}{--}}}

\newcommand{\PYCBCHIGHMASSPTERRES}[1]{\IfEqCase{#1}{{GW190413A}{$0.02$}{GW190719A}{$0.18$}{GW190620A}{--}{GW190514A}{$0.04$}{GW190731A}{$0.04$}{GW190503A}{$0.00$}{GW190602A}{$0.00$}{GW190929A}{--}{GW190517A}{$0.00$}{GW190915A}{$0.00$}{GW190425A}{--}{GW190512A}{$0.00$}{GW190630A}{--}{GW190521A}{--}{GW190413B}{$0.02$}{GW190924A}{$0.00$}{GW190930A}{$0.01$}{GW190706A}{$0.00$}{GW190408A}{$0.00$}{GW190909A}{--}{GW190728A}{$0.00$}{GW190426A}{--}{GW190412A}{$0.00$}{GW190720A}{$0.00$}{GW190521B}{$0.00$}{GW190910A}{--}{GW190803A}{$0.01$}{GW190519A}{$0.00$}{GW190708A}{--}{GW190527A}{--}{GW190513A}{$0.00$}{GW190424A}{--}{GW190727A}{$0.00$}{GW190814A}{--}{GW190707A}{$0.00$}{GW190828A}{$0.00$}{GW190828B}{$0.00$}{GW190701A}{--}{GW190421A}{$0.00$}}}

\newcommand{\PYCBCHIGHMASSPASTRO}[1]{\IfEqCase{#1}{{GW190413A}{$0.98$}{GW190719A}{$0.82$}{GW190620A}{~}{GW190514A}{$0.96$}{GW190731A}{$0.96$}{GW190503A}{$1.00$}{GW190602A}{$1.00$}{GW190929A}{--}{GW190517A}{$1.00$}{GW190915A}{$1.00$}{GW190425A}{~}{GW190512A}{$1.00$}{GW190630A}{~}{GW190521A}{--}{GW190413B}{$0.98$}{GW190924A}{$1.00$}{GW190930A}{$0.99$}{GW190706A}{$1.00$}{GW190408A}{$1.00$}{GW190909A}{--}{GW190728A}{$1.00$}{GW190426A}{--}{GW190412A}{$1.00$}{GW190720A}{$1.00$}{GW190521B}{$1.00$}{GW190910A}{~}{GW190803A}{$0.99$}{GW190519A}{$1.00$}{GW190708A}{~}{GW190527A}{--}{GW190513A}{$1.00$}{GW190424A}{~}{GW190727A}{$1.00$}{GW190814A}{~}{GW190707A}{$1.00$}{GW190828A}{$1.00$}{GW190828B}{$1.00$}{GW190701A}{--}{GW190421A}{$1.00$}}}

\newcommand{\PYCBCALLSKYPTERRES}[1]{\IfEqCase{#1}{{GW190413A}{--}{GW190719A}{--}{GW190620A}{--}{GW190514A}{--}{GW190731A}{--}{GW190503A}{$0.00$}{GW190602A}{--}{GW190929A}{--}{GW190517A}{$0.00$}{GW190915A}{$0.00$}{GW190425A}{--}{GW190512A}{$0.00$}{GW190630A}{--}{GW190521A}{$0.07$}{GW190413B}{--}{GW190924A}{$0.00$}{GW190930A}{$0.00$}{GW190706A}{$0.00$}{GW190408A}{$0.00$}{GW190909A}{--}{GW190728A}{$0.00$}{GW190426A}{--}{GW190412A}{$0.00$}{GW190720A}{$0.00$}{GW190521B}{$0.00$}{GW190910A}{--}{GW190803A}{--}{GW190519A}{$0.00$}{GW190708A}{--}{GW190527A}{--}{GW190513A}{$0.00$}{GW190424A}{--}{GW190727A}{$0.00$}{GW190814A}{--}{GW190707A}{$0.00$}{GW190828A}{$0.00$}{GW190828B}{$0.00$}{GW190701A}{--}{GW190421A}{$0.11$}}}

\newcommand{\PYCBCALLSKYPASTRO}[1]{\IfEqCase{#1}{{GW190413A}{--}{GW190719A}{--}{GW190620A}{~}{GW190514A}{--}{GW190731A}{--}{GW190503A}{$1.00$}{GW190602A}{--}{GW190929A}{--}{GW190517A}{$1.00$}{GW190915A}{$1.00$}{GW190425A}{~}{GW190512A}{$1.00$}{GW190630A}{~}{GW190521A}{$0.93$}{GW190413B}{--}{GW190924A}{$1.00$}{GW190930A}{$1.00$}{GW190706A}{$1.00$}{GW190408A}{$1.00$}{GW190909A}{--}{GW190728A}{$1.00$}{GW190426A}{--}{GW190412A}{$1.00$}{GW190720A}{$1.00$}{GW190521B}{$1.00$}{GW190910A}{~}{GW190803A}{--}{GW190519A}{$1.00$}{GW190708A}{~}{GW190527A}{--}{GW190513A}{$1.00$}{GW190424A}{~}{GW190727A}{$1.00$}{GW190814A}{~}{GW190707A}{$1.00$}{GW190828A}{$1.00$}{GW190828B}{$1.00$}{GW190701A}{--}{GW190421A}{$0.89$}}}

\newcommand{\GSTLALALLSKYPTERRES}[1]{\IfEqCase{#1}{{GW190413A}{--}{GW190719A}{--}{GW190620A}{$0.00$}{GW190514A}{--}{GW190731A}{$0.03$}{GW190503A}{$0.00$}{GW190602A}{$0.00$}{GW190929A}{$0.00$}{GW190517A}{$0.00$}{GW190915A}{$0.00$}{GW190425A}{--}{GW190512A}{$0.00$}{GW190630A}{$0.00$}{GW190521A}{$0.00$}{GW190413B}{$0.05$}{GW190924A}{$0.00$}{GW190930A}{$0.08$}{GW190706A}{$0.00$}{GW190408A}{$0.00$}{GW190909A}{$0.11$}{GW190728A}{$0.00$}{GW190426A}{--}{GW190412A}{$0.00$}{GW190720A}{$0.00$}{GW190521B}{$0.00$}{GW190910A}{$0.00$}{GW190803A}{$0.01$}{GW190519A}{$0.00$}{GW190708A}{$0.00$}{GW190527A}{$0.01$}{GW190513A}{$0.00$}{GW190424A}{$0.09$}{GW190727A}{$0.00$}{GW190814A}{$0.00$}{GW190707A}{$0.00$}{GW190828A}{$0.00$}{GW190828B}{$0.00$}{GW190701A}{$0.00$}{GW190421A}{$0.00$}}}

\newcommand{\GSTLALALLSKYPASTRO}[1]{\IfEqCase{#1}{{GW190413A}{--}{GW190719A}{--}{GW190620A}{$1.00$}{GW190514A}{--}{GW190731A}{$0.97$}{GW190503A}{$1.00$}{GW190602A}{$1.00$}{GW190929A}{$1.00$}{GW190517A}{$1.00$}{GW190915A}{$1.00$}{GW190425A}{--}{GW190512A}{$1.00$}{GW190630A}{$1.00$}{GW190521A}{$1.00$}{GW190413B}{$0.95$}{GW190924A}{$1.00$}{GW190930A}{$0.92$}{GW190706A}{$1.00$}{GW190408A}{$1.00$}{GW190909A}{$0.89$}{GW190728A}{$1.00$}{GW190426A}{--}{GW190412A}{$1.00$}{GW190720A}{$1.00$}{GW190521B}{$1.00$}{GW190910A}{$1.00$}{GW190803A}{$0.99$}{GW190519A}{$1.00$}{GW190708A}{$1.00$}{GW190527A}{$0.99$}{GW190513A}{$1.00$}{GW190424A}{$0.91$}{GW190727A}{$1.00$}{GW190814A}{$1.00$}{GW190707A}{$1.00$}{GW190828A}{$1.00$}{GW190828B}{$1.00$}{GW190701A}{$1.00$}{GW190421A}{$1.00$}}}

\newcommand{\MINFAR}[1]{\IfEqCase{#1}{{GW190413A}{$7.2 \times 10^{-2}$}{GW190719A}{$1.6 \times 10^{0}$}{GW190620A}{$2.9 \times 10^{-3}$}{GW190514A}{$5.3 \times 10^{-1}$}{GW190731A}{$2.1 \times 10^{-1}$}{GW190503A}{$1.0 \times 10^{-5}$}{GW190602A}{$1.1 \times 10^{-5}$}{GW190929A}{$2.0 \times 10^{-2}$}{GW190517A}{$5.7 \times 10^{-5}$}{GW190915A}{$1.0 \times 10^{-5}$}{GW190425A}{$7.5 \times 10^{-4}$}{GW190512A}{$1.0 \times 10^{-5}$}{GW190630A}{$1.0 \times 10^{-5}$}{GW190521A}{$2.0 \times 10^{-4}$}{GW190413B}{$4.4 \times 10^{-2}$}{GW190924A}{$1.0 \times 10^{-5}$}{GW190930A}{$3.3 \times 10^{-2}$}{GW190706A}{$1.0 \times 10^{-5}$}{GW190408A}{$1.0 \times 10^{-5}$}{GW190909A}{$1.1 \times 10^{0}$}{GW190728A}{$1.0 \times 10^{-5}$}{GW190426A}{$1.4 \times 10^{0}$}{GW190412A}{$1.0 \times 10^{-5}$}{GW190720A}{$1.0 \times 10^{-5}$}{GW190521B}{$1.0 \times 10^{-5}$}{GW190910A}{$1.9 \times 10^{-5}$}{GW190803A}{$2.7 \times 10^{-2}$}{GW190519A}{$1.0 \times 10^{-5}$}{GW190708A}{$2.8 \times 10^{-5}$}{GW190527A}{$6.2 \times 10^{-2}$}{GW190513A}{$1.0 \times 10^{-5}$}{GW190424A}{$7.8 \times 10^{-1}$}{GW190727A}{$1.0 \times 10^{-5}$}{GW190814A}{$1.0 \times 10^{-5}$}{GW190707A}{$1.0 \times 10^{-5}$}{GW190828A}{$1.0 \times 10^{-5}$}{GW190828B}{$1.0 \times 10^{-5}$}{GW190701A}{$1.1 \times 10^{-2}$}{GW190421A}{$7.7 \times 10^{-4}$}}}

\newcommand{\CWBALLSKYFAR}[1]{\IfEqCase{#1}{{GW190413A}{--}{GW190719A}{--}{GW190620A}{~}{GW190514A}{--}{GW190731A}{--}{GW190503A}{$1.8 \times 10^{-3}$}{GW190602A}{$1.5 \times 10^{-2}$}{GW190929A}{--}{GW190517A}{$6.5 \times 10^{-3}$}{GW190915A}{$<$ $1.0 \times 10^{-3}$}{GW190425A}{~}{GW190512A}{$8.8 \times 10^{-1}$}{GW190630A}{~}{GW190521A}{$2.0 \times 10^{-4}$}{GW190413B}{--}{GW190924A}{--}{GW190930A}{--}{GW190706A}{$<$ $1.0 \times 10^{-3}$}{GW190408A}{$<$ $9.5 \times 10^{-4}$}{GW190909A}{--}{GW190728A}{--}{GW190426A}{--}{GW190412A}{$<$ $9.5 \times 10^{-4}$}{GW190720A}{--}{GW190521B}{$<$ $1.0 \times 10^{-4}$}{GW190910A}{~}{GW190803A}{--}{GW190519A}{$3.1 \times 10^{-4}$}{GW190708A}{~}{GW190527A}{--}{GW190513A}{--}{GW190424A}{~}{GW190727A}{$8.8 \times 10^{-2}$}{GW190814A}{~}{GW190707A}{--}{GW190828A}{$<$ $9.6 \times 10^{-4}$}{GW190828B}{--}{GW190701A}{$5.5 \times 10^{-1}$}{GW190421A}{$3.0 \times 10^{-1}$}}}

\newcommand{\CWBALLSKYIFAR}[1]{\IfEqCase{#1}{{GW190413A}{--}{GW190719A}{--}{GW190620A}{~}{GW190514A}{--}{GW190731A}{--}{GW190503A}{2.7}{GW190602A}{1.8}{GW190929A}{--}{GW190517A}{2.2}{GW190915A}{3.0}{GW190425A}{~}{GW190512A}{0.1}{GW190630A}{~}{GW190521A}{3.7}{GW190413B}{--}{GW190924A}{--}{GW190930A}{--}{GW190706A}{3.0}{GW190408A}{3.0}{GW190909A}{--}{GW190728A}{--}{GW190426A}{--}{GW190412A}{3.0}{GW190720A}{--}{GW190521B}{4.0}{GW190910A}{~}{GW190803A}{--}{GW190519A}{3.5}{GW190708A}{~}{GW190527A}{--}{GW190513A}{--}{GW190424A}{~}{GW190727A}{1.1}{GW190814A}{~}{GW190707A}{--}{GW190828A}{3.0}{GW190828B}{--}{GW190701A}{0.3}{GW190421A}{0.5}}}

\newcommand{\CWBALLSKYSNR}[1]{\IfEqCase{#1}{{GW190413A}{--}{GW190719A}{--}{GW190620A}{~}{GW190514A}{--}{GW190731A}{--}{GW190503A}{11.5}{GW190602A}{11.1}{GW190929A}{--}{GW190517A}{10.7}{GW190915A}{12.3}{GW190425A}{~}{GW190512A}{10.7}{GW190630A}{~}{GW190521A}{14.4}{GW190413B}{--}{GW190924A}{--}{GW190930A}{--}{GW190706A}{12.7}{GW190408A}{14.8}{GW190909A}{--}{GW190728A}{--}{GW190426A}{--}{GW190412A}{19.7}{GW190720A}{--}{GW190521B}{24.7}{GW190910A}{~}{GW190803A}{--}{GW190519A}{14.0}{GW190708A}{~}{GW190527A}{--}{GW190513A}{--}{GW190424A}{~}{GW190727A}{11.4}{GW190814A}{~}{GW190707A}{--}{GW190828A}{16.6}{GW190828B}{--}{GW190701A}{10.2}{GW190421A}{9.3}}}

\newcommand{\PYCBCHIGHMASSFAR}[1]{\IfEqCase{#1}{{GW190413A}{$7.2 \times 10^{-2}$}{GW190719A}{$1.6 \times 10^{0}$}{GW190620A}{~}{GW190514A}{$5.3 \times 10^{-1}$}{GW190731A}{$2.8 \times 10^{-1}$}{GW190503A}{$<$ $7.9 \times 10^{-5}$}{GW190602A}{$1.5 \times 10^{-2}$}{GW190929A}{--}{GW190517A}{$<$ $5.7 \times 10^{-5}$}{GW190915A}{$<$ $3.3 \times 10^{-5}$}{GW190425A}{~}{GW190512A}{$<$ $5.7 \times 10^{-5}$}{GW190630A}{~}{GW190521A}{--}{GW190413B}{$4.4 \times 10^{-2}$}{GW190924A}{$<$ $3.3 \times 10^{-5}$}{GW190930A}{$3.3 \times 10^{-2}$}{GW190706A}{$<$ $4.6 \times 10^{-5}$}{GW190408A}{$<$ $7.9 \times 10^{-5}$}{GW190909A}{--}{GW190728A}{$<$ $3.7 \times 10^{-5}$}{GW190426A}{--}{GW190412A}{$<$ $7.9 \times 10^{-5}$}{GW190720A}{$<$ $3.7 \times 10^{-5}$}{GW190521B}{$<$ $5.7 \times 10^{-5}$}{GW190910A}{~}{GW190803A}{$2.7 \times 10^{-2}$}{GW190519A}{$<$ $5.7 \times 10^{-5}$}{GW190708A}{~}{GW190527A}{--}{GW190513A}{$<$ $5.7 \times 10^{-5}$}{GW190424A}{~}{GW190727A}{$<$ $3.7 \times 10^{-5}$}{GW190814A}{~}{GW190707A}{$<$ $4.6 \times 10^{-5}$}{GW190828A}{$<$ $3.3 \times 10^{-5}$}{GW190828B}{$<$ $3.3 \times 10^{-5}$}{GW190701A}{--}{GW190421A}{$6.6 \times 10^{-3}$}}}

\newcommand{\PYCBCHIGHMASSIFAR}[1]{\IfEqCase{#1}{{GW190413A}{1.1}{GW190719A}{-0.2}{GW190620A}{~}{GW190514A}{0.3}{GW190731A}{0.6}{GW190503A}{4.1}{GW190602A}{1.8}{GW190929A}{--}{GW190517A}{4.2}{GW190915A}{4.5}{GW190425A}{~}{GW190512A}{4.2}{GW190630A}{~}{GW190521A}{--}{GW190413B}{1.4}{GW190924A}{4.5}{GW190930A}{1.5}{GW190706A}{4.3}{GW190408A}{4.1}{GW190909A}{--}{GW190728A}{4.4}{GW190426A}{--}{GW190412A}{4.1}{GW190720A}{4.4}{GW190521B}{4.2}{GW190910A}{~}{GW190803A}{1.6}{GW190519A}{4.2}{GW190708A}{~}{GW190527A}{--}{GW190513A}{4.2}{GW190424A}{~}{GW190727A}{4.4}{GW190814A}{~}{GW190707A}{4.3}{GW190828A}{4.5}{GW190828B}{4.5}{GW190701A}{--}{GW190421A}{2.2}}}

\newcommand{\PYCBCHIGHMASSSNR}[1]{\IfEqCase{#1}{{GW190413A}{8.6}{GW190719A}{8.0}{GW190620A}{~}{GW190514A}{8.3}{GW190731A}{8.2}{GW190503A}{12.2}{GW190602A}{11.4}{GW190929A}{--}{GW190517A}{10.2}{GW190915A}{12.7}{GW190425A}{~}{GW190512A}{12.2}{GW190630A}{~}{GW190521A}{--}{GW190413B}{9.0}{GW190924A}{12.4}{GW190930A}{9.8}{GW190706A}{12.3}{GW190408A}{13.6}{GW190909A}{--}{GW190728A}{13.4}{GW190426A}{--}{GW190412A}{17.8}{GW190720A}{10.5}{GW190521B}{24.0}{GW190910A}{~}{GW190803A}{8.6}{GW190519A}{13.0}{GW190708A}{~}{GW190527A}{--}{GW190513A}{11.9}{GW190424A}{~}{GW190727A}{11.8}{GW190814A}{~}{GW190707A}{12.8}{GW190828A}{15.3}{GW190828B}{10.8}{GW190701A}{--}{GW190421A}{10.2}}}

\newcommand{\PYCBCALLSKYFAR}[1]{\IfEqCase{#1}{{GW190413A}{--}{GW190719A}{--}{GW190620A}{~}{GW190514A}{--}{GW190731A}{--}{GW190503A}{$3.7 \times 10^{-2}$}{GW190602A}{--}{GW190929A}{--}{GW190517A}{$1.8 \times 10^{-2}$}{GW190915A}{$8.6 \times 10^{-4}$}{GW190425A}{~}{GW190512A}{$3.8 \times 10^{-5}$}{GW190630A}{~}{GW190521A}{$1.1 \times 10^{0}$}{GW190413B}{--}{GW190924A}{$<$ $6.3 \times 10^{-5}$}{GW190930A}{$3.4 \times 10^{-2}$}{GW190706A}{$6.7 \times 10^{-5}$}{GW190408A}{$<$ $2.5 \times 10^{-5}$}{GW190909A}{--}{GW190728A}{$<$ $1.6 \times 10^{-5}$}{GW190426A}{--}{GW190412A}{$<$ $3.1 \times 10^{-5}$}{GW190720A}{$<$ $2.0 \times 10^{-5}$}{GW190521B}{$<$ $1.8 \times 10^{-5}$}{GW190910A}{~}{GW190803A}{--}{GW190519A}{$<$ $1.8 \times 10^{-5}$}{GW190708A}{~}{GW190527A}{--}{GW190513A}{$3.7 \times 10^{-4}$}{GW190424A}{~}{GW190727A}{$3.5 \times 10^{-3}$}{GW190814A}{~}{GW190707A}{$<$ $1.0 \times 10^{-5}$}{GW190828A}{$<$ $1.5 \times 10^{-5}$}{GW190828B}{$5.8 \times 10^{-5}$}{GW190701A}{--}{GW190421A}{$1.9 \times 10^{0}$}}}

\newcommand{\PYCBCALLSKYIFAR}[1]{\IfEqCase{#1}{{GW190413A}{--}{GW190719A}{--}{GW190620A}{~}{GW190514A}{--}{GW190731A}{--}{GW190503A}{1.4}{GW190602A}{--}{GW190929A}{--}{GW190517A}{1.8}{GW190915A}{3.1}{GW190425A}{~}{GW190512A}{4.4}{GW190630A}{~}{GW190521A}{-0.0}{GW190413B}{--}{GW190924A}{4.2}{GW190930A}{1.5}{GW190706A}{4.2}{GW190408A}{4.6}{GW190909A}{--}{GW190728A}{4.8}{GW190426A}{--}{GW190412A}{4.5}{GW190720A}{4.7}{GW190521B}{4.8}{GW190910A}{~}{GW190803A}{--}{GW190519A}{4.8}{GW190708A}{~}{GW190527A}{--}{GW190513A}{3.4}{GW190424A}{~}{GW190727A}{2.5}{GW190814A}{~}{GW190707A}{5.0}{GW190828A}{4.8}{GW190828B}{4.2}{GW190701A}{--}{GW190421A}{-0.3}}}

\newcommand{\PYCBCALLSKYSNR}[1]{\IfEqCase{#1}{{GW190413A}{--}{GW190719A}{--}{GW190620A}{~}{GW190514A}{--}{GW190731A}{--}{GW190503A}{12.2}{GW190602A}{--}{GW190929A}{--}{GW190517A}{10.4}{GW190915A}{13.0}{GW190425A}{~}{GW190512A}{12.2}{GW190630A}{~}{GW190521A}{12.6}{GW190413B}{--}{GW190924A}{12.5}{GW190930A}{9.7}{GW190706A}{11.7}{GW190408A}{13.5}{GW190909A}{--}{GW190728A}{13.4}{GW190426A}{--}{GW190412A}{17.9}{GW190720A}{10.6}{GW190521B}{24.0}{GW190910A}{~}{GW190803A}{--}{GW190519A}{13.0}{GW190708A}{~}{GW190527A}{--}{GW190513A}{11.8}{GW190424A}{~}{GW190727A}{11.5}{GW190814A}{~}{GW190707A}{12.8}{GW190828A}{15.3}{GW190828B}{10.8}{GW190701A}{--}{GW190421A}{10.2}}}

\newcommand{\GSTLALALLSKYFAR}[1]{\IfEqCase{#1}{{GW190413A}{--}{GW190719A}{--}{GW190620A}{$2.9 \times 10^{-3}$$^\dagger$}{GW190514A}{--}{GW190731A}{$2.1 \times 10^{-1}$}{GW190503A}{$<$ $1.0 \times 10^{-5}$}{GW190602A}{$1.1 \times 10^{-5}$}{GW190929A}{$2.0 \times 10^{-2}$}{GW190517A}{$9.6 \times 10^{-4}$}{GW190915A}{$<$ $1.0 \times 10^{-5}$}{GW190425A}{$7.5 \times 10^{-4}$$^\dagger$}{GW190512A}{$<$ $1.0 \times 10^{-5}$}{GW190630A}{$<$ $1.0 \times 10^{-5}$}{GW190521A}{$1.2 \times 10^{-3}$}{GW190413B}{$3.8 \times 10^{-1}$}{GW190924A}{$<$ $1.0 \times 10^{-5}$}{GW190930A}{$5.8 \times 10^{-1}$}{GW190706A}{$<$ $1.0 \times 10^{-5}$}{GW190408A}{$<$ $1.0 \times 10^{-5}$}{GW190909A}{$1.1 \times 10^{0}$}{GW190728A}{$<$ $1.0 \times 10^{-5}$}{GW190426A}{$1.4 \times 10^{0}$}{GW190412A}{$<$ $1.0 \times 10^{-5}$}{GW190720A}{$<$ $1.0 \times 10^{-5}$}{GW190521B}{$<$ $1.0 \times 10^{-5}$}{GW190910A}{$1.9 \times 10^{-5}$$^\dagger$}{GW190803A}{$3.2 \times 10^{-2}$}{GW190519A}{$<$ $1.0 \times 10^{-5}$}{GW190708A}{$2.8 \times 10^{-5}$$^\dagger$}{GW190527A}{$6.2 \times 10^{-2}$}{GW190513A}{$<$ $1.0 \times 10^{-5}$}{GW190424A}{$7.8 \times 10^{-1}$$^\dagger$}{GW190727A}{$<$ $1.0 \times 10^{-5}$}{GW190814A}{$<$ $1.0 \times 10^{-5}$}{GW190707A}{$<$ $1.0 \times 10^{-5}$}{GW190828A}{$<$ $1.0 \times 10^{-5}$}{GW190828B}{$<$ $1.0 \times 10^{-5}$}{GW190701A}{$1.1 \times 10^{-2}$}{GW190421A}{$7.7 \times 10^{-4}$}}}

\newcommand{\GSTLALALLSKYIFAR}[1]{\IfEqCase{#1}{{GW190413A}{--}{GW190719A}{--}{GW190620A}{2.5$^\dagger$}{GW190514A}{--}{GW190731A}{0.7}{GW190503A}{5.0}{GW190602A}{4.9}{GW190929A}{1.7}{GW190517A}{3.0}{GW190915A}{5.0}{GW190425A}{3.1$^\dagger$}{GW190512A}{5.0}{GW190630A}{5.0}{GW190521A}{2.9}{GW190413B}{0.4}{GW190924A}{5.0}{GW190930A}{0.2}{GW190706A}{5.0}{GW190408A}{5.0}{GW190909A}{-0.0}{GW190728A}{5.0}{GW190426A}{-0.2}{GW190412A}{5.0}{GW190720A}{5.0}{GW190521B}{5.0}{GW190910A}{4.7$^\dagger$}{GW190803A}{1.5}{GW190519A}{5.0}{GW190708A}{4.5$^\dagger$}{GW190527A}{1.2}{GW190513A}{5.0}{GW190424A}{0.1$^\dagger$}{GW190727A}{5.0}{GW190814A}{5.0}{GW190707A}{5.0}{GW190828A}{5.0}{GW190828B}{5.0}{GW190701A}{2.0}{GW190421A}{3.1}}}

\newcommand{\GSTLALALLSKYSNR}[1]{\IfEqCase{#1}{{GW190413A}{--}{GW190719A}{--}{GW190620A}{10.9}{GW190514A}{--}{GW190731A}{8.5}{GW190503A}{12.1}{GW190602A}{12.1}{GW190929A}{9.9}{GW190517A}{10.6}{GW190915A}{13.1}{GW190425A}{13.0}{GW190512A}{12.3}{GW190630A}{15.6}{GW190521A}{14.7}{GW190413B}{10.0}{GW190924A}{13.2}{GW190930A}{10.0}{GW190706A}{12.3}{GW190408A}{14.7}{GW190909A}{8.5}{GW190728A}{13.6}{GW190426A}{10.1}{GW190412A}{18.9}{GW190720A}{11.7}{GW190521B}{24.4}{GW190910A}{13.4}{GW190803A}{9.0}{GW190519A}{12.0}{GW190708A}{13.1}{GW190527A}{8.9}{GW190513A}{12.3}{GW190424A}{10.0}{GW190727A}{12.3}{GW190814A}{22.2}{GW190707A}{13.0}{GW190828A}{16.0}{GW190828B}{11.1}{GW190701A}{11.6}{GW190421A}{10.6}}}
\newcommand{\loglikelihoodminus}[1]{\IfEqCase{#1}{{GW190930A}{8.4}{GW190929A}{10.9}{GW190924A}{8.6}{GW190915A}{8.1}{GW190910A}{4.8}{GW190909A}{4.5}{GW190828B}{5.3}{GW190828A}{5.0}{GW190814A}{4.9}{GW190803A}{4.4}{GW190731A}{4.0}{GW190728A}{48007.6}{GW190727A}{5.1}{GW190720A}{9.4}{GW190719A}{4.2}{GW190708A}{4.8}{GW190707A}{7.1}{GW190706A}{5.2}{GW190701A}{3.9}{GW190630A}{5.3}{GW190620A}{5.0}{GW190602A}{4.3}{GW190527A}{6.7}{GW190521B}{5.9}{GW190521A}{11.1}{GW190519A}{17.8}{GW190517A}{5.9}{GW190514A}{4.6}{GW190513A}{4.7}{GW190512A}{5.5}{GW190503A}{4.4}{GW190426A}{5.6}{GW190425A}{5.7}{GW190424A}{3.9}{GW190421A}{4.0}{GW190413B}{5.7}{GW190413A}{4.8}{GW190412A}{10.1}{GW190408A}{5.0}}}
\newcommand{\loglikelihoodmed}[1]{\IfEqCase{#1}{{GW190930A}{-15934.9}{GW190929A}{-11962.2}{GW190924A}{-97031.8}{GW190915A}{-2809.6}{GW190910A}{93.5}{GW190909A}{25.8}{GW190828B}{43.6}{GW190828A}{121.4}{GW190814A}{298.6}{GW190803A}{28.7}{GW190731A}{29.8}{GW190728A}{64.1}{GW190727A}{58.4}{GW190720A}{-23904.6}{GW190719A}{27.0}{GW190708A}{76.7}{GW190707A}{-15883.0}{GW190706A}{72.6}{GW190701A}{55.9}{GW190630A}{114.8}{GW190620A}{65.2}{GW190602A}{73.6}{GW190527A}{22.9}{GW190521B}{322.2}{GW190521A}{-11913.6}{GW190519A}{111.0}{GW190517A}{48.8}{GW190514A}{25.9}{GW190513A}{74.3}{GW190512A}{67.5}{GW190503A}{68.5}{GW190426A}{-389547.0}{GW190425A}{-500483.9}{GW190424A}{45.7}{GW190421A}{48.8}{GW190413B}{42.7}{GW190413A}{28.4}{GW190412A}{-22827.3}{GW190408A}{108.8}}}
\newcommand{\loglikelihoodplus}[1]{\IfEqCase{#1}{{GW190930A}{15973.8}{GW190929A}{12007.8}{GW190924A}{97091.0}{GW190915A}{2897.5}{GW190910A}{4.0}{GW190909A}{3.6}{GW190828B}{4.0}{GW190828A}{4.2}{GW190814A}{3.0}{GW190803A}{2.7}{GW190731A}{2.5}{GW190728A}{13.4}{GW190727A}{5.2}{GW190720A}{23953.2}{GW190719A}{2.8}{GW190708A}{3.3}{GW190707A}{15964.2}{GW190706A}{4.0}{GW190701A}{2.6}{GW190630A}{3.8}{GW190620A}{4.1}{GW190602A}{3.3}{GW190527A}{3.0}{GW190521B}{4.7}{GW190521A}{12013.1}{GW190519A}{9.5}{GW190517A}{4.4}{GW190514A}{2.7}{GW190513A}{4.2}{GW190512A}{3.8}{GW190503A}{3.4}{GW190426A}{4.6}{GW190425A}{4.5}{GW190424A}{2.8}{GW190421A}{2.6}{GW190413B}{3.6}{GW190413A}{4.0}{GW190412A}{23002.9}{GW190408A}{3.7}}}
\newcommand{\chieffminus}[1]{\IfEqCase{#1}{{GW190930A}{0.15}{GW190929A}{0.33}{GW190924A}{0.09}{GW190915A}{0.25}{GW190910A}{0.18}{GW190909A}{0.36}{GW190828B}{0.16}{GW190828A}{0.16}{GW190814A}{0.06}{GW190803A}{0.27}{GW190731A}{0.24}{GW190728A}{0.07}{GW190727A}{0.25}{GW190720A}{0.12}{GW190719A}{0.31}{GW190708A}{0.08}{GW190707A}{0.08}{GW190706A}{0.29}{GW190701A}{0.29}{GW190630A}{0.13}{GW190620A}{0.25}{GW190602A}{0.24}{GW190527A}{0.28}{GW190521B}{0.13}{GW190521A}{0.39}{GW190519A}{0.22}{GW190517A}{0.19}{GW190514A}{0.32}{GW190513A}{0.17}{GW190512A}{0.13}{GW190503A}{0.26}{GW190426A}{0.30}{GW190425A}{0.05}{GW190424A}{0.22}{GW190421A}{0.27}{GW190413B}{0.29}{GW190413A}{0.34}{GW190412A}{0.11}{GW190408A}{0.19}}}
\newcommand{\chieffmed}[1]{\IfEqCase{#1}{{GW190930A}{0.14}{GW190929A}{0.01}{GW190924A}{0.03}{GW190915A}{0.02}{GW190910A}{0.02}{GW190909A}{-0.06}{GW190828B}{0.08}{GW190828A}{0.19}{GW190814A}{0.00}{GW190803A}{-0.03}{GW190731A}{0.06}{GW190728A}{0.12}{GW190727A}{0.11}{GW190720A}{0.18}{GW190719A}{0.32}{GW190708A}{0.02}{GW190707A}{-0.05}{GW190706A}{0.28}{GW190701A}{-0.07}{GW190630A}{0.10}{GW190620A}{0.33}{GW190602A}{0.07}{GW190527A}{0.11}{GW190521B}{0.09}{GW190521A}{0.03}{GW190519A}{0.31}{GW190517A}{0.52}{GW190514A}{-0.19}{GW190513A}{0.11}{GW190512A}{0.03}{GW190503A}{-0.03}{GW190426A}{-0.03}{GW190425A}{0.06}{GW190424A}{0.13}{GW190421A}{-0.06}{GW190413B}{-0.03}{GW190413A}{-0.01}{GW190412A}{0.25}{GW190408A}{-0.03}}}
\newcommand{\chieffplus}[1]{\IfEqCase{#1}{{GW190930A}{0.31}{GW190929A}{0.34}{GW190924A}{0.30}{GW190915A}{0.20}{GW190910A}{0.18}{GW190909A}{0.37}{GW190828B}{0.16}{GW190828A}{0.15}{GW190814A}{0.06}{GW190803A}{0.24}{GW190731A}{0.24}{GW190728A}{0.20}{GW190727A}{0.26}{GW190720A}{0.14}{GW190719A}{0.29}{GW190708A}{0.10}{GW190707A}{0.10}{GW190706A}{0.26}{GW190701A}{0.23}{GW190630A}{0.12}{GW190620A}{0.22}{GW190602A}{0.25}{GW190527A}{0.28}{GW190521B}{0.10}{GW190521A}{0.32}{GW190519A}{0.20}{GW190517A}{0.19}{GW190514A}{0.29}{GW190513A}{0.28}{GW190512A}{0.12}{GW190503A}{0.20}{GW190426A}{0.32}{GW190425A}{0.11}{GW190424A}{0.22}{GW190421A}{0.22}{GW190413B}{0.25}{GW190413A}{0.29}{GW190412A}{0.08}{GW190408A}{0.14}}}
\newcommand{\totalmasssourceminus}[1]{\IfEqCase{#1}{{GW190930A}{1.5}{GW190929A}{25.2}{GW190924A}{1.0}{GW190915A}{6.4}{GW190910A}{9.1}{GW190909A}{17.6}{GW190828B}{4.4}{GW190828A}{4.8}{GW190814A}{0.9}{GW190803A}{9.0}{GW190731A}{11.3}{GW190728A}{1.3}{GW190727A}{8.0}{GW190720A}{2.3}{GW190719A}{10.7}{GW190708A}{1.8}{GW190707A}{1.3}{GW190706A}{13.9}{GW190701A}{9.5}{GW190630A}{4.8}{GW190620A}{13.1}{GW190602A}{15.6}{GW190527A}{9.8}{GW190521B}{4.8}{GW190521A}{23.5}{GW190519A}{14.8}{GW190517A}{9.6}{GW190514A}{10.8}{GW190513A}{5.9}{GW190512A}{3.5}{GW190503A}{8.3}{GW190426A}{1.5}{GW190425A}{0.1}{GW190424A}{10.7}{GW190421A}{9.2}{GW190413B}{11.9}{GW190413A}{9.7}{GW190412A}{3.7}{GW190408A}{3.0}}}
\newcommand{\totalmasssourcemed}[1]{\IfEqCase{#1}{{GW190930A}{20.3}{GW190929A}{104.3}{GW190924A}{13.9}{GW190915A}{59.9}{GW190910A}{79.6}{GW190909A}{75.0}{GW190828B}{34.4}{GW190828A}{58.0}{GW190814A}{25.8}{GW190803A}{64.5}{GW190731A}{70.1}{GW190728A}{20.6}{GW190727A}{67.1}{GW190720A}{21.5}{GW190719A}{57.8}{GW190708A}{30.9}{GW190707A}{20.1}{GW190706A}{104.1}{GW190701A}{94.3}{GW190630A}{59.1}{GW190620A}{92.1}{GW190602A}{116.3}{GW190527A}{59.1}{GW190521B}{74.7}{GW190521A}{163.9}{GW190519A}{106.6}{GW190517A}{63.5}{GW190514A}{67.2}{GW190513A}{53.9}{GW190512A}{35.9}{GW190503A}{71.7}{GW190426A}{7.2}{GW190425A}{3.4}{GW190424A}{72.6}{GW190421A}{72.9}{GW190413B}{78.8}{GW190413A}{58.6}{GW190412A}{38.4}{GW190408A}{43.0}}}
\newcommand{\totalmasssourceplus}[1]{\IfEqCase{#1}{{GW190930A}{8.9}{GW190929A}{34.9}{GW190924A}{5.1}{GW190915A}{7.5}{GW190910A}{9.3}{GW190909A}{55.9}{GW190828B}{5.4}{GW190828A}{7.7}{GW190814A}{1.0}{GW190803A}{12.6}{GW190731A}{15.8}{GW190728A}{4.5}{GW190727A}{11.7}{GW190720A}{4.3}{GW190719A}{18.3}{GW190708A}{2.5}{GW190707A}{1.9}{GW190706A}{20.2}{GW190701A}{12.1}{GW190630A}{4.6}{GW190620A}{18.5}{GW190602A}{19.0}{GW190527A}{21.3}{GW190521B}{7.0}{GW190521A}{39.2}{GW190519A}{13.5}{GW190517A}{9.6}{GW190514A}{18.7}{GW190513A}{8.6}{GW190512A}{3.8}{GW190503A}{9.4}{GW190426A}{3.5}{GW190425A}{0.3}{GW190424A}{13.3}{GW190421A}{13.4}{GW190413B}{17.4}{GW190413A}{13.3}{GW190412A}{3.8}{GW190408A}{4.2}}}
\newcommand{\chipminus}[1]{\IfEqCase{#1}{{GW190930A}{0.24}{GW190929A}{0.45}{GW190924A}{0.18}{GW190915A}{0.39}{GW190910A}{0.32}{GW190909A}{0.38}{GW190828B}{0.23}{GW190828A}{0.31}{GW190814A}{0.03}{GW190803A}{0.33}{GW190731A}{0.30}{GW190728A}{0.20}{GW190727A}{0.36}{GW190720A}{0.22}{GW190719A}{0.30}{GW190708A}{0.24}{GW190707A}{0.23}{GW190706A}{0.28}{GW190701A}{0.31}{GW190630A}{0.23}{GW190620A}{0.28}{GW190602A}{0.31}{GW190527A}{0.34}{GW190521B}{0.29}{GW190521A}{0.44}{GW190519A}{0.29}{GW190517A}{0.29}{GW190514A}{0.34}{GW190513A}{0.22}{GW190512A}{0.17}{GW190503A}{0.29}{GW190426A}{0.00}{GW190425A}{0.27}{GW190424A}{0.38}{GW190421A}{0.36}{GW190413B}{0.41}{GW190413A}{0.31}{GW190412A}{0.16}{GW190408A}{0.31}}}
\newcommand{\chipmed}[1]{\IfEqCase{#1}{{GW190930A}{0.34}{GW190929A}{0.59}{GW190924A}{0.24}{GW190915A}{0.55}{GW190910A}{0.40}{GW190909A}{0.52}{GW190828B}{0.31}{GW190828A}{0.43}{GW190814A}{0.04}{GW190803A}{0.43}{GW190731A}{0.39}{GW190728A}{0.29}{GW190727A}{0.47}{GW190720A}{0.33}{GW190719A}{0.43}{GW190708A}{0.29}{GW190707A}{0.29}{GW190706A}{0.39}{GW190701A}{0.42}{GW190630A}{0.32}{GW190620A}{0.43}{GW190602A}{0.41}{GW190527A}{0.44}{GW190521B}{0.40}{GW190521A}{0.68}{GW190519A}{0.44}{GW190517A}{0.49}{GW190514A}{0.47}{GW190513A}{0.30}{GW190512A}{0.22}{GW190503A}{0.38}{GW190426A}{0.00}{GW190425A}{0.34}{GW190424A}{0.52}{GW190421A}{0.48}{GW190413B}{0.56}{GW190413A}{0.41}{GW190412A}{0.30}{GW190408A}{0.39}}}
\newcommand{\chipplus}[1]{\IfEqCase{#1}{{GW190930A}{0.40}{GW190929A}{0.32}{GW190924A}{0.40}{GW190915A}{0.36}{GW190910A}{0.39}{GW190909A}{0.39}{GW190828B}{0.38}{GW190828A}{0.36}{GW190814A}{0.04}{GW190803A}{0.42}{GW190731A}{0.46}{GW190728A}{0.37}{GW190727A}{0.40}{GW190720A}{0.43}{GW190719A}{0.37}{GW190708A}{0.43}{GW190707A}{0.39}{GW190706A}{0.39}{GW190701A}{0.41}{GW190630A}{0.32}{GW190620A}{0.37}{GW190602A}{0.42}{GW190527A}{0.43}{GW190521B}{0.32}{GW190521A}{0.26}{GW190519A}{0.34}{GW190517A}{0.30}{GW190514A}{0.39}{GW190513A}{0.39}{GW190512A}{0.36}{GW190503A}{0.41}{GW190426A}{0.00}{GW190425A}{0.43}{GW190424A}{0.38}{GW190421A}{0.39}{GW190413B}{0.37}{GW190413A}{0.41}{GW190412A}{0.19}{GW190408A}{0.38}}}
\newcommand{\spinoneyminus}[1]{\IfEqCase{#1}{{GW190930A}{0.47}{GW190929A}{0.71}{GW190924A}{0.35}{GW190915A}{0.68}{GW190910A}{0.48}{GW190909A}{0.66}{GW190828B}{0.41}{GW190828A}{0.51}{GW190814A}{0.04}{GW190803A}{0.58}{GW190731A}{0.52}{GW190728A}{0.37}{GW190727A}{0.61}{GW190720A}{0.49}{GW190719A}{0.55}{GW190708A}{0.43}{GW190707A}{0.39}{GW190706A}{0.50}{GW190701A}{0.52}{GW190630A}{0.36}{GW190620A}{0.53}{GW190602A}{0.52}{GW190527A}{0.61}{GW190521B}{0.44}{GW190521A}{0.76}{GW190519A}{0.55}{GW190517A}{0.58}{GW190514A}{0.56}{GW190513A}{0.41}{GW190512A}{0.29}{GW190503A}{0.48}{GW190426A}{0.00}{GW190425A}{0.48}{GW190424A}{0.64}{GW190421A}{0.58}{GW190413B}{0.70}{GW190413A}{0.54}{GW190412A}{0.39}{GW190408A}{0.48}}}
\newcommand{\spinoneymed}[1]{\IfEqCase{#1}{{GW190930A}{0.002}{GW190929A}{0.005}{GW190924A}{0.0009}{GW190915A}{0.00}{GW190910A}{0.0008}{GW190909A}{-0.01}{GW190828B}{0.0010}{GW190828A}{0.00}{GW190814A}{0.0008}{GW190803A}{0.0007}{GW190731A}{0.0007}{GW190728A}{0.004}{GW190727A}{0.0002}{GW190720A}{0.0009}{GW190719A}{0.004}{GW190708A}{0.00}{GW190707A}{0.00}{GW190706A}{0.00}{GW190701A}{0.003}{GW190630A}{0.0002}{GW190620A}{-0.01}{GW190602A}{0.00}{GW190527A}{0.00}{GW190521B}{0.00}{GW190521A}{0.0003}{GW190519A}{0.00}{GW190517A}{-0.01}{GW190514A}{0.00}{GW190513A}{0.0005}{GW190512A}{0.00}{GW190503A}{0.00}{GW190426A}{0.00}{GW190425A}{0.003}{GW190424A}{0.00}{GW190421A}{0.0008}{GW190413B}{0.00}{GW190413A}{0.003}{GW190412A}{0.06}{GW190408A}{0.001}}}
\newcommand{\spinoneyplus}[1]{\IfEqCase{#1}{{GW190930A}{0.48}{GW190929A}{0.70}{GW190924A}{0.36}{GW190915A}{0.68}{GW190910A}{0.50}{GW190909A}{0.64}{GW190828B}{0.43}{GW190828A}{0.53}{GW190814A}{0.04}{GW190803A}{0.55}{GW190731A}{0.54}{GW190728A}{0.39}{GW190727A}{0.59}{GW190720A}{0.44}{GW190719A}{0.55}{GW190708A}{0.41}{GW190707A}{0.38}{GW190706A}{0.51}{GW190701A}{0.53}{GW190630A}{0.37}{GW190620A}{0.55}{GW190602A}{0.54}{GW190527A}{0.59}{GW190521B}{0.45}{GW190521A}{0.75}{GW190519A}{0.54}{GW190517A}{0.57}{GW190514A}{0.59}{GW190513A}{0.41}{GW190512A}{0.29}{GW190503A}{0.49}{GW190426A}{0.00}{GW190425A}{0.48}{GW190424A}{0.63}{GW190421A}{0.60}{GW190413B}{0.69}{GW190413A}{0.55}{GW190412A}{0.33}{GW190408A}{0.49}}}
\newcommand{\finalmassdetminus}[1]{\IfEqCase{#1}{{GW190930A}{0.9}{GW190929A}{24.5}{GW190924A}{0.8}{GW190915A}{7.4}{GW190910A}{7.1}{GW190909A}{21.3}{GW190828B}{4.2}{GW190828A}{5.2}{GW190814A}{1.0}{GW190803A}{10.9}{GW190731A}{12.8}{GW190728A}{0.7}{GW190727A}{9.8}{GW190720A}{1.2}{GW190719A}{14.1}{GW190708A}{0.7}{GW190707A}{0.5}{GW190706A}{23.7}{GW190701A}{13.4}{GW190630A}{3.3}{GW190620A}{16.2}{GW190602A}{18.3}{GW190527A}{9.5}{GW190521B}{4.8}{GW190521A}{30.4}{GW190519A}{15.4}{GW190517A}{6.4}{GW190514A}{13.9}{GW190513A}{6.7}{GW190512A}{2.8}{GW190503A}{10.8}{GW190424A}{10.0}{GW190421A}{11.3}{GW190413B}{16.7}{GW190413A}{14.0}{GW190412A}{4.7}{GW190408A}{3.4}}}
\newcommand{\finalmassdetmed}[1]{\IfEqCase{#1}{{GW190930A}{22.1}{GW190929A}{144.3}{GW190924A}{14.8}{GW190915A}{74.8}{GW190910A}{97.0}{GW190909A}{114.5}{GW190828B}{42.7}{GW190828A}{75.7}{GW190814A}{26.9}{GW190803A}{95.8}{GW190731A}{104.6}{GW190728A}{22.7}{GW190727A}{99.2}{GW190720A}{23.7}{GW190719A}{90.0}{GW190708A}{34.4}{GW190707A}{22.1}{GW190706A}{171.1}{GW190701A}{124.0}{GW190630A}{66.3}{GW190620A}{130.3}{GW190602A}{163.8}{GW190527A}{80.3}{GW190521B}{88.0}{GW190521A}{256.6}{GW190519A}{146.8}{GW190517A}{79.8}{GW190514A}{108.3}{GW190513A}{70.6}{GW190512A}{43.5}{GW190503A}{87.6}{GW190424A}{96.0}{GW190421A}{103.9}{GW190413B}{129.8}{GW190413A}{89.6}{GW190412A}{42.9}{GW190408A}{53.0}}}
\newcommand{\finalmassdetplus}[1]{\IfEqCase{#1}{{GW190930A}{10.8}{GW190929A}{36.4}{GW190924A}{5.9}{GW190915A}{7.9}{GW190910A}{9.3}{GW190909A}{92.0}{GW190828B}{6.6}{GW190828A}{6.0}{GW190814A}{1.1}{GW190803A}{13.1}{GW190731A}{12.8}{GW190728A}{5.5}{GW190727A}{10.7}{GW190720A}{5.2}{GW190719A}{22.5}{GW190708A}{2.7}{GW190707A}{1.9}{GW190706A}{20.0}{GW190701A}{15.1}{GW190630A}{4.2}{GW190620A}{17.7}{GW190602A}{20.7}{GW190527A}{51.0}{GW190521B}{4.3}{GW190521A}{36.6}{GW190519A}{14.7}{GW190517A}{8.8}{GW190514A}{16.6}{GW190513A}{11.5}{GW190512A}{4.0}{GW190503A}{10.2}{GW190424A}{13.0}{GW190421A}{14.1}{GW190413B}{16.4}{GW190413A}{16.3}{GW190412A}{4.6}{GW190408A}{3.2}}}
\newcommand{\phioneminus}[1]{\IfEqCase{#1}{{GW190930A}{2.79}{GW190929A}{2.78}{GW190924A}{2.82}{GW190915A}{2.86}{GW190910A}{2.79}{GW190909A}{2.98}{GW190828B}{2.78}{GW190828A}{2.84}{GW190814A}{2.65}{GW190803A}{2.80}{GW190731A}{2.83}{GW190728A}{2.76}{GW190727A}{2.83}{GW190720A}{2.81}{GW190719A}{2.77}{GW190708A}{2.90}{GW190707A}{2.85}{GW190706A}{2.84}{GW190701A}{2.77}{GW190630A}{2.80}{GW190620A}{2.89}{GW190602A}{2.85}{GW190527A}{2.88}{GW190521B}{2.87}{GW190521A}{2.79}{GW190519A}{2.84}{GW190517A}{2.90}{GW190514A}{2.82}{GW190513A}{2.82}{GW190512A}{2.84}{GW190503A}{2.83}{GW190426A}{0.00}{GW190425A}{2.73}{GW190424A}{2.84}{GW190421A}{2.82}{GW190413B}{2.83}{GW190413A}{2.76}{GW190412A}{2.36}{GW190408A}{2.77}}}
\newcommand{\phionemed}[1]{\IfEqCase{#1}{{GW190930A}{3.10}{GW190929A}{3.09}{GW190924A}{3.09}{GW190915A}{3.17}{GW190910A}{3.12}{GW190909A}{3.26}{GW190828B}{3.09}{GW190828A}{3.16}{GW190814A}{2.97}{GW190803A}{3.12}{GW190731A}{3.12}{GW190728A}{3.05}{GW190727A}{3.13}{GW190720A}{3.12}{GW190719A}{3.09}{GW190708A}{3.21}{GW190707A}{3.16}{GW190706A}{3.15}{GW190701A}{3.07}{GW190630A}{3.13}{GW190620A}{3.23}{GW190602A}{3.18}{GW190527A}{3.16}{GW190521B}{3.17}{GW190521A}{3.14}{GW190519A}{3.15}{GW190517A}{3.23}{GW190514A}{3.18}{GW190513A}{3.11}{GW190512A}{3.15}{GW190503A}{3.15}{GW190426A}{0.00}{GW190425A}{3.05}{GW190424A}{3.16}{GW190421A}{3.13}{GW190413B}{3.16}{GW190413A}{3.06}{GW190412A}{2.69}{GW190408A}{3.08}}}
\newcommand{\phioneplus}[1]{\IfEqCase{#1}{{GW190930A}{2.86}{GW190929A}{2.90}{GW190924A}{2.86}{GW190915A}{2.79}{GW190910A}{2.85}{GW190909A}{2.71}{GW190828B}{2.87}{GW190828A}{2.82}{GW190814A}{2.95}{GW190803A}{2.85}{GW190731A}{2.88}{GW190728A}{2.92}{GW190727A}{2.83}{GW190720A}{2.84}{GW190719A}{2.86}{GW190708A}{2.79}{GW190707A}{2.83}{GW190706A}{2.83}{GW190701A}{2.88}{GW190630A}{2.83}{GW190620A}{2.75}{GW190602A}{2.77}{GW190527A}{2.83}{GW190521B}{2.80}{GW190521A}{2.80}{GW190519A}{2.83}{GW190517A}{2.76}{GW190514A}{2.77}{GW190513A}{2.86}{GW190512A}{2.84}{GW190503A}{2.81}{GW190426A}{0.00}{GW190425A}{2.90}{GW190424A}{2.78}{GW190421A}{2.82}{GW190413B}{2.79}{GW190413A}{2.90}{GW190412A}{3.20}{GW190408A}{2.87}}}
\newcommand{\phitwominus}[1]{\IfEqCase{#1}{{GW190930A}{2.89}{GW190929A}{2.80}{GW190924A}{2.83}{GW190915A}{2.82}{GW190910A}{2.83}{GW190909A}{2.79}{GW190828B}{2.84}{GW190828A}{2.83}{GW190814A}{2.74}{GW190803A}{2.82}{GW190731A}{2.88}{GW190728A}{2.85}{GW190727A}{2.83}{GW190720A}{2.82}{GW190719A}{2.74}{GW190708A}{2.76}{GW190707A}{2.79}{GW190706A}{2.81}{GW190701A}{2.97}{GW190630A}{2.80}{GW190620A}{2.84}{GW190602A}{2.84}{GW190527A}{2.79}{GW190521B}{2.82}{GW190521A}{2.86}{GW190519A}{2.79}{GW190517A}{2.82}{GW190514A}{2.89}{GW190513A}{2.80}{GW190512A}{2.75}{GW190503A}{2.80}{GW190426A}{0.00}{GW190425A}{2.84}{GW190424A}{2.84}{GW190421A}{2.81}{GW190413B}{2.97}{GW190413A}{2.83}{GW190412A}{2.75}{GW190408A}{2.80}}}
\newcommand{\phitwomed}[1]{\IfEqCase{#1}{{GW190930A}{3.21}{GW190929A}{3.12}{GW190924A}{3.15}{GW190915A}{3.16}{GW190910A}{3.14}{GW190909A}{3.09}{GW190828B}{3.16}{GW190828A}{3.13}{GW190814A}{3.07}{GW190803A}{3.14}{GW190731A}{3.21}{GW190728A}{3.15}{GW190727A}{3.12}{GW190720A}{3.15}{GW190719A}{3.08}{GW190708A}{3.05}{GW190707A}{3.09}{GW190706A}{3.12}{GW190701A}{3.27}{GW190630A}{3.12}{GW190620A}{3.14}{GW190602A}{3.15}{GW190527A}{3.06}{GW190521B}{3.15}{GW190521A}{3.17}{GW190519A}{3.12}{GW190517A}{3.14}{GW190514A}{3.19}{GW190513A}{3.14}{GW190512A}{3.07}{GW190503A}{3.14}{GW190426A}{0.00}{GW190425A}{3.15}{GW190424A}{3.15}{GW190421A}{3.14}{GW190413B}{3.28}{GW190413A}{3.16}{GW190412A}{3.09}{GW190408A}{3.11}}}
\newcommand{\phitwoplus}[1]{\IfEqCase{#1}{{GW190930A}{2.74}{GW190929A}{2.83}{GW190924A}{2.82}{GW190915A}{2.80}{GW190910A}{2.82}{GW190909A}{2.86}{GW190828B}{2.84}{GW190828A}{2.84}{GW190814A}{2.86}{GW190803A}{2.87}{GW190731A}{2.77}{GW190728A}{2.78}{GW190727A}{2.88}{GW190720A}{2.84}{GW190719A}{2.90}{GW190708A}{2.90}{GW190707A}{2.88}{GW190706A}{2.88}{GW190701A}{2.70}{GW190630A}{2.84}{GW190620A}{2.81}{GW190602A}{2.79}{GW190527A}{2.91}{GW190521B}{2.82}{GW190521A}{2.82}{GW190519A}{2.84}{GW190517A}{2.83}{GW190514A}{2.75}{GW190513A}{2.85}{GW190512A}{2.89}{GW190503A}{2.81}{GW190426A}{0.00}{GW190425A}{2.83}{GW190424A}{2.83}{GW190421A}{2.81}{GW190413B}{2.72}{GW190413A}{2.82}{GW190412A}{2.85}{GW190408A}{2.88}}}
\newcommand{\phionetwominus}[1]{\IfEqCase{#1}{{GW190930A}{2.84}{GW190929A}{2.79}{GW190924A}{2.83}{GW190915A}{2.89}{GW190910A}{2.78}{GW190909A}{2.85}{GW190828B}{2.83}{GW190828A}{2.63}{GW190814A}{2.71}{GW190803A}{2.82}{GW190731A}{2.73}{GW190728A}{2.79}{GW190727A}{2.83}{GW190720A}{2.74}{GW190719A}{2.75}{GW190708A}{2.66}{GW190707A}{2.79}{GW190706A}{2.80}{GW190701A}{2.82}{GW190630A}{2.74}{GW190620A}{2.78}{GW190602A}{2.78}{GW190527A}{2.77}{GW190521B}{2.78}{GW190521A}{3.09}{GW190519A}{2.92}{GW190517A}{2.91}{GW190514A}{2.83}{GW190513A}{2.84}{GW190512A}{2.80}{GW190503A}{2.70}{GW190426A}{0.00}{GW190425A}{2.87}{GW190424A}{2.85}{GW190421A}{2.83}{GW190413B}{2.67}{GW190413A}{2.84}{GW190412A}{3.10}{GW190408A}{2.77}}}
\newcommand{\phionetwomed}[1]{\IfEqCase{#1}{{GW190930A}{3.17}{GW190929A}{3.11}{GW190924A}{3.17}{GW190915A}{3.18}{GW190910A}{3.15}{GW190909A}{3.17}{GW190828B}{3.16}{GW190828A}{2.94}{GW190814A}{3.01}{GW190803A}{3.12}{GW190731A}{3.03}{GW190728A}{3.13}{GW190727A}{3.13}{GW190720A}{3.09}{GW190719A}{3.13}{GW190708A}{3.08}{GW190707A}{3.13}{GW190706A}{3.11}{GW190701A}{3.15}{GW190630A}{3.06}{GW190620A}{3.16}{GW190602A}{3.09}{GW190527A}{3.08}{GW190521B}{3.16}{GW190521A}{3.35}{GW190519A}{3.23}{GW190517A}{3.22}{GW190514A}{3.15}{GW190513A}{3.15}{GW190512A}{3.14}{GW190503A}{3.02}{GW190426A}{0.00}{GW190425A}{3.18}{GW190424A}{3.15}{GW190421A}{3.17}{GW190413B}{3.00}{GW190413A}{3.15}{GW190412A}{3.44}{GW190408A}{3.11}}}
\newcommand{\phionetwoplus}[1]{\IfEqCase{#1}{{GW190930A}{2.77}{GW190929A}{2.87}{GW190924A}{2.80}{GW190915A}{2.78}{GW190910A}{2.80}{GW190909A}{2.82}{GW190828B}{2.78}{GW190828A}{3.01}{GW190814A}{2.95}{GW190803A}{2.83}{GW190731A}{2.88}{GW190728A}{2.82}{GW190727A}{2.86}{GW190720A}{2.84}{GW190719A}{2.80}{GW190708A}{2.83}{GW190707A}{2.81}{GW190706A}{2.86}{GW190701A}{2.81}{GW190630A}{2.89}{GW190620A}{2.77}{GW190602A}{2.84}{GW190527A}{2.89}{GW190521B}{2.73}{GW190521A}{2.68}{GW190519A}{2.71}{GW190517A}{2.77}{GW190514A}{2.80}{GW190513A}{2.79}{GW190512A}{2.82}{GW190503A}{2.94}{GW190426A}{0.00}{GW190425A}{2.76}{GW190424A}{2.84}{GW190421A}{2.81}{GW190413B}{2.95}{GW190413A}{2.82}{GW190412A}{2.54}{GW190408A}{2.84}}}
\newcommand{\raminus}[1]{\IfEqCase{#1}{{GW190930A}{5.23224}{GW190929A}{3.05568}{GW190924A}{0.14434}{GW190915A}{0.08865}{GW190910A}{2.79654}{GW190909A}{1.21034}{GW190828B}{0.25804}{GW190828A}{0.19728}{GW190814A}{0.02832}{GW190803A}{0.32018}{GW190731A}{2.05429}{GW190728A}{3.93966}{GW190727A}{0.28275}{GW190720A}{5.03310}{GW190719A}{1.97394}{GW190708A}{2.51716}{GW190707A}{1.47850}{GW190706A}{0.11777}{GW190701A}{0.02938}{GW190630A}{3.00747}{GW190620A}{3.79912}{GW190602A}{0.16873}{GW190527A}{4.81181}{GW190521B}{0.49181}{GW190521A}{3.83925}{GW190519A}{3.53823}{GW190517A}{0.11026}{GW190514A}{2.80859}{GW190513A}{0.16674}{GW190512A}{0.34154}{GW190503A}{0.07308}{GW190426A}{5.11703}{GW190425A}{1.14713}{GW190424A}{2.87584}{GW190421A}{1.90797}{GW190413B}{0.19894}{GW190413A}{2.38305}{GW190412A}{0.06390}{GW190408A}{0.23336}}}
\newcommand{\ramed}[1]{\IfEqCase{#1}{{GW190930A}{5.56800}{GW190929A}{4.57328}{GW190924A}{2.28446}{GW190915A}{3.41338}{GW190910A}{3.71091}{GW190909A}{1.53780}{GW190828B}{2.46668}{GW190828A}{2.55563}{GW190814A}{0.22230}{GW190803A}{1.64028}{GW190731A}{3.03870}{GW190728A}{5.47833}{GW190727A}{1.82934}{GW190720A}{5.19166}{GW190719A}{2.73410}{GW190708A}{2.97596}{GW190707A}{3.54979}{GW190706A}{2.59111}{GW190701A}{0.66145}{GW190630A}{5.86181}{GW190620A}{4.24971}{GW190602A}{1.30570}{GW190527A}{5.13405}{GW190521B}{4.87951}{GW190521A}{3.88408}{GW190519A}{3.58629}{GW190517A}{4.07092}{GW190514A}{3.58258}{GW190513A}{0.89431}{GW190512A}{4.37140}{GW190503A}{1.65900}{GW190426A}{5.27391}{GW190425A}{1.62833}{GW190424A}{3.13318}{GW190421A}{3.50423}{GW190413B}{2.70054}{GW190413A}{2.53690}{GW190412A}{3.81244}{GW190408A}{6.08868}}}
\newcommand{\raplus}[1]{\IfEqCase{#1}{{GW190930A}{0.45017}{GW190929A}{0.94968}{GW190924A}{0.18201}{GW190915A}{0.07354}{GW190910A}{1.06677}{GW190909A}{4.24495}{GW190828B}{3.42033}{GW190828A}{3.27852}{GW190814A}{0.16914}{GW190803A}{1.69538}{GW190731A}{0.37410}{GW190728A}{0.52075}{GW190727A}{4.30388}{GW190720A}{0.96175}{GW190719A}{3.20570}{GW190708A}{2.84449}{GW190707A}{2.17972}{GW190706A}{3.23517}{GW190701A}{0.02994}{GW190630A}{0.12425}{GW190620A}{0.37483}{GW190602A}{0.27462}{GW190527A}{0.80861}{GW190521B}{0.59768}{GW190521A}{2.35925}{GW190519A}{2.66341}{GW190517A}{1.75963}{GW190514A}{1.66161}{GW190513A}{4.13656}{GW190512A}{0.17389}{GW190503A}{0.07931}{GW190426A}{0.85181}{GW190425A}{3.12911}{GW190424A}{2.81897}{GW190421A}{0.15768}{GW190413B}{1.98722}{GW190413A}{0.89215}{GW190412A}{0.03095}{GW190408A}{0.08337}}}
\newcommand{\phijlminus}[1]{\IfEqCase{#1}{{GW190930A}{2.81}{GW190929A}{3.08}{GW190924A}{2.69}{GW190915A}{2.76}{GW190910A}{2.83}{GW190909A}{2.84}{GW190828B}{2.87}{GW190828A}{3.02}{GW190814A}{1.87}{GW190803A}{2.70}{GW190731A}{2.69}{GW190728A}{2.87}{GW190727A}{2.85}{GW190720A}{2.86}{GW190719A}{2.89}{GW190708A}{2.84}{GW190707A}{2.89}{GW190706A}{2.70}{GW190701A}{2.57}{GW190630A}{2.93}{GW190620A}{3.11}{GW190602A}{2.78}{GW190527A}{2.80}{GW190521B}{2.82}{GW190521A}{2.81}{GW190519A}{2.86}{GW190517A}{2.16}{GW190514A}{2.83}{GW190513A}{2.69}{GW190512A}{2.82}{GW190503A}{3.64}{GW190426A}{1.43}{GW190425A}{2.88}{GW190424A}{2.83}{GW190421A}{2.83}{GW190413B}{3.08}{GW190413A}{2.65}{GW190412A}{3.64}{GW190408A}{2.54}}}
\newcommand{\phijlmed}[1]{\IfEqCase{#1}{{GW190930A}{3.14}{GW190929A}{3.37}{GW190924A}{3.02}{GW190915A}{3.39}{GW190910A}{3.15}{GW190909A}{3.16}{GW190828B}{3.19}{GW190828A}{3.35}{GW190814A}{2.28}{GW190803A}{3.05}{GW190731A}{3.03}{GW190728A}{3.19}{GW190727A}{3.21}{GW190720A}{3.15}{GW190719A}{3.21}{GW190708A}{3.17}{GW190707A}{3.21}{GW190706A}{3.03}{GW190701A}{3.05}{GW190630A}{3.23}{GW190620A}{3.51}{GW190602A}{3.12}{GW190527A}{3.13}{GW190521B}{3.17}{GW190521A}{3.14}{GW190519A}{3.17}{GW190517A}{2.38}{GW190514A}{3.15}{GW190513A}{3.00}{GW190512A}{3.16}{GW190503A}{3.90}{GW190426A}{1.70}{GW190425A}{3.23}{GW190424A}{3.16}{GW190421A}{3.14}{GW190413B}{3.36}{GW190413A}{2.98}{GW190412A}{3.82}{GW190408A}{2.90}}}
\newcommand{\phijlplus}[1]{\IfEqCase{#1}{{GW190930A}{2.86}{GW190929A}{2.65}{GW190924A}{2.94}{GW190915A}{2.29}{GW190910A}{2.84}{GW190909A}{2.81}{GW190828B}{2.77}{GW190828A}{2.59}{GW190814A}{3.50}{GW190803A}{2.90}{GW190731A}{2.89}{GW190728A}{2.76}{GW190727A}{2.79}{GW190720A}{2.84}{GW190719A}{2.78}{GW190708A}{2.80}{GW190707A}{2.77}{GW190706A}{2.92}{GW190701A}{2.72}{GW190630A}{2.76}{GW190620A}{2.40}{GW190602A}{2.83}{GW190527A}{2.85}{GW190521B}{2.86}{GW190521A}{2.80}{GW190519A}{2.79}{GW190517A}{3.65}{GW190514A}{2.83}{GW190513A}{2.95}{GW190512A}{2.80}{GW190503A}{2.15}{GW190426A}{1.19}{GW190425A}{2.76}{GW190424A}{2.82}{GW190421A}{2.84}{GW190413B}{2.68}{GW190413A}{2.97}{GW190412A}{2.23}{GW190408A}{3.00}}}
\newcommand{\tilttwominus}[1]{\IfEqCase{#1}{{GW190930A}{0.90}{GW190929A}{1.09}{GW190924A}{1.01}{GW190915A}{1.03}{GW190910A}{1.01}{GW190909A}{1.23}{GW190828B}{0.96}{GW190828A}{0.84}{GW190814A}{1.02}{GW190803A}{1.12}{GW190731A}{1.02}{GW190728A}{0.85}{GW190727A}{0.97}{GW190720A}{0.90}{GW190719A}{0.81}{GW190708A}{1.00}{GW190707A}{1.18}{GW190706A}{0.88}{GW190701A}{1.14}{GW190630A}{0.87}{GW190620A}{0.78}{GW190602A}{0.97}{GW190527A}{1.01}{GW190521B}{0.85}{GW190521A}{1.02}{GW190519A}{0.77}{GW190517A}{0.67}{GW190514A}{1.21}{GW190513A}{0.94}{GW190512A}{0.99}{GW190503A}{1.06}{GW190426A}{0.00}{GW190425A}{0.87}{GW190424A}{0.96}{GW190421A}{1.10}{GW190413B}{1.18}{GW190413A}{1.13}{GW190412A}{0.90}{GW190408A}{1.06}}}
\newcommand{\tilttwomed}[1]{\IfEqCase{#1}{{GW190930A}{1.26}{GW190929A}{1.53}{GW190924A}{1.42}{GW190915A}{1.56}{GW190910A}{1.53}{GW190909A}{1.74}{GW190828B}{1.32}{GW190828A}{1.19}{GW190814A}{1.60}{GW190803A}{1.63}{GW190731A}{1.42}{GW190728A}{1.17}{GW190727A}{1.37}{GW190720A}{1.25}{GW190719A}{1.11}{GW190708A}{1.43}{GW190707A}{1.76}{GW190706A}{1.19}{GW190701A}{1.74}{GW190630A}{1.23}{GW190620A}{1.05}{GW190602A}{1.38}{GW190527A}{1.41}{GW190521B}{1.27}{GW190521A}{1.59}{GW190519A}{1.05}{GW190517A}{0.91}{GW190514A}{1.94}{GW190513A}{1.32}{GW190512A}{1.41}{GW190503A}{1.59}{GW190426A}{0.00}{GW190425A}{1.41}{GW190424A}{1.37}{GW190421A}{1.73}{GW190413B}{1.70}{GW190413A}{1.65}{GW190412A}{1.32}{GW190408A}{1.59}}}
\newcommand{\tilttwoplus}[1]{\IfEqCase{#1}{{GW190930A}{1.21}{GW190929A}{1.13}{GW190924A}{1.16}{GW190915A}{1.09}{GW190910A}{1.04}{GW190909A}{1.01}{GW190828B}{1.20}{GW190828A}{1.25}{GW190814A}{1.00}{GW190803A}{1.05}{GW190731A}{1.16}{GW190728A}{1.29}{GW190727A}{1.20}{GW190720A}{1.25}{GW190719A}{1.31}{GW190708A}{1.13}{GW190707A}{0.92}{GW190706A}{1.29}{GW190701A}{0.97}{GW190630A}{1.18}{GW190620A}{1.32}{GW190602A}{1.15}{GW190527A}{1.20}{GW190521B}{1.13}{GW190521A}{1.05}{GW190519A}{1.26}{GW190517A}{1.30}{GW190514A}{0.88}{GW190513A}{1.19}{GW190512A}{1.11}{GW190503A}{1.07}{GW190426A}{3.14}{GW190425A}{0.94}{GW190424A}{1.16}{GW190421A}{0.99}{GW190413B}{1.00}{GW190413A}{1.04}{GW190412A}{1.12}{GW190408A}{1.03}}}
\newcommand{\costhetajnminus}[1]{\IfEqCase{#1}{{GW190930A}{1.54}{GW190929A}{0.76}{GW190924A}{1.67}{GW190915A}{0.50}{GW190910A}{0.88}{GW190909A}{1.06}{GW190828B}{0.71}{GW190828A}{0.36}{GW190814A}{1.35}{GW190803A}{1.39}{GW190731A}{1.33}{GW190728A}{1.30}{GW190727A}{0.98}{GW190720A}{0.20}{GW190719A}{0.92}{GW190708A}{1.17}{GW190707A}{0.43}{GW190706A}{1.14}{GW190701A}{0.41}{GW190630A}{1.29}{GW190620A}{0.60}{GW190602A}{0.82}{GW190527A}{1.25}{GW190521B}{1.05}{GW190521A}{1.35}{GW190519A}{0.81}{GW190517A}{0.34}{GW190514A}{1.02}{GW190513A}{1.65}{GW190512A}{0.91}{GW190503A}{0.23}{GW190426A}{0.84}{GW190425A}{1.43}{GW190424A}{1.02}{GW190421A}{0.80}{GW190413B}{0.63}{GW190413A}{1.35}{GW190412A}{0.35}{GW190408A}{1.65}}}
\newcommand{\costhetajnmed}[1]{\IfEqCase{#1}{{GW190930A}{0.59}{GW190929A}{-0.13}{GW190924A}{0.74}{GW190915A}{-0.45}{GW190910A}{-0.05}{GW190909A}{0.12}{GW190828B}{-0.25}{GW190828A}{-0.62}{GW190814A}{0.65}{GW190803A}{0.44}{GW190731A}{0.37}{GW190728A}{0.33}{GW190727A}{0.02}{GW190720A}{-0.79}{GW190719A}{-0.04}{GW190708A}{0.20}{GW190707A}{-0.55}{GW190706A}{0.20}{GW190701A}{0.84}{GW190630A}{0.34}{GW190620A}{-0.36}{GW190602A}{-0.14}{GW190527A}{0.30}{GW190521B}{0.09}{GW190521A}{0.41}{GW190519A}{-0.01}{GW190517A}{-0.64}{GW190514A}{0.07}{GW190513A}{0.70}{GW190512A}{-0.04}{GW190503A}{-0.75}{GW190426A}{-0.13}{GW190425A}{0.47}{GW190424A}{0.05}{GW190421A}{-0.17}{GW190413B}{-0.33}{GW190413A}{0.41}{GW190412A}{0.75}{GW190408A}{0.70}}}
\newcommand{\costhetajnplus}[1]{\IfEqCase{#1}{{GW190930A}{0.39}{GW190929A}{1.01}{GW190924A}{0.25}{GW190915A}{1.32}{GW190910A}{0.96}{GW190909A}{0.82}{GW190828B}{1.20}{GW190828A}{1.57}{GW190814A}{0.16}{GW190803A}{0.53}{GW190731A}{0.59}{GW190728A}{0.65}{GW190727A}{0.94}{GW190720A}{1.68}{GW190719A}{1.01}{GW190708A}{0.78}{GW190707A}{1.51}{GW190706A}{0.75}{GW190701A}{0.15}{GW190630A}{0.62}{GW190620A}{1.27}{GW190602A}{1.10}{GW190527A}{0.66}{GW190521B}{0.87}{GW190521A}{0.55}{GW190519A}{0.81}{GW190517A}{1.38}{GW190514A}{0.89}{GW190513A}{0.27}{GW190512A}{0.99}{GW190503A}{0.48}{GW190426A}{1.09}{GW190425A}{0.50}{GW190424A}{0.91}{GW190421A}{1.12}{GW190413B}{1.27}{GW190413A}{0.55}{GW190412A}{0.14}{GW190408A}{0.28}}}
\newcommand{\spintwominus}[1]{\IfEqCase{#1}{{GW190930A}{0.37}{GW190929A}{0.44}{GW190924A}{0.32}{GW190915A}{0.43}{GW190910A}{0.33}{GW190909A}{0.43}{GW190828B}{0.38}{GW190828A}{0.37}{GW190814A}{0.46}{GW190803A}{0.40}{GW190731A}{0.40}{GW190728A}{0.35}{GW190727A}{0.41}{GW190720A}{0.45}{GW190719A}{0.49}{GW190708A}{0.28}{GW190707A}{0.28}{GW190706A}{0.44}{GW190701A}{0.40}{GW190630A}{0.34}{GW190620A}{0.50}{GW190602A}{0.45}{GW190527A}{0.45}{GW190521B}{0.37}{GW190521A}{0.52}{GW190519A}{0.48}{GW190517A}{0.52}{GW190514A}{0.48}{GW190513A}{0.39}{GW190512A}{0.32}{GW190503A}{0.40}{GW190426A}{0.009}{GW190425A}{0.25}{GW190424A}{0.42}{GW190421A}{0.42}{GW190413B}{0.45}{GW190413A}{0.41}{GW190412A}{0.43}{GW190408A}{0.32}}}
\newcommand{\spintwomed}[1]{\IfEqCase{#1}{{GW190930A}{0.42}{GW190929A}{0.49}{GW190924A}{0.35}{GW190915A}{0.48}{GW190910A}{0.37}{GW190909A}{0.49}{GW190828B}{0.42}{GW190828A}{0.41}{GW190814A}{0.52}{GW190803A}{0.45}{GW190731A}{0.45}{GW190728A}{0.39}{GW190727A}{0.45}{GW190720A}{0.51}{GW190719A}{0.55}{GW190708A}{0.30}{GW190707A}{0.31}{GW190706A}{0.49}{GW190701A}{0.44}{GW190630A}{0.38}{GW190620A}{0.56}{GW190602A}{0.50}{GW190527A}{0.50}{GW190521B}{0.42}{GW190521A}{0.58}{GW190519A}{0.54}{GW190517A}{0.58}{GW190514A}{0.54}{GW190513A}{0.43}{GW190512A}{0.36}{GW190503A}{0.44}{GW190426A}{0.009}{GW190425A}{0.28}{GW190424A}{0.47}{GW190421A}{0.46}{GW190413B}{0.50}{GW190413A}{0.45}{GW190412A}{0.49}{GW190408A}{0.36}}}
\newcommand{\spintwoplus}[1]{\IfEqCase{#1}{{GW190930A}{0.49}{GW190929A}{0.45}{GW190924A}{0.51}{GW190915A}{0.46}{GW190910A}{0.51}{GW190909A}{0.45}{GW190828B}{0.49}{GW190828A}{0.46}{GW190814A}{0.41}{GW190803A}{0.49}{GW190731A}{0.48}{GW190728A}{0.50}{GW190727A}{0.46}{GW190720A}{0.43}{GW190719A}{0.40}{GW190708A}{0.50}{GW190707A}{0.52}{GW190706A}{0.45}{GW190701A}{0.48}{GW190630A}{0.46}{GW190620A}{0.40}{GW190602A}{0.44}{GW190527A}{0.45}{GW190521B}{0.39}{GW190521A}{0.38}{GW190519A}{0.41}{GW190517A}{0.38}{GW190514A}{0.42}{GW190513A}{0.48}{GW190512A}{0.51}{GW190503A}{0.48}{GW190426A}{0.03}{GW190425A}{0.51}{GW190424A}{0.47}{GW190421A}{0.47}{GW190413B}{0.44}{GW190413A}{0.48}{GW190412A}{0.44}{GW190408A}{0.53}}}
\newcommand{\massonedetminus}[1]{\IfEqCase{#1}{{GW190930A}{2.6}{GW190929A}{28.9}{GW190924A}{2.2}{GW190915A}{7.7}{GW190910A}{6.2}{GW190909A}{17.5}{GW190828B}{8.7}{GW190828A}{4.7}{GW190814A}{1.0}{GW190803A}{9.0}{GW190731A}{10.5}{GW190728A}{2.5}{GW190727A}{8.2}{GW190720A}{3.5}{GW190719A}{16.4}{GW190708A}{2.4}{GW190707A}{1.8}{GW190706A}{17.6}{GW190701A}{10.9}{GW190630A}{6.4}{GW190620A}{15.4}{GW190602A}{15.7}{GW190527A}{11.0}{GW190521B}{5.4}{GW190521A}{20.8}{GW190519A}{12.9}{GW190517A}{8.4}{GW190514A}{10.8}{GW190513A}{12.3}{GW190512A}{6.8}{GW190503A}{9.9}{GW190426A}{2.5}{GW190425A}{0.4}{GW190424A}{8.0}{GW190421A}{8.8}{GW190413B}{14.6}{GW190413A}{12.0}{GW190412A}{6.2}{GW190408A}{4.0}}}
\newcommand{\massonedetmed}[1]{\IfEqCase{#1}{{GW190930A}{14.2}{GW190929A}{111.3}{GW190924A}{9.9}{GW190915A}{46.0}{GW190910A}{56.3}{GW190909A}{73.0}{GW190828B}{31.1}{GW190828A}{43.9}{GW190814A}{24.4}{GW190803A}{57.6}{GW190731A}{64.4}{GW190728A}{14.4}{GW190727A}{58.8}{GW190720A}{15.7}{GW190719A}{60.3}{GW190708A}{20.6}{GW190707A}{13.4}{GW190706A}{112.9}{GW190701A}{74.1}{GW190630A}{41.4}{GW190620A}{84.6}{GW190602A}{101.7}{GW190527A}{52.0}{GW190521B}{51.9}{GW190521A}{153.4}{GW190519A}{95.4}{GW190517A}{50.5}{GW190514A}{64.9}{GW190513A}{49.0}{GW190512A}{29.4}{GW190503A}{55.2}{GW190426A}{6.2}{GW190425A}{2.1}{GW190424A}{56.1}{GW190421A}{61.1}{GW190413B}{80.7}{GW190413A}{55.3}{GW190412A}{34.6}{GW190408A}{31.5}}}
\newcommand{\massonedetplus}[1]{\IfEqCase{#1}{{GW190930A}{14.3}{GW190929A}{39.3}{GW190924A}{7.8}{GW190915A}{11.6}{GW190910A}{8.8}{GW190909A}{84.2}{GW190828B}{8.8}{GW190828A}{7.5}{GW190814A}{1.2}{GW190803A}{14.2}{GW190731A}{13.4}{GW190728A}{8.4}{GW190727A}{12.8}{GW190720A}{7.7}{GW190719A}{26.5}{GW190708A}{5.7}{GW190707A}{3.7}{GW190706A}{20.5}{GW190701A}{15.1}{GW190630A}{8.3}{GW190620A}{20.0}{GW190602A}{19.9}{GW190527A}{32.9}{GW190521B}{7.1}{GW190521A}{45.9}{GW190519A}{14.9}{GW190517A}{14.7}{GW190514A}{17.6}{GW190513A}{12.5}{GW190512A}{6.5}{GW190503A}{10.6}{GW190426A}{4.2}{GW190425A}{0.6}{GW190424A}{14.9}{GW190421A}{14.7}{GW190413B}{19.0}{GW190413A}{17.1}{GW190412A}{5.5}{GW190408A}{6.3}}}
\newcommand{\massratiominus}[1]{\IfEqCase{#1}{{GW190930A}{0.46}{GW190929A}{0.16}{GW190924A}{0.37}{GW190915A}{0.26}{GW190910A}{0.23}{GW190909A}{0.39}{GW190828B}{0.16}{GW190828A}{0.23}{GW190814A}{0.009}{GW190803A}{0.31}{GW190731A}{0.31}{GW190728A}{0.38}{GW190727A}{0.32}{GW190720A}{0.30}{GW190719A}{0.29}{GW190708A}{0.28}{GW190707A}{0.27}{GW190706A}{0.25}{GW190701A}{0.30}{GW190630A}{0.22}{GW190620A}{0.27}{GW190602A}{0.33}{GW190527A}{0.32}{GW190521B}{0.21}{GW190521A}{0.34}{GW190519A}{0.19}{GW190517A}{0.29}{GW190514A}{0.33}{GW190513A}{0.18}{GW190512A}{0.18}{GW190503A}{0.23}{GW190426A}{0.15}{GW190425A}{0.25}{GW190424A}{0.29}{GW190421A}{0.30}{GW190413B}{0.31}{GW190413A}{0.28}{GW190412A}{0.06}{GW190408A}{0.25}}}
\newcommand{\massratiomed}[1]{\IfEqCase{#1}{{GW190930A}{0.64}{GW190929A}{0.30}{GW190924A}{0.57}{GW190915A}{0.69}{GW190910A}{0.82}{GW190909A}{0.62}{GW190828B}{0.42}{GW190828A}{0.82}{GW190814A}{0.112}{GW190803A}{0.75}{GW190731A}{0.72}{GW190728A}{0.66}{GW190727A}{0.79}{GW190720A}{0.59}{GW190719A}{0.58}{GW190708A}{0.75}{GW190707A}{0.72}{GW190706A}{0.58}{GW190701A}{0.76}{GW190630A}{0.68}{GW190620A}{0.62}{GW190602A}{0.71}{GW190527A}{0.64}{GW190521B}{0.78}{GW190521A}{0.75}{GW190519A}{0.61}{GW190517A}{0.68}{GW190514A}{0.75}{GW190513A}{0.50}{GW190512A}{0.54}{GW190503A}{0.65}{GW190426A}{0.25}{GW190425A}{0.67}{GW190424A}{0.81}{GW190421A}{0.79}{GW190413B}{0.69}{GW190413A}{0.69}{GW190412A}{0.28}{GW190408A}{0.76}}}
\newcommand{\massratioplus}[1]{\IfEqCase{#1}{{GW190930A}{0.30}{GW190929A}{0.52}{GW190924A}{0.36}{GW190915A}{0.27}{GW190910A}{0.15}{GW190909A}{0.33}{GW190828B}{0.38}{GW190828A}{0.15}{GW190814A}{0.008}{GW190803A}{0.22}{GW190731A}{0.25}{GW190728A}{0.29}{GW190727A}{0.18}{GW190720A}{0.36}{GW190719A}{0.37}{GW190708A}{0.21}{GW190707A}{0.24}{GW190706A}{0.34}{GW190701A}{0.21}{GW190630A}{0.27}{GW190620A}{0.33}{GW190602A}{0.25}{GW190527A}{0.32}{GW190521B}{0.19}{GW190521A}{0.23}{GW190519A}{0.28}{GW190517A}{0.27}{GW190514A}{0.21}{GW190513A}{0.42}{GW190512A}{0.37}{GW190503A}{0.29}{GW190426A}{0.41}{GW190425A}{0.29}{GW190424A}{0.17}{GW190421A}{0.18}{GW190413B}{0.28}{GW190413A}{0.28}{GW190412A}{0.12}{GW190408A}{0.21}}}
\newcommand{\spinoneminus}[1]{\IfEqCase{#1}{{GW190930A}{0.35}{GW190929A}{0.54}{GW190924A}{0.21}{GW190915A}{0.49}{GW190910A}{0.30}{GW190909A}{0.52}{GW190828B}{0.26}{GW190828A}{0.40}{GW190814A}{0.03}{GW190803A}{0.37}{GW190731A}{0.34}{GW190728A}{0.28}{GW190727A}{0.42}{GW190720A}{0.35}{GW190719A}{0.53}{GW190708A}{0.20}{GW190707A}{0.21}{GW190706A}{0.48}{GW190701A}{0.36}{GW190630A}{0.23}{GW190620A}{0.50}{GW190602A}{0.34}{GW190527A}{0.43}{GW190521B}{0.28}{GW190521A}{0.63}{GW190519A}{0.50}{GW190517A}{0.35}{GW190514A}{0.46}{GW190513A}{0.28}{GW190512A}{0.16}{GW190503A}{0.31}{GW190426A}{0.14}{GW190425A}{0.25}{GW190424A}{0.47}{GW190421A}{0.41}{GW190413B}{0.52}{GW190413A}{0.36}{GW190412A}{0.22}{GW190408A}{0.31}}}
\newcommand{\spinonemed}[1]{\IfEqCase{#1}{{GW190930A}{0.39}{GW190929A}{0.64}{GW190924A}{0.24}{GW190915A}{0.55}{GW190910A}{0.34}{GW190909A}{0.58}{GW190828B}{0.28}{GW190828A}{0.44}{GW190814A}{0.03}{GW190803A}{0.41}{GW190731A}{0.37}{GW190728A}{0.32}{GW190727A}{0.46}{GW190720A}{0.40}{GW190719A}{0.62}{GW190708A}{0.22}{GW190707A}{0.24}{GW190706A}{0.55}{GW190701A}{0.40}{GW190630A}{0.26}{GW190620A}{0.61}{GW190602A}{0.38}{GW190527A}{0.47}{GW190521B}{0.31}{GW190521A}{0.73}{GW190519A}{0.60}{GW190517A}{0.86}{GW190514A}{0.52}{GW190513A}{0.30}{GW190512A}{0.17}{GW190503A}{0.34}{GW190426A}{0.14}{GW190425A}{0.27}{GW190424A}{0.53}{GW190421A}{0.46}{GW190413B}{0.58}{GW190413A}{0.40}{GW190412A}{0.44}{GW190408A}{0.34}}}
\newcommand{\spinoneplus}[1]{\IfEqCase{#1}{{GW190930A}{0.40}{GW190929A}{0.32}{GW190924A}{0.43}{GW190915A}{0.39}{GW190910A}{0.50}{GW190909A}{0.37}{GW190828B}{0.43}{GW190828A}{0.45}{GW190814A}{0.05}{GW190803A}{0.51}{GW190731A}{0.54}{GW190728A}{0.37}{GW190727A}{0.47}{GW190720A}{0.40}{GW190719A}{0.34}{GW190708A}{0.52}{GW190707A}{0.47}{GW190706A}{0.39}{GW190701A}{0.50}{GW190630A}{0.37}{GW190620A}{0.34}{GW190602A}{0.51}{GW190527A}{0.47}{GW190521B}{0.42}{GW190521A}{0.25}{GW190519A}{0.33}{GW190517A}{0.13}{GW190514A}{0.43}{GW190513A}{0.51}{GW190512A}{0.44}{GW190503A}{0.51}{GW190426A}{0.40}{GW190425A}{0.51}{GW190424A}{0.42}{GW190421A}{0.47}{GW190413B}{0.38}{GW190413A}{0.51}{GW190412A}{0.16}{GW190408A}{0.47}}}
\newcommand{\costiltoneminus}[1]{\IfEqCase{#1}{{GW190930A}{1.08}{GW190929A}{0.83}{GW190924A}{0.97}{GW190915A}{0.82}{GW190910A}{0.93}{GW190909A}{0.78}{GW190828B}{0.99}{GW190828A}{1.09}{GW190814A}{0.90}{GW190803A}{0.80}{GW190731A}{0.99}{GW190728A}{1.13}{GW190727A}{1.02}{GW190720A}{0.97}{GW190719A}{1.02}{GW190708A}{0.88}{GW190707A}{0.68}{GW190706A}{0.99}{GW190701A}{0.70}{GW190630A}{1.02}{GW190620A}{0.84}{GW190602A}{0.99}{GW190527A}{1.02}{GW190521B}{0.97}{GW190521A}{0.89}{GW190519A}{0.74}{GW190517A}{0.37}{GW190514A}{0.50}{GW190513A}{1.10}{GW190512A}{0.93}{GW190503A}{0.75}{GW190426A}{0.00}{GW190425A}{0.65}{GW190424A}{1.00}{GW190421A}{0.75}{GW190413B}{0.76}{GW190413A}{0.90}{GW190412A}{0.47}{GW190408A}{0.73}}}
\newcommand{\costiltonemed}[1]{\IfEqCase{#1}{{GW190930A}{0.47}{GW190929A}{0.02}{GW190924A}{0.19}{GW190915A}{0.06}{GW190910A}{0.08}{GW190909A}{-0.10}{GW190828B}{0.26}{GW190828A}{0.51}{GW190814A}{0.01}{GW190803A}{-0.07}{GW190731A}{0.16}{GW190728A}{0.49}{GW190727A}{0.30}{GW190720A}{0.54}{GW190719A}{0.66}{GW190708A}{0.07}{GW190707A}{-0.19}{GW190706A}{0.66}{GW190701A}{-0.22}{GW190630A}{0.29}{GW190620A}{0.68}{GW190602A}{0.18}{GW190527A}{0.28}{GW190521B}{0.17}{GW190521A}{0.05}{GW190519A}{0.65}{GW190517A}{0.83}{GW190514A}{-0.45}{GW190513A}{0.41}{GW190512A}{0.07}{GW190503A}{-0.16}{GW190426A}{-1.00}{GW190425A}{0.26}{GW190424A}{0.36}{GW190421A}{-0.15}{GW190413B}{-0.06}{GW190413A}{0.01}{GW190412A}{0.70}{GW190408A}{-0.17}}}
\newcommand{\costiltoneplus}[1]{\IfEqCase{#1}{{GW190930A}{0.49}{GW190929A}{0.65}{GW190924A}{0.76}{GW190915A}{0.73}{GW190910A}{0.79}{GW190909A}{0.96}{GW190828B}{0.63}{GW190828A}{0.44}{GW190814A}{0.87}{GW190803A}{0.91}{GW190731A}{0.74}{GW190728A}{0.47}{GW190727A}{0.61}{GW190720A}{0.42}{GW190719A}{0.31}{GW190708A}{0.80}{GW190707A}{0.97}{GW190706A}{0.31}{GW190701A}{1.01}{GW190630A}{0.63}{GW190620A}{0.29}{GW190602A}{0.72}{GW190527A}{0.65}{GW190521B}{0.71}{GW190521A}{0.78}{GW190519A}{0.32}{GW190517A}{0.16}{GW190514A}{1.08}{GW190513A}{0.53}{GW190512A}{0.82}{GW190503A}{0.96}{GW190426A}{2.00}{GW190425A}{0.61}{GW190424A}{0.56}{GW190421A}{0.94}{GW190413B}{0.82}{GW190413A}{0.87}{GW190412A}{0.20}{GW190408A}{0.94}}}
\newcommand{\finalmasssourceminus}[1]{\IfEqCase{#1}{{GW190930A}{1.5}{GW190929A}{25.3}{GW190924A}{1.0}{GW190915A}{6.0}{GW190910A}{8.6}{GW190909A}{16.8}{GW190828B}{4.5}{GW190828A}{4.3}{GW190814A}{0.9}{GW190803A}{8.5}{GW190731A}{10.8}{GW190728A}{1.3}{GW190727A}{7.5}{GW190720A}{2.2}{GW190719A}{10.2}{GW190708A}{1.8}{GW190707A}{1.3}{GW190706A}{13.5}{GW190701A}{8.9}{GW190630A}{4.6}{GW190620A}{12.1}{GW190602A}{14.9}{GW190527A}{9.3}{GW190521B}{4.4}{GW190521A}{22.4}{GW190519A}{13.8}{GW190517A}{8.9}{GW190514A}{10.4}{GW190513A}{5.8}{GW190512A}{3.5}{GW190503A}{7.7}{GW190424A}{10.1}{GW190421A}{8.7}{GW190413B}{11.4}{GW190413A}{9.2}{GW190412A}{3.8}{GW190408A}{2.8}}}
\newcommand{\finalmasssourcemed}[1]{\IfEqCase{#1}{{GW190930A}{19.4}{GW190929A}{101.5}{GW190924A}{13.3}{GW190915A}{57.2}{GW190910A}{75.8}{GW190909A}{72.0}{GW190828B}{33.1}{GW190828A}{54.9}{GW190814A}{25.6}{GW190803A}{61.7}{GW190731A}{67.0}{GW190728A}{19.6}{GW190727A}{63.8}{GW190720A}{20.4}{GW190719A}{54.9}{GW190708A}{29.5}{GW190707A}{19.2}{GW190706A}{99.0}{GW190701A}{90.2}{GW190630A}{56.4}{GW190620A}{87.2}{GW190602A}{110.9}{GW190527A}{56.4}{GW190521B}{71.0}{GW190521A}{156.3}{GW190519A}{101.0}{GW190517A}{59.3}{GW190514A}{64.5}{GW190513A}{51.6}{GW190512A}{34.5}{GW190503A}{68.6}{GW190424A}{68.9}{GW190421A}{69.7}{GW190413B}{75.5}{GW190413A}{56.0}{GW190412A}{37.3}{GW190408A}{41.1}}}
\newcommand{\finalmasssourceplus}[1]{\IfEqCase{#1}{{GW190930A}{9.2}{GW190929A}{33.6}{GW190924A}{5.2}{GW190915A}{7.1}{GW190910A}{8.5}{GW190909A}{54.9}{GW190828B}{5.5}{GW190828A}{7.2}{GW190814A}{1.1}{GW190803A}{11.8}{GW190731A}{14.6}{GW190728A}{4.7}{GW190727A}{10.9}{GW190720A}{4.5}{GW190719A}{17.3}{GW190708A}{2.5}{GW190707A}{1.9}{GW190706A}{18.3}{GW190701A}{11.3}{GW190630A}{4.4}{GW190620A}{16.8}{GW190602A}{17.7}{GW190527A}{20.2}{GW190521B}{6.5}{GW190521A}{36.8}{GW190519A}{12.4}{GW190517A}{9.1}{GW190514A}{17.9}{GW190513A}{8.2}{GW190512A}{3.8}{GW190503A}{8.8}{GW190424A}{12.4}{GW190421A}{12.5}{GW190413B}{16.4}{GW190413A}{12.5}{GW190412A}{3.9}{GW190408A}{3.9}}}
\newcommand{\phaseminus}[1]{\IfEqCase{#1}{{GW190930A}{2.91}{GW190929A}{2.79}{GW190924A}{2.77}{GW190915A}{2.96}{GW190910A}{2.81}{GW190909A}{2.80}{GW190828B}{2.90}{GW190828A}{2.89}{GW190814A}{2.76}{GW190803A}{2.85}{GW190731A}{2.83}{GW190728A}{2.80}{GW190727A}{2.88}{GW190720A}{2.81}{GW190719A}{2.84}{GW190708A}{2.83}{GW190707A}{2.91}{GW190706A}{2.45}{GW190701A}{2.60}{GW190630A}{3.46}{GW190620A}{2.68}{GW190602A}{2.80}{GW190527A}{2.85}{GW190521B}{1.76}{GW190521A}{2.87}{GW190519A}{2.80}{GW190517A}{2.79}{GW190514A}{2.81}{GW190513A}{2.78}{GW190512A}{2.87}{GW190503A}{2.88}{GW190426A}{2.79}{GW190425A}{2.82}{GW190424A}{2.86}{GW190421A}{2.89}{GW190413B}{2.82}{GW190413A}{2.81}{GW190412A}{1.89}{GW190408A}{2.79}}}
\newcommand{\phasemed}[1]{\IfEqCase{#1}{{GW190930A}{3.22}{GW190929A}{3.08}{GW190924A}{3.08}{GW190915A}{3.30}{GW190910A}{3.15}{GW190909A}{3.15}{GW190828B}{3.23}{GW190828A}{3.17}{GW190814A}{3.13}{GW190803A}{3.14}{GW190731A}{3.15}{GW190728A}{3.11}{GW190727A}{3.21}{GW190720A}{3.12}{GW190719A}{3.15}{GW190708A}{3.14}{GW190707A}{3.24}{GW190706A}{3.01}{GW190701A}{2.89}{GW190630A}{3.82}{GW190620A}{2.97}{GW190602A}{3.13}{GW190527A}{3.13}{GW190521B}{2.03}{GW190521A}{3.12}{GW190519A}{3.13}{GW190517A}{3.10}{GW190514A}{3.13}{GW190513A}{3.08}{GW190512A}{3.18}{GW190503A}{3.16}{GW190426A}{3.09}{GW190425A}{3.12}{GW190424A}{3.16}{GW190421A}{3.20}{GW190413B}{3.11}{GW190413A}{3.12}{GW190412A}{2.15}{GW190408A}{3.11}}}
\newcommand{\phaseplus}[1]{\IfEqCase{#1}{{GW190930A}{2.77}{GW190929A}{2.89}{GW190924A}{2.88}{GW190915A}{2.70}{GW190910A}{2.84}{GW190909A}{2.82}{GW190828B}{2.73}{GW190828A}{2.85}{GW190814A}{2.81}{GW190803A}{2.81}{GW190731A}{2.80}{GW190728A}{2.86}{GW190727A}{2.76}{GW190720A}{2.85}{GW190719A}{2.83}{GW190708A}{2.83}{GW190707A}{2.76}{GW190706A}{2.65}{GW190701A}{3.10}{GW190630A}{2.15}{GW190620A}{2.97}{GW190602A}{2.84}{GW190527A}{2.81}{GW190521B}{3.94}{GW190521A}{2.87}{GW190519A}{2.83}{GW190517A}{2.88}{GW190514A}{2.82}{GW190513A}{2.90}{GW190512A}{2.82}{GW190503A}{2.84}{GW190426A}{2.85}{GW190425A}{2.87}{GW190424A}{2.81}{GW190421A}{2.77}{GW190413B}{2.84}{GW190413A}{2.84}{GW190412A}{3.84}{GW190408A}{2.86}}}
\newcommand{\radiatedenergyminus}[1]{\IfEqCase{#1}{{GW190930A}{0.2}{GW190929A}{1.4}{GW190924A}{0.1}{GW190915A}{0.7}{GW190910A}{0.7}{GW190909A}{1.4}{GW190828B}{0.2}{GW190828A}{0.5}{GW190814A}{0.007}{GW190803A}{0.8}{GW190731A}{1.1}{GW190728A}{0.2}{GW190727A}{0.9}{GW190720A}{0.2}{GW190719A}{1.1}{GW190708A}{0.2}{GW190707A}{0.09}{GW190706A}{2.1}{GW190701A}{1.1}{GW190630A}{0.5}{GW190620A}{1.9}{GW190602A}{1.9}{GW190527A}{1.0}{GW190521B}{0.7}{GW190521A}{2.4}{GW190519A}{1.7}{GW190517A}{1.2}{GW190514A}{0.8}{GW190513A}{0.6}{GW190512A}{0.3}{GW190503A}{1.0}{GW190424A}{0.9}{GW190421A}{0.9}{GW190413B}{1.1}{GW190413A}{0.8}{GW190412A}{0.1}{GW190408A}{0.3}}}
\newcommand{\radiatedenergymed}[1]{\IfEqCase{#1}{{GW190930A}{0.9}{GW190929A}{2.7}{GW190924A}{0.6}{GW190915A}{2.7}{GW190910A}{3.8}{GW190909A}{3.0}{GW190828B}{1.2}{GW190828A}{3.1}{GW190814A}{0.2}{GW190803A}{2.9}{GW190731A}{3.2}{GW190728A}{1.0}{GW190727A}{3.3}{GW190720A}{1.0}{GW190719A}{2.9}{GW190708A}{1.4}{GW190707A}{0.9}{GW190706A}{5.3}{GW190701A}{4.1}{GW190630A}{2.8}{GW190620A}{4.9}{GW190602A}{5.4}{GW190527A}{2.7}{GW190521B}{3.7}{GW190521A}{7.6}{GW190519A}{5.6}{GW190517A}{4.1}{GW190514A}{2.7}{GW190513A}{2.2}{GW190512A}{1.5}{GW190503A}{3.1}{GW190424A}{3.6}{GW190421A}{3.3}{GW190413B}{3.4}{GW190413A}{2.6}{GW190412A}{1.1}{GW190408A}{1.9}}}
\newcommand{\radiatedenergyplus}[1]{\IfEqCase{#1}{{GW190930A}{0.1}{GW190929A}{2.8}{GW190924A}{0.06}{GW190915A}{0.7}{GW190910A}{0.9}{GW190909A}{2.2}{GW190828B}{0.3}{GW190828A}{0.7}{GW190814A}{0.006}{GW190803A}{0.9}{GW190731A}{1.2}{GW190728A}{0.09}{GW190727A}{1.1}{GW190720A}{0.1}{GW190719A}{1.7}{GW190708A}{0.1}{GW190707A}{0.08}{GW190706A}{2.3}{GW190701A}{1.1}{GW190630A}{0.5}{GW190620A}{2.0}{GW190602A}{1.8}{GW190527A}{1.5}{GW190521B}{0.6}{GW190521A}{2.9}{GW190519A}{1.7}{GW190517A}{1.3}{GW190514A}{1.1}{GW190513A}{1.1}{GW190512A}{0.3}{GW190503A}{0.9}{GW190424A}{1.2}{GW190421A}{1.0}{GW190413B}{1.1}{GW190413A}{1.0}{GW190412A}{0.2}{GW190408A}{0.3}}}
\newcommand{\masstwodetminus}[1]{\IfEqCase{#1}{{GW190930A}{3.9}{GW190929A}{16.4}{GW190924A}{2.1}{GW190915A}{8.5}{GW190910A}{9.1}{GW190909A}{23.3}{GW190828B}{2.7}{GW190828A}{6.5}{GW190814A}{0.09}{GW190803A}{13.5}{GW190731A}{16.4}{GW190728A}{3.0}{GW190727A}{13.8}{GW190720A}{2.6}{GW190719A}{12.7}{GW190708A}{3.1}{GW190707A}{1.9}{GW190706A}{27.0}{GW190701A}{17.5}{GW190630A}{5.5}{GW190620A}{19.7}{GW190602A}{27.8}{GW190527A}{12.3}{GW190521B}{7.6}{GW190521A}{39.7}{GW190519A}{16.9}{GW190517A}{9.5}{GW190514A}{15.9}{GW190513A}{6.0}{GW190512A}{2.9}{GW190503A}{10.6}{GW190426A}{0.5}{GW190425A}{0.3}{GW190424A}{10.3}{GW190421A}{13.2}{GW190413B}{20.0}{GW190413A}{11.5}{GW190412A}{1.0}{GW190408A}{4.6}}}
\newcommand{\masstwodetmed}[1]{\IfEqCase{#1}{{GW190930A}{9.1}{GW190929A}{33.9}{GW190924A}{5.6}{GW190915A}{31.9}{GW190910A}{45.9}{GW190909A}{47.1}{GW190828B}{13.3}{GW190828A}{36.1}{GW190814A}{2.72}{GW190803A}{42.7}{GW190731A}{45.6}{GW190728A}{9.5}{GW190727A}{46.0}{GW190720A}{9.2}{GW190719A}{34.5}{GW190708A}{15.5}{GW190707A}{9.7}{GW190706A}{66.6}{GW190701A}{56.2}{GW190630A}{28.0}{GW190620A}{53.1}{GW190602A}{71.5}{GW190527A}{32.8}{GW190521B}{40.5}{GW190521A}{114.8}{GW190519A}{59.3}{GW190517A}{34.4}{GW190514A}{48.0}{GW190513A}{24.7}{GW190512A}{15.8}{GW190503A}{36.2}{GW190426A}{1.6}{GW190425A}{1.4}{GW190424A}{44.8}{GW190421A}{47.8}{GW190413B}{55.2}{GW190413A}{38.0}{GW190412A}{9.6}{GW190408A}{23.7}}}
\newcommand{\masstwodetplus}[1]{\IfEqCase{#1}{{GW190930A}{1.8}{GW190929A}{37.2}{GW190924A}{1.5}{GW190915A}{7.0}{GW190910A}{7.0}{GW190909A}{20.9}{GW190828B}{4.9}{GW190828A}{4.6}{GW190814A}{0.08}{GW190803A}{9.8}{GW190731A}{11.8}{GW190728A}{1.8}{GW190727A}{8.4}{GW190720A}{2.4}{GW190719A}{13.3}{GW190708A}{2.0}{GW190707A}{1.4}{GW190706A}{23.9}{GW190701A}{11.7}{GW190630A}{5.5}{GW190620A}{17.1}{GW190602A}{18.6}{GW190527A}{19.4}{GW190521B}{5.7}{GW190521A}{26.0}{GW190519A}{16.1}{GW190517A}{8.2}{GW190514A}{11.1}{GW190513A}{10.6}{GW190512A}{4.7}{GW190503A}{10.4}{GW190426A}{0.9}{GW190425A}{0.3}{GW190424A}{8.3}{GW190421A}{9.4}{GW190413B}{14.9}{GW190413A}{10.7}{GW190412A}{1.7}{GW190408A}{3.6}}}
\newcommand{\masstwosourceminus}[1]{\IfEqCase{#1}{{GW190930A}{3.3}{GW190929A}{10.6}{GW190924A}{1.9}{GW190915A}{6.1}{GW190910A}{7.2}{GW190909A}{12.7}{GW190828B}{2.1}{GW190828A}{4.8}{GW190814A}{0.09}{GW190803A}{8.2}{GW190731A}{9.5}{GW190728A}{2.6}{GW190727A}{8.4}{GW190720A}{2.2}{GW190719A}{7.2}{GW190708A}{2.7}{GW190707A}{1.7}{GW190706A}{13.3}{GW190701A}{12.0}{GW190630A}{5.1}{GW190620A}{12.3}{GW190602A}{17.4}{GW190527A}{8.1}{GW190521B}{6.4}{GW190521A}{23.1}{GW190519A}{11.1}{GW190517A}{7.3}{GW190514A}{8.8}{GW190513A}{4.1}{GW190512A}{2.5}{GW190503A}{8.0}{GW190426A}{0.5}{GW190425A}{0.3}{GW190424A}{7.7}{GW190421A}{8.8}{GW190413B}{10.8}{GW190413A}{6.7}{GW190412A}{0.9}{GW190408A}{3.6}}}
\newcommand{\masstwosourcemed}[1]{\IfEqCase{#1}{{GW190930A}{7.8}{GW190929A}{24.1}{GW190924A}{5.0}{GW190915A}{24.4}{GW190910A}{35.6}{GW190909A}{28.3}{GW190828B}{10.2}{GW190828A}{26.2}{GW190814A}{2.59}{GW190803A}{27.3}{GW190731A}{28.8}{GW190728A}{8.1}{GW190727A}{29.4}{GW190720A}{7.8}{GW190719A}{20.8}{GW190708A}{13.2}{GW190707A}{8.4}{GW190706A}{38.2}{GW190701A}{40.8}{GW190630A}{23.7}{GW190620A}{35.5}{GW190602A}{47.8}{GW190527A}{22.6}{GW190521B}{32.8}{GW190521A}{69.0}{GW190519A}{40.5}{GW190517A}{25.3}{GW190514A}{28.4}{GW190513A}{18.0}{GW190512A}{12.6}{GW190503A}{28.4}{GW190426A}{1.5}{GW190425A}{1.4}{GW190424A}{31.8}{GW190421A}{31.9}{GW190413B}{31.8}{GW190413A}{23.7}{GW190412A}{8.3}{GW190408A}{18.4}}}
\newcommand{\masstwosourceplus}[1]{\IfEqCase{#1}{{GW190930A}{1.7}{GW190929A}{19.3}{GW190924A}{1.4}{GW190915A}{5.6}{GW190910A}{6.3}{GW190909A}{13.4}{GW190828B}{3.6}{GW190828A}{4.6}{GW190814A}{0.08}{GW190803A}{7.8}{GW190731A}{9.7}{GW190728A}{1.7}{GW190727A}{7.1}{GW190720A}{2.3}{GW190719A}{9.0}{GW190708A}{2.0}{GW190707A}{1.4}{GW190706A}{14.6}{GW190701A}{8.7}{GW190630A}{5.2}{GW190620A}{12.2}{GW190602A}{14.3}{GW190527A}{10.5}{GW190521B}{5.4}{GW190521A}{22.7}{GW190519A}{11.0}{GW190517A}{7.0}{GW190514A}{9.3}{GW190513A}{7.7}{GW190512A}{3.6}{GW190503A}{7.7}{GW190426A}{0.8}{GW190425A}{0.3}{GW190424A}{7.6}{GW190421A}{8.0}{GW190413B}{11.7}{GW190413A}{7.3}{GW190412A}{1.6}{GW190408A}{3.3}}}
\newcommand{\decminus}[1]{\IfEqCase{#1}{{GW190930A}{0.66804}{GW190929A}{1.09453}{GW190924A}{0.31095}{GW190915A}{0.43700}{GW190910A}{0.78759}{GW190909A}{1.37456}{GW190828B}{0.42413}{GW190828A}{0.45822}{GW190814A}{0.12765}{GW190803A}{0.76320}{GW190731A}{0.54299}{GW190728A}{1.45903}{GW190727A}{0.53050}{GW190720A}{1.79814}{GW190719A}{1.52271}{GW190708A}{1.12280}{GW190707A}{0.66130}{GW190706A}{1.13851}{GW190701A}{0.08561}{GW190630A}{0.88066}{GW190620A}{1.15741}{GW190602A}{0.22445}{GW190527A}{0.63740}{GW190521B}{0.61963}{GW190521A}{0.40205}{GW190519A}{1.22807}{GW190517A}{0.23138}{GW190514A}{1.30610}{GW190513A}{1.20492}{GW190512A}{0.07267}{GW190503A}{0.08741}{GW190426A}{1.53579}{GW190425A}{0.89977}{GW190424A}{1.08941}{GW190421A}{0.52647}{GW190413B}{0.10088}{GW190413A}{1.21171}{GW190412A}{0.03938}{GW190408A}{0.33259}}}
\newcommand{\decmed}[1]{\IfEqCase{#1}{{GW190930A}{0.64489}{GW190929A}{0.12643}{GW190924A}{0.16308}{GW190915A}{0.64820}{GW190910A}{-0.19299}{GW190909A}{0.45798}{GW190828B}{-0.70225}{GW190828A}{-0.38234}{GW190814A}{-0.43746}{GW190803A}{0.56728}{GW190731A}{-0.84452}{GW190728A}{0.14876}{GW190727A}{-0.69575}{GW190720A}{0.61861}{GW190719A}{0.60387}{GW190708A}{0.31278}{GW190707A}{-0.26771}{GW190706A}{0.49342}{GW190701A}{-0.11450}{GW190630A}{-0.17794}{GW190620A}{0.40157}{GW190602A}{-0.60585}{GW190527A}{-0.67031}{GW190521B}{0.31697}{GW190521A}{-0.79351}{GW190519A}{0.62138}{GW190517A}{-0.77834}{GW190514A}{0.75342}{GW190513A}{0.66794}{GW190512A}{-0.46498}{GW190503A}{-0.88291}{GW190426A}{0.90282}{GW190425A}{-0.13006}{GW190424A}{-0.00051}{GW190421A}{-0.82328}{GW190413B}{-0.53598}{GW190413A}{0.45042}{GW190412A}{0.63309}{GW190408A}{0.92018}}}
\newcommand{\decplus}[1]{\IfEqCase{#1}{{GW190930A}{0.44937}{GW190929A}{0.92641}{GW190924A}{0.26435}{GW190915A}{0.49993}{GW190910A}{0.99066}{GW190909A}{0.80671}{GW190828B}{1.28711}{GW190828A}{1.25309}{GW190814A}{0.03175}{GW190803A}{0.63137}{GW190731A}{1.11036}{GW190728A}{0.43956}{GW190727A}{1.58271}{GW190720A}{0.05969}{GW190719A}{0.55220}{GW190708A}{0.86640}{GW190707A}{1.42731}{GW190706A}{0.43610}{GW190701A}{0.08803}{GW190630A}{0.78734}{GW190620A}{0.77559}{GW190602A}{0.57550}{GW190527A}{0.97030}{GW190521B}{0.25837}{GW190521A}{1.54032}{GW190519A}{0.40124}{GW190517A}{0.96906}{GW190514A}{0.60827}{GW190513A}{0.37574}{GW190512A}{0.32502}{GW190503A}{0.10637}{GW190426A}{0.61722}{GW190425A}{0.96811}{GW190424A}{1.08246}{GW190421A}{0.53382}{GW190413B}{1.13504}{GW190413A}{0.90354}{GW190412A}{0.02643}{GW190408A}{0.08290}}}
\newcommand{\psiminus}[1]{\IfEqCase{#1}{{GW190930A}{1.82}{GW190929A}{1.43}{GW190924A}{1.79}{GW190915A}{1.85}{GW190910A}{2.23}{GW190909A}{1.42}{GW190828B}{1.30}{GW190828A}{2.05}{GW190814A}{0.32}{GW190803A}{2.82}{GW190731A}{2.78}{GW190728A}{1.92}{GW190727A}{2.89}{GW190720A}{1.90}{GW190719A}{1.84}{GW190708A}{2.83}{GW190707A}{1.83}{GW190706A}{2.03}{GW190701A}{1.87}{GW190630A}{1.25}{GW190620A}{1.54}{GW190602A}{2.89}{GW190527A}{2.77}{GW190521B}{1.24}{GW190521A}{1.42}{GW190519A}{2.65}{GW190517A}{2.20}{GW190514A}{2.82}{GW190513A}{2.03}{GW190512A}{2.81}{GW190503A}{2.51}{GW190426A}{1.40}{GW190425A}{1.46}{GW190424A}{2.78}{GW190421A}{2.77}{GW190413B}{2.84}{GW190413A}{2.79}{GW190412A}{2.36}{GW190408A}{2.73}}}
\newcommand{\psimed}[1]{\IfEqCase{#1}{{GW190930A}{2.02}{GW190929A}{1.62}{GW190924A}{2.00}{GW190915A}{2.06}{GW190910A}{3.19}{GW190909A}{1.60}{GW190828B}{1.45}{GW190828A}{2.21}{GW190814A}{0.39}{GW190803A}{3.18}{GW190731A}{3.13}{GW190728A}{2.16}{GW190727A}{3.16}{GW190720A}{2.09}{GW190719A}{2.05}{GW190708A}{3.14}{GW190707A}{2.05}{GW190706A}{2.19}{GW190701A}{2.03}{GW190630A}{1.81}{GW190620A}{1.82}{GW190602A}{3.13}{GW190527A}{3.10}{GW190521B}{1.73}{GW190521A}{1.59}{GW190519A}{3.30}{GW190517A}{2.37}{GW190514A}{3.18}{GW190513A}{2.26}{GW190512A}{3.14}{GW190503A}{3.12}{GW190426A}{1.58}{GW190425A}{1.62}{GW190424A}{3.15}{GW190421A}{3.11}{GW190413B}{3.10}{GW190413A}{3.15}{GW190412A}{2.56}{GW190408A}{3.14}}}
\newcommand{\psiplus}[1]{\IfEqCase{#1}{{GW190930A}{3.53}{GW190929A}{1.36}{GW190924A}{3.51}{GW190915A}{3.52}{GW190910A}{2.39}{GW190909A}{1.40}{GW190828B}{1.52}{GW190828A}{3.56}{GW190814A}{2.62}{GW190803A}{2.75}{GW190731A}{2.84}{GW190728A}{3.57}{GW190727A}{2.83}{GW190720A}{3.47}{GW190719A}{3.54}{GW190708A}{2.87}{GW190707A}{3.56}{GW190706A}{3.34}{GW190701A}{3.64}{GW190630A}{3.38}{GW190620A}{3.56}{GW190602A}{2.94}{GW190527A}{2.81}{GW190521B}{3.41}{GW190521A}{1.38}{GW190519A}{2.15}{GW190517A}{3.46}{GW190514A}{2.79}{GW190513A}{3.41}{GW190512A}{2.71}{GW190503A}{2.59}{GW190426A}{1.36}{GW190425A}{1.38}{GW190424A}{2.77}{GW190421A}{2.85}{GW190413B}{2.87}{GW190413A}{2.74}{GW190412A}{0.44}{GW190408A}{2.70}}}
\newcommand{\totalmassdetminus}[1]{\IfEqCase{#1}{{GW190930A}{1.0}{GW190929A}{26.3}{GW190924A}{0.7}{GW190915A}{8.1}{GW190910A}{7.8}{GW190909A}{22.9}{GW190828B}{4.0}{GW190828A}{5.9}{GW190814A}{1.0}{GW190803A}{11.9}{GW190731A}{14.3}{GW190728A}{0.7}{GW190727A}{10.9}{GW190720A}{1.2}{GW190719A}{15.5}{GW190708A}{0.8}{GW190707A}{0.5}{GW190706A}{27.7}{GW190701A}{14.8}{GW190630A}{3.5}{GW190620A}{18.4}{GW190602A}{20.6}{GW190527A}{10.3}{GW190521B}{5.4}{GW190521A}{34.6}{GW190519A}{17.9}{GW190517A}{7.3}{GW190514A}{15.1}{GW190513A}{6.7}{GW190512A}{2.8}{GW190503A}{11.8}{GW190426A}{1.6}{GW190425A}{0.08}{GW190424A}{10.9}{GW190421A}{12.4}{GW190413B}{18.0}{GW190413A}{15.3}{GW190412A}{4.6}{GW190408A}{3.8}}}
\newcommand{\totalmassdetmed}[1]{\IfEqCase{#1}{{GW190930A}{23.2}{GW190929A}{148.8}{GW190924A}{15.5}{GW190915A}{78.3}{GW190910A}{101.9}{GW190909A}{119.7}{GW190828B}{44.4}{GW190828A}{79.9}{GW190814A}{27.1}{GW190803A}{100.3}{GW190731A}{109.7}{GW190728A}{23.9}{GW190727A}{104.4}{GW190720A}{24.9}{GW190719A}{94.9}{GW190708A}{36.1}{GW190707A}{23.1}{GW190706A}{180.3}{GW190701A}{129.7}{GW190630A}{69.6}{GW190620A}{137.6}{GW190602A}{171.8}{GW190527A}{84.1}{GW190521B}{92.6}{GW190521A}{269.4}{GW190519A}{155.1}{GW190517A}{85.4}{GW190514A}{112.9}{GW190513A}{73.6}{GW190512A}{45.3}{GW190503A}{91.6}{GW190426A}{7.8}{GW190425A}{3.50}{GW190424A}{101.1}{GW190421A}{108.7}{GW190413B}{135.4}{GW190413A}{93.7}{GW190412A}{44.2}{GW190408A}{55.5}}}
\newcommand{\totalmassdetplus}[1]{\IfEqCase{#1}{{GW190930A}{10.5}{GW190929A}{38.6}{GW190924A}{5.7}{GW190915A}{8.4}{GW190910A}{10.4}{GW190909A}{95.3}{GW190828B}{6.4}{GW190828A}{6.9}{GW190814A}{1.1}{GW190803A}{14.1}{GW190731A}{14.3}{GW190728A}{5.3}{GW190727A}{11.9}{GW190720A}{5.0}{GW190719A}{24.4}{GW190708A}{2.5}{GW190707A}{1.8}{GW190706A}{23.3}{GW190701A}{16.4}{GW190630A}{4.2}{GW190620A}{20.1}{GW190602A}{23.2}{GW190527A}{53.7}{GW190521B}{4.8}{GW190521A}{39.8}{GW190519A}{16.7}{GW190517A}{9.6}{GW190514A}{17.8}{GW190513A}{12.7}{GW190512A}{3.9}{GW190503A}{11.2}{GW190426A}{3.7}{GW190425A}{0.3}{GW190424A}{14.4}{GW190421A}{15.3}{GW190413B}{17.9}{GW190413A}{17.8}{GW190412A}{4.5}{GW190408A}{3.5}}}
\newcommand{\thetajnminus}[1]{\IfEqCase{#1}{{GW190930A}{0.74}{GW190929A}{1.20}{GW190924A}{0.57}{GW190915A}{1.52}{GW190910A}{1.19}{GW190909A}{1.12}{GW190828B}{1.51}{GW190828A}{1.92}{GW190814A}{0.24}{GW190803A}{0.87}{GW190731A}{0.93}{GW190728A}{1.02}{GW190727A}{1.28}{GW190720A}{2.01}{GW190719A}{1.34}{GW190708A}{1.15}{GW190707A}{1.87}{GW190706A}{1.06}{GW190701A}{0.42}{GW190630A}{0.97}{GW190620A}{1.52}{GW190602A}{1.43}{GW190527A}{0.99}{GW190521B}{1.20}{GW190521A}{0.87}{GW190519A}{0.94}{GW190517A}{1.52}{GW190514A}{1.23}{GW190513A}{0.58}{GW190512A}{1.30}{GW190503A}{0.57}{GW190426A}{1.43}{GW190425A}{0.85}{GW190424A}{1.26}{GW190421A}{1.44}{GW190413B}{1.55}{GW190413A}{0.87}{GW190412A}{0.25}{GW190408A}{0.59}}}
\newcommand{\thetajnmed}[1]{\IfEqCase{#1}{{GW190930A}{0.94}{GW190929A}{1.70}{GW190924A}{0.74}{GW190915A}{2.03}{GW190910A}{1.62}{GW190909A}{1.45}{GW190828B}{1.83}{GW190828A}{2.24}{GW190814A}{0.86}{GW190803A}{1.12}{GW190731A}{1.19}{GW190728A}{1.23}{GW190727A}{1.55}{GW190720A}{2.48}{GW190719A}{1.61}{GW190708A}{1.37}{GW190707A}{2.15}{GW190706A}{1.37}{GW190701A}{0.58}{GW190630A}{1.22}{GW190620A}{1.94}{GW190602A}{1.71}{GW190527A}{1.26}{GW190521B}{1.48}{GW190521A}{1.15}{GW190519A}{1.58}{GW190517A}{2.26}{GW190514A}{1.50}{GW190513A}{0.79}{GW190512A}{1.61}{GW190503A}{2.41}{GW190426A}{1.70}{GW190425A}{1.08}{GW190424A}{1.52}{GW190421A}{1.74}{GW190413B}{1.90}{GW190413A}{1.15}{GW190412A}{0.72}{GW190408A}{0.79}}}
\newcommand{\thetajnplus}[1]{\IfEqCase{#1}{{GW190930A}{1.90}{GW190929A}{0.97}{GW190924A}{2.04}{GW190915A}{0.78}{GW190910A}{1.13}{GW190909A}{1.34}{GW190828B}{1.02}{GW190828A}{0.70}{GW190814A}{1.48}{GW190803A}{1.73}{GW190731A}{1.66}{GW190728A}{1.66}{GW190727A}{1.31}{GW190720A}{0.50}{GW190719A}{1.25}{GW190708A}{1.54}{GW190707A}{0.77}{GW190706A}{1.43}{GW190701A}{0.55}{GW190630A}{1.60}{GW190620A}{0.90}{GW190602A}{1.17}{GW190527A}{1.55}{GW190521B}{1.37}{GW190521A}{1.65}{GW190519A}{0.95}{GW190517A}{0.64}{GW190514A}{1.33}{GW190513A}{2.02}{GW190512A}{1.22}{GW190503A}{0.52}{GW190426A}{1.19}{GW190425A}{1.77}{GW190424A}{1.36}{GW190421A}{1.13}{GW190413B}{0.95}{GW190413A}{1.64}{GW190412A}{0.44}{GW190408A}{2.03}}}
\newcommand{\redshiftminus}[1]{\IfEqCase{#1}{{GW190930A}{0.06}{GW190929A}{0.17}{GW190924A}{0.04}{GW190915A}{0.10}{GW190910A}{0.10}{GW190909A}{0.33}{GW190828B}{0.10}{GW190828A}{0.15}{GW190814A}{0.010}{GW190803A}{0.24}{GW190731A}{0.26}{GW190728A}{0.07}{GW190727A}{0.22}{GW190720A}{0.06}{GW190719A}{0.29}{GW190708A}{0.07}{GW190707A}{0.07}{GW190706A}{0.27}{GW190701A}{0.12}{GW190630A}{0.07}{GW190620A}{0.20}{GW190602A}{0.17}{GW190527A}{0.20}{GW190521B}{0.10}{GW190521A}{0.28}{GW190519A}{0.14}{GW190517A}{0.14}{GW190514A}{0.31}{GW190513A}{0.13}{GW190512A}{0.10}{GW190503A}{0.11}{GW190426A}{0.03}{GW190425A}{0.02}{GW190424A}{0.19}{GW190421A}{0.21}{GW190413B}{0.30}{GW190413A}{0.24}{GW190412A}{0.03}{GW190408A}{0.10}}}
\newcommand{\redshiftmed}[1]{\IfEqCase{#1}{{GW190930A}{0.15}{GW190929A}{0.38}{GW190924A}{0.12}{GW190915A}{0.30}{GW190910A}{0.28}{GW190909A}{0.62}{GW190828B}{0.30}{GW190828A}{0.38}{GW190814A}{0.05}{GW190803A}{0.55}{GW190731A}{0.55}{GW190728A}{0.18}{GW190727A}{0.55}{GW190720A}{0.16}{GW190719A}{0.64}{GW190708A}{0.18}{GW190707A}{0.16}{GW190706A}{0.71}{GW190701A}{0.37}{GW190630A}{0.18}{GW190620A}{0.49}{GW190602A}{0.47}{GW190527A}{0.44}{GW190521B}{0.24}{GW190521A}{0.64}{GW190519A}{0.44}{GW190517A}{0.34}{GW190514A}{0.67}{GW190513A}{0.37}{GW190512A}{0.27}{GW190503A}{0.27}{GW190426A}{0.08}{GW190425A}{0.03}{GW190424A}{0.39}{GW190421A}{0.49}{GW190413B}{0.71}{GW190413A}{0.59}{GW190412A}{0.15}{GW190408A}{0.29}}}
\newcommand{\redshiftplus}[1]{\IfEqCase{#1}{{GW190930A}{0.06}{GW190929A}{0.49}{GW190924A}{0.04}{GW190915A}{0.11}{GW190910A}{0.16}{GW190909A}{0.41}{GW190828B}{0.10}{GW190828A}{0.10}{GW190814A}{0.009}{GW190803A}{0.26}{GW190731A}{0.31}{GW190728A}{0.05}{GW190727A}{0.21}{GW190720A}{0.12}{GW190719A}{0.33}{GW190708A}{0.06}{GW190707A}{0.07}{GW190706A}{0.32}{GW190701A}{0.11}{GW190630A}{0.10}{GW190620A}{0.23}{GW190602A}{0.25}{GW190527A}{0.34}{GW190521B}{0.07}{GW190521A}{0.28}{GW190519A}{0.25}{GW190517A}{0.24}{GW190514A}{0.33}{GW190513A}{0.13}{GW190512A}{0.09}{GW190503A}{0.11}{GW190426A}{0.04}{GW190425A}{0.01}{GW190424A}{0.23}{GW190421A}{0.19}{GW190413B}{0.31}{GW190413A}{0.29}{GW190412A}{0.03}{GW190408A}{0.06}}}
\newcommand{\iotaminus}[1]{\IfEqCase{#1}{{GW190930A}{0.73}{GW190929A}{1.24}{GW190924A}{0.56}{GW190915A}{1.73}{GW190910A}{1.19}{GW190909A}{1.11}{GW190828B}{1.55}{GW190828A}{1.97}{GW190814A}{0.27}{GW190803A}{0.85}{GW190731A}{0.90}{GW190728A}{1.01}{GW190727A}{1.26}{GW190720A}{2.01}{GW190719A}{1.34}{GW190708A}{1.14}{GW190707A}{1.88}{GW190706A}{1.04}{GW190701A}{0.43}{GW190630A}{0.98}{GW190620A}{1.69}{GW190602A}{1.53}{GW190527A}{0.97}{GW190521B}{1.19}{GW190521A}{0.87}{GW190519A}{0.93}{GW190517A}{1.40}{GW190514A}{1.22}{GW190513A}{0.58}{GW190512A}{1.29}{GW190503A}{0.57}{GW190426A}{1.43}{GW190425A}{0.85}{GW190424A}{1.26}{GW190421A}{1.48}{GW190413B}{1.61}{GW190413A}{0.86}{GW190412A}{0.35}{GW190408A}{0.59}}}
\newcommand{\iotamed}[1]{\IfEqCase{#1}{{GW190930A}{0.94}{GW190929A}{1.70}{GW190924A}{0.74}{GW190915A}{2.13}{GW190910A}{1.62}{GW190909A}{1.47}{GW190828B}{1.84}{GW190828A}{2.27}{GW190814A}{0.85}{GW190803A}{1.09}{GW190731A}{1.15}{GW190728A}{1.23}{GW190727A}{1.53}{GW190720A}{2.47}{GW190719A}{1.61}{GW190708A}{1.37}{GW190707A}{2.15}{GW190706A}{1.33}{GW190701A}{0.59}{GW190630A}{1.24}{GW190620A}{2.03}{GW190602A}{1.79}{GW190527A}{1.24}{GW190521B}{1.46}{GW190521A}{1.15}{GW190519A}{1.59}{GW190517A}{2.21}{GW190514A}{1.48}{GW190513A}{0.79}{GW190512A}{1.61}{GW190503A}{2.43}{GW190426A}{1.70}{GW190425A}{1.09}{GW190424A}{1.52}{GW190421A}{1.76}{GW190413B}{1.97}{GW190413A}{1.13}{GW190412A}{0.84}{GW190408A}{0.78}}}
\newcommand{\iotaplus}[1]{\IfEqCase{#1}{{GW190930A}{1.89}{GW190929A}{1.03}{GW190924A}{2.04}{GW190915A}{0.74}{GW190910A}{1.14}{GW190909A}{1.30}{GW190828B}{1.01}{GW190828A}{0.67}{GW190814A}{1.51}{GW190803A}{1.76}{GW190731A}{1.71}{GW190728A}{1.66}{GW190727A}{1.34}{GW190720A}{0.49}{GW190719A}{1.27}{GW190708A}{1.55}{GW190707A}{0.78}{GW190706A}{1.49}{GW190701A}{0.57}{GW190630A}{1.55}{GW190620A}{0.85}{GW190602A}{1.11}{GW190527A}{1.57}{GW190521B}{1.40}{GW190521A}{1.65}{GW190519A}{0.92}{GW190517A}{0.66}{GW190514A}{1.37}{GW190513A}{2.03}{GW190512A}{1.22}{GW190503A}{0.51}{GW190426A}{1.19}{GW190425A}{1.77}{GW190424A}{1.38}{GW190421A}{1.11}{GW190413B}{0.88}{GW190413A}{1.66}{GW190412A}{0.38}{GW190408A}{2.06}}}
\newcommand{\spinonexminus}[1]{\IfEqCase{#1}{{GW190930A}{0.44}{GW190929A}{0.71}{GW190924A}{0.35}{GW190915A}{0.67}{GW190910A}{0.51}{GW190909A}{0.63}{GW190828B}{0.42}{GW190828A}{0.52}{GW190814A}{0.04}{GW190803A}{0.57}{GW190731A}{0.56}{GW190728A}{0.37}{GW190727A}{0.58}{GW190720A}{0.43}{GW190719A}{0.55}{GW190708A}{0.42}{GW190707A}{0.39}{GW190706A}{0.53}{GW190701A}{0.52}{GW190630A}{0.36}{GW190620A}{0.55}{GW190602A}{0.52}{GW190527A}{0.59}{GW190521B}{0.44}{GW190521A}{0.74}{GW190519A}{0.54}{GW190517A}{0.57}{GW190514A}{0.57}{GW190513A}{0.42}{GW190512A}{0.30}{GW190503A}{0.49}{GW190426A}{0.00}{GW190425A}{0.50}{GW190424A}{0.64}{GW190421A}{0.59}{GW190413B}{0.68}{GW190413A}{0.53}{GW190412A}{0.33}{GW190408A}{0.47}}}
\newcommand{\spinonexmed}[1]{\IfEqCase{#1}{{GW190930A}{0.002}{GW190929A}{0.007}{GW190924A}{0.0001}{GW190915A}{0.00}{GW190910A}{0.00}{GW190909A}{0.002}{GW190828B}{0.00}{GW190828A}{0.00}{GW190814A}{0.00}{GW190803A}{0.00}{GW190731A}{0.0007}{GW190728A}{0.0008}{GW190727A}{0.002}{GW190720A}{0.003}{GW190719A}{0.004}{GW190708A}{0.004}{GW190707A}{0.003}{GW190706A}{0.00}{GW190701A}{0.00}{GW190630A}{0.00}{GW190620A}{0.00}{GW190602A}{0.00}{GW190527A}{0.002}{GW190521B}{0.001}{GW190521A}{-0.02}{GW190519A}{0.004}{GW190517A}{0.0009}{GW190514A}{0.00007}{GW190513A}{0.0006}{GW190512A}{0.0010}{GW190503A}{0.00}{GW190426A}{0.00}{GW190425A}{0.00}{GW190424A}{0.00}{GW190421A}{0.00005}{GW190413B}{0.00}{GW190413A}{0.0005}{GW190412A}{-0.02}{GW190408A}{0.003}}}
\newcommand{\spinonexplus}[1]{\IfEqCase{#1}{{GW190930A}{0.47}{GW190929A}{0.69}{GW190924A}{0.36}{GW190915A}{0.66}{GW190910A}{0.51}{GW190909A}{0.68}{GW190828B}{0.43}{GW190828A}{0.51}{GW190814A}{0.04}{GW190803A}{0.55}{GW190731A}{0.51}{GW190728A}{0.40}{GW190727A}{0.58}{GW190720A}{0.45}{GW190719A}{0.57}{GW190708A}{0.47}{GW190707A}{0.42}{GW190706A}{0.53}{GW190701A}{0.52}{GW190630A}{0.36}{GW190620A}{0.53}{GW190602A}{0.53}{GW190527A}{0.59}{GW190521B}{0.45}{GW190521A}{0.75}{GW190519A}{0.55}{GW190517A}{0.57}{GW190514A}{0.60}{GW190513A}{0.43}{GW190512A}{0.28}{GW190503A}{0.50}{GW190426A}{0.00}{GW190425A}{0.47}{GW190424A}{0.63}{GW190421A}{0.60}{GW190413B}{0.71}{GW190413A}{0.53}{GW190412A}{0.39}{GW190408A}{0.49}}}
\newcommand{\chirpmassdetminus}[1]{\IfEqCase{#1}{{GW190930A}{0.2}{GW190929A}{15.4}{GW190924A}{0.03}{GW190915A}{3.9}{GW190910A}{3.6}{GW190909A}{12.4}{GW190828B}{0.7}{GW190828A}{2.8}{GW190814A}{0.02}{GW190803A}{6.1}{GW190731A}{8.2}{GW190728A}{0.08}{GW190727A}{5.7}{GW190720A}{0.1}{GW190719A}{6.6}{GW190708A}{0.2}{GW190707A}{0.09}{GW190706A}{17.5}{GW190701A}{8.1}{GW190630A}{1.5}{GW190620A}{11.2}{GW190602A}{13.7}{GW190527A}{5.5}{GW190521B}{3.0}{GW190521A}{17.6}{GW190519A}{10.3}{GW190517A}{3.4}{GW190514A}{7.7}{GW190513A}{2.5}{GW190512A}{0.8}{GW190503A}{6.0}{GW190426A}{0.01}{GW190425A}{0.0006}{GW190424A}{4.8}{GW190421A}{6.0}{GW190413B}{9.8}{GW190413A}{6.6}{GW190412A}{0.2}{GW190408A}{1.7}}}
\newcommand{\chirpmassdetmed}[1]{\IfEqCase{#1}{{GW190930A}{9.8}{GW190929A}{52.2}{GW190924A}{6.44}{GW190915A}{33.1}{GW190910A}{43.9}{GW190909A}{49.8}{GW190828B}{17.4}{GW190828A}{34.5}{GW190814A}{6.41}{GW190803A}{42.7}{GW190731A}{46.6}{GW190728A}{10.1}{GW190727A}{44.7}{GW190720A}{10.4}{GW190719A}{38.7}{GW190708A}{15.5}{GW190707A}{9.89}{GW190706A}{75.1}{GW190701A}{55.5}{GW190630A}{29.4}{GW190620A}{57.5}{GW190602A}{72.9}{GW190527A}{34.9}{GW190521B}{39.8}{GW190521A}{114.8}{GW190519A}{65.1}{GW190517A}{35.9}{GW190514A}{48.1}{GW190513A}{29.5}{GW190512A}{18.6}{GW190503A}{38.6}{GW190426A}{2.60}{GW190425A}{1.49}{GW190424A}{43.4}{GW190421A}{46.6}{GW190413B}{57.0}{GW190413A}{39.4}{GW190412A}{15.2}{GW190408A}{23.7}}}
\newcommand{\chirpmassdetplus}[1]{\IfEqCase{#1}{{GW190930A}{0.2}{GW190929A}{19.9}{GW190924A}{0.04}{GW190915A}{3.3}{GW190910A}{4.6}{GW190909A}{32.2}{GW190828B}{0.6}{GW190828A}{2.9}{GW190814A}{0.02}{GW190803A}{6.3}{GW190731A}{6.8}{GW190728A}{0.09}{GW190727A}{5.3}{GW190720A}{0.2}{GW190719A}{9.2}{GW190708A}{0.3}{GW190707A}{0.1}{GW190706A}{11.0}{GW190701A}{7.3}{GW190630A}{1.6}{GW190620A}{9.0}{GW190602A}{10.8}{GW190527A}{21.7}{GW190521B}{2.2}{GW190521A}{15.2}{GW190519A}{7.7}{GW190517A}{4.0}{GW190514A}{7.5}{GW190513A}{5.6}{GW190512A}{0.9}{GW190503A}{5.3}{GW190426A}{0.01}{GW190425A}{0.0008}{GW190424A}{6.0}{GW190421A}{6.6}{GW190413B}{8.6}{GW190413A}{7.7}{GW190412A}{0.2}{GW190408A}{1.4}}}
\newcommand{\cosiotaminus}[1]{\IfEqCase{#1}{{GW190930A}{1.54}{GW190929A}{0.79}{GW190924A}{1.67}{GW190915A}{0.43}{GW190910A}{0.88}{GW190909A}{1.03}{GW190828B}{0.69}{GW190828A}{0.33}{GW190814A}{1.37}{GW190803A}{1.42}{GW190731A}{1.37}{GW190728A}{1.30}{GW190727A}{1.00}{GW190720A}{0.20}{GW190719A}{0.93}{GW190708A}{1.18}{GW190707A}{0.43}{GW190706A}{1.19}{GW190701A}{0.44}{GW190630A}{1.26}{GW190620A}{0.52}{GW190602A}{0.75}{GW190527A}{1.27}{GW190521B}{1.07}{GW190521A}{1.35}{GW190519A}{0.79}{GW190517A}{0.37}{GW190514A}{1.05}{GW190513A}{1.65}{GW190512A}{0.91}{GW190503A}{0.22}{GW190426A}{0.84}{GW190425A}{1.42}{GW190424A}{1.02}{GW190421A}{0.78}{GW190413B}{0.57}{GW190413A}{1.37}{GW190412A}{0.32}{GW190408A}{1.67}}}
\newcommand{\cosiotamed}[1]{\IfEqCase{#1}{{GW190930A}{0.59}{GW190929A}{-0.13}{GW190924A}{0.74}{GW190915A}{-0.53}{GW190910A}{-0.05}{GW190909A}{0.10}{GW190828B}{-0.27}{GW190828A}{-0.65}{GW190814A}{0.66}{GW190803A}{0.47}{GW190731A}{0.41}{GW190728A}{0.34}{GW190727A}{0.04}{GW190720A}{-0.78}{GW190719A}{-0.04}{GW190708A}{0.20}{GW190707A}{-0.55}{GW190706A}{0.24}{GW190701A}{0.83}{GW190630A}{0.32}{GW190620A}{-0.45}{GW190602A}{-0.22}{GW190527A}{0.32}{GW190521B}{0.11}{GW190521A}{0.41}{GW190519A}{-0.02}{GW190517A}{-0.59}{GW190514A}{0.09}{GW190513A}{0.71}{GW190512A}{-0.04}{GW190503A}{-0.76}{GW190426A}{-0.13}{GW190425A}{0.46}{GW190424A}{0.05}{GW190421A}{-0.19}{GW190413B}{-0.39}{GW190413A}{0.42}{GW190412A}{0.67}{GW190408A}{0.71}}}
\newcommand{\cosiotaplus}[1]{\IfEqCase{#1}{{GW190930A}{0.39}{GW190929A}{1.03}{GW190924A}{0.24}{GW190915A}{1.45}{GW190910A}{0.96}{GW190909A}{0.84}{GW190828B}{1.23}{GW190828A}{1.60}{GW190814A}{0.17}{GW190803A}{0.51}{GW190731A}{0.56}{GW190728A}{0.64}{GW190727A}{0.93}{GW190720A}{1.68}{GW190719A}{1.00}{GW190708A}{0.77}{GW190707A}{1.51}{GW190706A}{0.72}{GW190701A}{0.16}{GW190630A}{0.65}{GW190620A}{1.39}{GW190602A}{1.18}{GW190527A}{0.64}{GW190521B}{0.85}{GW190521A}{0.55}{GW190519A}{0.81}{GW190517A}{1.29}{GW190514A}{0.88}{GW190513A}{0.27}{GW190512A}{0.99}{GW190503A}{0.47}{GW190426A}{1.09}{GW190425A}{0.51}{GW190424A}{0.91}{GW190421A}{1.15}{GW190413B}{1.32}{GW190413A}{0.54}{GW190412A}{0.22}{GW190408A}{0.27}}}
\newcommand{\comovingdistminus}[1]{\IfEqCase{#1}{{GW190930A}{257}{GW190929A}{650}{GW190924A}{184}{GW190915A}{401}{GW190910A}{396}{GW190909A}{1128}{GW190828B}{395}{GW190828A}{567}{GW190814A}{41}{GW190803A}{824}{GW190731A}{906}{GW190728A}{285}{GW190727A}{774}{GW190720A}{254}{GW190719A}{965}{GW190708A}{307}{GW190707A}{295}{GW190706A}{859}{GW190701A}{438}{GW190630A}{289}{GW190620A}{723}{GW190602A}{621}{GW190527A}{725}{GW190521B}{409}{GW190521A}{943}{GW190519A}{518}{GW190517A}{538}{GW190514A}{1034}{GW190513A}{484}{GW190512A}{378}{GW190503A}{431}{GW190426A}{143}{GW190425A}{67}{GW190424A}{717}{GW190421A}{759}{GW190413B}{957}{GW190413A}{829}{GW190412A}{134}{GW190408A}{399}}}
\newcommand{\comovingdistmed}[1]{\IfEqCase{#1}{{GW190930A}{658}{GW190929A}{1541}{GW190924A}{507}{GW190915A}{1244}{GW190910A}{1139}{GW190909A}{2329}{GW190828B}{1228}{GW190828A}{1539}{GW190814A}{229}{GW190803A}{2108}{GW190731A}{2120}{GW190728A}{742}{GW190727A}{2119}{GW190720A}{678}{GW190719A}{2399}{GW190708A}{748}{GW190707A}{667}{GW190706A}{2594}{GW190701A}{1498}{GW190630A}{752}{GW190620A}{1893}{GW190602A}{1832}{GW190527A}{1729}{GW190521B}{1003}{GW190521A}{2390}{GW190519A}{1754}{GW190517A}{1389}{GW190514A}{2475}{GW190513A}{1501}{GW190512A}{1120}{GW190503A}{1133}{GW190426A}{344}{GW190425A}{151}{GW190424A}{1578}{GW190421A}{1923}{GW190413B}{2603}{GW190413A}{2232}{GW190412A}{640}{GW190408A}{1198}}}
\newcommand{\comovingdistplus}[1]{\IfEqCase{#1}{{GW190930A}{258}{GW190929A}{1537}{GW190924A}{174}{GW190915A}{403}{GW190910A}{588}{GW190909A}{1139}{GW190828B}{359}{GW190828A}{342}{GW190814A}{37}{GW190803A}{778}{GW190731A}{925}{GW190728A}{182}{GW190727A}{627}{GW190720A}{472}{GW190719A}{915}{GW190708A}{235}{GW190707A}{272}{GW190706A}{867}{GW190701A}{400}{GW190630A}{379}{GW190620A}{725}{GW190602A}{784}{GW190527A}{1068}{GW190521B}{253}{GW190521A}{795}{GW190519A}{816}{GW190517A}{816}{GW190514A}{915}{GW190513A}{453}{GW190512A}{330}{GW190503A}{406}{GW190426A}{154}{GW190425A}{64}{GW190424A}{755}{GW190421A}{598}{GW190413B}{831}{GW190413A}{856}{GW190412A}{105}{GW190408A}{240}}}
\newcommand{\spintwoyminus}[1]{\IfEqCase{#1}{{GW190930A}{0.54}{GW190929A}{0.57}{GW190924A}{0.48}{GW190915A}{0.61}{GW190910A}{0.52}{GW190909A}{0.61}{GW190828B}{0.54}{GW190828A}{0.50}{GW190814A}{0.61}{GW190803A}{0.57}{GW190731A}{0.58}{GW190728A}{0.51}{GW190727A}{0.59}{GW190720A}{0.55}{GW190719A}{0.55}{GW190708A}{0.44}{GW190707A}{0.47}{GW190706A}{0.51}{GW190701A}{0.58}{GW190630A}{0.48}{GW190620A}{0.56}{GW190602A}{0.60}{GW190527A}{0.61}{GW190521B}{0.53}{GW190521A}{0.68}{GW190519A}{0.55}{GW190517A}{0.54}{GW190514A}{0.60}{GW190513A}{0.54}{GW190512A}{0.51}{GW190503A}{0.57}{GW190426A}{0.00}{GW190425A}{0.48}{GW190424A}{0.60}{GW190421A}{0.59}{GW190413B}{0.59}{GW190413A}{0.57}{GW190412A}{0.57}{GW190408A}{0.52}}}
\newcommand{\spintwoymed}[1]{\IfEqCase{#1}{{GW190930A}{0.00}{GW190929A}{0.0008}{GW190924A}{0.00}{GW190915A}{0.00}{GW190910A}{0.00}{GW190909A}{0.003}{GW190828B}{0.00}{GW190828A}{0.0004}{GW190814A}{0.005}{GW190803A}{0.0002}{GW190731A}{0.00}{GW190728A}{0.00}{GW190727A}{0.0009}{GW190720A}{0.00}{GW190719A}{0.003}{GW190708A}{0.003}{GW190707A}{0.002}{GW190706A}{0.001}{GW190701A}{-0.01}{GW190630A}{0.0006}{GW190620A}{0.00010}{GW190602A}{0.00}{GW190527A}{0.004}{GW190521B}{0.00}{GW190521A}{0.00}{GW190519A}{0.001}{GW190517A}{0.00}{GW190514A}{0.00}{GW190513A}{0.00}{GW190512A}{0.002}{GW190503A}{0.00}{GW190426A}{0.00}{GW190425A}{0.00}{GW190424A}{0.00}{GW190421A}{0.00}{GW190413B}{-0.01}{GW190413A}{0.00}{GW190412A}{0.004}{GW190408A}{0.0008}}}
\newcommand{\spintwoyplus}[1]{\IfEqCase{#1}{{GW190930A}{0.51}{GW190929A}{0.59}{GW190924A}{0.49}{GW190915A}{0.60}{GW190910A}{0.52}{GW190909A}{0.57}{GW190828B}{0.54}{GW190828A}{0.50}{GW190814A}{0.61}{GW190803A}{0.58}{GW190731A}{0.55}{GW190728A}{0.50}{GW190727A}{0.56}{GW190720A}{0.56}{GW190719A}{0.55}{GW190708A}{0.46}{GW190707A}{0.48}{GW190706A}{0.53}{GW190701A}{0.55}{GW190630A}{0.49}{GW190620A}{0.54}{GW190602A}{0.58}{GW190527A}{0.59}{GW190521B}{0.52}{GW190521A}{0.69}{GW190519A}{0.55}{GW190517A}{0.55}{GW190514A}{0.60}{GW190513A}{0.55}{GW190512A}{0.55}{GW190503A}{0.58}{GW190426A}{0.00}{GW190425A}{0.48}{GW190424A}{0.60}{GW190421A}{0.59}{GW190413B}{0.62}{GW190413A}{0.56}{GW190412A}{0.58}{GW190408A}{0.53}}}
\newcommand{\tiltoneminus}[1]{\IfEqCase{#1}{{GW190930A}{0.79}{GW190929A}{0.72}{GW190924A}{1.05}{GW190915A}{0.85}{GW190910A}{0.97}{GW190909A}{1.14}{GW190828B}{0.83}{GW190828A}{0.72}{GW190814A}{1.08}{GW190803A}{1.06}{GW190731A}{0.96}{GW190728A}{0.77}{GW190727A}{0.85}{GW190720A}{0.72}{GW190719A}{0.59}{GW190708A}{0.98}{GW190707A}{1.08}{GW190706A}{0.61}{GW190701A}{1.12}{GW190630A}{0.87}{GW190620A}{0.59}{GW190602A}{0.94}{GW190527A}{0.89}{GW190521B}{0.90}{GW190521A}{0.93}{GW190519A}{0.60}{GW190517A}{0.42}{GW190514A}{1.15}{GW190513A}{0.81}{GW190512A}{1.03}{GW190503A}{1.10}{GW190426A}{0.00}{GW190425A}{0.80}{GW190424A}{0.80}{GW190421A}{1.06}{GW190413B}{0.93}{GW190413A}{1.06}{GW190412A}{0.35}{GW190408A}{1.06}}}
\newcommand{\tiltonemed}[1]{\IfEqCase{#1}{{GW190930A}{1.08}{GW190929A}{1.55}{GW190924A}{1.38}{GW190915A}{1.51}{GW190910A}{1.49}{GW190909A}{1.67}{GW190828B}{1.31}{GW190828A}{1.04}{GW190814A}{1.56}{GW190803A}{1.64}{GW190731A}{1.41}{GW190728A}{1.06}{GW190727A}{1.26}{GW190720A}{1.00}{GW190719A}{0.85}{GW190708A}{1.50}{GW190707A}{1.77}{GW190706A}{0.85}{GW190701A}{1.79}{GW190630A}{1.28}{GW190620A}{0.82}{GW190602A}{1.39}{GW190527A}{1.29}{GW190521B}{1.40}{GW190521A}{1.52}{GW190519A}{0.87}{GW190517A}{0.59}{GW190514A}{2.03}{GW190513A}{1.14}{GW190512A}{1.50}{GW190503A}{1.73}{GW190426A}{0.00}{GW190425A}{1.31}{GW190424A}{1.20}{GW190421A}{1.72}{GW190413B}{1.63}{GW190413A}{1.56}{GW190412A}{0.80}{GW190408A}{1.74}}}
\newcommand{\tiltoneplus}[1]{\IfEqCase{#1}{{GW190930A}{1.14}{GW190929A}{0.97}{GW190924A}{1.09}{GW190915A}{0.92}{GW190910A}{1.09}{GW190909A}{0.99}{GW190828B}{1.08}{GW190828A}{1.15}{GW190814A}{1.11}{GW190803A}{0.99}{GW190731A}{1.13}{GW190728A}{1.20}{GW190727A}{1.11}{GW190720A}{1.01}{GW190719A}{1.09}{GW190708A}{1.00}{GW190707A}{0.86}{GW190706A}{1.06}{GW190701A}{0.95}{GW190630A}{1.11}{GW190620A}{0.91}{GW190602A}{1.12}{GW190527A}{1.11}{GW190521B}{1.09}{GW190521A}{1.04}{GW190519A}{0.79}{GW190517A}{0.50}{GW190514A}{0.78}{GW190513A}{1.18}{GW190512A}{1.11}{GW190503A}{0.97}{GW190426A}{3.14}{GW190425A}{0.66}{GW190424A}{1.06}{GW190421A}{0.97}{GW190413B}{0.91}{GW190413A}{1.11}{GW190412A}{0.54}{GW190408A}{0.94}}}
\newcommand{\spintwozminus}[1]{\IfEqCase{#1}{{GW190930A}{0.41}{GW190929A}{0.55}{GW190924A}{0.36}{GW190915A}{0.51}{GW190910A}{0.36}{GW190909A}{0.62}{GW190828B}{0.42}{GW190828A}{0.35}{GW190814A}{0.53}{GW190803A}{0.55}{GW190731A}{0.47}{GW190728A}{0.38}{GW190727A}{0.45}{GW190720A}{0.53}{GW190719A}{0.46}{GW190708A}{0.33}{GW190707A}{0.36}{GW190706A}{0.45}{GW190701A}{0.54}{GW190630A}{0.31}{GW190620A}{0.47}{GW190602A}{0.46}{GW190527A}{0.49}{GW190521B}{0.32}{GW190521A}{0.54}{GW190519A}{0.43}{GW190517A}{0.46}{GW190514A}{0.59}{GW190513A}{0.41}{GW190512A}{0.33}{GW190503A}{0.51}{GW190426A}{0.03}{GW190425A}{0.18}{GW190424A}{0.44}{GW190421A}{0.53}{GW190413B}{0.55}{GW190413A}{0.54}{GW190412A}{0.44}{GW190408A}{0.37}}}
\newcommand{\spintwozmed}[1]{\IfEqCase{#1}{{GW190930A}{0.08}{GW190929A}{0.008}{GW190924A}{0.02}{GW190915A}{0.003}{GW190910A}{0.006}{GW190909A}{-0.04}{GW190828B}{0.06}{GW190828A}{0.11}{GW190814A}{-0.01}{GW190803A}{-0.01}{GW190731A}{0.03}{GW190728A}{0.11}{GW190727A}{0.04}{GW190720A}{0.11}{GW190719A}{0.16}{GW190708A}{0.03}{GW190707A}{-0.03}{GW190706A}{0.11}{GW190701A}{-0.04}{GW190630A}{0.09}{GW190620A}{0.21}{GW190602A}{0.05}{GW190527A}{0.04}{GW190521B}{0.09}{GW190521A}{-0.01}{GW190519A}{0.21}{GW190517A}{0.29}{GW190514A}{-0.13}{GW190513A}{0.06}{GW190512A}{0.03}{GW190503A}{0.00}{GW190426A}{0.00}{GW190425A}{0.03}{GW190424A}{0.06}{GW190421A}{-0.04}{GW190413B}{-0.03}{GW190413A}{-0.02}{GW190412A}{0.07}{GW190408A}{0.00}}}
\newcommand{\spintwozplus}[1]{\IfEqCase{#1}{{GW190930A}{0.50}{GW190929A}{0.57}{GW190924A}{0.48}{GW190915A}{0.47}{GW190910A}{0.39}{GW190909A}{0.51}{GW190828B}{0.54}{GW190828A}{0.48}{GW190814A}{0.52}{GW190803A}{0.47}{GW190731A}{0.54}{GW190728A}{0.48}{GW190727A}{0.53}{GW190720A}{0.54}{GW190719A}{0.61}{GW190708A}{0.39}{GW190707A}{0.34}{GW190706A}{0.62}{GW190701A}{0.42}{GW190630A}{0.44}{GW190620A}{0.57}{GW190602A}{0.56}{GW190527A}{0.60}{GW190521B}{0.36}{GW190521A}{0.53}{GW190519A}{0.56}{GW190517A}{0.53}{GW190514A}{0.44}{GW190513A}{0.54}{GW190512A}{0.45}{GW190503A}{0.46}{GW190426A}{0.03}{GW190425A}{0.30}{GW190424A}{0.52}{GW190421A}{0.41}{GW190413B}{0.49}{GW190413A}{0.48}{GW190412A}{0.57}{GW190408A}{0.38}}}
\newcommand{\massonesourceminus}[1]{\IfEqCase{#1}{{GW190930A}{2.3}{GW190929A}{33.2}{GW190924A}{2.0}{GW190915A}{6.4}{GW190910A}{6.1}{GW190909A}{13.3}{GW190828B}{7.2}{GW190828A}{4.0}{GW190814A}{1.0}{GW190803A}{7.0}{GW190731A}{9.0}{GW190728A}{2.2}{GW190727A}{6.2}{GW190720A}{3.0}{GW190719A}{10.3}{GW190708A}{2.3}{GW190707A}{1.7}{GW190706A}{16.2}{GW190701A}{8.0}{GW190630A}{5.6}{GW190620A}{12.7}{GW190602A}{13.0}{GW190527A}{9.0}{GW190521B}{4.8}{GW190521A}{18.9}{GW190519A}{12.0}{GW190517A}{7.6}{GW190514A}{8.2}{GW190513A}{9.2}{GW190512A}{5.8}{GW190503A}{8.1}{GW190426A}{2.3}{GW190425A}{0.3}{GW190424A}{7.3}{GW190421A}{6.9}{GW190413B}{10.7}{GW190413A}{8.1}{GW190412A}{5.1}{GW190408A}{3.4}}}
\newcommand{\massonesourcemed}[1]{\IfEqCase{#1}{{GW190930A}{12.3}{GW190929A}{80.8}{GW190924A}{8.9}{GW190915A}{35.3}{GW190910A}{43.9}{GW190909A}{45.8}{GW190828B}{24.1}{GW190828A}{32.1}{GW190814A}{23.2}{GW190803A}{37.3}{GW190731A}{41.5}{GW190728A}{12.3}{GW190727A}{38.0}{GW190720A}{13.4}{GW190719A}{36.5}{GW190708A}{17.6}{GW190707A}{11.6}{GW190706A}{67.0}{GW190701A}{53.9}{GW190630A}{35.1}{GW190620A}{57.1}{GW190602A}{69.1}{GW190527A}{36.5}{GW190521B}{42.2}{GW190521A}{95.3}{GW190519A}{66.0}{GW190517A}{37.4}{GW190514A}{39.0}{GW190513A}{35.7}{GW190512A}{23.3}{GW190503A}{43.3}{GW190426A}{5.7}{GW190425A}{2.0}{GW190424A}{40.5}{GW190421A}{41.3}{GW190413B}{47.5}{GW190413A}{34.7}{GW190412A}{30.1}{GW190408A}{24.6}}}
\newcommand{\massonesourceplus}[1]{\IfEqCase{#1}{{GW190930A}{12.4}{GW190929A}{33.0}{GW190924A}{7.0}{GW190915A}{9.5}{GW190910A}{7.6}{GW190909A}{52.7}{GW190828B}{7.0}{GW190828A}{5.8}{GW190814A}{1.1}{GW190803A}{10.6}{GW190731A}{12.2}{GW190728A}{7.2}{GW190727A}{9.5}{GW190720A}{6.7}{GW190719A}{18.0}{GW190708A}{4.7}{GW190707A}{3.3}{GW190706A}{14.6}{GW190701A}{11.8}{GW190630A}{6.9}{GW190620A}{16.0}{GW190602A}{15.7}{GW190527A}{16.4}{GW190521B}{5.9}{GW190521A}{28.7}{GW190519A}{10.7}{GW190517A}{11.7}{GW190514A}{14.7}{GW190513A}{9.5}{GW190512A}{5.3}{GW190503A}{9.2}{GW190426A}{3.9}{GW190425A}{0.6}{GW190424A}{11.1}{GW190421A}{10.4}{GW190413B}{13.5}{GW190413A}{12.6}{GW190412A}{4.7}{GW190408A}{5.1}}}
\newcommand{\geocenttimeminus}[1]{\IfEqCase{#1}{{GW190930A}{0.02}{GW190929A}{0.02}{GW190924A}{0.008}{GW190915A}{0.003}{GW190910A}{0.0}{GW190909A}{0.0}{GW190828B}{0.0}{GW190828A}{0.0}{GW190814A}{0.0009}{GW190803A}{0.0}{GW190731A}{0.0}{GW190728A}{0.03}{GW190727A}{0.0}{GW190720A}{0.01}{GW190719A}{0.0}{GW190708A}{0.0}{GW190707A}{0.03}{GW190706A}{0.0000002}{GW190701A}{0.0}{GW190630A}{0.0000002}{GW190620A}{0.0000002}{GW190602A}{0.0}{GW190527A}{0.0}{GW190521B}{0.0}{GW190521A}{0.04}{GW190519A}{0.0}{GW190517A}{0.0}{GW190514A}{0.0}{GW190513A}{0.010}{GW190512A}{0.0}{GW190503A}{0.0}{GW190426A}{0.03}{GW190425A}{0.009}{GW190424A}{0.0}{GW190421A}{0.0}{GW190413B}{0.0}{GW190413A}{0.0}{GW190412A}{0.001}{GW190408A}{0.0}}}
\newcommand{\geocenttimemed}[1]{\IfEqCase{#1}{{GW190930A}{1253885759.2}{GW190929A}{1253755327.5}{GW190924A}{1253326744.8}{GW190915A}{1252627040.7}{GW190910A}{1252150105.3}{GW190909A}{1252064527.7}{GW190828B}{1251010527.9}{GW190828A}{1251009263.8}{GW190814A}{1249852257.0}{GW190803A}{1248834439.9}{GW190731A}{1248617394.6}{GW190728A}{1248331528.6}{GW190727A}{1248242632.0}{GW190720A}{1247616534.7}{GW190719A}{1247608532.9}{GW190708A}{1246663515.4}{GW190707A}{1246527224.2}{GW190706A}{1246487219.3}{GW190701A}{1246048404.6}{GW190630A}{1245955943.2}{GW190620A}{1245035079.3}{GW190602A}{1243533585.1}{GW190527A}{1242984073.8}{GW190521B}{1242459857.5}{GW190521A}{1242442967.4}{GW190519A}{1242315362.4}{GW190517A}{1242107479.8}{GW190514A}{1241852074.8}{GW190513A}{1241816086.8}{GW190512A}{1241719652.4}{GW190503A}{1240944862.3}{GW190426A}{1240327333.4}{GW190425A}{1240215503.0}{GW190424A}{1240164426.1}{GW190421A}{1239917954.2}{GW190413B}{1239198206.7}{GW190413A}{1239168612.5}{GW190412A}{1239082262.2}{GW190408A}{1238782700.3}}}
\newcommand{\geocenttimeplus}[1]{\IfEqCase{#1}{{GW190930A}{0.002}{GW190929A}{0.03}{GW190924A}{0.008}{GW190915A}{0.002}{GW190910A}{0.0}{GW190909A}{0.0}{GW190828B}{0.0}{GW190828A}{0.0000005}{GW190814A}{0.004}{GW190803A}{0.0}{GW190731A}{0.0}{GW190728A}{0.0010}{GW190727A}{0.0}{GW190720A}{0.01}{GW190719A}{0.05}{GW190708A}{0.0}{GW190707A}{0.009}{GW190706A}{0.0}{GW190701A}{0.0000005}{GW190630A}{0.0}{GW190620A}{0.0}{GW190602A}{0.0}{GW190527A}{0.0}{GW190521B}{0.0}{GW190521A}{0.01}{GW190519A}{0.0}{GW190517A}{0.0}{GW190514A}{0.0}{GW190513A}{0.0}{GW190512A}{0.0}{GW190503A}{0.0}{GW190426A}{0.02}{GW190425A}{0.03}{GW190424A}{0.0}{GW190421A}{0.0}{GW190413B}{0.0}{GW190413A}{0.0}{GW190412A}{0.007}{GW190408A}{0.0}}}
\newcommand{\costilttwominus}[1]{\IfEqCase{#1}{{GW190930A}{1.09}{GW190929A}{0.93}{GW190924A}{1.00}{GW190915A}{0.90}{GW190910A}{0.88}{GW190909A}{0.75}{GW190828B}{1.06}{GW190828A}{1.14}{GW190814A}{0.83}{GW190803A}{0.83}{GW190731A}{0.99}{GW190728A}{1.16}{GW190727A}{1.04}{GW190720A}{1.11}{GW190719A}{1.19}{GW190708A}{0.97}{GW190707A}{0.71}{GW190706A}{1.16}{GW190701A}{0.74}{GW190630A}{1.08}{GW190620A}{1.22}{GW190602A}{1.01}{GW190527A}{1.02}{GW190521B}{1.03}{GW190521A}{0.86}{GW190519A}{1.17}{GW190517A}{1.21}{GW190514A}{0.59}{GW190513A}{1.05}{GW190512A}{0.98}{GW190503A}{0.87}{GW190426A}{0.00}{GW190425A}{0.87}{GW190424A}{1.02}{GW190421A}{0.76}{GW190413B}{0.77}{GW190413A}{0.82}{GW190412A}{1.01}{GW190408A}{0.85}}}
\newcommand{\costilttwomed}[1]{\IfEqCase{#1}{{GW190930A}{0.31}{GW190929A}{0.04}{GW190924A}{0.15}{GW190915A}{0.02}{GW190910A}{0.04}{GW190909A}{-0.17}{GW190828B}{0.24}{GW190828A}{0.37}{GW190814A}{-0.03}{GW190803A}{-0.06}{GW190731A}{0.15}{GW190728A}{0.39}{GW190727A}{0.20}{GW190720A}{0.31}{GW190719A}{0.44}{GW190708A}{0.14}{GW190707A}{-0.19}{GW190706A}{0.37}{GW190701A}{-0.17}{GW190630A}{0.33}{GW190620A}{0.50}{GW190602A}{0.19}{GW190527A}{0.16}{GW190521B}{0.29}{GW190521A}{-0.02}{GW190519A}{0.50}{GW190517A}{0.61}{GW190514A}{-0.36}{GW190513A}{0.25}{GW190512A}{0.16}{GW190503A}{-0.02}{GW190426A}{-1.00}{GW190425A}{0.16}{GW190424A}{0.20}{GW190421A}{-0.16}{GW190413B}{-0.13}{GW190413A}{-0.08}{GW190412A}{0.25}{GW190408A}{-0.02}}}
\newcommand{\costilttwoplus}[1]{\IfEqCase{#1}{{GW190930A}{0.63}{GW190929A}{0.86}{GW190924A}{0.77}{GW190915A}{0.85}{GW190910A}{0.83}{GW190909A}{1.04}{GW190828B}{0.69}{GW190828A}{0.57}{GW190814A}{0.87}{GW190803A}{0.93}{GW190731A}{0.77}{GW190728A}{0.56}{GW190727A}{0.72}{GW190720A}{0.62}{GW190719A}{0.51}{GW190708A}{0.77}{GW190707A}{1.02}{GW190706A}{0.58}{GW190701A}{1.00}{GW190630A}{0.60}{GW190620A}{0.47}{GW190602A}{0.73}{GW190527A}{0.76}{GW190521B}{0.62}{GW190521A}{0.86}{GW190519A}{0.46}{GW190517A}{0.36}{GW190514A}{1.11}{GW190513A}{0.68}{GW190512A}{0.75}{GW190503A}{0.88}{GW190426A}{2.00}{GW190425A}{0.70}{GW190424A}{0.72}{GW190421A}{0.96}{GW190413B}{1.00}{GW190413A}{0.95}{GW190412A}{0.67}{GW190408A}{0.88}}}
\newcommand{\finalspinminus}[1]{\IfEqCase{#1}{{GW190930A}{0.06}{GW190929A}{0.31}{GW190924A}{0.05}{GW190915A}{0.11}{GW190910A}{0.07}{GW190909A}{0.20}{GW190828B}{0.08}{GW190828A}{0.07}{GW190814A}{0.02}{GW190803A}{0.11}{GW190731A}{0.13}{GW190728A}{0.04}{GW190727A}{0.10}{GW190720A}{0.05}{GW190719A}{0.17}{GW190708A}{0.04}{GW190707A}{0.04}{GW190706A}{0.18}{GW190701A}{0.13}{GW190630A}{0.07}{GW190620A}{0.15}{GW190602A}{0.14}{GW190527A}{0.16}{GW190521B}{0.07}{GW190521A}{0.16}{GW190519A}{0.13}{GW190517A}{0.07}{GW190514A}{0.15}{GW190513A}{0.12}{GW190512A}{0.07}{GW190503A}{0.12}{GW190424A}{0.09}{GW190421A}{0.11}{GW190413B}{0.12}{GW190413A}{0.13}{GW190412A}{0.06}{GW190408A}{0.07}}}
\newcommand{\finalspinmed}[1]{\IfEqCase{#1}{{GW190930A}{0.72}{GW190929A}{0.66}{GW190924A}{0.67}{GW190915A}{0.70}{GW190910A}{0.70}{GW190909A}{0.66}{GW190828B}{0.65}{GW190828A}{0.75}{GW190814A}{0.28}{GW190803A}{0.68}{GW190731A}{0.70}{GW190728A}{0.71}{GW190727A}{0.73}{GW190720A}{0.72}{GW190719A}{0.78}{GW190708A}{0.69}{GW190707A}{0.66}{GW190706A}{0.78}{GW190701A}{0.66}{GW190630A}{0.70}{GW190620A}{0.79}{GW190602A}{0.70}{GW190527A}{0.71}{GW190521B}{0.72}{GW190521A}{0.71}{GW190519A}{0.79}{GW190517A}{0.87}{GW190514A}{0.63}{GW190513A}{0.68}{GW190512A}{0.65}{GW190503A}{0.66}{GW190424A}{0.74}{GW190421A}{0.67}{GW190413B}{0.68}{GW190413A}{0.68}{GW190412A}{0.67}{GW190408A}{0.67}}}
\newcommand{\finalspinplus}[1]{\IfEqCase{#1}{{GW190930A}{0.07}{GW190929A}{0.20}{GW190924A}{0.05}{GW190915A}{0.09}{GW190910A}{0.08}{GW190909A}{0.15}{GW190828B}{0.08}{GW190828A}{0.06}{GW190814A}{0.02}{GW190803A}{0.10}{GW190731A}{0.10}{GW190728A}{0.04}{GW190727A}{0.10}{GW190720A}{0.06}{GW190719A}{0.11}{GW190708A}{0.04}{GW190707A}{0.03}{GW190706A}{0.09}{GW190701A}{0.09}{GW190630A}{0.05}{GW190620A}{0.08}{GW190602A}{0.10}{GW190527A}{0.12}{GW190521B}{0.05}{GW190521A}{0.12}{GW190519A}{0.07}{GW190517A}{0.05}{GW190514A}{0.11}{GW190513A}{0.14}{GW190512A}{0.07}{GW190503A}{0.09}{GW190424A}{0.09}{GW190421A}{0.10}{GW190413B}{0.10}{GW190413A}{0.12}{GW190412A}{0.05}{GW190408A}{0.06}}}
\newcommand{\luminositydistanceminus}[1]{\IfEqCase{#1}{{GW190930A}{0.32}{GW190929A}{1.05}{GW190924A}{0.22}{GW190915A}{0.61}{GW190910A}{0.58}{GW190909A}{2.22}{GW190828B}{0.60}{GW190828A}{0.93}{GW190814A}{0.05}{GW190803A}{1.58}{GW190731A}{1.72}{GW190728A}{0.37}{GW190727A}{1.50}{GW190720A}{0.32}{GW190719A}{2.00}{GW190708A}{0.39}{GW190707A}{0.37}{GW190706A}{1.93}{GW190701A}{0.73}{GW190630A}{0.37}{GW190620A}{1.31}{GW190602A}{1.12}{GW190527A}{1.24}{GW190521B}{0.57}{GW190521A}{1.95}{GW190519A}{0.92}{GW190517A}{0.84}{GW190514A}{2.17}{GW190513A}{0.80}{GW190512A}{0.55}{GW190503A}{0.63}{GW190426A}{0.16}{GW190425A}{0.07}{GW190424A}{1.16}{GW190421A}{1.38}{GW190413B}{2.12}{GW190413A}{1.66}{GW190412A}{0.17}{GW190408A}{0.60}}}
\newcommand{\luminositydistancemed}[1]{\IfEqCase{#1}{{GW190930A}{0.76}{GW190929A}{2.13}{GW190924A}{0.57}{GW190915A}{1.62}{GW190910A}{1.46}{GW190909A}{3.77}{GW190828B}{1.60}{GW190828A}{2.13}{GW190814A}{0.24}{GW190803A}{3.27}{GW190731A}{3.30}{GW190728A}{0.87}{GW190727A}{3.30}{GW190720A}{0.79}{GW190719A}{3.94}{GW190708A}{0.88}{GW190707A}{0.77}{GW190706A}{4.42}{GW190701A}{2.06}{GW190630A}{0.89}{GW190620A}{2.81}{GW190602A}{2.69}{GW190527A}{2.49}{GW190521B}{1.24}{GW190521A}{3.92}{GW190519A}{2.53}{GW190517A}{1.86}{GW190514A}{4.13}{GW190513A}{2.06}{GW190512A}{1.43}{GW190503A}{1.45}{GW190426A}{0.37}{GW190425A}{0.16}{GW190424A}{2.20}{GW190421A}{2.88}{GW190413B}{4.45}{GW190413A}{3.55}{GW190412A}{0.74}{GW190408A}{1.55}}}
\newcommand{\luminositydistanceplus}[1]{\IfEqCase{#1}{{GW190930A}{0.36}{GW190929A}{3.65}{GW190924A}{0.22}{GW190915A}{0.71}{GW190910A}{1.03}{GW190909A}{3.27}{GW190828B}{0.62}{GW190828A}{0.66}{GW190814A}{0.04}{GW190803A}{1.95}{GW190731A}{2.39}{GW190728A}{0.26}{GW190727A}{1.54}{GW190720A}{0.69}{GW190719A}{2.59}{GW190708A}{0.33}{GW190707A}{0.38}{GW190706A}{2.59}{GW190701A}{0.76}{GW190630A}{0.56}{GW190620A}{1.68}{GW190602A}{1.79}{GW190527A}{2.48}{GW190521B}{0.40}{GW190521A}{2.19}{GW190519A}{1.83}{GW190517A}{1.62}{GW190514A}{2.65}{GW190513A}{0.88}{GW190512A}{0.55}{GW190503A}{0.69}{GW190426A}{0.18}{GW190425A}{0.07}{GW190424A}{1.58}{GW190421A}{1.37}{GW190413B}{2.48}{GW190413A}{2.27}{GW190412A}{0.14}{GW190408A}{0.40}}}
\newcommand{\spinonezminus}[1]{\IfEqCase{#1}{{GW190930A}{0.29}{GW190929A}{0.43}{GW190924A}{0.24}{GW190915A}{0.41}{GW190910A}{0.34}{GW190909A}{0.56}{GW190828B}{0.23}{GW190828A}{0.31}{GW190814A}{0.05}{GW190803A}{0.46}{GW190731A}{0.33}{GW190728A}{0.27}{GW190727A}{0.33}{GW190720A}{0.29}{GW190719A}{0.44}{GW190708A}{0.25}{GW190707A}{0.29}{GW190706A}{0.38}{GW190701A}{0.51}{GW190630A}{0.21}{GW190620A}{0.39}{GW190602A}{0.34}{GW190527A}{0.36}{GW190521B}{0.23}{GW190521A}{0.58}{GW190519A}{0.37}{GW190517A}{0.36}{GW190514A}{0.54}{GW190513A}{0.24}{GW190512A}{0.25}{GW190503A}{0.44}{GW190426A}{0.51}{GW190425A}{0.12}{GW190424A}{0.35}{GW190421A}{0.49}{GW190413B}{0.48}{GW190413A}{0.52}{GW190412A}{0.23}{GW190408A}{0.42}}}
\newcommand{\spinonezmed}[1]{\IfEqCase{#1}{{GW190930A}{0.15}{GW190929A}{0.008}{GW190924A}{0.02}{GW190915A}{0.02}{GW190910A}{0.01}{GW190909A}{-0.03}{GW190828B}{0.06}{GW190828A}{0.20}{GW190814A}{0.0001}{GW190803A}{-0.01}{GW190731A}{0.03}{GW190728A}{0.13}{GW190727A}{0.10}{GW190720A}{0.20}{GW190719A}{0.36}{GW190708A}{0.009}{GW190707A}{-0.03}{GW190706A}{0.33}{GW190701A}{-0.05}{GW190630A}{0.05}{GW190620A}{0.37}{GW190602A}{0.04}{GW190527A}{0.09}{GW190521B}{0.04}{GW190521A}{0.02}{GW190519A}{0.35}{GW190517A}{0.67}{GW190514A}{-0.18}{GW190513A}{0.09}{GW190512A}{0.005}{GW190503A}{-0.03}{GW190426A}{-0.03}{GW190425A}{0.06}{GW190424A}{0.15}{GW190421A}{-0.04}{GW190413B}{-0.02}{GW190413A}{0.001}{GW190412A}{0.30}{GW190408A}{-0.03}}}
\newcommand{\spinonezplus}[1]{\IfEqCase{#1}{{GW190930A}{0.41}{GW190929A}{0.45}{GW190924A}{0.39}{GW190915A}{0.40}{GW190910A}{0.39}{GW190909A}{0.53}{GW190828B}{0.24}{GW190828A}{0.41}{GW190814A}{0.04}{GW190803A}{0.39}{GW190731A}{0.45}{GW190728A}{0.30}{GW190727A}{0.48}{GW190720A}{0.29}{GW190719A}{0.43}{GW190708A}{0.24}{GW190707A}{0.20}{GW190706A}{0.43}{GW190701A}{0.34}{GW190630A}{0.28}{GW190620A}{0.42}{GW190602A}{0.46}{GW190527A}{0.50}{GW190521B}{0.32}{GW190521A}{0.53}{GW190519A}{0.37}{GW190517A}{0.25}{GW190514A}{0.39}{GW190513A}{0.46}{GW190512A}{0.21}{GW190503A}{0.31}{GW190426A}{0.36}{GW190425A}{0.18}{GW190424A}{0.46}{GW190421A}{0.40}{GW190413B}{0.40}{GW190413A}{0.44}{GW190412A}{0.12}{GW190408A}{0.26}}}
\newcommand{\chirpmasssourceminus}[1]{\IfEqCase{#1}{{GW190930A}{0.5}{GW190929A}{8.2}{GW190924A}{0.2}{GW190915A}{2.7}{GW190910A}{4.1}{GW190909A}{7.5}{GW190828B}{1.0}{GW190828A}{2.1}{GW190814A}{0.06}{GW190803A}{4.1}{GW190731A}{5.2}{GW190728A}{0.3}{GW190727A}{3.7}{GW190720A}{0.8}{GW190719A}{4.0}{GW190708A}{0.6}{GW190707A}{0.5}{GW190706A}{7.0}{GW190701A}{4.9}{GW190630A}{2.1}{GW190620A}{6.5}{GW190602A}{8.5}{GW190527A}{4.2}{GW190521B}{2.5}{GW190521A}{10.6}{GW190519A}{7.1}{GW190517A}{4.0}{GW190514A}{4.8}{GW190513A}{1.9}{GW190512A}{1.0}{GW190503A}{4.2}{GW190426A}{0.08}{GW190425A}{0.02}{GW190424A}{4.6}{GW190421A}{4.2}{GW190413B}{5.4}{GW190413A}{4.1}{GW190412A}{0.3}{GW190408A}{1.2}}}
\newcommand{\chirpmasssourcemed}[1]{\IfEqCase{#1}{{GW190930A}{8.5}{GW190929A}{35.8}{GW190924A}{5.8}{GW190915A}{25.3}{GW190910A}{34.3}{GW190909A}{30.9}{GW190828B}{13.3}{GW190828A}{25.0}{GW190814A}{6.09}{GW190803A}{27.3}{GW190731A}{29.5}{GW190728A}{8.6}{GW190727A}{28.6}{GW190720A}{8.9}{GW190719A}{23.5}{GW190708A}{13.2}{GW190707A}{8.5}{GW190706A}{42.7}{GW190701A}{40.3}{GW190630A}{24.9}{GW190620A}{38.3}{GW190602A}{49.1}{GW190527A}{24.3}{GW190521B}{32.1}{GW190521A}{69.2}{GW190519A}{44.5}{GW190517A}{26.6}{GW190514A}{28.5}{GW190513A}{21.6}{GW190512A}{14.6}{GW190503A}{30.2}{GW190426A}{2.41}{GW190425A}{1.44}{GW190424A}{31.0}{GW190421A}{31.2}{GW190413B}{33.0}{GW190413A}{24.6}{GW190412A}{13.3}{GW190408A}{18.3}}}
\newcommand{\chirpmasssourceplus}[1]{\IfEqCase{#1}{{GW190930A}{0.5}{GW190929A}{14.9}{GW190924A}{0.2}{GW190915A}{3.2}{GW190910A}{4.1}{GW190909A}{17.2}{GW190828B}{1.2}{GW190828A}{3.4}{GW190814A}{0.06}{GW190803A}{5.7}{GW190731A}{7.1}{GW190728A}{0.5}{GW190727A}{5.3}{GW190720A}{0.5}{GW190719A}{6.5}{GW190708A}{0.9}{GW190707A}{0.6}{GW190706A}{10.0}{GW190701A}{5.4}{GW190630A}{2.1}{GW190620A}{8.3}{GW190602A}{9.1}{GW190527A}{9.1}{GW190521B}{3.2}{GW190521A}{17.0}{GW190519A}{6.4}{GW190517A}{4.0}{GW190514A}{7.9}{GW190513A}{3.8}{GW190512A}{1.3}{GW190503A}{4.2}{GW190426A}{0.08}{GW190425A}{0.02}{GW190424A}{5.8}{GW190421A}{5.9}{GW190413B}{8.2}{GW190413A}{5.5}{GW190412A}{0.4}{GW190408A}{1.9}}}
\newcommand{\symmetricmassratiominus}[1]{\IfEqCase{#1}{{GW190930A}{0.11}{GW190929A}{0.07}{GW190924A}{0.09}{GW190915A}{0.03}{GW190910A}{0.01}{GW190909A}{0.09}{GW190828B}{0.04}{GW190828A}{0.01}{GW190814A}{0.006}{GW190803A}{0.03}{GW190731A}{0.04}{GW190728A}{0.07}{GW190727A}{0.03}{GW190720A}{0.06}{GW190719A}{0.06}{GW190708A}{0.03}{GW190707A}{0.03}{GW190706A}{0.05}{GW190701A}{0.03}{GW190630A}{0.03}{GW190620A}{0.04}{GW190602A}{0.04}{GW190527A}{0.06}{GW190521B}{0.01}{GW190521A}{0.04}{GW190519A}{0.03}{GW190517A}{0.04}{GW190514A}{0.04}{GW190513A}{0.04}{GW190512A}{0.03}{GW190503A}{0.03}{GW190426A}{0.08}{GW190425A}{0.03}{GW190424A}{0.02}{GW190421A}{0.03}{GW190413B}{0.04}{GW190413A}{0.04}{GW190412A}{0.02}{GW190408A}{0.02}}}
\newcommand{\symmetricmassratiomed}[1]{\IfEqCase{#1}{{GW190930A}{0.24}{GW190929A}{0.18}{GW190924A}{0.23}{GW190915A}{0.242}{GW190910A}{0.248}{GW190909A}{0.24}{GW190828B}{0.21}{GW190828A}{0.248}{GW190814A}{0.090}{GW190803A}{0.245}{GW190731A}{0.243}{GW190728A}{0.24}{GW190727A}{0.247}{GW190720A}{0.23}{GW190719A}{0.23}{GW190708A}{0.245}{GW190707A}{0.244}{GW190706A}{0.23}{GW190701A}{0.246}{GW190630A}{0.241}{GW190620A}{0.24}{GW190602A}{0.243}{GW190527A}{0.24}{GW190521B}{0.246}{GW190521A}{0.245}{GW190519A}{0.24}{GW190517A}{0.241}{GW190514A}{0.245}{GW190513A}{0.22}{GW190512A}{0.23}{GW190503A}{0.24}{GW190426A}{0.16}{GW190425A}{0.240}{GW190424A}{0.247}{GW190421A}{0.247}{GW190413B}{0.241}{GW190413A}{0.241}{GW190412A}{0.17}{GW190408A}{0.245}}}
\newcommand{\symmetricmassratioplus}[1]{\IfEqCase{#1}{{GW190930A}{0.01}{GW190929A}{0.07}{GW190924A}{0.02}{GW190915A}{0.008}{GW190910A}{0.002}{GW190909A}{0.01}{GW190828B}{0.04}{GW190828A}{0.002}{GW190814A}{0.005}{GW190803A}{0.005}{GW190731A}{0.007}{GW190728A}{0.01}{GW190727A}{0.003}{GW190720A}{0.02}{GW190719A}{0.02}{GW190708A}{0.005}{GW190707A}{0.006}{GW190706A}{0.02}{GW190701A}{0.004}{GW190630A}{0.009}{GW190620A}{0.01}{GW190602A}{0.007}{GW190527A}{0.01}{GW190521B}{0.004}{GW190521A}{0.005}{GW190519A}{0.01}{GW190517A}{0.009}{GW190514A}{0.005}{GW190513A}{0.03}{GW190512A}{0.02}{GW190503A}{0.01}{GW190426A}{0.08}{GW190425A}{0.010}{GW190424A}{0.003}{GW190421A}{0.003}{GW190413B}{0.008}{GW190413A}{0.008}{GW190412A}{0.03}{GW190408A}{0.005}}}
\newcommand{\spintwoxminus}[1]{\IfEqCase{#1}{{GW190930A}{0.52}{GW190929A}{0.60}{GW190924A}{0.49}{GW190915A}{0.59}{GW190910A}{0.52}{GW190909A}{0.58}{GW190828B}{0.55}{GW190828A}{0.50}{GW190814A}{0.62}{GW190803A}{0.58}{GW190731A}{0.57}{GW190728A}{0.46}{GW190727A}{0.55}{GW190720A}{0.58}{GW190719A}{0.56}{GW190708A}{0.46}{GW190707A}{0.46}{GW190706A}{0.53}{GW190701A}{0.55}{GW190630A}{0.49}{GW190620A}{0.54}{GW190602A}{0.60}{GW190527A}{0.61}{GW190521B}{0.51}{GW190521A}{0.66}{GW190519A}{0.57}{GW190517A}{0.53}{GW190514A}{0.60}{GW190513A}{0.53}{GW190512A}{0.50}{GW190503A}{0.55}{GW190426A}{0.00}{GW190425A}{0.47}{GW190424A}{0.59}{GW190421A}{0.59}{GW190413B}{0.59}{GW190413A}{0.58}{GW190412A}{0.57}{GW190408A}{0.53}}}
\newcommand{\spintwoxmed}[1]{\IfEqCase{#1}{{GW190930A}{0.00}{GW190929A}{0.0009}{GW190924A}{0.00}{GW190915A}{0.00}{GW190910A}{0.0003}{GW190909A}{0.0003}{GW190828B}{0.00}{GW190828A}{0.002}{GW190814A}{-0.01}{GW190803A}{0.006}{GW190731A}{0.003}{GW190728A}{0.001}{GW190727A}{0.002}{GW190720A}{0.002}{GW190719A}{0.00}{GW190708A}{0.002}{GW190707A}{0.0004}{GW190706A}{0.003}{GW190701A}{0.003}{GW190630A}{0.00}{GW190620A}{0.00006}{GW190602A}{0.00}{GW190527A}{0.005}{GW190521B}{0.00}{GW190521A}{0.002}{GW190519A}{0.0006}{GW190517A}{0.001}{GW190514A}{0.001}{GW190513A}{0.00}{GW190512A}{0.0009}{GW190503A}{0.001}{GW190426A}{0.00}{GW190425A}{0.0006}{GW190424A}{0.00}{GW190421A}{0.00}{GW190413B}{0.0007}{GW190413A}{0.00}{GW190412A}{-0.01}{GW190408A}{0.002}}}
\newcommand{\spintwoxplus}[1]{\IfEqCase{#1}{{GW190930A}{0.52}{GW190929A}{0.59}{GW190924A}{0.48}{GW190915A}{0.60}{GW190910A}{0.54}{GW190909A}{0.57}{GW190828B}{0.54}{GW190828A}{0.51}{GW190814A}{0.59}{GW190803A}{0.57}{GW190731A}{0.58}{GW190728A}{0.48}{GW190727A}{0.55}{GW190720A}{0.57}{GW190719A}{0.56}{GW190708A}{0.43}{GW190707A}{0.46}{GW190706A}{0.54}{GW190701A}{0.56}{GW190630A}{0.48}{GW190620A}{0.56}{GW190602A}{0.60}{GW190527A}{0.59}{GW190521B}{0.51}{GW190521A}{0.69}{GW190519A}{0.55}{GW190517A}{0.53}{GW190514A}{0.59}{GW190513A}{0.54}{GW190512A}{0.49}{GW190503A}{0.58}{GW190426A}{0.00}{GW190425A}{0.47}{GW190424A}{0.59}{GW190421A}{0.58}{GW190413B}{0.60}{GW190413A}{0.57}{GW190412A}{0.56}{GW190408A}{0.53}}}
\newcommand{\networkoptimalsnrminus}[1]{\IfEqCase{#1}{{GW190814A}{1.7}{GW190426A}{1.8}{GW190425A}{1.7}}}
\newcommand{\networkoptimalsnrmed}[1]{\IfEqCase{#1}{{GW190814A}{24.7}{GW190426A}{8.3}{GW190425A}{12.0}}}
\newcommand{\networkoptimalsnrplus}[1]{\IfEqCase{#1}{{GW190814A}{1.7}{GW190426A}{1.8}{GW190425A}{1.7}}}
\newcommand{\networkmatchedfiltersnrminus}[1]{\IfEqCase{#1}{{GW190814A}{0.2}{GW190426A}{0.6}{GW190425A}{0.4}{GW190412A}{0.4}}}
\newcommand{\networkmatchedfiltersnrmed}[1]{\IfEqCase{#1}{{GW190814A}{24.9}{GW190426A}{8.7}{GW190425A}{12.4}{GW190412A}{19.0}}}
\newcommand{\networkmatchedfiltersnrplus}[1]{\IfEqCase{#1}{{GW190814A}{0.1}{GW190426A}{0.5}{GW190425A}{0.3}{GW190412A}{0.2}}}
\newcommand{\logpriorminus}[1]{\IfEqCase{#1}{{GW190426A}{10.5}{GW190425A}{8.6}}}
\newcommand{\logpriormed}[1]{\IfEqCase{#1}{{GW190426A}{161.3}{GW190425A}{98.4}}}
\newcommand{\logpriorplus}[1]{\IfEqCase{#1}{{GW190426A}{8.6}{GW190425A}{6.7}}}
\newcommand{\PEpercentBNS}[1]{\IfEqCase{#1}{{GW190930A}{0}{GW190929A}{0}{GW190924A}{0}{GW190915A}{0}{GW190910A}{0}{GW190909A}{0}{GW190828B}{0}{GW190828A}{0}{GW190814A}{0}{GW190803A}{0}{GW190731A}{0}{GW190728A}{0}{GW190727A}{0}{GW190720A}{0}{GW190719A}{0}{GW190708A}{0}{GW190707A}{0}{GW190706A}{0}{GW190701A}{0}{GW190630A}{0}{GW190620A}{0}{GW190602A}{0}{GW190527A}{0}{GW190521B}{0}{GW190521A}{0}{GW190519A}{0}{GW190517A}{0}{GW190514A}{0}{GW190513A}{0}{GW190512A}{0}{GW190503A}{0}{GW190426A}{1}{GW190425A}{100}{GW190424A}{0}{GW190421A}{0}{GW190413B}{0}{GW190413A}{0}{GW190412A}{0}{GW190408A}{0}}}
\newcommand{\PEpercentNSBH}[1]{\IfEqCase{#1}{{GW190930A}{0}{GW190929A}{0}{GW190924A}{4}{GW190915A}{0}{GW190910A}{0}{GW190909A}{0}{GW190828B}{0}{GW190828A}{0}{GW190814A}{100}{GW190803A}{0}{GW190731A}{0}{GW190728A}{0}{GW190727A}{0}{GW190720A}{0}{GW190719A}{0}{GW190708A}{0}{GW190707A}{0}{GW190706A}{0}{GW190701A}{0}{GW190630A}{0}{GW190620A}{0}{GW190602A}{0}{GW190527A}{0}{GW190521B}{0}{GW190521A}{0}{GW190519A}{0}{GW190517A}{0}{GW190514A}{0}{GW190513A}{0}{GW190512A}{0}{GW190503A}{0}{GW190426A}{99}{GW190425A}{0}{GW190424A}{0}{GW190421A}{0}{GW190413B}{0}{GW190413A}{0}{GW190412A}{0}{GW190408A}{0}}}
\newcommand{\PEpercentBBH}[1]{\IfEqCase{#1}{{GW190930A}{100}{GW190929A}{100}{GW190924A}{96}{GW190915A}{100}{GW190910A}{100}{GW190909A}{100}{GW190828B}{100}{GW190828A}{100}{GW190814A}{0}{GW190803A}{100}{GW190731A}{100}{GW190728A}{100}{GW190727A}{100}{GW190720A}{100}{GW190719A}{100}{GW190708A}{100}{GW190707A}{100}{GW190706A}{100}{GW190701A}{100}{GW190630A}{100}{GW190620A}{100}{GW190602A}{100}{GW190527A}{100}{GW190521B}{100}{GW190521A}{100}{GW190519A}{100}{GW190517A}{100}{GW190514A}{100}{GW190513A}{100}{GW190512A}{100}{GW190503A}{100}{GW190426A}{0}{GW190425A}{0}{GW190424A}{100}{GW190421A}{100}{GW190413B}{100}{GW190413A}{100}{GW190412A}{100}{GW190408A}{100}}}
\newcommand{\PEpercentMassGap}[1]{\IfEqCase{#1}{{GW190930A}{0}{GW190929A}{0}{GW190924A}{0}{GW190915A}{0}{GW190910A}{0}{GW190909A}{0}{GW190828B}{0}{GW190828A}{0}{GW190814A}{0}{GW190803A}{0}{GW190731A}{0}{GW190728A}{0}{GW190727A}{0}{GW190720A}{0}{GW190719A}{0}{GW190708A}{0}{GW190707A}{0}{GW190706A}{0}{GW190701A}{0}{GW190630A}{0}{GW190620A}{0}{GW190602A}{0}{GW190527A}{0}{GW190521B}{0}{GW190521A}{0}{GW190519A}{0}{GW190517A}{0}{GW190514A}{0}{GW190513A}{0}{GW190512A}{0}{GW190503A}{0}{GW190426A}{0}{GW190425A}{0}{GW190424A}{0}{GW190421A}{0}{GW190413B}{0}{GW190413A}{0}{GW190412A}{0}{GW190408A}{0}}}

    \newcommand{\luminositydistancemostsecond}{GW190706A}

\newcommand{\fractionChiEffNegativeExtremeLowerLimit}{\ensuremath{7}}
\newcommand{\fractionChiEffNegativeLow}{\ensuremath{12}}

\newcommand{\fractionChiEffNegativeHigh}{\ensuremath{44}}

\newcommand{\fractionChiEffNegative}{\ensuremath{0.27}}
\newcommand{\fractionChiEffNegativeErrorHigh}{\ensuremath{0.17}}
\newcommand{\fractionChiEffNegativeErrorLow}{\ensuremath{0.15}}

\newcommand{\fractionChiEffPositive}{\ensuremath{0.67}}
\newcommand{\fractionChiEffPositiveErrorHigh}{\ensuremath{0.16}}
\newcommand{\fractionChiEffPositiveErrorLow}{\ensuremath{0.16}}

\newcommand{\fractionChiEffVanishing}{\ensuremath{0.05}}
\newcommand{\fractionChiEffVanishingErrorHigh}{\ensuremath{0.02}}
\newcommand{\fractionChiEffVanishingErrorLow}{\ensuremath{0.01}}

\newcommand{\fractionChiEffDynamicalLow}{\ensuremath{0.25}}
\newcommand{\fractionChiEffDynamicalHigh}{\ensuremath{0.93}}

\newcommand{\fractionChiEffNegativeVarMin}{\ensuremath{0.26}}

\newcommand{\fractionChiEffNegativeErrorHighVarMin}{\ensuremath{0.17}}
\newcommand{\fractionChiEffNegativeErrorLowVarMin}{\ensuremath{0.21}}

\newcommand{\fractionChiEffPositiveVarMin}{\ensuremath{0.65}}
\newcommand{\fractionChiEffPositiveErrorHighVarMin}{\ensuremath{0.17}}
\newcommand{\fractionChiEffPositiveErrorLowVarMin}{\ensuremath{0.16}}

\newcommand{\FractionOfSmallChiPSamples}{\ensuremath{0.02\%}}

\newcommand{\medianMuEff}{\ensuremath{0.06}}
\newcommand{\errorMuEffHigh}{\ensuremath{0.05}}
\newcommand{\errorMuEffLow}{\ensuremath{0.05}}

\newcommand{\medianSigmaEff}{\ensuremath{0.12}}
\newcommand{\errorSigmaEffHigh}{\ensuremath{0.06}}
\newcommand{\errorSigmaEffLow}{\ensuremath{0.04}}

\newcommand{\percentMinChiLessThanZero}{\ensuremath{99\%}}

\newcommand{\mmaxModelB}{\ensuremath{40.8^{+11.8}_{-4.4} }}
\newcommand{\rateOTwo}{\ensuremath{53.2^{+55.8}_{-28.2}}}
\newcommand{\BNSrate}{\ensuremath{320^{+490}_{-240}}}

\newcommand{\result}[1]{\textcolor{black}{#1}}

\newcommand{\mmax}{m_\mathrm{max}}
\newcommand{\mmin}{m_\mathrm{min}}
\newcommand{\bbhevnt}{BBH events}
\newcommand{\bbhsys}{BBH systems}
\newcommand{\appd}{posterior population distribution}
\newcommand{\oppd}{posterior predictive distribution}
\newcommand{\antialigned}{anti-aligned}
\newcommand{\chieff}{effective inspiral spin parameter}
\newcommand{\chip}{effective precession spin parameter}
\newcommand{\astrophysical}{\textit{astrophysical}}
\newcommand{\observed}{\textit{observed}}
\newcommand{\truncated}{\textsc{Truncated}}
\newcommand{\ppsn}{\textsc{Power Law + Peak}}
\newcommand{\tapered}{\textsc{Broken Power Law}}
\newcommand{\multipeak}{\textsc{Multi Peak}}
\newcommand{\ModelH}{\textsc{Multi Spin}}
\newcommand{\DefineRemark}[2]{%
  \expandafter\newcommand\csname rmk-#1\endcsname{#2}%
}
\newcommand{\Remark}[1]{\csname rmk-#1\endcsname}
\DefineRemark{gw190521}{GW190521}

\mathchardef\mhyphen="2D

\begin{document}

\title{Population properties of compact objects from the second LIGO--Virgo Gravitational-Wave Transient Catalog}

\AuthorCollaborationLimit=3000  


\author{R.~Abbott}
\affiliation{LIGO, California Institute of Technology, Pasadena, CA 91125, USA}
\author{T.~D.~Abbott}
\affiliation{Louisiana State University, Baton Rouge, LA 70803, USA}
\author{S.~Abraham}
\affiliation{Inter-University Centre for Astronomy and Astrophysics, Pune 411007, India}
\author{F.~Acernese}
\affiliation{Dipartimento di Farmacia, Universit\`a di Salerno, I-84084 Fisciano, Salerno, Italy  }
\affiliation{INFN, Sezione di Napoli, Complesso Universitario di Monte S.Angelo, I-80126 Napoli, Italy  }
\author{K.~Ackley}
\affiliation{OzGrav, School of Physics \& Astronomy, Monash University, Clayton 3800, Victoria, Australia}
\author{A.~Adams}
\affiliation{Christopher Newport University, Newport News, VA 23606, USA}
\author{C.~Adams}
\affiliation{LIGO Livingston Observatory, Livingston, LA 70754, USA}
\author{R.~X.~Adhikari}
\affiliation{LIGO, California Institute of Technology, Pasadena, CA 91125, USA}
\author{V.~B.~Adya}
\affiliation{OzGrav, Australian National University, Canberra, Australian Capital Territory 0200, Australia}
\author{C.~Affeldt}
\affiliation{Max Planck Institute for Gravitational Physics (Albert Einstein Institute), D-30167 Hannover, Germany}
\affiliation{Leibniz Universit\"at Hannover, D-30167 Hannover, Germany}
% ...

\collaboration{The LIGO Scientific Collaboration and the Virgo Collaboration}



\begin{abstract}
We report on the population of the 47 compact binary mergers detected with a false-alarm rate <$1\,\mathrm{yr}^{-1}$ in the second LIGO--Virgo Gravitational-Wave~Transient~Catalog, GWTC-2. We observe several characteristics of the merging binary~black~hole (BBH) population not discernible until now. First, the primary mass spectrum contains structure beyond a power-law with a sharp high-mass cut-off; it is more consistent with a \textit{broken}~power~law with a break at $\unit[\BPLNoAugNoEvolutionmbreak]{M_\odot}$, or a power~law with a Gaussian feature peaking at $\unit[\peakNoAugNoEvolutionmpp]{M_\odot}$~(90\% credible interval). While the primary mass distribution must extend to $\result{\sim65\,M_\odot}$~or beyond, only $\peakNoAugNoEvolutionFractionAbovePrimaryFortyFive\%$~of systems have primary masses greater than $45\,M_\odot$. Second, we find that a fraction of BBH~systems have component spins misaligned with the orbital angular momentum, giving rise to precession of the orbital plane. Moreover, \fractionChiEffNegativeLow\%~to~\fractionChiEffNegativeHigh\% of BBH systems have spins tilted by more than $90^\circ$, giving rise to a negative effective inspiral spin parameter~$\chi_\mathrm{eff}$. \added{Under the assumption that such systems can only be formed by dynamical interactions, we infer that between~25\% and~93\% of BBH with non-vanishing $|\chi_\mathrm{eff}|>0.01$ are dynamically assembled.} Third, we estimate merger rates, finding $\mathcal{R}_\text{BBH} = \unit[\peakNoAugNoEvolutionrate]{Gpc^{-3}\,yr^{-1}}$ for BBH and $\mathcal{R}_\text{BNS}= \unit[\BNSrate]{Gpc^{-3}\,yr^{-1}}$ for binary neutron stars. We find that the BBH rate likely increases with redshift ($\peakNoAugEvolutionLambdaPGrZero\%$ credibility), but not faster than the star-formation rate ($\peakNoAugEvolutionMDPercentileInLambda\%$ credibility). Additionally, we examine recent exceptional events in the context of our population models, finding that the asymmetric masses of GW190412 and the high component masses of GW190521 are consistent with our models, but the low secondary mass of GW190814 makes it an outlier.
\end{abstract}
\keywords{gravitational waves}

\section{Introduction}
We analyze the population properties of black holes (BHs) and neutron stars (NSs) in compact binary systems using data from the LIGO--Virgo Gravitational-wave Transient Catalog~2~\citep[GWTC-2;][]{O3acatalog}.
The GWTC-2 catalog combines observations from the first two observing runs~\citep[O1 and O2;][]{GWTC1} and the first half of the third observing run~\citep[O3a;][]{O3acatalog} of the Advanced LIGO~\citep{aLIGO} and Advanced Virgo~\citep{aVirgo} gravitational-wave observatories.
With the $\result{39}$ additional candidates from O3a, we have more than quadrupled the number of events from O1 and O2, published in the first LIGO--Virgo Transient Catalog~\citep[GWTC-1;][]{GWTC1}.
Counting only events with false alarm rate (FAR) $<\unit[1]{yr^{-1}}$ (as opposed to the less conservative FAR threshold of $<\unit[2]{yr^{-1}}$ in GWTC-2), the new combined catalog includes: two binary NS (BNS) events, \result{44} confident binary black hole (BBH) events, and one NS--BH (NSBH) candidate, which may be a BBH---a topic we discuss below.
\added{We define \bbhevnt{} as systems where both masses are above $3\,M_\odot$ at 90\% credibility.}
These 47 events are listed in Table~\ref{tab:events}.
Our chosen FAR threshold ensures a relatively pure sample with only $\sim 1$ noise event (see Section~\ref{data}) and excludes three marginal triggers presented in GWTC-2.
Two of these excluded events are BBH candidates (\NAME{GW190719A}{} and \NAME{GW190909A}{}) and one is an NSBH candidate (\NAME{GW190426A}).

The latest observations include a number of individually remarkable events, which invite theoretical speculation while providing challenges to existing models.
The observation of high-mass binaries such as \NAME{GW190521A}~\citep{GW190521}, which has a primary mass $m_1 \sim 85\,M_\odot$, is in tension with the sharp mass cutoff $\mmax = \mmaxModelB \ M_\odot$ inferred from the GWTC-1 detections, forcing us to reconsider models for the distribution of black hole (BH) masses in binary systems~\citep{GW190521, Abbott:GW190521_implications}.
Here and in the following, the primary mass $m_1$ refers to the bigger of the two component masses in the binary, while the secondary mass $m_2$ refers to the smaller of the two.
Along with \NAME{GW190521A}{}, \NAME{GW190602A}{} and \NAME{GW190519A}{} also have primary masses above $45 \ M_\odot$ at $>99\%$ credibility. 
These high-mass binaries are interesting from a theoretical perspective since they occur in the predicted pair-instability gap~\citep{1967PhRvL..18..379B, 1964ApJS....9..201F, Heger2002,Heger2003,Woosley2015,2016A&A...594A..97B}.
Additionally, GWTC-2 includes the first systems with confidently asymmetric component masses, including \NAME{GW190412A}{} with mass ratio $m_2/ m_1  \equiv q = \massratiomed{GW190412A}_{-\massratiominus{GW190412A}}^{+\massratioplus{GW190412A}}$~\citep{GW190412} and \NAME{GW190814A}{}~\citep{GW190814}, with $q=\massratiomed{GW190814A}_{-\massratiominus{GW190814A}}^{+\massratioplus{GW190814A}}$.
Furthermore, the secondary mass of \NAME{GW190814A}{}, $m_2 = \masstwosourcemed{GW190814A}_{-\masstwosourceminus{GW190814A}}^{+\masstwosourceplus{GW190814A}}\ M_\odot$, is near the purported NS--BH gap~\citep{Bailyn1998,Ozel2011,Farr2011}, posing a challenge to our understanding of binary formation~\citep{GW190814, Zevin:2020gma, Mandel:2020cig}.
We can gain insight into these exceptional events by studying them in the context of the larger population of compact binaries~\citep{Fishbach:2019ckx}.

In particular, studying the enlarged population of \bbhevnt{} enables us to investigate how compact binaries form.\footnote{\added{For the sake of brevity, we refer throughout to "BBHs" when we really mean "merging BBHs." It is possible that non-merging BBHs have different properties from those that merge.}}
Several evolutionary channels have been proposed to explain the origin of the compact binary mergers observed with Advanced LIGO and Advanced Virgo.
The isolated binary channel may occur via common envelope evolution (e.g., \citealt{1998ApJ...506..780B,1998A&A...332..173P,2002ApJ...572..407B, 2015ApJ...806..263D}), the remnants of Population III stars (e.g., \citealt{Madau:2001sc, Inayoshi:2017mrs}), or chemically homogeneous stellar evolution (e.g., \citealt{2016A&A...588A..50M,2016MNRAS.460.3545D,2016MNRAS.458.2634M}). Evolution via common envelope predicts \bbhsys{} with component masses up to $\sim{}50$ M$_\odot$, mass ratios in the range $0.3 \lesssim q < 1$, and nearly aligned spins~\citep{2000ApJ...541..319K, 2010CQGra..27k4007M,2013ApJ...779...72D,2018MNRAS.474.2959G, 2017PASA...34...58E,  2020arXiv200411866O}.

In the chemically homogeneous scenario, \bbhsys{} are expected to form with nearly equal mass components, in addition to aligned spins and component masses in the range $\sim 20$--$ 50 \ M_\odot$.
In isolated formation scenarios, component BHs form via stellar collapse, and are thus not expected to occur within the pair-instability mass gap, $\sim 50$--$120 \ M_\odot$, but may populate either side of the gap.

Alternatively, BBH mergers could form via dynamical interactions in young stellar clusters, globular clusters, or nuclear star clusters (e.g., \citealt{1993Natur.364..421K,1993Natur.364..423S,2000ApJ...528L..17P}), triple systems (e.g. \citealt{2014ApJ...781...45A,2016MNRAS.463.2443K,2017ApJ...841...77A,2020arXiv201013669V}) or the disks of active galactic nuclei (e.g. \citealt{2012MNRAS.425..460M,2017ApJ...835..165B,2017MNRAS.464..946S,2019MNRAS.488...47F}).
Dynamical formation in dense stellar clusters typically produces an isotropic distribution of spin directions (e.g. \citealt{Vitale:2015tea,Rodriguez2016}), and may enable hierarchical mergers characterized by relatively high-mass binaries (e.g. \citealt{2016ApJ...831..187A,2016MNRAS.459.3432M,mckernan_constraining_2018,Rodriguez2018,2020ApJ...894..133A}) and large BH spins~\citep{Fishbach:2017dwv, 2017PhRvD..95l4046G}.
Finally, \bbhsys{} might originate as part of a primordial BH population in the early Universe \citep{1974MNRAS.168..399C,2016PhRvD..94h3504C}.
Most primordial BH models predict low spins and isotropic spin orientation~\citep{2019JCAP...08..022F}.

Before we continue, we summarize key questions addressed in the previous analysis of GWTC-1 data~\citep{O2pop} and how the inclusion of O3a events affects our findings:
\begin{enumerate}
    \item \emph{Are there \bbhsys{} with component masses $\gtrsim 45 \ M_\odot$?}
    Following O1 and O2, we inferred that 99\% of \bbhsys{} have primary mass below $m_{99\%} \approx 45 \ M_\odot$. 
    Moreover, this limit was consistent with a sharp cutoff at $\mmax = \mmaxModelB \ M_\odot$,
    hypothesized to correspond to the lower edge of the pair-instability mass gap expected from supernova theory~\citep{Woosley2015,Heger2002,Heger2003,Fishbach2017,Talbot2018}.
    While the GWTC-2 events remain consistent with $\peakNoAugNoEvolutionFractionBelowPrimaryFortyFive$ \% of \bbhsys{} having primary masses below $45 \ M_\odot$ (inferred using the the \ppsn{} mass model described in Section~\ref{models}, or ``Model C'' from~\citealp{O2pop}), high-mass detections such as \NAME{GW190521A}{}, \NAME{GW190602A}{} and \NAME{GW190519A}{} imply a non-zero rate of BBH mergers beyond the $\sim45 \ M_\odot$ mass limit. We infer that the merger rate for systems with primary masses in the range $ 45 \ M_\odot < m_1 < 100 \ M_\odot$ is $\peakNoAugNoEvolutionRateAbovePrimaryFortyFive \ \mathrm{Gpc}^{-3} \ \mathrm{yr}^{-1}$, consistent with estimates inferred from GWTC-1~\citep{Fishbach:2019ckx}.
    \item \emph{Is there a preference for nearly equal component masses?}
    All of the GWTC-1 observations are consistent with equal-mass binaries, with mass ratios $q\equiv m_2/m_1 = 1$.
    In O3a, however, we detected two events with mass ratios bounded confidently away from unity: \NAME{GW190814A}{}  and \NAME{GW190412A}{}, though, most binaries are consistent with equal-mass.
    The NSBH candidate \NAME{GW190426A}{}, if real, also has unequal component masses $q=\massratiomed{GW190426A}^{+\massratioplus{GW190426A}}_{-\massratiominus{GW190426A}}$, but is above the FAR threshold for this work.
    \item \emph{Does the merger rate evolve with redshift?}
    From GWTC-1 we inferred that the BBH merger rate is {$\rateOTwo \ \mathrm{Gpc}^{-3} \ \mathrm{yr}^{-1}$}, assuming a rate density that is uniform in comoving volume.
    Allowing for a rate that evolves with redshift~\citep{Fishbach2018}, we found that the local merger rate is $\mathcal{R}_\text{BBH}(z = 0) = 19.7^{+ 57.3}_{-15.9} \ \mathrm{Gpc}^{-3} \ \mathrm{yr}^{-1}$, and that, while still consistent with a uniform-in-comoving volume model, the rate density is likely increasing with redshift with {93\%} credibility~\citep{O2pop}. 
    With GWTC-2, we are able to more tightly bound the BBH merger rate at {$\mathcal{R}_\text{BBH} = \unit[\peakNoAugNoEvolutionrate]{Gpc^{-3}\,yr^{-1}}$} (assuming the \ppsn{} mass model \added{and a constant-in-comoving-volume merger rate}), as well as its evolution with redshift.
    The data remain consistent with a merger rate that does not evolve with redshift, but prefer a rate that increases with redshift (\peakNoAugEvolutionLambdaPGrZero\% credibility). 
    Using the \textsc{Power Law} redshift evolution model of Section~\ref{models}, we find that the merger rate evolves slower than the naive expectation of $(1+z)^{2.7}$ from the \added{local ($z \lesssim 1$)} star formation rate~\citep[SFR;][]{MadauDickinson} at \peakNoAugEvolutionMDPercentileInLambda\% credibility.
    \item \emph{How fast do black holes spin?}
    From GWTC-1, we inferred that the BH spin component aligned with the orbital angular momentum is typically small~\citep{2016PhRvD..94f4035A,Farr:2017uvj, BFarrSpin, Tiwari:2018qch,O2pop, Wysocki2019, 2019JCAP...08..022F,Roulet:2019, Miller2020,Roulet:2020}. Among the GWTC-1 events, GW151226 is the only event to exhibit unambiguous signs of spin~\citep{GW151226,Vitale2017,Kimball}, \added{while a few other events, including GW170729, show a mild preference for spin~\citep{2019PhRvD.100j4015C}}.
    We were unable to determine if this typical smallness was because the spin vectors are misaligned, because the spin magnitudes are small, or both, although the GWTC-1 data weakly disfavors the scenario in which all spins are perfectly aligned~\citep{Farr:2017uvj, Tiwari:2018qch, O2pop, Biscoveanu:2020are}.
    With additional data, we can now say confidently that some BBH systems have spins misaligned with the orbital angular momentum. 
    A nonzero fraction of systems have measurable in-plane spin components, leading them to undergo precession of the orbital plane. 
    Additionally, \result{\fractionChiEffNegativeLow}\% to \result{\fractionChiEffNegativeHigh}\% of \bbhsys{} merge with a negative \chieff{} $\chi_\mathrm{eff}$, see Eq.~\eqref{eq:chi_eff} below, implying that some component spins are tilted by more than $90^\circ$ relative to the orbital angular momentum axis. We refer to spins tilted more than $90^\circ$ as \emph{\antialigned{} spins}.\footnote{Our definition of \antialigned{} does not imply \textit{perfect} anti-alignment (tilt angles of exactly $180^\circ$) as the phrase is sometimes used to mean in the context of waveform modelling or numerical relativity.}
    While some events identified in O3a have individually measurable nonzero spin, they occur infrequently, consistent with our previous estimates.
    We identify \result{nine} out of \result{44} BBH events in GWTC-2 with a positive \chieff{} that excludes zero at greater than 95\% credibility.\footnote{This result is obtained using a prior informed by the full population of GWTC-2 events. In particular, we employ the \textsc{Gaussian} model described in Section~\ref{models}.}
    These \result{nine} events include both massive \bbhsys{} like \result{\NAME{GW190519A}{}}, with a source  primary mass $m_1 = \massonesourcemed{GW190519A}^{+\massonesourceplus{GW190519A}}_{-\massonesourceminus{GW190519A}}\ {M_\odot}$ (90\% credibility, uniform in redshifted mass prior) and less massive \bbhsys{} like \result{\NAME{GW190728A}{}}, with $m_1 = \massonesourcemed{GW190728A}^{+\massonesourceplus{GW190728A}}_{-\massonesourceminus{GW190728A}}\ {M_\odot}$.
    No individual BBH events are observed with confidently negative \chieff{}.
    \item \emph{What is the minimum black hole mass?}
    Using GWTC-1, we previously inferred that, if there is a low-mass cut-off in the BBH mass spectrum, it is somewhere below $9\ M_\odot$~\citep{O2pop}. 
    With GWTC-2, we tighten the constraints on the minimum BH mass in a BBH system, finding $\mmin < \truncatedNoAugNoEvolutionMminUpperNinety \ M_\odot$ (90\% credibility). Furthermore, we find that if the BH mass spectrum extends down to $3 \ M_\odot$, it likely turns over at $\sim \peakNoAugNoEvolutionTurnoverMass \ M_\odot$.
    This suggests that there may be a dearth of systems between NS and BH masses~\citep{Fishbach:2020ryj}.
    However, the O3a observation of GW190814~\citep{GW190814}, with a secondary mass $m_2 = \masstwosourcemed{GW190814A}^{+\masstwosourceplus{GW190814A}}_{-\masstwosourceminus{GW190814A}}\ {M_\odot}$, complicates this picture.
    If the secondary mass is a BH, it would indicate that the BH mass spectrum extends below $3 \ M_\odot$, to much lower masses than exhibited by the Galactic \added{X-ray binary} population~(\citealt{Bailyn1998, Ozel2011, Farr2011}, but see also \citealt{Kreidberg:2012ud, Thompson, Mandel:2020cig}).
    Alternatively, the secondary object in GW190814 could be an NS \replaced{
    Although, unless it were significantly spinning, this may be in tension with constraints in the maximum NS mass}{, but it would likely have to be significantly spinning to satisfy constraints on the maximum NS mass}~\citep{GW190814,Most:2020bba,Tan:2020ics,Essick:2020ghc,Tews:2020ylw, Zhang:2020zsc}.
    Either way, we find that GW190814 is an outlier with respect to the BBH population and the models we consider in this work. 
    Unless stated otherwise, when presenting results on the BBH population, we exclude GW190814.
\end{enumerate}

The remainder of this paper is organized as follows.
In Section~\ref{data} we describe the data used in this study and detail how we select events for analysis.
In Section~\ref{models}, we provide a high-level overview of our models for the distributions of binary mass, spins, and redshift.
In Section~\ref{method}, we describe the hierarchical method used to fit population models to data.
In Section~\ref{discussion}, we present key results and discuss the astrophysical implications.
Concluding remarks are provided in Section~\ref{sec:conclusion}.
The Appendix provides \added{additional details regarding the statistical method (Appendix~\ref{Appendix:xi}) } and descriptions of models (Appendix~\ref{details},~\ref{sec:spindetails},~\ref{Appendix:redshift}), outlier analyses and model checking (Appendix~\ref{Appendix:massmodelchecking}), and other supplementary material, including a discussion of gravitational lensing (Appendix~\ref{other}).
\added{
  The data release for this paper is available online in \cite{P2000434}.
}

\section{Data and event selection}\label{data}
Searches for gravitational wave transients in the O3a data identified $\result{39}$ candidate events with FAR below $\unit[2]{yr^{-1}}$~\citep{O3acatalog}.
\added{
  This FAR cut excludes two BBH candidates and one low-significance NSBH candidate.
  It is unlikely the results of our BBH analyses would differ qualitatively with the inclusion of these two additional \bbhevnt{} because they are typical of other, more confident GWTC-2 detections, and including them would lead to a modest 5\% increase in the catalog size.
}
The event list was collated by combining results from several pipelines searching for compact binary mergers using archival data. 
The search pipelines include \texttt{GstLAL}~\citep{Sachdev:2019vvd, Hanna:2019ezx,Messick:2016aqy} and \texttt{PyCBC}~\citep{PyCBC1,PyCBC2,PyCBC3,PyCBC4,PyCBC5,PyCBC6}, which use template-based matched filtering techniques, and \texttt{cWB}~\citep{Klimenko:2004qh,Klimenko:2015ypf}, which uses a wavelet-based search that does not assume a physically parameterized signal model.
These pipelines, along with two additional pipelines, \texttt{MBTA}~\citep{MBTA}, and \texttt{SPIIR}~\citep{spiir}, recovered most of the event candidates in low-latency.

For the population studies presented here, the event list is further restricted to the \result{36} events with
$\text{FAR}<\unit[1]{yr^{-1}}$ as a means to increase the purity of the sample. 
This FAR threshold excludes the lower-significance triggers \NAME{GW190426A}{} (a potential NSBH or BBH candidate), \NAME{GW190719A}{} and \NAME{GW190909A}{} that appear in \cite{O3acatalog}.
At this FAR threshold, we expect $\sim$1 noise event, and therefore a contamination fraction of $\lesssim 3\%$.\footnote{In this estimate, we do not include a trial factor penalty for the fact that we look for candidates with multiple pipelines.}
In this work we assume that all of the event candidates that meet this FAR threshold are of astrophysical origin.
For the population analysis of the GWTC-1 \bbhevnt{}, the selection criteria used for inclusion in the study is the FAR and the probability $p_\text{astro}$ that the source is of astrophysical origin. This value was only computed for BBHs with $\text{FAR}<\unit[2]{yr^{-1}}$ in \cite{O3acatalog}, so we
simplify our criterion to only select on FAR. The set of GWTC-1 events that pass this FAR threshold is identical to the previous set chosen by FAR and $p_\text{astro}$.
Therefore, while the selection criteria here are different from our GWTC-1 analysis, the analyzed events are consistent.

In addition to the \result{36} events from O3a which passed the FAR threshold, the 11 detections presented
in GWTC-1~\citep{GWTC1} are also included in the event list used here to infer properties of the underlying population.
All \result{47} events used in this analysis are tabulated in Table~\ref{tab:events}.
Of these systems, \result{44} have both masses confidently in the BH mass range, with $m_1 \geq m_2 > 3 \ M_\odot$.
Unless stated otherwise, we restrict BBH population results to these \result{44} (see also Appendix~\ref{Appendix:GW190814}).
Since our statistical framework relies on accurately quantifying the selection effects of our search, we only include events identified in GWTC-2, for which we have measured the search sensitivity; see Appendix~\ref{Appendix:xi}.
In particular, the event list does not include candidates identified by independent analyses~\citep{Zackay:2019btq, Venumadhav:2019lyq, Venumadhav:2019tad,
Zackay:2019tzo, Nitz:2018imz, Nitz:2019hdf, Magee:2019vmb} of the publicly released LIGO and Virgo data~\citep{Abbott:2019ebz,Trovato:2019liz}.
\cite{Galaudage} and \cite{Roulet:2020} suggest that our results are unlikely to change significantly with the inclusion of these events, and in the future it may be possible to include events from independent groups using a unified framework for the calculation of significance~\citep{Ashton,Pratten:2020ruz, Roulet:2020}.

Parameter estimation results for each candidate event~\citep{O3acatalog} were obtained using the \texttt{LALInference}~\citep{lalinference,lalsuite}, \texttt{RIFT}~\citep{Lange:2018pyp, Wysocki_2019}, and \texttt{Bilby}~\citep{bilby,bilby_gwtc1} pipelines, the latter employing the \texttt{dynesty} nested sampling tool~\citep{dynesty}.
The parameter estimation pipelines use Bayesian sampling methods to produce fair draws from the posterior distribution function of the source parameters, conditioned on the data and given a models for the signal and noise~\citep{GW150914PE}.

For the BBH events previously published in GWTC-1, we use the published posterior samples inferred using the \textsc{IMRPhenomPv2}~\citep{Hannam:2013oca,Khan:2015jqa,Husa:2015iqa, Bohe} and \textsc{SEOBNRv3}~\citep{Pan:2013rra,Taracchini:2013rva,Babak:2016tgq} waveform models.
\result{For the new BBH events of GWTC-2, we use waveform approximants that include higher-order multipole content, including  the \textsc{IMRPhenomPv3HM}~\citep{Khan2020_IMRPhenomPv3HM}, \textsc{NRSur7dq4}~\citep{Varma2019_NRSur7dq4} and \textsc{SEOBNRv4PHM}~\citep{Bohe:2016gbl,Ossokine:2020kjp}.}
For all events, we average over these waveform families, in contrast to our previously published parameter estimation results for GW190521, which highlighted results from one waveform, \textsc{NRSur7dq4}~\citep{GW190521}.
A more complete description of the parameter estimation methods and waveform models used can be found in Section~V of~\cite{O3acatalog}.

\begin{table*}[t]
\renewcommand{\arraystretch}{0.95}
    \centering
    \begin{tabular}{l  l  l  r }
        \hline
        {\bf Event} & $\boldsymbol{m_1} [M_{\odot}]$ & $\boldsymbol{m_2} [M_{\odot}]$ & {\bf FAR} [yr$^{-1}$] \\
        \hline\hline
        GW150914 &  $\massonesourcemedA{GW150914}^{+\massonesourceplusA{GW150914}}_{-\massonesourceminusA{GW150914}}$ & $\masstwosourcemedA{GW150914}^{+\masstwosourceplusA{GW150914}}_{-\masstwosourceminusA{GW150914}}$ & {$<$ $1.0 \times 10^{-7}$} \\
        GW151012 & $\massonesourcemedA{GW151012}^{+\massonesourceplusA{GW151012}}_{-\massonesourceminusA{GW151012}}$ & $\masstwosourcemedA{GW151012}^{+\masstwosourceplusA{GW151012}}_{-\masstwosourceminusA{GW151012}}$ & $7.9 \times 10^{-3}$ \\
	GW151226 & $\massonesourcemedA{GW151226}^{+\massonesourceplusA{GW151226}}_{-\massonesourceminusA{GW151226}}$ & $\masstwosourcemedA{GW151226}^{+\masstwosourceplusA{GW151226}}_{-\masstwosourceminusA{GW151226}}$  & {$<$ $1.0 \times 10^{-7}$}  \\        
                GW170104 &  $\massonesourcemedA{GW170104}^{+\massonesourceplusA{GW170104}}_{-\massonesourceminusA{GW170104}}$ & $\masstwosourcemedA{GW170104}^{+\masstwosourceplusA{GW170104}}_{-\masstwosourceminusA{GW170104}}$ & {$<$ $1.0 \times 10^{-7}$}  \\
        GW170608 &  $\massonesourcemedA{GW170608}^{+\massonesourceplusA{GW170608}}_{-\massonesourceminusA{GW170608}}$ & $\masstwosourcemedA{GW170608}^{+\masstwosourceplusA{GW170608}}_{-\masstwosourceminusA{GW170608}}$ & {$<$ $1.0 \times 10^{-7}$}  \\
        GW170729 &  $\massonesourcemedA{GW170729}^{+\massonesourceplusA{GW170729}}_{-\massonesourceminusA{GW170729}}$ & $\masstwosourcemedA{GW170729}^{+\masstwosourceplusA{GW170729}}_{-\masstwosourceminusA{GW170729}}$ & {$2.0 \times 10^{-2}$}  \\
        GW170809 &  $\massonesourcemedA{GW170809}^{+\massonesourceplusA{GW170809}}_{-\massonesourceminusA{GW170809}}$ & $\masstwosourcemedA{GW170809}^{+\masstwosourceplusA{GW170809}}_{-\masstwosourceminusA{GW170809}}$ & {$<$ $1.0 \times 10^{-7}$}  \\
        GW170814 &  $\massonesourcemedA{GW170814}^{+\massonesourceplusA{GW170814}}_{-\massonesourceminusA{GW170814}}$ & $\masstwosourcemedA{GW170814}^{+\masstwosourceplusA{GW170814}}_{-\masstwosourceminusA{GW170814}}$ & {$<$ $1.0 \times 10^{-7}$}  \\
        \hspace{-2.1mm}*\,GW170817 & $\massonesourcemedA{GW170817}^{+\massonesourceplusA{GW170817}}_{-\massonesourceminusA{GW170817}}$ & $\masstwosourcemedA{GW170817}^{+\masstwosourceplusA{GW170817}}_{-\masstwosourceminusA{GW170817}}$ & {$<$ $1.0 \times 10^{-7}$}  \\
        GW170818 & ${35.4}^{+7.5}_{-4.7}$ & ${26.7}^{+4.3}_{-5.2}$ & {$4.2 \times 10^{-5}$}  \\
        GW170823 &  $\massonesourcemedA{GW170823}^{+\massonesourceplusA{GW170823}}_{-\massonesourceminusA{GW170823}}$ & $\masstwosourcemedA{GW170823}^{+\masstwosourceplusA{GW170823}}_{-\masstwosourceminusA{GW170823}}$ & {$<$ $1.0 \times 10^{-7}$}  \\
        \NAME{GW190408A} & $\massonesourcemed{GW190408A}^{+\massonesourceplus{GW190408A}}_{-\massonesourceminus{GW190408A}}$ & $\masstwosourcemed{GW190408A}^{+\masstwosourceplus{GW190408A}}_{-\masstwosourceminus{GW190408A}}$ & $<$ \MINFAR{GW190408A} \\
        \NAME{GW190412A} & $\massonesourcemed{GW190412A}^{+\massonesourceplus{GW190412A}}_{-\massonesourceminus{GW190412A}}$ & $\masstwosourcemed{GW190412A}^{+\masstwosourceplus{GW190412A}}_{-\masstwosourceminus{GW190412A}}$ & $<$ \MINFAR{GW190412A} \\
        \NAME{GW190413A} & $\massonesourcemed{GW190413A}^{+\massonesourceplus{GW190413A}}_{-\massonesourceminus{GW190413A}}$ & $\masstwosourcemed{GW190413A}^{+\masstwosourceplus{GW190413A}}_{-\masstwosourceminus{GW190413A}}$ & \MINFAR{GW190413A} \\
        \NAME{GW190413B} & $\massonesourcemed{GW190413B}^{+\massonesourceplus{GW190413B}}_{-\massonesourceminus{GW190413B}}$ & $\masstwosourcemed{GW190413B}^{+\masstwosourceplus{GW190413B}}_{-\masstwosourceminus{GW190413B}}$ & \MINFAR{GW190413B} \\
        \NAME{GW190421A} & $\massonesourcemed{GW190421A}^{+\massonesourceplus{GW190421A}}_{-\massonesourceminus{GW190421A}}$ & $\masstwosourcemed{GW190421A}^{+\masstwosourceplus{GW190421A}}_{-\masstwosourceminus{GW190421A}}$ & \MINFAR{GW190421A} \\
        \NAME{GW190424A} & $\massonesourcemed{GW190424A}^{+\massonesourceplus{GW190424A}}_{-\massonesourceminus{GW190424A}}$ & $\masstwosourcemed{GW190424A}^{+\masstwosourceplus{GW190424A}}_{-\masstwosourceminus{GW190424A}}$ & \MINFAR{GW190424A} \\
        \hspace{-2.1mm}*\,\NAME{GW190425A} & $\massonesourcemed{GW190425A}^{+\massonesourceplus{GW190425A}}_{-\massonesourceminus{GW190425A}}$ & $\masstwosourcemed{GW190425A}^{+\masstwosourceplus{GW190425A}}_{-\masstwosourceminus{GW190425A}}$ & \MINFAR{GW190425A} \\
        \NAME{GW190503A} & $\massonesourcemed{GW190503A}^{+\massonesourceplus{GW190503A}}_{-\massonesourceminus{GW190503A}}$ & $\masstwosourcemed{GW190503A}^{+\masstwosourceplus{GW190503A}}_{-\masstwosourceminus{GW190503A}}$ & $<$ \MINFAR{GW190503A} \\
        \NAME{GW190512A} & $\massonesourcemed{GW190512A}^{+\massonesourceplus{GW190512A}}_{-\massonesourceminus{GW190512A}}$ & $\masstwosourcemed{GW190512A}^{+\masstwosourceplus{GW190512A}}_{-\masstwosourceminus{GW190512A}}$ & $<$ \MINFAR{GW190512A} \\
        \NAME{GW190513A} & $\massonesourcemed{GW190513A}^{+\massonesourceplus{GW190513A}}_{-\massonesourceminus{GW190513A}}$ & $\masstwosourcemed{GW190513A}^{+\masstwosourceplus{GW190513A}}_{-\masstwosourceminus{GW190513A}}$ & $<$ \MINFAR{GW190513A} \\
        \NAME{GW190514A} & $\massonesourcemed{GW190514A}^{+\massonesourceplus{GW190514A}}_{-\massonesourceminus{GW190514A}}$ & $\masstwosourcemed{GW190514A}^{+\masstwosourceplus{GW190514A}}_{-\masstwosourceminus{GW190514A}}$ & \MINFAR{GW190514A} \\
        \NAME{GW190517A} & $\massonesourcemed{GW190517A}^{+\massonesourceplus{GW190517A}}_{-\massonesourceminus{GW190517A}}$ & $\masstwosourcemed{GW190517A}^{+\masstwosourceplus{GW190517A}}_{-\masstwosourceminus{GW190517A}}$ & \MINFAR{GW190517A} \\
        \NAME{GW190519A} & $\massonesourcemed{GW190519A}^{+\massonesourceplus{GW190519A}}_{-\massonesourceminus{GW190519A}}$ & $\masstwosourcemed{GW190519A}^{+\masstwosourceplus{GW190519A}}_{-\masstwosourceminus{GW190519A}}$ & $<$ \MINFAR{GW190519A} \\
        \NAME{GW190521A} & $\massonesourcemed{GW190521A}^{+\massonesourceplus{GW190521A}}_{-\massonesourceminus{GW190521A}}$ & $\masstwosourcemed{GW190521A}^{+\masstwosourceplus{GW190521A}}_{-\masstwosourceminus{GW190521A}}$ & \MINFAR{GW190521A} \\
        \NAME{GW190521B} & $\massonesourcemed{GW190521B}^{+\massonesourceplus{GW190521B}}_{-\massonesourceminus{GW190521B}}$ & $\masstwosourcemed{GW190521B}^{+\masstwosourceplus{GW190521B}}_{-\masstwosourceminus{GW190521B}}$ & $<$ \MINFAR{GW190521B} \\
        \NAME{GW190527A} & $\massonesourcemed{GW190527A}^{+\massonesourceplus{GW190527A}}_{-\massonesourceminus{GW190527A}}$ & $\masstwosourcemed{GW190527A}^{+\masstwosourceplus{GW190527A}}_{-\masstwosourceminus{GW190527A}}$ & \MINFAR{GW190527A} \\
        \NAME{GW190602A} & $\massonesourcemed{GW190602A}^{+\massonesourceplus{GW190602A}}_{-\massonesourceminus{GW190602A}}$ & $\masstwosourcemed{GW190602A}^{+\masstwosourceplus{GW190602A}}_{-\masstwosourceminus{GW190602A}}$ & \MINFAR{GW190602A} \\
        \NAME{GW190620A} & $\massonesourcemed{GW190620A}^{+\massonesourceplus{GW190620A}}_{-\massonesourceminus{GW190620A}}$ & $\masstwosourcemed{GW190620A}^{+\masstwosourceplus{GW190620A}}_{-\masstwosourceminus{GW190620A}}$ & \MINFAR{GW190620A} \\
        \NAME{GW190630A} & $\massonesourcemed{GW190630A}^{+\massonesourceplus{GW190630A}}_{-\massonesourceminus{GW190630A}}$ & $\masstwosourcemed{GW190630A}^{+\masstwosourceplus{GW190630A}}_{-\masstwosourceminus{GW190630A}}$ & $<$ \MINFAR{GW190630A} \\
        \NAME{GW190701A} & $\massonesourcemed{GW190701A}^{+\massonesourceplus{GW190701A}}_{-\massonesourceminus{GW190701A}}$ & $\masstwosourcemed{GW190701A}^{+\masstwosourceplus{GW190701A}}_{-\masstwosourceminus{GW190701A}}$ & \MINFAR{GW190701A} \\
        \NAME{GW190706A} & $\massonesourcemed{GW190706A}^{+\massonesourceplus{GW190706A}}_{-\massonesourceminus{GW190706A}}$ & $\masstwosourcemed{GW190706A}^{+\masstwosourceplus{GW190706A}}_{-\masstwosourceminus{GW190706A}}$ & $<$ \MINFAR{GW190706A} \\
        \NAME{GW190707A} & $\massonesourcemed{GW190707A}^{+\massonesourceplus{GW190707A}}_{-\massonesourceminus{GW190707A}}$ & $\masstwosourcemed{GW190707A}^{+\masstwosourceplus{GW190707A}}_{-\masstwosourceminus{GW190707A}}$ & $<$ \MINFAR{GW190707A} \\
        \NAME{GW190708A} & $\massonesourcemed{GW190708A}^{+\massonesourceplus{GW190708A}}_{-\massonesourceminus{GW190708A}}$ & $\masstwosourcemed{GW190708A}^{+\masstwosourceplus{GW190708A}}_{-\masstwosourceminus{GW190708A}}$ & \MINFAR{GW190708A} \\
        \NAME{GW190720A} & $\massonesourcemed{GW190720A}^{+\massonesourceplus{GW190720A}}_{-\massonesourceminus{GW190720A}}$ & $\masstwosourcemed{GW190720A}^{+\masstwosourceplus{GW190720A}}_{-\masstwosourceminus{GW190720A}}$ & $<$ \MINFAR{GW190720A} \\
        \NAME{GW190727A} & $\massonesourcemed{GW190727A}^{+\massonesourceplus{GW190727A}}_{-\massonesourceminus{GW190727A}}$ & $\masstwosourcemed{GW190727A}^{+\masstwosourceplus{GW190727A}}_{-\masstwosourceminus{GW190727A}}$ & $<$ \MINFAR{GW190727A} \\
        \NAME{GW190728A} & $\massonesourcemed{GW190728A}^{+\massonesourceplus{GW190728A}}_{-\massonesourceminus{GW190728A}}$ & $\masstwosourcemed{GW190728A}^{+\masstwosourceplus{GW190728A}}_{-\masstwosourceminus{GW190728A}}$ & $<$ \MINFAR{GW190728A} \\
        \NAME{GW190731A} & $\massonesourcemed{GW190731A}^{+\massonesourceplus{GW190731A}}_{-\massonesourceminus{GW190731A}}$ & $\masstwosourcemed{GW190731A}^{+\masstwosourceplus{GW190731A}}_{-\masstwosourceminus{GW190731A}}$ & \MINFAR{GW190731A} \\
        \NAME{GW190803A} & $\massonesourcemed{GW190803A}^{+\massonesourceplus{GW190803A}}_{-\massonesourceminus{GW190803A}}$ & $\masstwosourcemed{GW190803A}^{+\masstwosourceplus{GW190803A}}_{-\masstwosourceminus{GW190803A}}$ & \MINFAR{GW190803A} \\
        \hspace{-2mm}\dag\,\NAME{GW190814A} & $\massonesourcemed{GW190814A}^{+\massonesourceplus{GW190814A}}_{-\massonesourceminus{GW190814A}}$ & $\masstwosourcemed{GW190814A}^{+\masstwosourceplus{GW190814A}}_{-\masstwosourceminus{GW190814A}}$ & $<$ \MINFAR{GW190814A} \\
        \NAME{GW190828A} & $\massonesourcemed{GW190828A}^{+\massonesourceplus{GW190828A}}_{-\massonesourceminus{GW190828A}}$ & $\masstwosourcemed{GW190828A}^{+\masstwosourceplus{GW190828A}}_{-\masstwosourceminus{GW190828A}}$ & $<$ \MINFAR{GW190828A} \\
        \NAME{GW190828B} & $\massonesourcemed{GW190828B}^{+\massonesourceplus{GW190828B}}_{-\massonesourceminus{GW190828B}}$ & $\masstwosourcemed{GW190828B}^{+\masstwosourceplus{GW190828B}}_{-\masstwosourceminus{GW190828B}}$ & $<$ \MINFAR{GW190828B} \\
        \NAME{GW190910A} & $\massonesourcemed{GW190910A}^{+\massonesourceplus{GW190910A}}_{-\massonesourceminus{GW190910A}}$ & $\masstwosourcemed{GW190910A}^{+\masstwosourceplus{GW190910A}}_{-\masstwosourceminus{GW190910A}}$& \MINFAR{GW190910A} \\
        \NAME{GW190915A} &$\massonesourcemed{GW190915A}^{+\massonesourceplus{GW190915A}}_{-\massonesourceminus{GW190915A}}$ & $\masstwosourcemed{GW190915A}^{+\masstwosourceplus{GW190915A}}_{-\masstwosourceminus{GW190915A}}$ & $<$ \MINFAR{GW190915A} \\
        \NAME{GW190924A} & $\massonesourcemed{GW190924A}^{+\massonesourceplus{GW190924A}}_{-\massonesourceminus{GW190924A}}$ & $\masstwosourcemed{GW190924A}^{+\masstwosourceplus{GW190924A}}_{-\masstwosourceminus{GW190924A}}$ & $<$ \MINFAR{GW190924A} \\
        \NAME{GW190929A} & $\massonesourcemed{GW190929A}^{+\massonesourceplus{GW190929A}}_{-\massonesourceminus{GW190929A}}$ & $\masstwosourcemed{GW190929A}^{+\masstwosourceplus{GW190929A}}_{-\masstwosourceminus{GW190929A}}$ & \MINFAR{GW190929A} \\
        \NAME{GW190930A} & $\massonesourcemed{GW190930A}^{+\massonesourceplus{GW190930A}}_{-\massonesourceminus{GW190930A}}$ & $\masstwosourcemed{GW190930A}^{+\masstwosourceplus{GW190930A}}_{-\masstwosourceminus{GW190930A}}$ & \MINFAR{GW190930A} \\
        \hline
        \end{tabular}
    \caption{
        A summary of the events included in this analysis.
        Asterisks (*) denote binaries in which both components lie in the NS mass range while a dagger (\dag) indicates a system in which the nature of the secondary component is unknown.
        Both of these classes are excluded from our analyses unless explicitly stated.
        From left to right, the columns show the event name, the 90\% credible interval for primary source mass (units of $M_\odot$), the 90\% credible interval for secondary 
        mass (units of $M_\odot$), and the minimum available FAR.
        For events detected in more than one CBC detection pipeline, we report the lowest of the available FAR estimate.
        These credible intervals are obtained using a prior that is uniform in redshifted component mass and comoving volume, as in Table 6 of~\citet{O3acatalog}.
    }
    \label{tab:events}
\end{table*}

\section{Population Models}\label{models}
\begin{figure*}
    \centering
    \includegraphics[width = \textwidth]{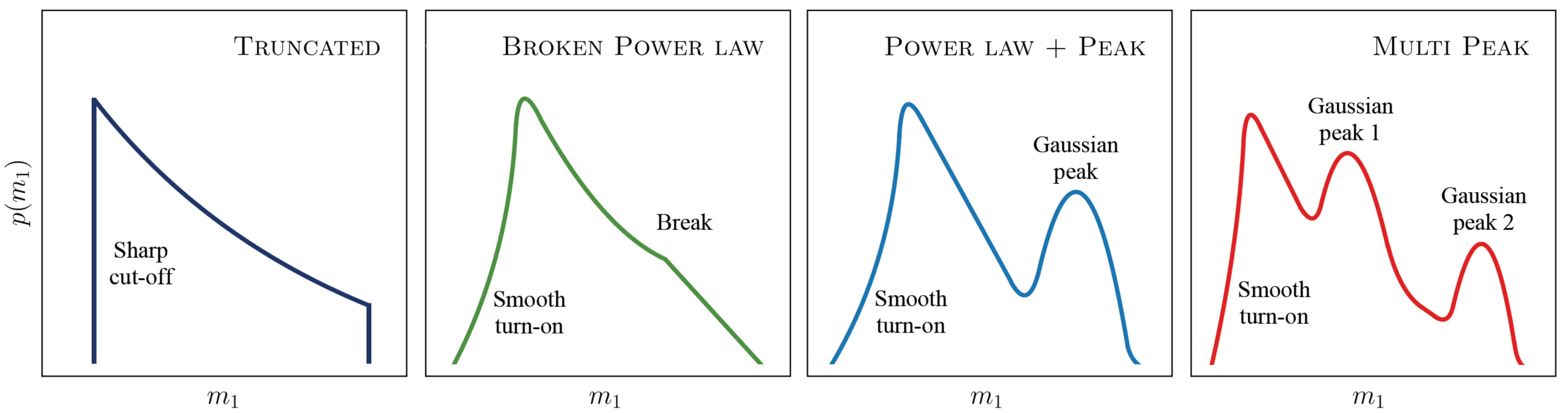}
     \caption{
     Graphical representations of the various mass distributions described in Section~\ref{mass}. \ModelH{}, a model of both mass and spin, is similar to the mass distribution of \ppsn{}, with a sharp lower mass cutoff rather than the smooth low mass turn-on.
     }
     \label{fig:thumbs}
\end{figure*}

In this section, we provide a high-level overview of the different population models used in this paper.
Each model is given a nickname indicated in boldface.
There are three categories of models: models for the shape of the mass distribution (Section~\ref{mass}), models for the spin distribution (Section~\ref{spin}), and models for the redshift distribution (Section~\ref{redshift}).
Readers interested in the astrophysical results may wish to skip ahead to Section~\ref{discussion}.
In Fig.~\ref{fig:thumbs}, we provide graphical representations of each mass model described below.
Additional details about each model are provided in Appendix~\ref{details}, including their mathematical definitions, lists of hyper-parameters, and their associated prior distributions.

\subsection{Black hole mass distribution}\label{mass}
\begin{itemize}
    \item \textbf{\truncated{}} (4~parameters; Appendix~\ref{truncated}).
    Our simplest mass model, the distribution of primary masses is a power-law with hard cut-offs at both low ($m_\mathrm{min}$) and high ($m_\mathrm{max}$) masses.
    The high-mass cut-off corresponds to the lower edge of the theorized pair-instability gap in the BH mass spectrum~\citep{Woosley2015,Heger2002,Heger2003}.
    The mass ratio distribution is also assumed to be a power law~\citep{Kovetz,Fishbach2017}.
    In~\cite{O2pop}, it is referred to as ``Model~B.''
    This model struggles to accommodate high-mass events like \NAME{GW190521A}, necessitating more complicated models.
    \item \textbf{\ppsn{}} (8~parameters; Appendix~\ref{ppsn}).
    Similar to the \truncated{} model, but with two modifications.
    At low masses we implement a smoothing function to avoid a hard cut-off.
    At high masses, we include a Gaussian peak, initially introduced to model a pile-up from pulsational pair-instability supernovae~\citep{Talbot2018}.
    This model is better able to accommodate the high-mass events that pose a challenge for the \truncated{} model.
    In~\cite{O2pop}, it is referred to as ``Model~C.''
    \item \textbf{\tapered} (7~parameters; Appendix~\ref{tapered}).
    The same as \truncated{} except, instead of a single power-law between $m_\mathrm{min}$ and $m_\mathrm{max}$, the model allows for a break in the power-law at some mass $m_\mathrm{break}$. 
    This model includes the low-mass smoothing function used in the \ppsn{} model.
    The high mass feature $m_\mathrm{break}$ may correspond to the onset of pair-instability, and the second power-law can be thought of as either a gradual tapering off, or a subpopulation of BHs within the pair-instability mass gap.
    This model accommodates the high-mass events that pose a challenge for the \truncated{} while providing an alternative to the \ppsn{} model.
    \item \textbf{\multipeak} (11~parameters; Appendix~\ref{Multi-Peak}).
    This phenomenological model is similar to \ppsn{} except we allow for two peaks.
    The resulting mass distribution can accommodate hierarchical BBH mergers~\citep{2002MNRAS.330..232C,2004ApJ...616..221G,2016ApJ...831..187A,Rodriguez2019}, in which second-generation mergers create a second high-mass peak in the BH mass spectrum.
    We use this model to test if GWTC-2 exhibits evidence for hierarchical mergers.
\end{itemize}

\subsection{Spin Distribution}\label{spin}
\begin{itemize}
    \item \textbf{\textsc{Default}} (4 parameters; Appendix~\ref{sec:default}).
    This parameterization for the component BH spin magnitudes and tilts was previously used to explore the spin distribution of compact binaries in GWTC-1~\citep{O2pop}. 
    The spin of each component BH in a binary is assumed to be independently drawn from the same underlying distribution.
    The dimensionless spin magnitude is described using a beta distribution as in~\cite{Wysocki2019}.
    The spin tilt distribution from~\cite{Talbot2017} is a mixture model comprising two components: an isotropic component designed to model dynamically assembled binaries, and a second component in which the spins are preferentially aligned with the orbital angular momentum, as expected for isolated field binaries; \added{the fraction of binaries in the purely-aligned sub-population is denoted $\zeta$}.\footnote{Throughout the paper, spin tilt is measured at a reference frequency of $\unit[20]{Hz}$ for all events except \NAME{GW190521A}{}, for which the spin tilt is measured at $\unit[11]{Hz}$ (see discussion in~\citealp{O3acatalog}). We verified that for \NAME{GW190521A}{}, the difference between the spin measurements at $\unit[20]{Hz}$ and $\unit[11]{Hz}$ are smaller than the systematic uncertainty between the waveform models.}
    For this latter component, the spin tilt angles are distributed as a truncated Gaussian, \added{with width $\sigma_t$}, that peaks when the BH spin is aligned to the orbital angular momentum.
    We use this model in concert with the mass models described above.
    \item \textbf{\textsc{Gaussian}} (5 parameters; Appendix~\ref{sec:Gaussian}). 
    While the \textsc{Default} spin model is physically inspired,
    this model, based on that of~\cite{Miller2020}, allows us to fit the distribution of phenomenological spin parameters $\chi_\mathrm{eff}$ (the \chieff{}, Eq.~\ref{eq:chi_eff}) and $\chi_\mathrm{p}$ (the \chip{}, Eq.~\ref{eq:chi_P}), assuming that their distribution is jointly described as a bivariate Gaussian.
    The ensemble properties of $\chi_\mathrm{eff}$ and $\chi_\mathrm{p}$ allow us to conclude that the BBHs in GWTC-2 exhibit general relativistic spin-induced precession of the orbital plane ($\chi_\mathrm{p} > 0$), and that some systems have component spins misaligned by more than $90^\circ$ ($\chi_\mathrm{eff} < 0$) relative to the orbital angular momentum.
    \item \textbf{\ModelH{}} (12 spin parameters, 10 mass parameters; Appendix~\ref{modelH}).
    This model allows for multiple subpopulations of \bbhsys{} with distinct mass and spin distributions.
    Specifically, this model assumes a \truncated{} power-law mass distribution with the additional presence of a 2-D Gaussian subpopulation in $m_1$ and $m_2$, truncated such that $m_1 \geq m_2$.
    While similar to the \ppsn{} mass model, there is no smooth turn on and the mass ratio distribution is allowed to differ between each subpopulation.
    Most importantly, the two subpopulations have independently parameterized \textsc{Default} spin distributions.
    We use this model to test whether the BBH spin distribution varies as a function of mass as expected if higher-mass systems are the products of hierarchical mergers.
\end{itemize}

\subsection{Redshift evolution}\label{redshift}
\begin{itemize}
    \item \textbf{\textsc{Non-Evolving}} (0 parameters).
    Our default model posits that the merger rate is uniform in comoving volume.
    \item \textbf{\textsc{Power-law Evolution}} (1 parameter; Appendix~\ref{Appendix:redshift}).
    Following~\cite{Fishbach2018}, the merger rate density is described by a power-law in $(1+z)$ where $z$ is redshift.
    Given the finite range of Advanced LIGO and Advanced Virgo to BBH mergers, we only expect to constrain the redshift evolution at redshifts $z \lesssim 1$~\citep{ObsProspects}.
    The farthest event in our analysis is likely \NAME{\luminositydistancemostsecond}{}, at redshift
$z = \redshiftmed{\luminositydistancemostsecond}_
{-\redshiftminus{\luminositydistancemostsecond}}^
{+\redshiftplus{\luminositydistancemostsecond}}$.
\end{itemize}

\section{Method}\label{method}
We adopt a hierarchical Bayesian approach, marginalizing over the properties of individual events to measure parameters of the population models described above; see, e.g.,~\citep{Thrane2019,Mandel2019,Vitale:2020aaz}.
Given data $\{d_i\}$ from $N_\mathrm{det}$ gravitational-wave detections, the likelihood of the data given population parameters $\Lambda$ is~\citep{Loredo2004,Mandel2019,Thrane2019}
\begin{equation}
    \label{eq:generic-likelihood}
    {\cal L}(\{d\}|\Lambda, N)
        \propto N^{N_\mathrm{det}} e^{-N\xi(\Lambda)} \prod_{i=1}^{N_\mathrm{det}} \int {\cal L}(d_i|\theta) \pi(\theta|\Lambda) d\theta.
\end{equation}
Here, $N$ is the total number of events expected during the observation period (including both resolvable and unresolvable signals).
Each event is described by a set of parameters $\theta$, the likelihood of which is $\mathcal{L}(d|\theta)$.
The conditional prior $\pi(\theta|\Lambda)$, meanwhile, is defined by the mass, spin, and redshift models described above in Sec.~\ref{models}.
It serves to quantify the predicted distribution of event parameters $\theta$ given the hyper-parameters $\Lambda$ of the population model.
An example of a hyper-parameter is the power-law index $\alpha$ governing primary mass distribution $\pi(m_1|\alpha)\propto m_1^{-\alpha}$ for the \truncated{} mass model.
Finally, $\xi(\Lambda)$ is the fraction of binaries that we expect to successfully detect for a population described by hyper-parameters $\Lambda$.
The procedure for calculating $\xi(\Lambda)$ is described in Appendix~\ref{Appendix:xi}.
One of our primary goals in this paper is to constrain the population hyper-parameters describing the distribution of gravitational-wave signals.
Given a log-uniform prior on $N$, one can marginalize Eq.~\eqref{eq:generic-likelihood} over $N$ to obtain~\citep{Fishbach2018,Mandel2019,Vitale:2020aaz}
\begin{equation}
    \label{eq:generic-likelihood-marginalized}
    {\cal L}(\{d\}|\Lambda)
        \propto \prod_{i=1}^{N_\mathrm{det}} \frac{\int {\cal L}(d_i|\theta) \pi(\theta|\Lambda) d\theta}{\xi(\Lambda)} .
\end{equation}

To evaluate the single-event likelihood $\mathcal{L}(d_i \mid \theta)$, we use posterior samples that are obtained using some default prior $\pi_\varnothing(\theta)$.
In this case, integrals over the likelihood can be replaced with weighted averages ``$\langle \ldots \rangle$'' over discrete samples.
Equation~\eqref{eq:generic-likelihood-marginalized}, for example, becomes
    \begin{equation}
    \mathcal{L}(\{d\}|\Lambda)
        \propto \prod_{i=1}^{N_\mathrm{det}} \frac{1}{\xi(\Lambda)} \bigg\langle \frac{\pi(\theta|\Lambda)}{\pi_\varnothing(\theta)}\bigg\rangle_\mathrm{samples},
    \end{equation}
where the factor of $\pi_\varnothing(\theta)$ serves to divide out the prior used for initial parameter estimation.
We evaluate the likelihoods for population models using the \texttt{emcee}, \texttt{dynesty}, and \texttt{stan} packages~\citep{emcee,dynesty,stan,pystan}.
The likelihoods are implemented in a variety of software including \texttt{GWPopulation}~\citep{gwpopulation} and \texttt{PopModels}~\citep{RITpop}.
The priors adopted for each of our hyper-parameters are described in Appendix~\ref{details}.

Throughout this paper, we find it useful to distinguish between the \astrophysical{} distribution of a parameter---the distribution as it is in nature---and the \observed{} distribution of a parameter---the distribution as it appears among detected events due to selection effects.
The \textit{\appd{}} for a given model represents our best guess for the \astrophysical{} distribution of some source parameter $\theta$, averaged over the posterior for population parameters $\Lambda$:
\begin{align}
    p_\Lambda(\theta) = \int d\Lambda \, 
    \pi(\theta|\Lambda) p(\Lambda | \{d\}) .
\end{align}
The subscript $\Lambda$ indicates that we have marginalised over population parameters.
Meanwhile, the \textit{\oppd{}} refers to the population-averaged distribution of source parameters $\theta$ {\it conditioned on detection}.

\added{
In Appendix~\ref{Appendix:massmodelchecking}, \ref{sec:spindetails} and \ref{Appendix:redshift}, we provide posterior predictive checks for our population models. These checks consist of comparing simulated sets of $N_\mathrm{obs}$ ``predicted" and ``observed" events. For every sample in the model hyper-posterior, we generate a set of predicted events by reweighting our found injections to the population model, and drawing $N_\mathrm{obs}$ synthetic events. For each hyper-posterior sample, we then generate a set of observed events by reweighting the single-event posteriors of the $N_\mathrm{obs}$ events to this population prior, and drawing one sample per event. Therefore, the  inferred distribution of observed events depends on the population model considered. The first example of such a posterior predictive check is shown in Fig.~\ref{fig:compare_cdfs}. We calculate the cumulative distribution function for each set of predicted and observed events in the model hyper-posterior, and summarize these curves with 90\% credibility bands.
The uncertainty in the predicted bands comes from uncertainty in the population hyper-posterior, as well as Poisson fluctuations, because each cumulative distribution curve consists of $N_\mathrm{obs}$ simulated events. 
}

\added{
  Phenomenological models such as we employ here can fail if their assumed form does not adequately represent reality or if priors are inappropriately chosen.
  Inferences from such models are inevitably prone to systematic error, given that the model is unlikely to be a perfect description of reality; for instance, the inferred local rate of BBH mergers is subject to a systematic error associated with our choice of model(s) for the BBH mass distribution.
  While such errors cannot be eliminated, we take steps to control and minimize their magnitude. 
  First, we carry out posterior predictive checks to make sure our models are consistent with the data.
  Next, where possible, we check for consistency between models with different assumptions, for example, looking for common features in the \ppsn{} and \tapered{} models.
  Finally, we carry out tests to make sure that our inferences are not artifacts of our model design, for example, by applying the models to simulated uninformative data.
  Additional information about these tests is provided in the Appendix.}

\section{Results and Discussion}\label{discussion}
\subsection{Mass Distribution}\label{mass_results}
In this subsection we report results obtained using the mass models described in Section~\ref{mass} (see Fig.~\ref{fig:thumbs}).\footnote{\added{In this study, we employ models in which the mass and redshift distributions factorize. This is a reasonable assumption for the $z \lesssim 1$ binaries in GWTC-2 and preliminary tests suggest our data are well-fit with this assumption. However, as more binaries are detected from ever greater distances, it will be interesting to test models that allow for the mass distribution to evolve with redshift.}}
We employ the \textsc{Default} spin model and the \textsc{Non-Evolving} redshift model.
The results shown here are marginalized over the hyperparameters of the spin distribution.

\begin{figure}
    \centering
    \includegraphics[width=0.48\textwidth]{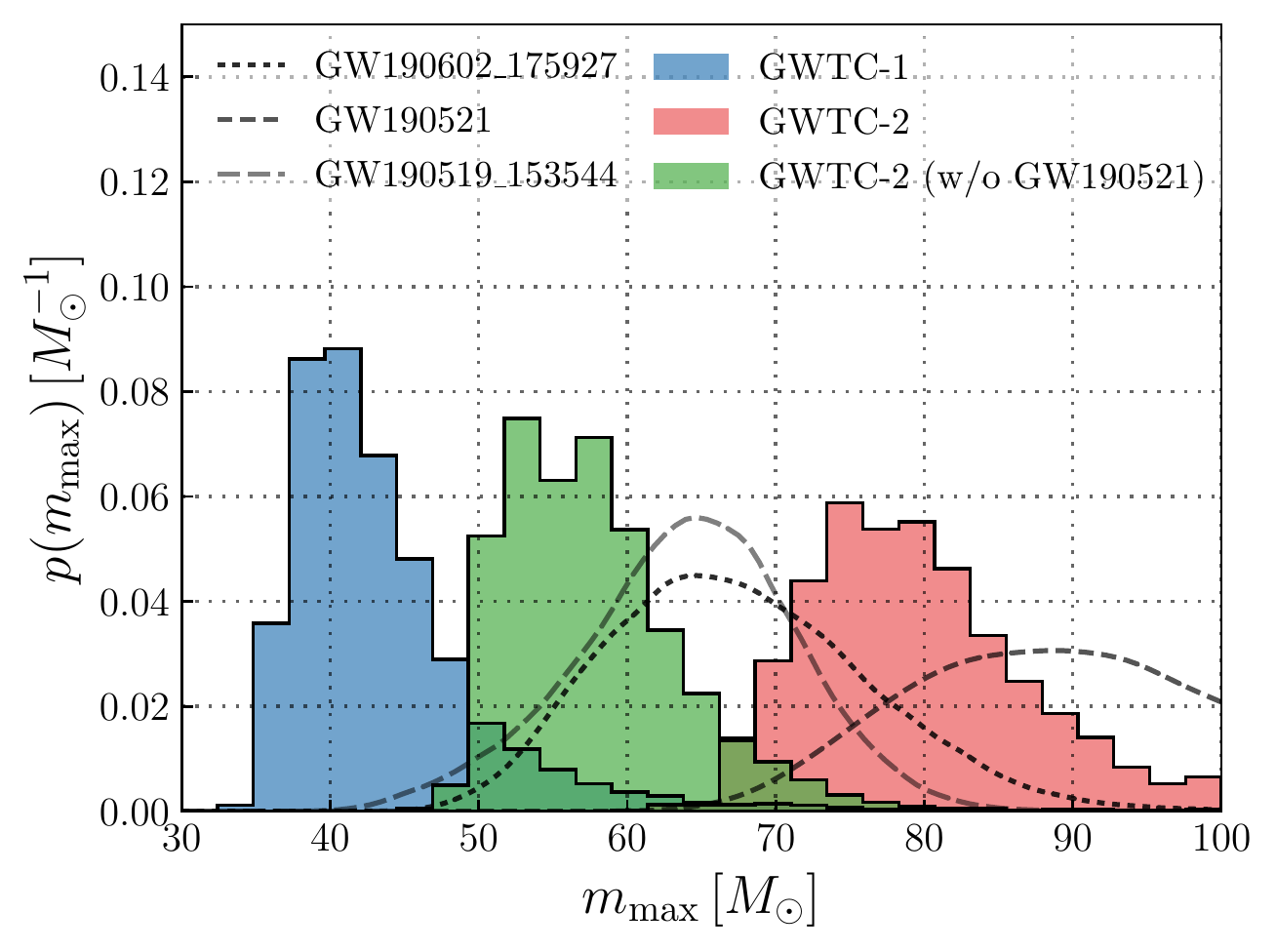}
    \caption{
    Posterior for the maximum mass using GWTC-1 and fit to the \truncated{} model (blue), compared to the posterior obtained by adding events from O3a data (red).
    The two distributions are inconsistent, suggesting the \truncated{} model is inadequate.
    The tension between GWTC-1 and GWTC-2 is somewhat alleviated by the exclusion of the high-mass event \NAME{GW190521A}{} (green).
    However, there remain several other high-mass events in O3a.
    The black dashed lines show primary mass posteriors for the three events in which $m_1>45 \ M_\odot$ at 99\% credibility (we employ a prior that is uniform in redshifted masses).
    These events cause a significant shift in the $\mmax$ posterior if we assume a simple power-law fits the data.
    }
    \label{fig:mmax_compare}
\end{figure}

\textbf{A truncated power law fails to fit the high-mass BBH events.}
The \truncated{} model applied in~\cite{O2pop} measured the sharp high-mass cutoff to be $\mmax = \mmaxModelB \ M_\odot$.
When we fit this model to GWTC-2 data, we obtain $\mmax = \truncatedNoAugNoEvolutionmmax \ M_\odot$, in significant tension ($>99\%$ credibility) with the GWTC-1 result; see Fig.~\ref{fig:mmax_compare}.
The \truncated{} model struggles to accommodate the high-mass systems of GWTC-2.
At 99\% credibility, three events of GWTC-2 have $m_1 > 45 \ M_\odot$ (regardless of whether we use a population-informed prior;~\citealp{Fishbach:2019ckx}).
This tension remains (at the $>93\%$ credibility level) even when we exclude the highest-mass event \NAME{GW190521A}{}~\citep{GW190521, Abbott:GW190521_implications}.
The poor fit of the \truncated{} model is further seen in the posterior predictive check of Fig.~\ref{fig:compare_cdfs} in Appendix~\ref{Appendix:massmodelchecking}, which shows that the \truncated{} model fails to capture the relative excess of observations with $m_1 \sim 30\ M_\odot$ compared to the number of events with $m_1 \gtrsim 45 \ M_\odot$. \added{Our fit to the \truncated{} model overpredicts the number of observations with $m_1 > 45 \ M_\odot$ relative to the number of observations with $30 M_\odot < m_1 < 45 \ M_\odot$ (98\% credibility).}

The \ppsn{}, \tapered{}, and \multipeak{} models provide better fits to the shape of the mass distribution, particularly at high masses.
Although our updated fit to the mass distribution extends to higher masses than the GWTC-1 fit, we find that $\peakNoAugNoEvolutionFractionBelowPrimaryFortyFive \%$ of BBH systems have primary masses below $45 \ M_\odot$ (\ppsn{} model), consistent with the GWTC-1 estimate that 99\% of primary masses lie below $\sim45 \ M_\odot$~\citep{O2pop, Kimball}.
In Table~\ref{tab:BF}, we provide log Bayes factors ($\log_{10}{\cal B}$) comparing the mass models; each Bayes factor is measured relative to the model with the highest Bayesian evidence: \ppsn{}. 
In each case we use the \textsc{Default} spin model.
For context, $\log_{10}{\cal B} > 1.5$ is often interpreted as a strong preference for one model over another, and $\log_{10}{\cal B} > 2$ as decisive evidence~\citep{Jeffreys1961}.

While Bayes factors depend on the choice of hyper-parameter priors, it is nonetheless possible to see that the \truncated{} model is disfavored compared to the more complicated models.
\added{This inference is driven in part by the fact that--in our posterior predictive checks--94\% of the time, the \truncated{} model overpredicts the number of detections with $m_1 > 50\ M_\odot$. See Appendix~\ref{Appendix:massmodelchecking} for information about the posterior predictive checks.}
Meanwhile, there is not a strong preference for \ppsn{} over \tapered{} or \multipeak.
We currently lack the resolving power to determine whether the deviations from \truncated{} are best modeled as a break, a Gaussian peak, or two Gaussian peaks.
As a further check, we carried out a follow-up analysis using a hybrid \tapered{} + \textsc{peak} model, which indicated only modest support for a peak on top of the \tapered{} distribution ($\log_{10}{\cal B} = \result{0.79}$).

\begin{table}[t]
  \centering
  \begin{tabular}{p{5.0cm}  c r }
    \hline
    {\bf Mass model} & ${\cal B}$ & \textbf{$\log_{10}{\cal B}$}  \\\hline\hline
    \ppsn{} & \peakNoAugNoEvolutionBF & \peakNoAugNoEvolutionLogBF \\
    \multipeak{} & \multipeakNoAugNoEvolutionBF & \multipeakNoAugNoEvolutionLogBF \\
    \tapered{} & \BPLNoAugNoEvolutionBF & \BPLNoAugNoEvolutionLogBF \\    
    \truncated{} & \truncatedNoAugNoEvolutionBF & \truncatedNoAugNoEvolutionLogBF \\
    \hline
    \ppsn{} ($\delta_m = 0$) & \peakNoAugNoEvolutionNoSmoothingBF & \peakNoAugNoEvolutionNoSmoothingLogBF \\
    \textsc{Broken Power Law + Peak} & \BPLPeakNoAugNoEvolutionBF & \BPLPeakNoAugNoEvolutionLogBF \\
    \tapered{} ($\delta_m = 0$) & \BPLNoAugNoEvolutionNoSmoothingBF & \BPLNoAugNoEvolutionNoSmoothingLogBF \\
    \ppsn{} ($\lambda_\text{peak} = 0$) & \peakNoAugNoEvolutionNoPeakBF & \peakNoAugNoEvolutionNoPeakLogBF \\
    \hline
  \end{tabular}
  \caption{
  Bayes factors for each mass model relative to the favored \ppsn{} model, which gives the highest Bayesian evidence for GWTC-2. For models that have a smooth turn on at low masses parameterized by $\delta_m$, we also compare the corresponding sub-model with a sharp minimum mass cutoff ($\delta_m = 0$). For the \ppsn{} model which includes a fraction $\lambda_\mathrm{peak}$ of systems in the Gaussian component, we compare the sub-model with $\lambda_\mathrm{peak} = 0$. GW190814 is excluded from this analysis.
  }
  \label{tab:BF}
\end{table}

\begin{table*}
  \centering
  \begin{tabular}{l c c c }
    \hline
    {\bf Events} & {\bf Mass model} & {$ \bf m_{1\%}$} ($M_\odot$) & {$ \bf m_{99\%}$} ($M_\odot$) \\\hline\hline
    \multirow{4}{*}{All confident \bbhevnt{} - \textit{excluding} GW190814 } & \truncated{} & \truncatedNoAugNoEvolutionMOnePercentile & \truncatedNoAugNoEvolutionMNinetyNinePercentile \\
    & \tapered{} & \BPLNoAugNoEvolutionMOnePercentile & \BPLNoAugNoEvolutionMNinetyNinePercentile  \\
    & \ppsn{} & \peakNoAugNoEvolutionMOnePercentile & \peakNoAugNoEvolutionMNinetyNinePercentile \\
    & \multipeak{} & \multipeakNoAugNoEvolutionMOnePercentile & \multipeakNoAugNoEvolutionMNinetyNinePercentile \\\hline
    \multirow{4}{*}{All confident \bbhevnt{} - \textit{including} GW190814}& \truncated{} & \truncatedAllNoEvolutionMOnePercentile & \truncatedAllNoEvolutionMNinetyNinePercentile \\
    & \tapered{} & \BPLAllNoEvolutionMOnePercentile & \BPLAllNoEvolutionMNinetyNinePercentile  \\
    & \ppsn{} & \peakAllNoEvolutionMOnePercentile & \peakAllNoEvolutionMNinetyNinePercentile \\
    & \multipeak{} & \multipeakAllNoEvolutionMOnePercentile & \multipeakAllNoEvolutionMNinetyNinePercentile  \\\hline
    \multirow{4}{*}{All confident \bbhevnt{} - \textit{excluding} \NAME{GW190521A}, GW190814} & \truncated{} & \truncatedNoMayNoEvolutionMOnePercentile & \truncatedNoMayNoEvolutionMNinetyNinePercentile \\
    & \tapered{} & \BPLNoMayNoEvolutionMOnePercentile & \BPLNoMayNoEvolutionMNinetyNinePercentile \\
    & \ppsn{} & \peakNoMayNoEvolutionMOnePercentile & \peakNoMayNoEvolutionMNinetyNinePercentile  \\
    & \multipeak{} & \multipeakNoMayNoEvolutionMOnePercentile & \multipeakNoMayNoEvolutionMNinetyNinePercentile \\\hline
  \end{tabular}
  \caption{
  The $m_{1\%}$ and $m_{99\%}$ credible intervals (90\%) for various mass models and combinations of events.
  These variables are defined such that, among the \textit{astrophysical} BBH population, $1\%$ of systems have primary masses $m_1\leq m_{1\%}$, while 99\% have primary masses $m_1\leq m_{99\%}$.
  The \ppsn{}, \multipeak{}, and \tapered{} models are preferred over the \truncated{} model.
  }
  \label{tab:m1m99}
\end{table*}

\begin{figure*}[t]
    \centering
    \includegraphics[width = 0.75\textwidth]{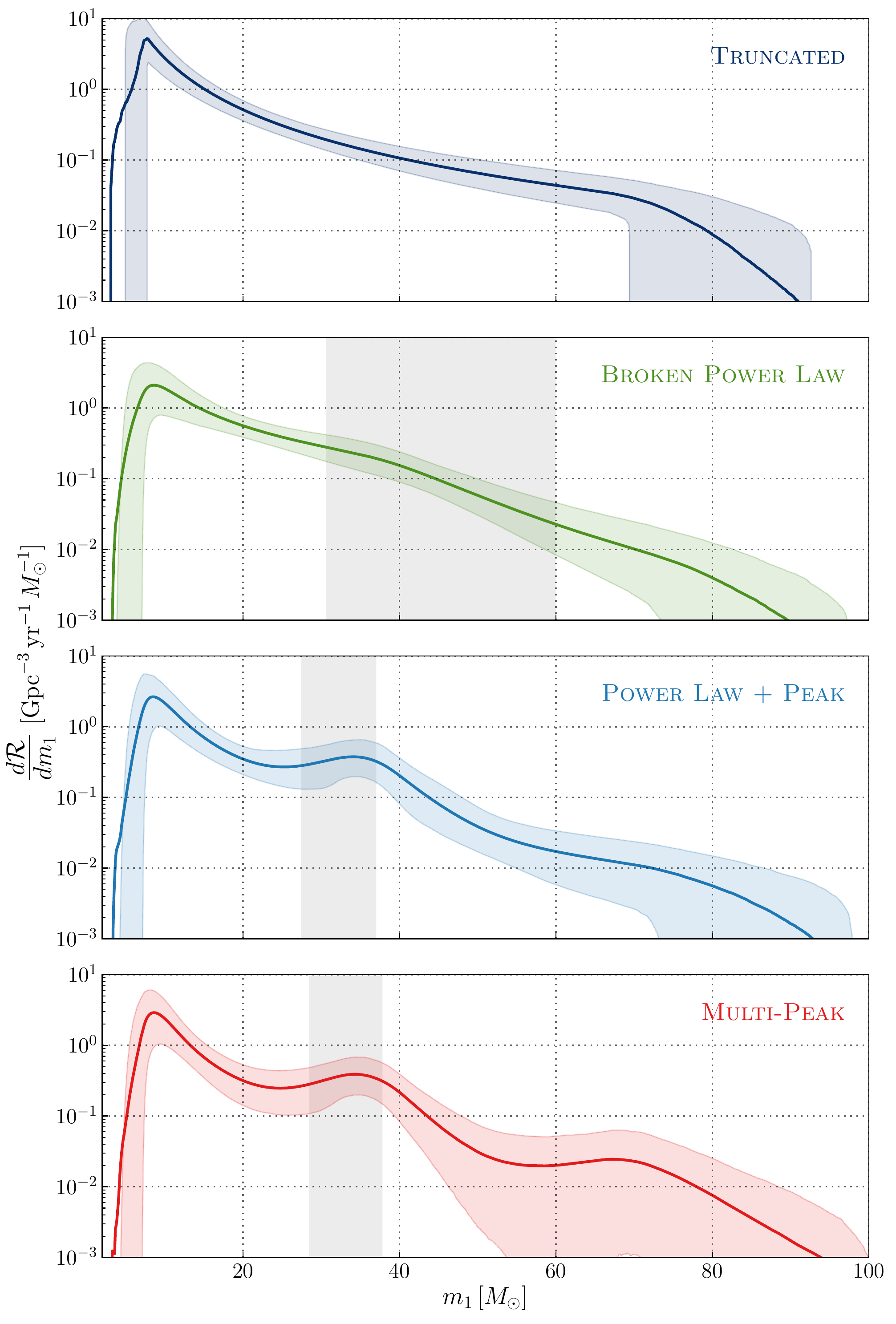}
    \caption{
    \textit{Astrophysical} primary BH mass distribution for the \truncated{}, \tapered{}, \ppsn{} and \multipeak{} models.
    The solid curve is the \appd{} (averaging over model uncertainty) while the shaded region shows the 90\% credible interval. While the median rate is always inside the credible region, the solid curve represents the \textit{mean}, which can be outside the credible region.
    Top (navy) is the \truncated{} model, second from the top (green) is the \tapered{} model, third from the top (blue) is for the \ppsn{} model, and bottom (red) is for the \multipeak{} model.
    The \truncated{} model is disfavored compared to the three latter models that predict a feature at $\sim 40 \ M_\odot$: a break in the mass spectrum in the \tapered{} model or additional Gaussian peaks in the \ppsn{} and \multipeak{} models.
    The vertical gray bands show 90\% credible bounds on the locations of these additional features.
    \label{fig:pm1_astrophysical}
}
\end{figure*}

\begin{figure*}
    \centering
    \includegraphics[width = 0.8\textwidth]{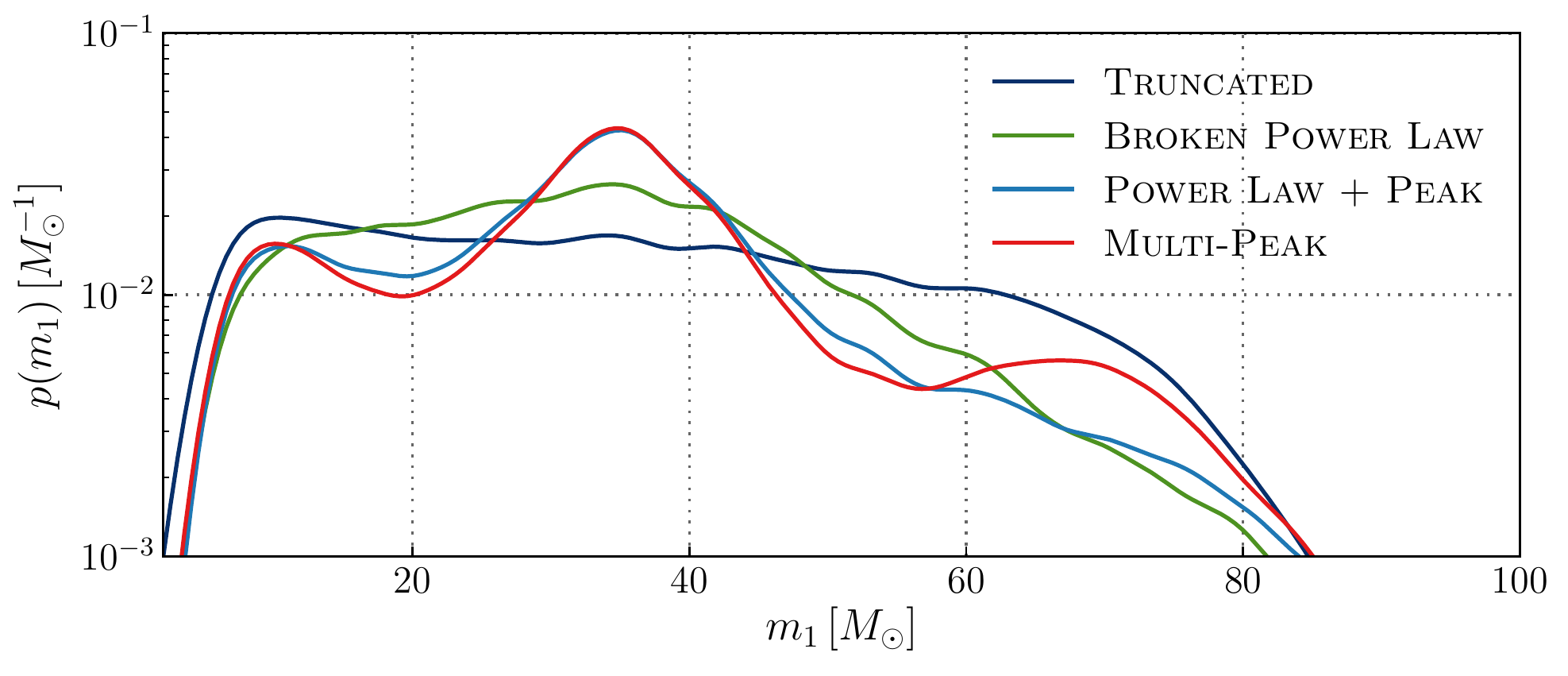}
    \caption{
    \textit{Observed} primary BH mass distributions predicted by each mass model. For each model, we average over the uncertainty in the hyper-parameter posterior.
    The \observed{} distribution describes the events successfully detected by LIGO--Virgo, preferentially favoring more massive systems relative to the astrophysical distribution due to selection effects.
    \label{fig:PPSN_observed_spectrum}
    }
\end{figure*}

\textbf{There are features in the black hole mass spectrum beyond a power-law.}
Figure~\ref{fig:pm1_astrophysical} shows the astrophysical merger rate density as a function of primary BH mass for the \truncated{}, \ppsn{}, \multipeak{} and \tapered{} models.
Figure~\ref{fig:PPSN_observed_spectrum}, meanwhile, shows the \oppd{} for primary masses, including selection effects.
Corner plots showing the constraints for the parameters in each model are available in Appendix~\ref{details}; see Figs.~\ref{fig:PPSN_corner}, \ref{fig:BPL_corner} and~\ref{fig:Multipeak_corner}. 
In Appendix~\ref{Appendix:massmodelchecking}, we show posterior predictive checks for each model (Fig.~\ref{fig:compare_cdfs} and Fig.~\ref{fig:cwb}), comparing mock observations predicted by the model to the empirical distribution inferred from GWTC-2.

\begin{figure}[h]
    \centering
    \includegraphics[width = 0.48\textwidth]{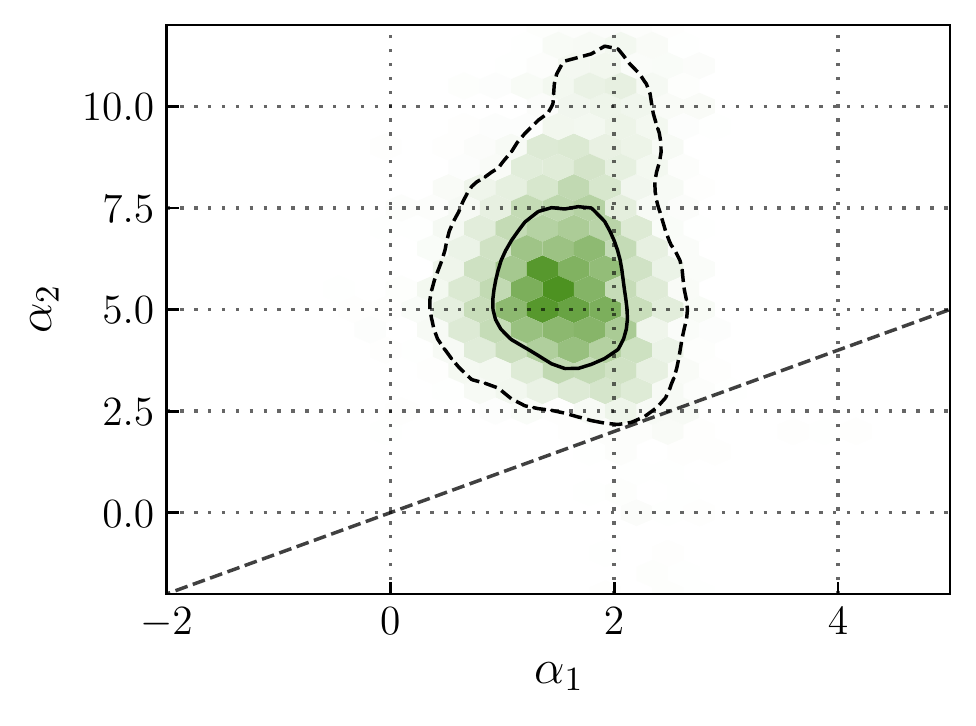}
    \caption{
    Constraints on the power-law indices governing the primary mass distribution within the \tapered{} model.
    The parameter $\alpha_1$ is the power-law index below the break, which is found to be $m_\mathrm{break} = \BPLNoAugNoEvolutionmbreak$ while $\alpha_2$ is the index above the break.
    The dashed and solid contours mark the central 50\% and 90\% posterior credible regions, respectively, under a flat prior on $\alpha_1$, $\alpha_2$ in the range $(-4, 12)$.
    We rule out with high confidence the hypothesis that $\alpha_1=\alpha_2$, indicated by the dashed diagonal line, finding $\alpha_2 > \alpha_1$ with \BPLNoAugNoEvolutionPAlphaOneLessAlphaTwoPercent\% credibility.
    }
    \label{fig:BPL_alpha1valpha2}
\end{figure}

We turn first to the \tapered{} model (second panel in Fig.~\ref{fig:pm1_astrophysical}), which is characterized by two spectral indices, $\alpha_1$ and $\alpha_2$, with $p(m_1) \propto m_1^{-\alpha_1}$ for $m_1<m_\mathrm{break}$, and $p(m_1) \propto m_1^{-\alpha_2}$ above the break.
We find the data prefer a break at $m_\mathrm{break} = \BPLNoAugNoEvolutionmbreak \ M_\odot$; 90\% credible bounds on the location of this break are denoted by the gray vertical band in the second panel of Fig.~\ref{fig:pm1_astrophysical}.
For masses above the break, the \tapered{} model prefers a significantly steeper power-law slope, from $\alpha_1 = \BPLNoAugNoEvolutionalphaOne$ before to $\alpha_2 = \BPLNoAugNoEvolutionalphaTwo$ after.
Figure~\ref{fig:BPL_alpha1valpha2} shows the joint posterior on $\alpha_1$ and $\alpha_2$.
We infer that $\alpha_2 > \alpha_1$ at credibility $\BPLNoAugNoEvolutionPAlphaOneLessAlphaTwoPercent\%$.
The break aligns with the cutoff $\mmax$ inferred with GWTC-1 data~\citep{O2pop}, and we speculate that the steep drop-off in the merger rate that occurs after $m_\mathrm{break}$ may be an imprint of PPSN, which are expected to become important for BH masses around $35 \, M_\odot$~\citep{Woosley2015,Heger2002,Heger2003}.

\begin{figure}[h]
    \centering
    \includegraphics[width = 0.48\textwidth]{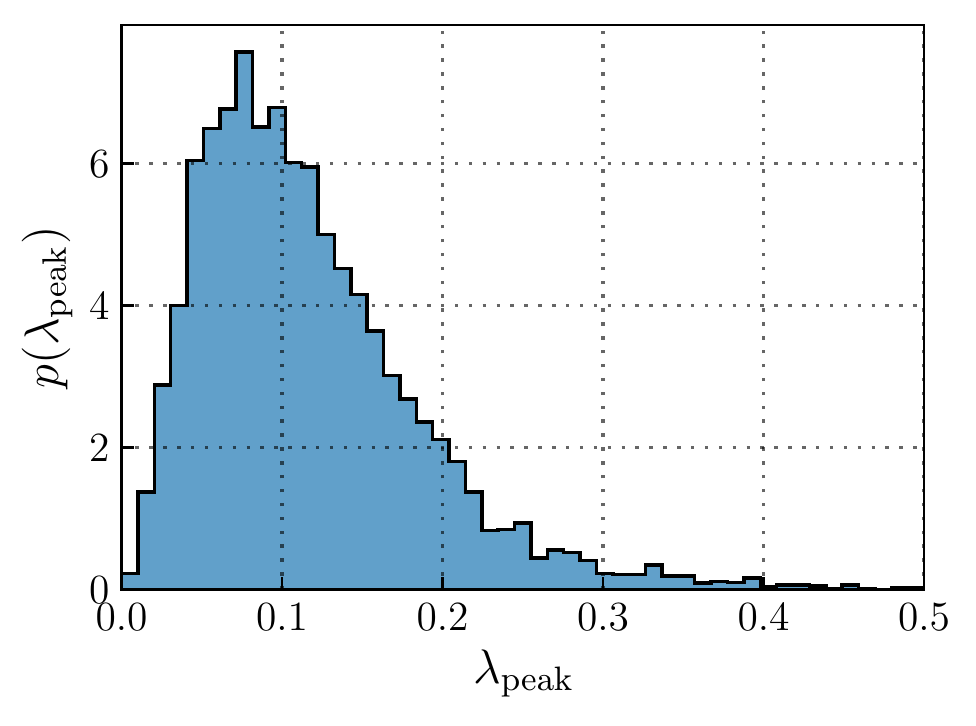}
    \caption{
    Posterior distribution on the fraction of binaries ($\lambda_\text{peak}$) in the Gaussian component of \ppsn{} model, under a flat prior on $\lambda_\text{peak}$; see Appendix~\ref{ppsn}.
    We find that $\lambda_\text{peak} = 0$ (which corresponds to no Gaussian peak) is disfavored, supporting the hypothesis that there is a feature in the BH primary mass spectrum.
    }
    \label{fig:zoomed_in_lambda}
\end{figure}

The \ppsn{} model (third panel of Fig.~\ref{fig:pm1_astrophysical}) produces a qualitatively similar fit to the one obtained from the \tapered{} model.
However, a key feature of the \ppsn{} model is the Gaussian peak at $\peakNoAugNoEvolutionmpp \ M_\odot$, denoted by the gray vertical band in the third panel of Fig.~\ref{fig:pm1_astrophysical}.
Evidence for a peak can be seen in Fig.~\ref{fig:zoomed_in_lambda}, which shows the posterior for $\lambda_\text{peak}$, the fraction of systems that belong to the Gaussian component.
We see that $\lambda_\text{peak}=0$ (pure power law) is disfavored.
It was envisioned~\citep{Talbot2018} that the power-law component of the \ppsn{} model would terminate in the vicinity of this peak to create a high-mass gap.
However, in order to accommodate the most massive binaries in GWTC-2, the power-law extends to values of $\mmax = \peakNoAugNoEvolutionmmax M_\odot$.

While the mass spectrum must extend to these high masses, we find that 99\% of primary BH masses lie below $m_{99\%} \sim 60 \ M_\odot$; see Table~\ref{tab:m1m99}.
The \astrophysical{} rate density at $\sim 80 M_\odot$ (the primary mass of \NAME{GW190521A}) is two orders of magnitude lower than the rate density at $\sim 40 M_\odot$.
However, because of selection effects, the \oppd{} skews to much higher masses, as seen in Fig.~\ref{fig:PPSN_observed_spectrum}, so that the probability of detecting at least one event with $m_1 \geq 80 M_\odot$ after observing 44 BBH events drawn from the \ppsn{} \oppd{} of Fig.~\ref{fig:PPSN_observed_spectrum} is high: $\peakNoAugNoEvolutionObsPrimaryMassGrEightyPercent \%$.

We cannot determine whether the high-mass events of GWTC-2 belong to a distinct subpopulation rather than a high-mass tail of the normal BBH population.
An additional subpopulation may be expected if high-mass BHs have a different origin from low mass ones; for example, if they are the products of hierarchical mergers~\citep{Fishbach:2017dwv,2017PhRvD..95l4046G,2019PhRvD.100j4015C, Doctor2019, Kimball}.
Using the \multipeak{} model (bottom panel in Fig.~\ref{fig:pm1_astrophysical}), which allows for a second high-mass Gaussian component at $m_1 >50 \ M_\odot$ in addition to the Gaussian component at $m_1 \lesssim 40 \ M_\odot$, we find that the addition of a second Gaussian peak is not preferred by the data.
The \multipeak{} model is mildly disfavored compared to \ppsn{}, with a $\log_{10}{\cal B} = \multipeakNoAugNoEvolutionLogBF$ (or Bayes factor ${\cal B}=\multipeakNoAugNoEvolutionBF$) for \multipeak{} relative to \ppsn{}.
\added{Motivated by the hypothesis of hierarchical mergers, we consider a variation of the \multipeak{} model, in which the location of the second peak is required to be at twice the value of the first peak; that is, $\mu_{m,2} = 2\mu_{m,1}$ (see Appendix~\ref{Multi-Peak}.   Studying the $(\mu_{m,1},\mu_{m,2})$ panel of the corner plot in Fig.~\ref{fig:Multipeak_corner}, we see that the data mildly prefer the second peak at $\approx 70 M_\odot$, consistent with twice the value of the first peak at $\approx 35 M_\odot$. This ``\textsc{Modified MultiPeak}'' is mildly preferred over the original version by a Bayes factor of $\sim 2$.}
A similar conclusion is found using the \ModelH{} model; as discussed in Section \ref{sec:results-spin}, we find hints, but no significant evidence for subpopulations with distinct spin distributions.
\added{Additional evidence for hierarchical mergers is presented in~\cite{gwtc2_hierarchical}.}

Within the framework of the \tapered{}, \ppsn{} and \multipeak{} models, the most massive event, \NAME{GW190521A}, appears to be a normal member of the BBH population in the context of the other GWTC-2 events (see Appendix~\ref{GW190521}).
The event \NAME{GW190521A}{} \textit{is} an outlier if we consider it in the context of GWTC-1 with the \truncated{} model, but we interpret this as a limitation of the \truncated{} model (see Fig.~\ref{fig:mmax_compare}; \citealp{Abbott:GW190521_implications}).

The GWTC-1 detections showed that BHs more massive than  $\sim 45 \ M_\odot$ merge relatively rarely, based on simple extrapolations from below $45 \ M_\odot$.
With GWTC-2, we are beginning to resolve the shape of the primary BH mass spectrum above $45 \ M_\odot$.
The implications are not yet clear, but there are intriguing possibilities.
One hypothesis is that the events with $m_1>45 \ M_\odot$ are simply the high-mass tail of the ordinary BBH population, and do not form through a distinct channel.
For example, if the lower edge of the PPSN gap may be modeled as a smooth tapering rather than a sharp cutoff, the feature at $m_\mathrm{break} = \BPLNoAugNoEvolutionmbreak$ may represent the onset of pair-instability.
This explanation may pose challenges to our understanding of stellar evolution since the pair-instability cutoff of BH masses at $\sim 40\ M_\odot$ is thought to be relatively abrupt~\citep{ 2002RvMP...74.1015W,2017ApJ...836..244W,Farmer_2019}, even though its precise location is uncertain~\citep{2020ApJ...888...76M,2017MNRAS.470.4739S,2018MNRAS.474.2959G,2020arXiv200405187V,Farmer:2020,Croon:2020oga,Marchant:2020haw}.
If the PPSN cutoff is indeed sharp and all observed BBH systems lie below the PPSN gap, the cutoff must occur at relatively high masses; in the \truncated{} model, $\mmax = \truncatedNoAugNoEvolutionmmax \ M_\odot$ (or, excluding the most massive event, \NAME{GW190521A}, $\mmax = \truncatedNoMayNoEvolutionmmax \ M_\odot$). This may have significant implications for nuclear~\citep{Farmer:2020} and particle~\citep{Croon:2020oga,2020arXiv201000254Z} physics.

Another hypothesis is that the events with $m_1>45 \ M_\odot$ constitute a distinct population, created, for example, from hierarchical mergers of lower mass binaries in globular clusters or galactic nuclei~\citep{Miller2002, 2016ApJ...831..187A,2016MNRAS.463.2443K, Rodriguez2018, Yang2018, 2020ApJ...891...47A}.
Alternatively, the high-mass gap might be populated from low-metallicity stellar mergers in young star clusters, the remnants of which can merge dynamically~\citep{2019arXiv191101434D,2019MNRAS.487.2947D}, BH growth through accretion~\citep{2019A&A...632L...8R,2020arXiv200911326R,2020arXiv200909156N,2020arXiv200909320S}, or Population III stellar remnants with large hydrogen envelopes~\citep{2020arXiv200801890T,Farrell:2020zju,2020arXiv200911447L}.

\begin{figure}[h]
    \centering
    \includegraphics[width = 0.48\textwidth]{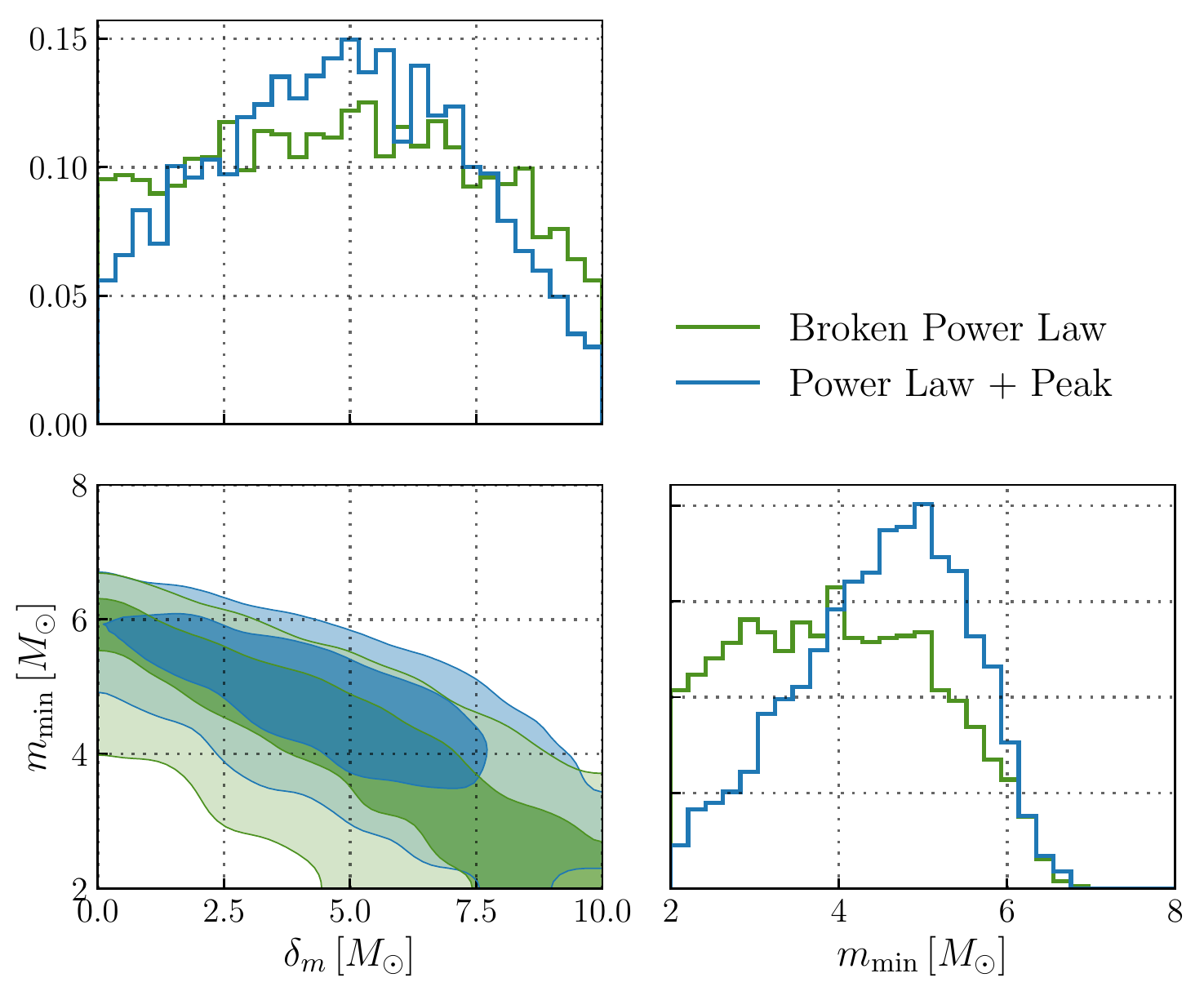}
    \caption{
    Posterior distribution for population parameters $m_\text{min}$, the minimum BH mass, and $\delta_m$, which controls the sharpness of the low-mass cut-off.
    A sharp cut-off corresponds to $\delta_m=0$. Analyzing the \result{44} BBH events (with the exclusion of \NAME{GW190814A}{}),
    both models exclude $(m_\text{min}=\unit[3]{M_\odot}, \delta_m=0)$ with $>99\%$ credibility, indicating that the rate drops off at low masses.
    To varying degrees, both models allow for $m_\text{min} \leq \unit[3]{M_\odot},\delta_m>0$, suggesting that the low-mass gap may not be empty.
    See Appendices~\ref{tapered} and~\ref{ppsn} for additional details about the $\delta_m, m_\text{min}$ parameters.
    }
    \label{fig:deltam_mmin}
\end{figure}

\textbf{The BBH primary mass distribution exhibits a global maximum between $\sim 4$ and $\sim 10 \ M_\odot$.}
Figure~\ref{fig:deltam_mmin} shows the joint posterior for the $\mmin$ and $\delta_m$ parameters inferred using the \ppsn{} and \tapered{} mass models, including only BBH events with $m_2 > 3 \ M_\odot$. 
Recall that, while the \truncated{} model has a sharp cutoff at $\mmin$, the remaining models implement a smooth turn-on of width $\delta_m$ above $\mmin$, causing the mass spectrum to peak and turn over between $\mmin$, and $\delta_m + \mmin$~\citep{Talbot2018}. 
Using both the \tapered{} model and the \ppsn{} model, we find that the primary mass spectrum does not decrease monotonically from $3 \ M_\odot$.
Rather, it turns over at $\peakNoAugNoEvolutionTurnoverMass \ M_\odot$ (\ppsn{} model) or $\BPLNoAugNoEvolutionTurnoverMass \ M_\odot$ (\tapered{} model).
In other words, the mass distribution must turn over at $m_1 > 3 \ M_\odot$, with $\peakNoAugNoEvolutionTurnoverMassPGrThree \%$ credibility (assuming the \ppsn{} model) or $\BPLNoAugNoEvolutionTurnoverMassPGrThree \%$ credibility (\tapered{} model).
As seen in Fig.~\ref{fig:deltam_mmin}, if the BH low-mass cutoff is sharp ($\delta_m = 0$), then $\mmin \gtrsim \result{4} \ M_\odot$. Conversely, if the BH mass spectrum extends below $\mmin \lesssim 4 \ M_\odot$, an extended turn-on $\delta_m \gtrsim \result{3} \ M_\odot$ is required.
These results support the existence of a low-mass gap (or dip) between $\sim 2.6 \ M_\odot$ (the secondary mass of \NAME{GW190814A}{}; the most massive component mass observed below $3 \ M_\odot$) and $\sim 6 \ M_\odot$, strengthening results from~\citet{Fishbach:2020ryj}, although we cannot determine whether the low-mass gap is empty.

\begin{figure*}
    \subfloat[Minimum primary mass distributions.]{
        \includegraphics[width=0.5\textwidth]{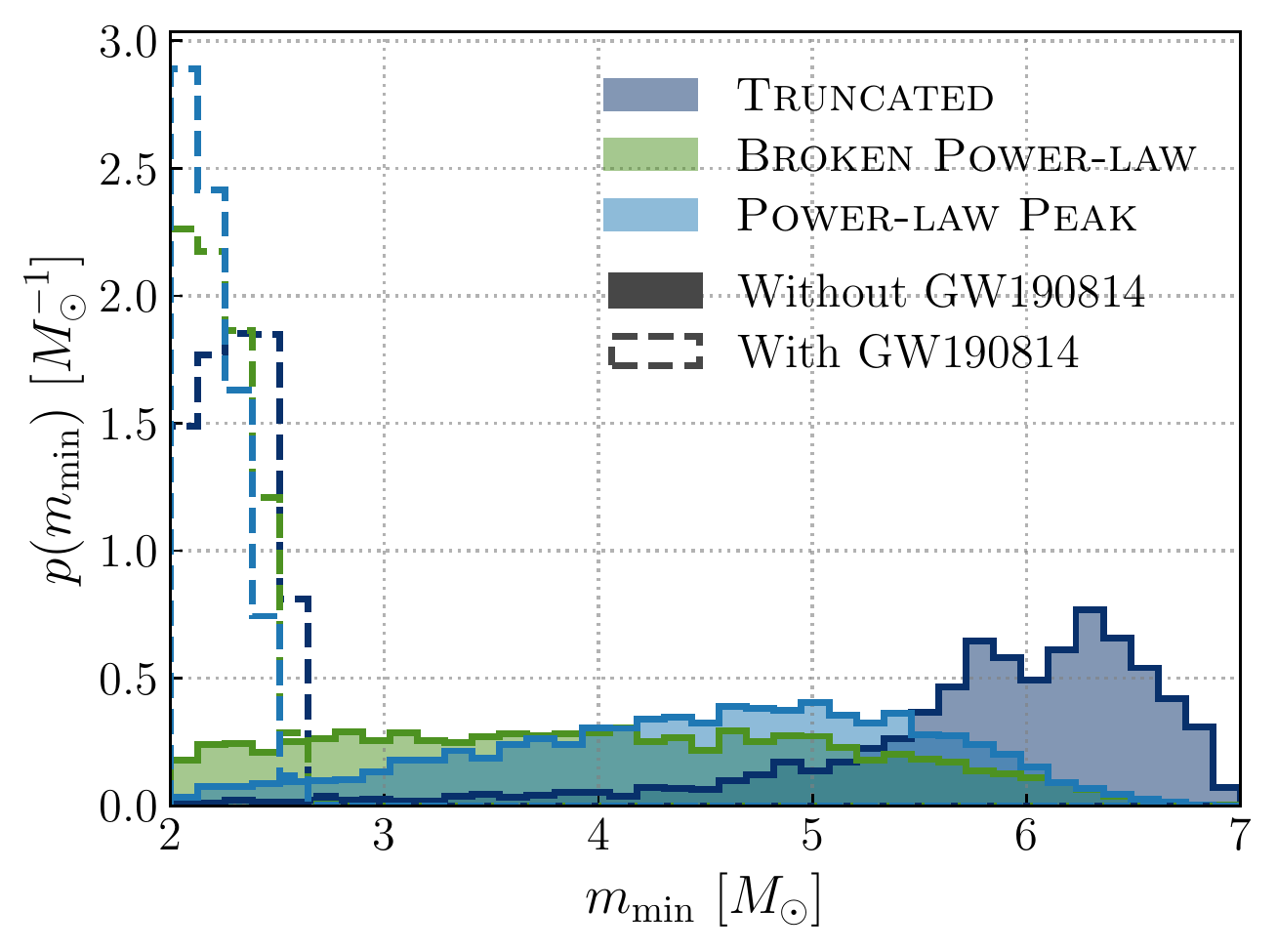}
        \label{fig:mmin_various_mass_models}
    }
    \subfloat[\ppsn{} mass distribution fit.]{
        \includegraphics[width=0.5\textwidth]{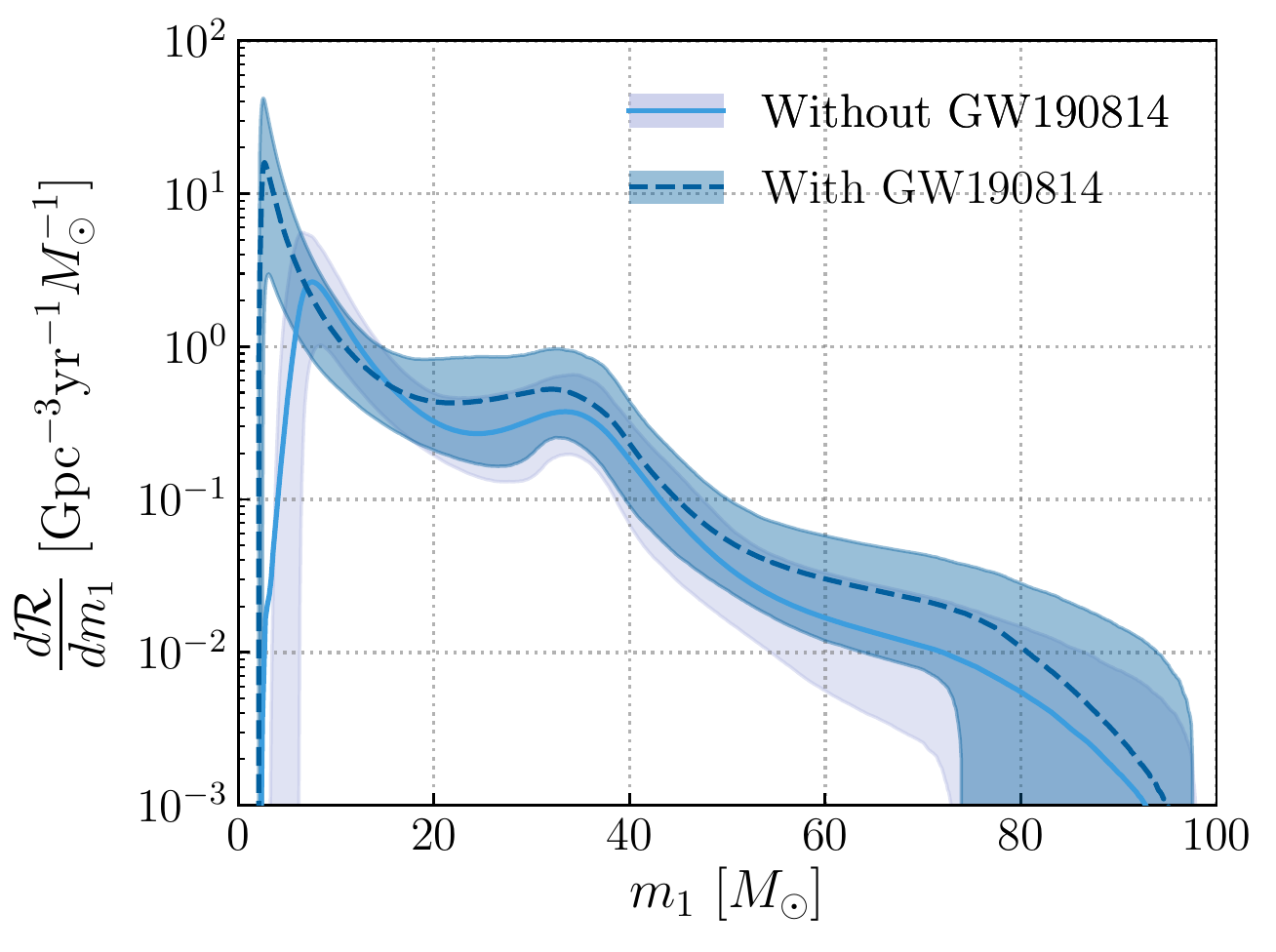}
        \label{fig:pm1_withlowmass}
    }
    \caption{
    (a) Posterior distribution for the minimum mass parameter in the \truncated{} (navy), \tapered{} (green) and \ppsn{} (blue) models.
    The solid bars show the posterior for $\mmin$ when fitting the models to the confident \bbhevnt{} excluding \NAME{GW190814A}{}, while the dashed lines show the fits to the \bbhevnt{} including \NAME{GW190814A}{}.
    The \truncated{} model is disfavored.
    (b) Distribution of primary masses inferred using the \ppsn{} model when including (navy, dashed line) and excluding (light blue, solid line) \NAME{GW190814A}{}. \added{Only the secondary mass of \NAME{GW190814A}{} is below $3\,M_\odot$.
      However, the primary and secondary mass distributions share a common $\mmin$ parameter in all models we consider.}
    The effect of \NAME{GW190814A}{} on the mass spectrum at low masses inferred using the \tapered{} model is similar.
    }
\end{figure*}

\replaced{Since our BBH mass distribution models are not designed to simultaneously fit systems above and below the low-mass gap}{Since our models do not permit additional features beyond a smooth turn on to a power law at low masses}, they struggle to accommodate \NAME{GW190814A}{} (with secondary mass at $m_2 = \masstwosourcemed{GW190814A}^{+\masstwosourceplus{GW190814A}}_{-\masstwosourceminus{GW190814A}}$). \deleted{This system occurs below the turnover mass inferred above from systems with both component masses above $3 \ M_\odot$.}
\added{We can see this by comparing the mass distribution inferred from the events with $m_1 \geq m_2 > 3\,M_\odot$, discussed above, to the distribution inferred with \NAME{GW190814A}{}.}
If \NAME{GW190814A}{} is a BBH system, the minimum BH mass must extend to $\mmin = \peakAllNoEvolutionmmin \ M_\odot$ (see the dashed histograms in Fig.~\ref{fig:mmin_various_mass_models}).
In Fig.~\ref{fig:pm1_withlowmass} we show how the inclusion/exclusion of \NAME{GW190814A}{} affects the shape of the primary mass distribution below $\lesssim 5 \ M_\odot$.
\added{We see that the two distributions are inconsistent at the low mass end, suggesting that there is a feature in the mass distribution between $\sim2.6\,M_\odot$ and $\sim 6\,M_\odot$ that our models cannot capture.}
This effect can also be seen in the $m_{1\%}$ values inferred with/without \NAME{GW190814A}{}, shown in Table~\ref{tab:m1m99}.
Assuming the \truncated{} and \ppsn{} models, the $\mmin$ posteriors inferred with \NAME{GW190814A}{} lie at the $\MMinWithAugPercentileInNoAugtruncated$ and $\MMinWithAugPercentileInNoAugpeak$ percentiles, respectively, of the $\mmin$ posteriors obtained {without} \NAME{GW190814A}{}.
Even using the \tapered{} model, which admits greater overlap in the 1-dimensional $\mmin$ posteriors inferred with/without \NAME{GW190814A}{}, the addition of \NAME{GW190814A}{} significantly shifts the two-dimensional $(\delta_m, \mmin)$ posterior and the inferred mass spectrum.
Assuming the \tapered{} model with \NAME{GW190814A}{}, the mass at which the mass spectrum turns over is shifted down to $\BPLAllNoEvolutionTurnoverMass \ M_\odot$, which is inconsistent with the turnover mass (the low-mass local maximum) inferred without \NAME{GW190814A}{}, $\BPLNoAugNoEvolutionTurnoverMass \ M_\odot$. 
This indicates a failure of our models to fit \NAME{GW190814A}{} together with the BBH systems of GWTC-2. 
This finding is supported by additional studies described in Appendix~\ref{Appendix:GW190814}.
Because \NAME{GW190814A}{} is a population outlier with respect to the BBH events of GWTC-2 \added{and our choice of models}, we exclude it from the analyses here unless otherwise indicated.

\textbf{The distribution of mass ratios is broad.}
The GWTC-1 events are all individually consistent with $q=1$. Describing the conditional mass-ratio distribution as a power law $p(q|m_1)\propto q^{\beta_q}$, a population analysis of GWTC-1 allowed $\beta_q = 12$ (our maximum prior bound), consistent with a mass-ratio distribution sharply peaked at equal-mass pairings~\citep{O2pop,Roulet:2019,Fishbach:picky}.
GWTC-2 saw the first detections of confidently asymmetric systems: \NAME{GW190412A}~\citep{GW190412} and GW190814~\citep{GW190814}.
Excluding \NAME{GW190814A}{}, our reconstruction of the mass ratio distribution is consistent with the results published in~\cite{GW190412}: $\beta_q = \peakNoAugNoEvolutionbeta$ for the \ppsn{} model and $\beta_q = \BPLNoAugNoEvolutionbeta$ for the \tapered{} model.
We rule out distributions that are sharply peaked around $q=1$, with $\beta_q < \peakNoAugNoEvolutionbetaqUpperNinety$ (\ppsn{}) and $\beta_q < \BPLNoAugNoEvolutionbetaqUpperNinety$ (\tapered{}) at 90\% credibility.
However, we also disfavor distributions that prefer unequal-mass pairings, with $\beta_q > 0$ at \peakNoAugNoEvolutionbetaqPGrZero\% (\ppsn{}) and \BPLNoAugNoEvolutionbetaqPGrZero\% (\tapered{})  credibility. We find that 90\% of systems in the underlying population have mass ratios $q > \peakNoAugNoEvolutionMassRatioTenPercentile$.

\subsection{Spin Distribution}
\label{sec:results-spin}
In this subsection, we highlight the results from the \textsc{Gaussian}, \textsc{Default}, and \ModelH{} models. We fix the redshift distribution to a \textsc{Non-Evolving} merger rate.
The \textsc{Gaussian} and \ModelH{} models assume the mass distributions described in Appendix~\ref{sec:Gaussian} and~\ref{modelH}, respectively. 
For the \textsc{Default} spin model, we employ the \ppsn{} mass model, simultaneously fitting the mass and spin distribution as in the previous subsection.

\textbf{We observe spin-induced general relativistic precession of the orbital plane.}
As two BHs merge, the morphology of the resulting gravitational waveform depends on their spins.
The spin-dependence of a gravitational waveform is determined in part by two phenomenological parameters.
First, the \chieff{} $\chi_\mathrm{eff}$ quantifies the spin components aligned with the orbital angular momentum~\citep{PhysRevD.64.124013}:
    \begin{align}\label{eq:chi_eff}
    \chi_\text{eff} = \frac{\chi_1\cos \theta_1 + q\, \chi_2 \cos \theta_2}{1+q} .
    \end{align}
Here, $\chi_1$ and $\chi_2$ are the dimensionless component spins, defined by $\chi_i = \lvert {c S_i}/({G m_i^2}) \rvert$ where $S_i$ is the spin angular momentum of component $i$, and $\theta_1$ and $\theta_2$ are the misalignment angles between the component spins and the orbital angular momentum.
Second, spins with components perpendicular to the orbital angular momentum drive relativistic precession of the orbital plane~\citep{Apostolatos}.
The effect is quantified by the \chip{}~\citep{Schmidt2012,Hannam:2013oca,Schmidt2015}
\begin{align}\label{eq:chi_P}
    \chi_\mathrm{p} = \max\bigg[
    \chi_1 \sin \theta_1 , 
    \left(\frac{4q+3}{4+3q}\right) q\, \chi_2 \sin \theta_2
    \bigg].
\end{align}
A non-zero value of $\chi_\mathrm{p}$ indicates the presence of relativistic spin-induced precession of the orbital plane.
Although the component spin tilts $\theta_1$ and $\theta_2$ appearing in Eqs.~\eqref{eq:chi_eff} and \eqref{eq:chi_P} generically evolve over the course of a binary inspiral, $\chi_\mathrm{eff}$ and $\chi_\mathrm{p}$ are themselves approximately conserved quantities~\citep{Kidder1995,Schmidt2015}.

\begin{figure*}
    \centering
    \includegraphics[width=0.48\textwidth]{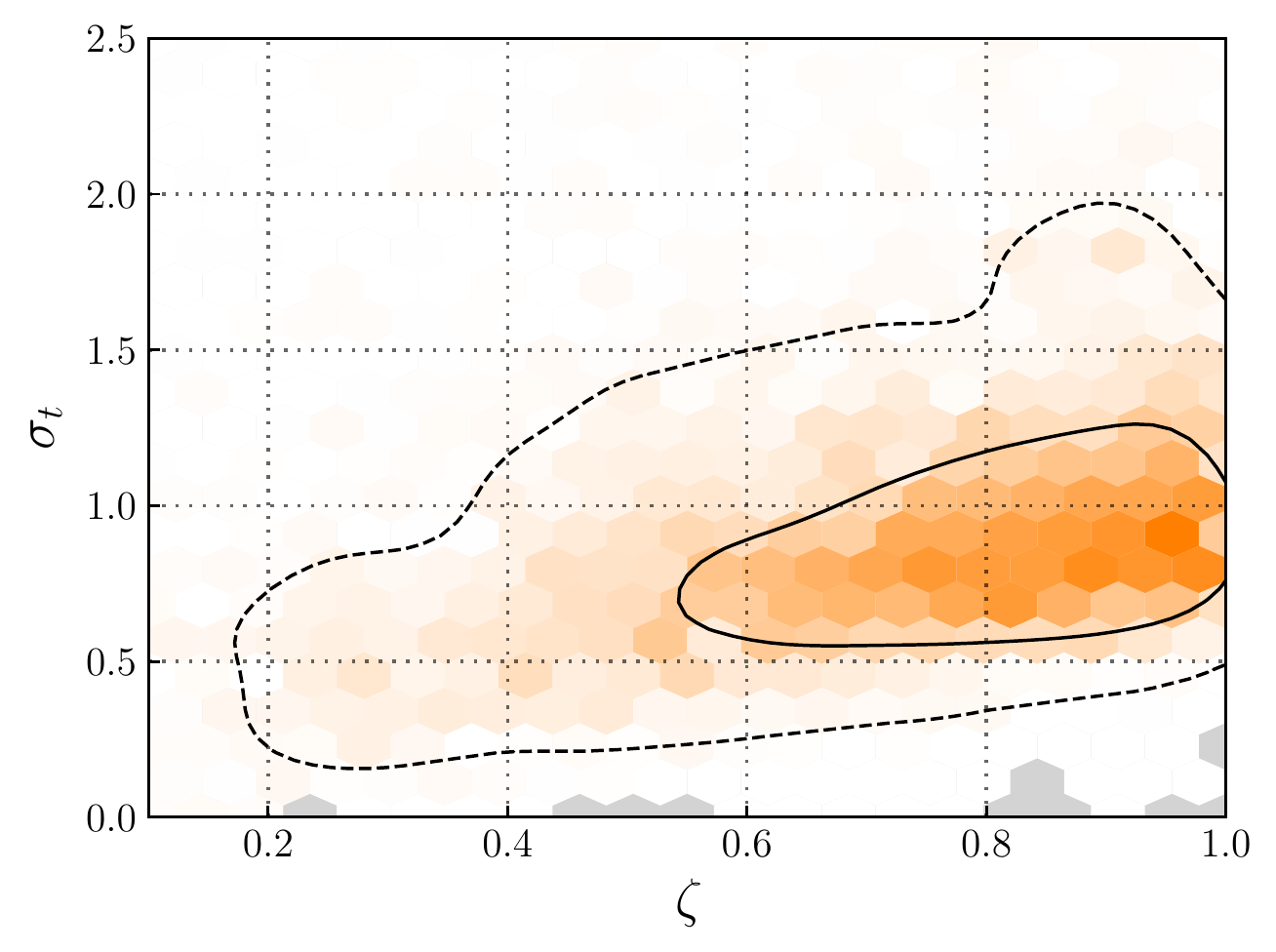}
    \hspace{5mm}
    \includegraphics[width=0.48\textwidth]{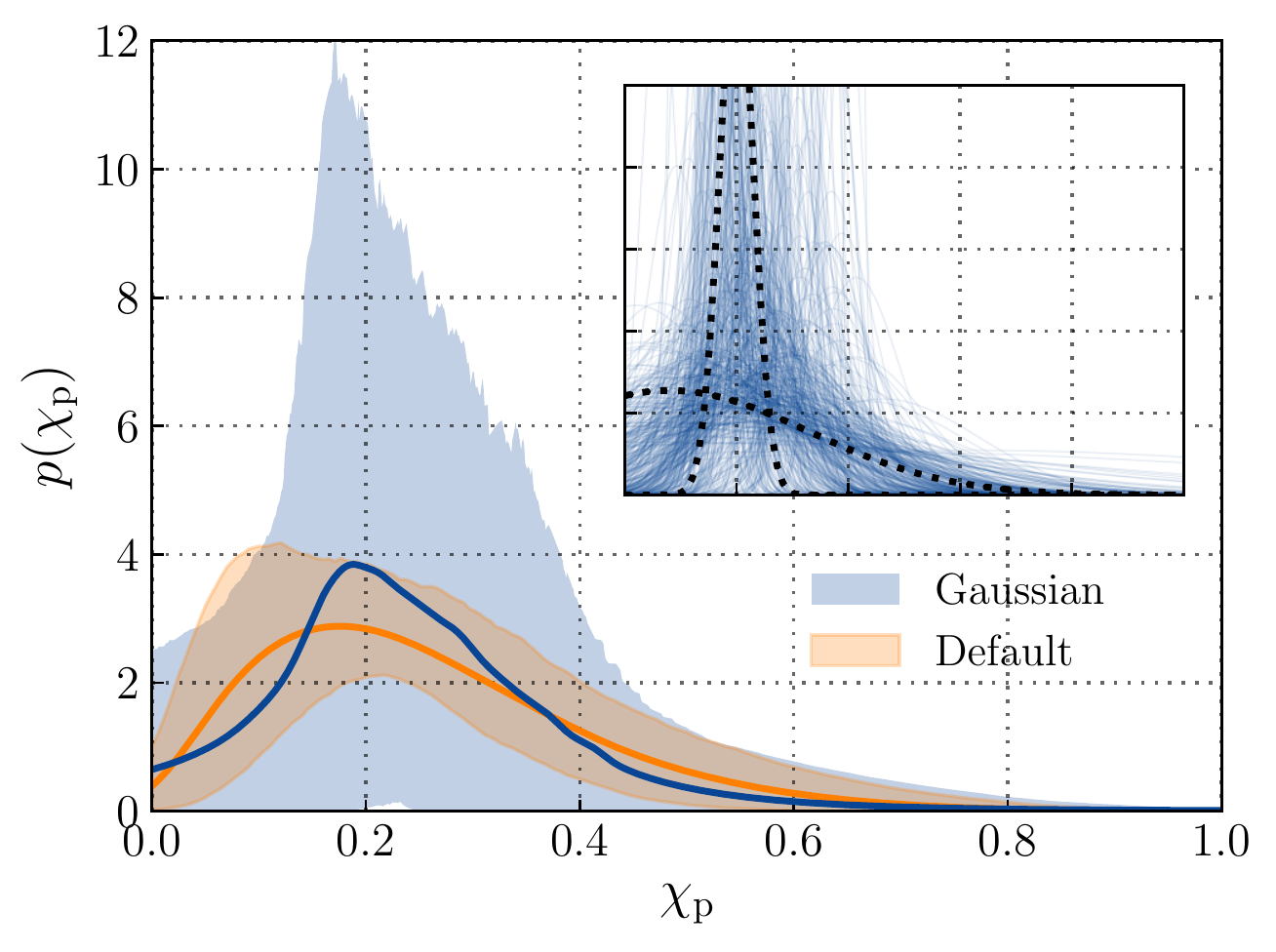}
    \caption{
    \deleted{\textit{Left}: Marginal posterior for the mean $\mu_p$ and standard deviation $\sigma_p$ of the $\chi_\mathrm{p}$ distribution for BBH mergers obtained using the \textsc{Gaussian} spin model (see Appendix~\ref{sec:Gaussian} for additional details).
    A population with perfectly aligned spins corresponds to $\mu_p = \sigma_p = 0$, which is excluded at $>99\%$ credibility, indicating the presence of in-plane spin components among the GWTC-2 BBH population.}
    \added{\textit{Left}: Joint posterior on the fraction $\zeta$ of BBHs with preferentially-aligned spins (versus isotropic spins) and the spread $\sigma_t$ of misalignment angles among this population obtained using the \textsc{Default} spin model (see Appendix~\ref{sec:default} for additional details).
      We rule out a population with perfectly aligned spins corresponding to $\zeta = 1$ and $\sigma_t = 0$.
      The gray shaded region represents the region of parameter space inaccessible to our analysis.
      This region is artificially excluded due to sampling uncertainties even when analyzing uninformative samples drawn from the spin tilt prior.
    The dashed and solid contours mark the central 50\% and 90\% posterior credible regions, respectively, assuming a flat prior on $\zeta$ and $\sigma_t$.}
    \textit{Right}: 
    Population predictive distributions for the \chip{} $\chi_\mathrm{p}$ of BBH systems obtained using the \textsc{Gaussian} (blue) and \textsc{Default} (orange) spin models.
    Shaded regions show the central 90\% credible bounds on $p(\chi_\mathrm{p})$ at a given spin value, while the solid lines show the median posterior prediction.
    The inset shows draws of the \textsc{Gaussian} $\chi_\mathrm{p}$ distributions implied by the posterior on $\mu_p$ and $\sigma_p$.
    Broadly, we see support for two possible morphologies, indicated schematically by the dashed black curves.
    GWTC-2 is compatible with a $\chi_\mathrm{p}$ distribution that is either broad, or one that is narrow and centered at $\mu_p\sim0.2$.
    }
    \label{fig:chiP-corner_combined}
\end{figure*}

The first unambiguous measurements of BH spin in gravitational-wave astronomy came from analyses of the BBH event GW151226.
This system had $\chi_\mathrm{eff}>0$ at 99\% credibility, with at least one of its components having spin magnitude $\chi>0.2$, and spin misalignment angles consistent with $\theta_1 = \theta_2 = 0$~\citep{GW151226}.
While analyses of GWTC-1 found no clear evidence for spin in the other events in GWTC-1~(\citealt{Miller2020}, but also see \citealt{Zackay:2019tzo} and \citealt{Huang2020}), GWTC-1 is collectively inconsistent with a population of non-spinning BHs, if one allows for both spinning and non-spinning subpopulations~\citep{Kimball}. 
Moreover, population analyses of GWTC-1 mildly disfavor the scenario in which all spins are perfectly aligned ($\theta_1 = \theta_2 = 0$), although the degree of misalignment is degenerate with the spin magnitude distribution~\citep{Farr:2017uvj,BFarrSpin,Tiwari:2018qch,O2pop,Wysocki_2019}.

In GWTC-2, additional \bbhevnt{} are observed with confidently positive \chieff{}.
No individual event is observed with confidently negative $\chi_\mathrm{eff}$~\citep{O3acatalog}.
Several events, including \NAME{GW190521A}~\citep{GW190521,Abbott:GW190521_implications} and \NAME{GW190412A}~\citep{GW190412}, show moderate evidence for non-zero $\chi_\mathrm{p}$, but no single event unambiguously exhibits spin-induced precession ~\citep{O3acatalog}.

\deleted{
Using the \textsc{Gaussian} model described in Sec.~\ref{models} (see also, Appendix~\ref{sec:Gaussian}), we obtain clear evidence for relativistic spin-induced precession among the population of \bbhevnt{} in GWTC-2.
In the left side of Fig.~\ref{fig:chiP-corner_combined}, we provide a joint posterior for the mean $\mu_p$ and standard deviation $\sigma_p$ of the $\chi_\mathrm{p}$ distribution.
We have marginalized over the parameters of the $\chi_\mathrm{eff}$ distribution as well as the covariance between $\chi_\mathrm{eff}$ and $\chi_\mathrm{p}$.
Although neither $\mu_p$ nor $\sigma_p$ are individually well-constrained, $\mu_p = \sigma_p = 0$ is ruled out at $> 99\%$ credibility; fewer than \FractionOfSmallChiPSamples{} of posterior samples occur at $\mu_p\leq 0.05$ and $\sigma_p\leq0.05$.
Since any non-zero $\mu_p$ \textit{or} $\sigma_p$ implies the existence of spin-induced precession, this result constitutes a clear observation of spin-induced precession of the orbital plane, although no individual system shows significant statistical evidence for $\chi_\mathrm{p}>0$.}

\begin{figure*}
    \centering
    \includegraphics[width=0.48\textwidth]{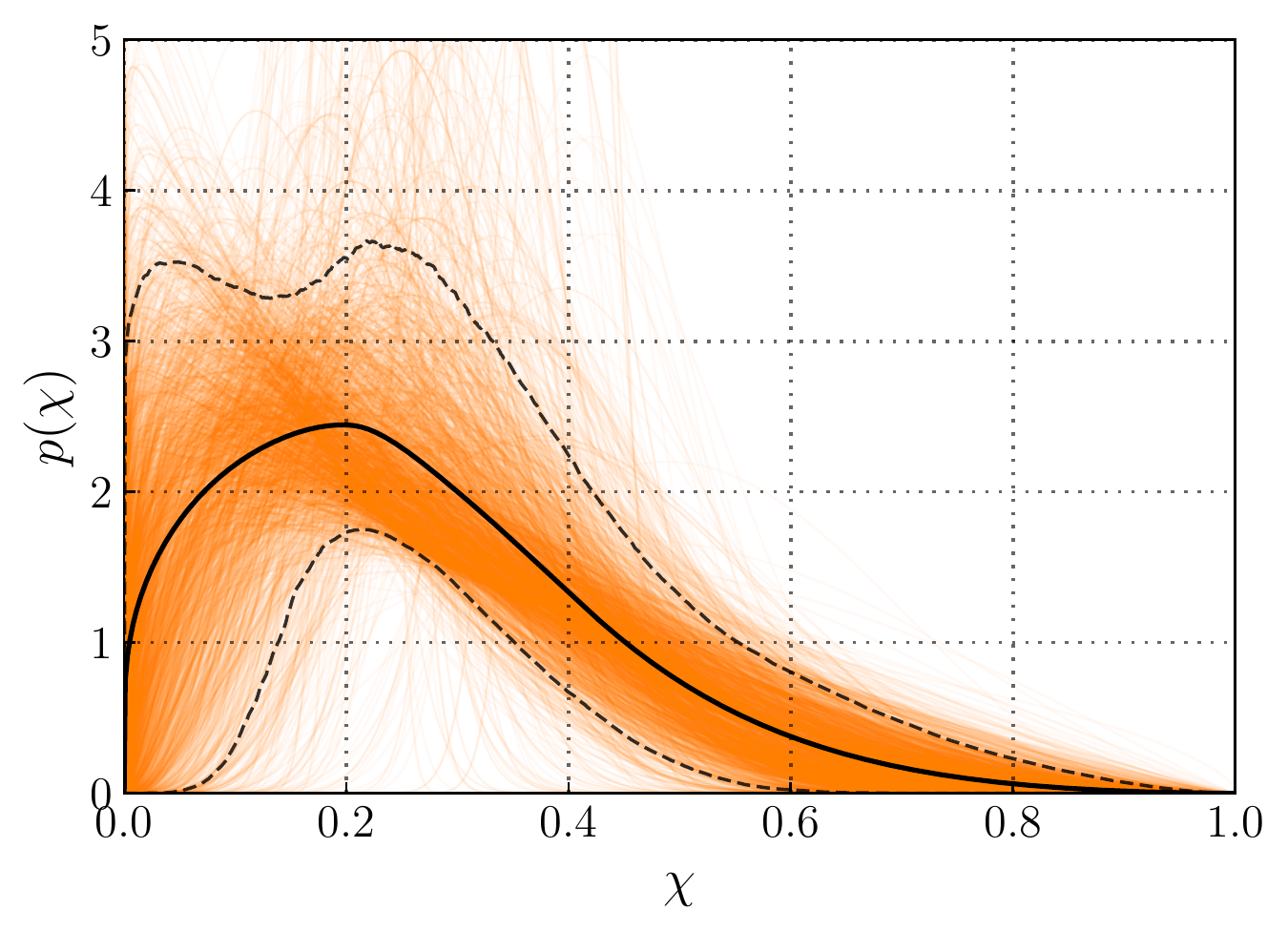}
    \hspace{5mm}
    \includegraphics[width = 0.48\textwidth]{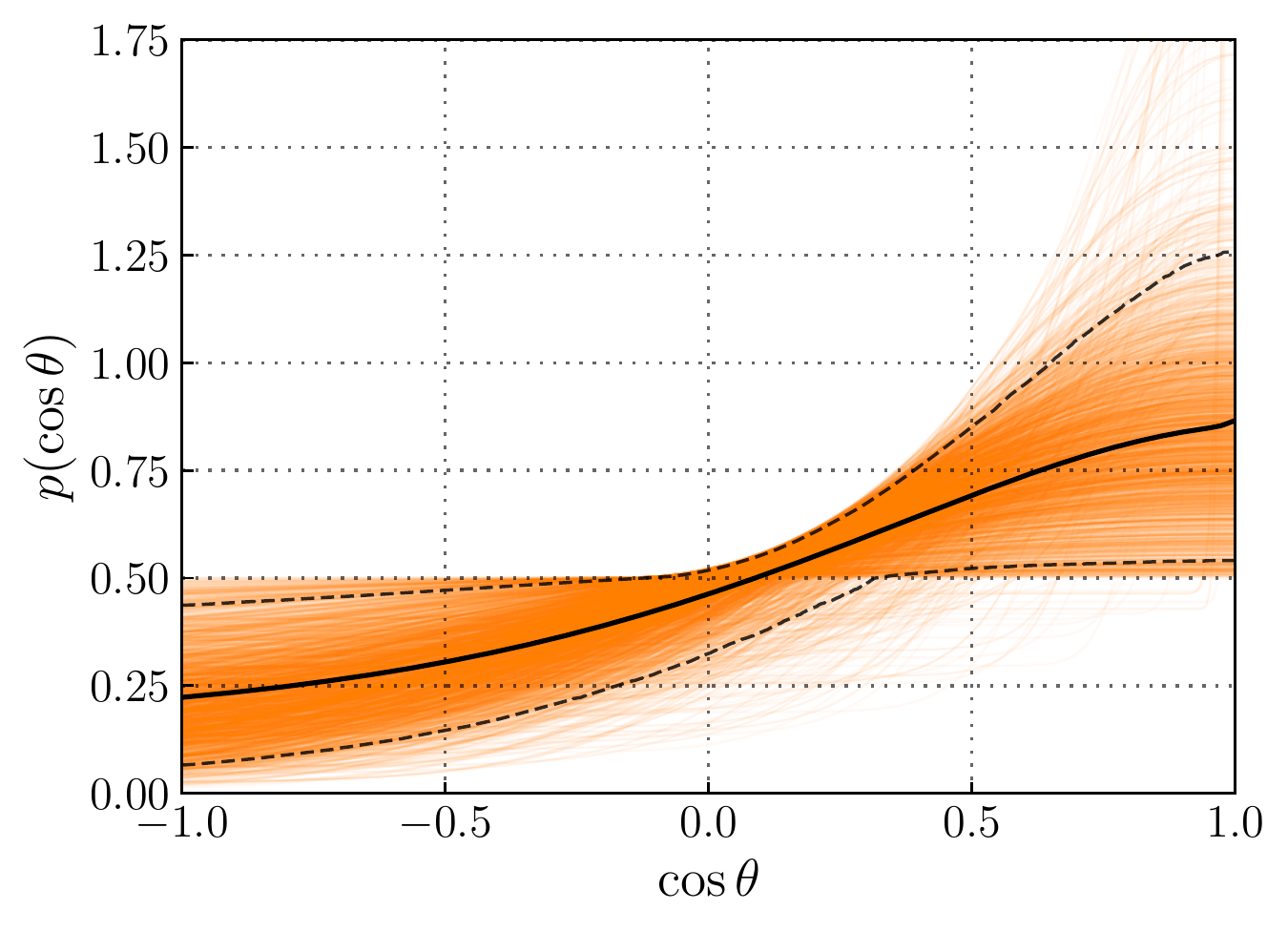}
     \caption{
     Reconstructions of the BH spin magnitude and tilt distributions.
     \textit{Left}: The distribution of dimensionless spin magnitude $\chi$ as inferred using the \textsc{Default} spin model (see Appendix~\ref{sec:default}).
     Light traces show individual draws from the \textsc{Default} posterior, while the solid black curve shows the \appd{} for $\chi$.
     Dashed lines mark the central 90\% quantiles.
     \textit{Right}: the reconstructed distribution of tilt angle $\cos \theta_{1,2}$ of BH component spins relative to the orbital angular momenta.
     An isotropic spin orientation, which corresponds to a uniform distribution in $\cos \theta_{1,2}$, is disfavored but not ruled out.
     The data do, however, rule out a highly peaked distribution at $\cos \theta_{1,2} = 1$.
     Rather, the data are consistent with a gently peaked distribution, with a modest preference for aligned spin ($\cos \theta_{1,2} > 0$).
     }
     \label{fig:default-spin-ppds}
\end{figure*}

\added{
Using the \textsc{Default} model described in Sec.~\ref{models} (see also Appendix~\ref{sec:default}), we obtain evidence for non-vanishing spin-orbit misalignment among the population of \bbhevnt{} in GWTC-2.
The \textsc{Default} model describes the distribution of spin-orbit misalignments as a mixture between two components: a component with isotropically-oriented spins and a preferentially-aligned component with $z = \cos \theta$ values centered at $z=0$ (perfect alignment) with a Gaussian spread of width $\sigma_t$.
In the left side of Fig.~\ref{fig:chiP-corner_combined}, we show the joint posterior on $\sigma_t$ and the fraction $\zeta$ of events in the preferentially-aligned subpopulation.
Perfect alignment corresponds to $\zeta = 1$ and $\sigma_t = 0$.
We see that this case is ruled out at $>99\%$ credibility.
Thus, either a non-zero fraction of \bbhevnt{} exhibit isotropically-oriented spins, \textit{or} BBH spins are preferentially aligned to their orbits but with a non-vanishing spread.
Either case constitutes an observation of in-plane spin components among the BBH population.
The shaded gray tiles in the left side of Fig.~\ref{fig:chiP-corner_combined} show the values of $\zeta$ and $\sigma_t$ that are artificially excluded by the prior and/or finite sampling effects; the true measurement using GWTC-2 lies well away from this artificial exclusion region.\footnote{In order to determine the artificial exclusion region, we generate prior samples for tilt angles $\theta_{1,2}$ conditioned on the measured values of mass ratio $q$ and spin magnitudes $\chi_{1,2}$. There are no prior samples in the gray tiles, which indicate that these tiles are artificially excluded due to finite sampling effects.}

In Fig.~\ref{fig:default-spin-ppds}, discussed further below, we plot the range of component spin magnitude and tilt angle distributions recovered using the \textsc{Default} model.
Although the data are consistent with tilt angle distributions that favor alignment, distributions that are highly peaked at $\cos \theta_{1,2} = 1$ are ruled out.
}

\added{
A similar conclusion regarding the presence of in-plane spin components may be drawn using \textsc{Gaussian} spin model, which imposes an entirely different parameterization for the BH spin distribution and makes different assumptions regarding their masses.
In particular, when measuring the mean $\mu_p$ and standard deviation $\sigma_p$ of the $\chi_{\rm p}$ distribution, the case $\mu_p = \sigma_p = 0$ is ruled out at $>99\%$ credibility; fewer than $1\%$ of posterior samples occur at $\mu_p\leq 0.05$ and $\sigma_p\leq0.05$.
Since any non-zero $\mu_p$ \textit{or} $\sigma_p$ implies the existence of spin-induced precession, this result supports the observation of spin misalignment seen in the \textsc{Default} model.
In
}
the right side of Fig.~\ref{fig:chiP-corner_combined},
\deleted{shows the range of allowed $\chi_\mathrm{p}$ distributions given by our posterior on $\mu_p$ and $\sigma_p$.}
the dark blue curve and shaded blue region mark the median and 90\% credible bound, respectively, on $p(\chi_\mathrm{p})$ as inferred by the \textsc{Gaussian} model.
\deleted{while the orange curve and shaded region show the inference from the \textsc{Default} model.}
While the blue region in this figure suggests a $\chi_\mathrm{p}$ distribution that peaks at $\sim 0.2$, there are in fact two morphologies preferred by the data according to the \textsc{Gaussian} model: the recovered $\chi_\mathrm{p}$ distribution is \textit{either} broad---or narrowly peaked at $\chi_\mathrm{p} \approx 0.2$.
This is illustrated by the inset, in which we plot an ensemble of distributions corresponding to individual draws from the $(\mu_p,\sigma_p)$ posterior;
the dashed black curves highlight traces representative of the two permitted morphologies.
\deleted{
We disfavor populations in which the $\chi_\mathrm{p}$ distribution is confined to very small (corresponding to nearly aligned spins) or very large (corresponding to preferentially in-plane spins) values. 
}

\added{
For comparison, the orange curve in the right side of Fig.~\ref{fig:chiP-corner_combined} shows the $\chi_{\rm p}$ distribution implied by the \textsc{Default}  results discussed above.
There are several potentially meaningful differences between the results from the \textsc{Gaussian} and \textsc{Default} models.
In particular, the \textsc{Default} model predicts $\chi_\mathrm{p}$ distributions that are generally broader and peaked at lower values.
This is due to additional physical constraints imposed by the \textsc{Default} spin model; component spins are presumed to preferentially cluster about $\theta=0$, an assumption that preferentially favors smaller $\chi_\mathrm{p}$ values.
Nevertheless, the two models agree well within statistical uncertainties,\footnote{For technical reasons, we do not have a Bayes factor to compare the \textsc{Default} and \textsc{Gaussian} spin models, though, this comparison is possible in principle.}
indicating that the identification of spin-induced precession is robust to the systematic modeling choices and prior uncertainties.
}

As mentioned above, \NAME{GW190521A}{} and \NAME{GW190412A}{} individually show mild evidence of precession, with $\chi_\mathrm{p}$ posteriors shifted away from their respective priors~\citep{GW190412, GW190521,Abbott:GW190521_implications}.
To verify that our population-level conclusions are not driven primarily by these two events, we have repeated the \textsc{Gaussian} analysis excluding \NAME{GW190521A}{} and \NAME{GW190412A}.
Our results again exclude $\mu_p = \sigma_p = 0$ at a similar level of confidence ($>99\%$ credibility).
This implies that the signature of precession observed here is due to the combined influence of many systems with only weakly measured $\chi_\mathrm{p}$, consistent with expectations from simulation studies~\citep{Fairhurst:2019srr,Wysocki_2019}.

The injection sets used to quantify search selection effects (see Appendix~\ref{Appendix:xi}) contain only events whose component spins are perfectly aligned with their orbital angular momenta.
The results in Fig.~\ref{fig:chiP-corner_combined} therefore \textit{do not} account for systematics possibly affecting our ability to detect events with \deleted{non-zero $\chi_\mathrm{p}$} \added{misaligned spins}.
The matched filter template banks adopted by the \texttt{GstLAL} and \texttt{PyCBC} search pipelines, for instance, are composed of purely aligned-spin waveforms, and so may have reduced sensitivity to events with high $\chi_\mathrm{p}$~\citep{Harry_2016,Calder_n_Bustillo_2017}.
Selection effects can, however, only \textit{decrease} the efficiency with which events with large in-plane spins are detected; incorporating such effects would further shift the posterior in Fig.~\ref{fig:chiP-corner_combined} \deleted{towards larger values of $\mu_p$ and/or $\sigma_p$ and further rule out $\mu_p = \sigma_p = 0$} \added{away from $\zeta=1$ and $\sigma_t = 0$ and/or more strongly rule out a delta function at $\chi_{\rm p} = 0$.}
Thus, the presence of in-plane spin components is robust to selection effects.
The specific preference for $\mu_p \approx 0.2$, though, may not be.
In the future, accurately characterizing the effects of in-plane spins on detection efficiency will be crucial in order to robustly determine the shape of the $\cos \theta$ and $\chi_\mathrm{p}$ distributions.

\deleted{
A similar conclusion regarding the presence of in-plane spin components may be drawn using \textsc{Default} spin model, which imposes an entirely different parameterization for the BH spin distribution and makes different assumptions regarding their masses.
In Fig.~\ref{fig:default-spin-ppds} we show the posteriors for the component spin magnitude and tilt angle distributions obtained using the \textsc{Default} model.
Together, these distributions themselves imply a posterior for the $\chi_\mathrm{p}$ distribution of BBH systems; this implied $\chi_\mathrm{p}$ distribution is shown in orange on the right side of Fig.~\ref{fig:chiP-corner_combined}.
As discussed further below, the \textsc{Default} model excludes perfect spin alignment (i.e. spin tilts identically zero) at >99\% credibility.
As a result, the corresponding $\chi_\mathrm{p}$ distribution excludes the case of vanishing $\chi_\mathrm{p}$.
The identification of spin-induced precession within the BBH population is therefore robust to the systematic differences exhibited by our \textsc{Default} and \textsc{Gaussian} models.
}

\deleted{
There are, though, several potentially meaningful differences between the results from the \textsc{Gaussian} and \textsc{Default} models.
In particular, the \textsc{Default} model predicts $\chi_\mathrm{p}$ distributions that are generally broader and peaked at lower values.
This is due to additional physical constraints imposed by the \textsc{Default} spin model; component spins are presumed to preferentially cluster about $\theta=0$, an assumption that preferentially favors smaller $\chi_\mathrm{p}$ values.
Nevertheless, the two models agree within statistical uncertainties
}

\begin{figure*}
    \centering
    \includegraphics[width = 0.48\textwidth]{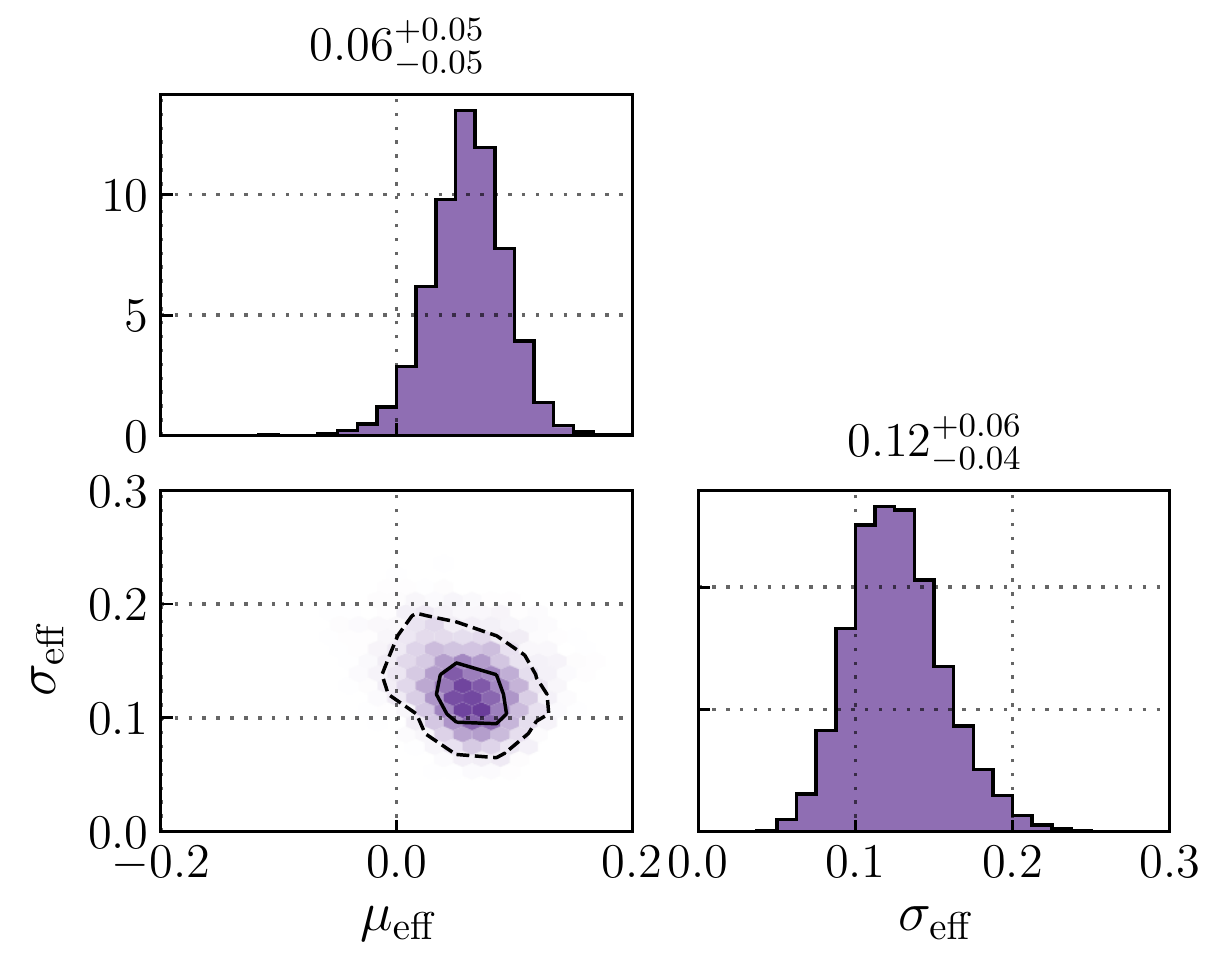}
    \hspace{5mm}
    \includegraphics[width = 0.48\textwidth]{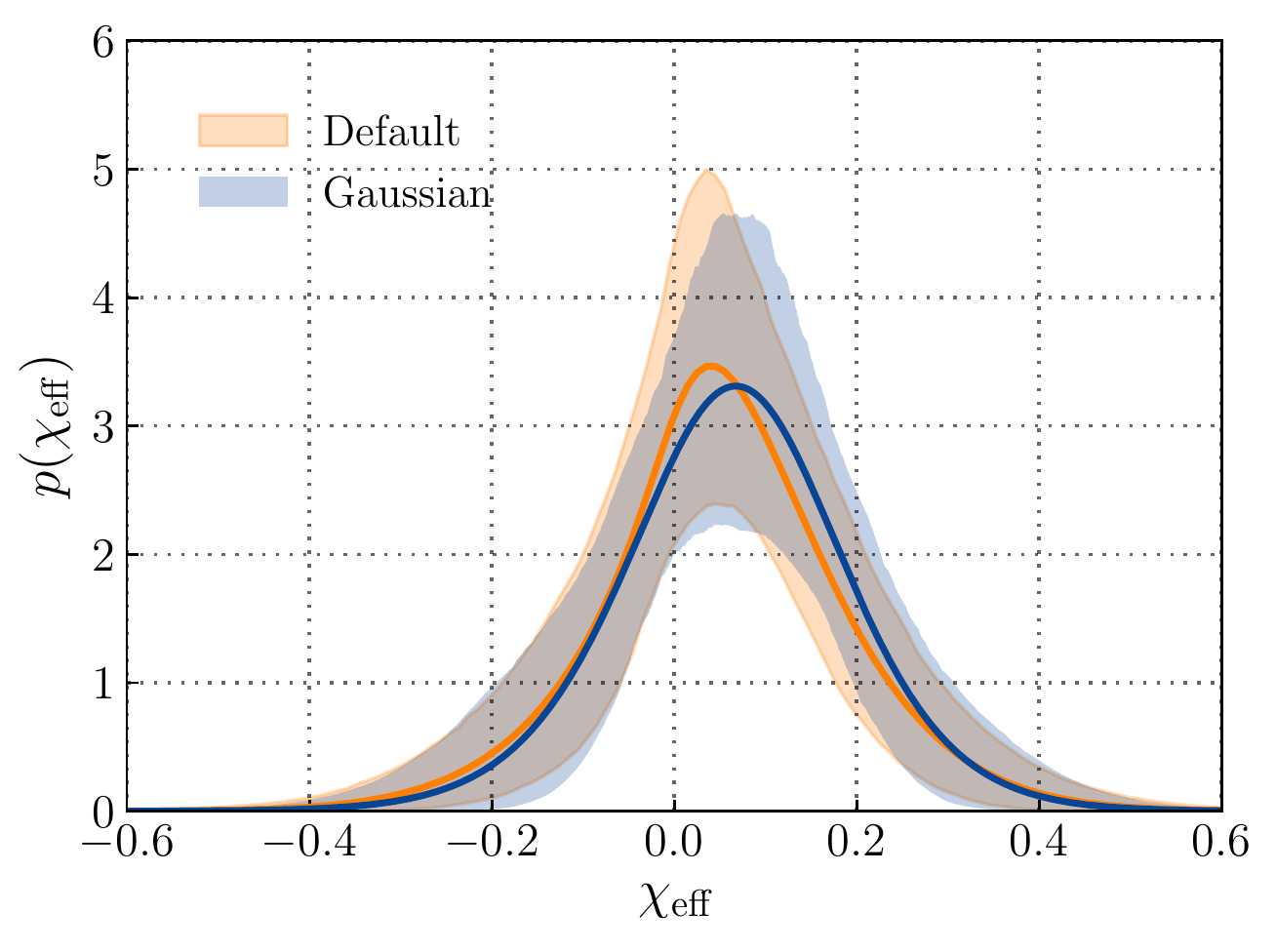}
    \caption{
    \textit{Left}: Posterior for the mean $\mu_\mathrm{eff}$ and standard deviation $\sigma_\mathrm{eff}$ of the BBH $\chi_\mathrm{eff}$ distribution, obtained using the \textsc{Gaussian} model described in Appendix~\ref{sec:Gaussian}.
    We marginalize over the parameters governing the distribution of the \chieff{} and the \chip{}.
    While we infer a $\chi_\mathrm{eff}$ distribution that is peaked at positive values, its measured width implies that a non-zero fraction of \bbhsys{} have negative $\chi_\mathrm{eff}$, implying component spins misaligned by $t_{1,2}>90^\circ$ relative to the orbital angular momentum.
    \textit{Right}: Population predictive distributions for the \chieff{} $\chi_\mathrm{eff}$ obtained with both the \textsc{Gaussian} and \textsc{Default} spin models.
    Shaded regions show the central 90\% credible bounds on $p(\chi_\mathrm{eff})$ and the solid lines show the median posterior prediction for the $\chi_\mathrm{eff}$ distribution.
    }
    \label{fig:chi_eff_cornerPlot}
\end{figure*}

\textbf{We observe \antialigned{} spin, which may suggest the presence of more than one binary formation channel.}
Using the \textsc{Gaussian} model described in Sec.~\ref{models}, we infer the presence of systems with negative \chieff{}: $\chi_\mathrm{eff} < 0$.
Thus, there exist \bbhsys{} with at least one component spin tilted by $\theta>90^\circ$ relative to the orbital angular momenta.
Figure~\ref{fig:chi_eff_cornerPlot} shows posteriors for the mean $\mu_\text{eff}$ and standard deviation $\sigma_\text{eff}$ of the $\chi_\mathrm{eff}$ distribution, marginalized over $\mu_p$, $\sigma_p$, and the covariance between the \chieff{} and the \chip{}.
With a peak at $\mu_\mathrm{eff} = {\medianMuEff{}^{+\errorMuEffHigh{}}_{-\errorMuEffLow{}}}$, we find that most systems have small but positive $\chi_\mathrm{eff}$, in agreement with the inference from GWTC-1~\citep{Miller2020,Roulet:2019}.
With GWTC-2, we can now also constrain the width of the $\chi_\mathrm{eff}$ distribution.
The result, $\sigma_\text{eff}={\medianSigmaEff{}^{+\errorSigmaEffHigh{}}_{-\errorSigmaEffLow{}}}$, requires that a nonzero fraction of BBH systems have $\chi_\text{eff}<0$.
Unlike the constraints on the $\chi_\mathrm{p}$ distribution presented above, the results for the presence of negative \chieff{} \textit{do} incorporate selection effects via the prescription described in Sec.~\ref{method}.

Analysis with the \textsc{Default} spin model is also suggestive of an anisotropic distribution of spin orientations.
In Fig.~\ref{fig:default-spin-ppds}, we plot the population distribution of $\cos \theta_{1,2} $ reconstructed using the \textsc{Default} model.
While the $\cos \theta_{1,2}$ distribution shows a preference for primarily aligned spins, with $\cos \theta_{1,2} > 0$, it also exhibits non-vanishing posterior support for $\cos \theta_{1,2} < 0$, indicating the presence of component spins misaligned by more than $90^\circ$.
The $\chi_\mathrm{eff}$ distribution inferred with the \textsc{Default} model closely matches the distribution inferred using the \textsc{Gaussian} model; compare the orange and blue bands in the right panel of Fig.~\ref{fig:chi_eff_cornerPlot}. The two models therefore agree on the fraction of systems with \antialigned{} component spins.

To further verify that the apparent presence of events with negative $\chi_\mathrm{eff}$ is physical and not an artifact of our choice of models, we repeat our inference of the \textsc{Gaussian} $\chi_\mathrm{eff}$ distribution, this time permitting the minimum allowed \chieff{} $\chi_\text{eff}^\text{min}$ (until now fixed to $\chi_\text{eff}^\text{min}=-1$) to vary as an additional hyper-parameter to be inferred from the data.
When fitting for $\chi_\mathrm{eff}^\mathrm{min}$ alongside $\mu_\mathrm{eff}$ and $\sigma_\mathrm{eff}$, we find that $\chi_\text{eff}^\text{min}$ is less than zero at \percentMinChiLessThanZero{} credibility (see Fig.~\ref{fig:chiMin_posterior} in the Appendix), confirming that the evidence for \antialigned{} spin is not an artifact of our parameterization. Allowing $\chi_\mathrm{eff}^\mathrm{min}$ to vary yields similar results for the implied $\chi_\mathrm{eff}$ distribution, and in particular, the fraction of systems with negative  $\chi_\mathrm{eff}$.

The presence of BBH systems with negative \chieff{} carries implications for the formation channels that give rise to stellar-mass BBH mergers.
BBHs born in the field from isolated stellar progenitors are predicted to contain components whose spins are nearly aligned with their orbital angular momenta, although sufficiently strong supernova kicks might produce modest misalignment~\citep{2017PhRvL.119a1101O,Stevenson,Gerosa2018,Rodriguez2016,Bavera2019}.
In contrast, binaries assembled dynamically in dense stellar environments are expected to have randomly oriented component spins, yielding positive or negative $\chi_\mathrm{eff}$ with equal probabilities~\citep{2000ApJ...541..319K,2010CQGra..27k4007M,Rodriguez2016,Zevin2017,Rodriguez2018,Doctor2019}.
\deleted{Measurements of BBH spin~\citep{2010CQGra..27k4007M,Stevenson,Fishbach:2017dwv,Talbot2017,Wysocki2019} provide one means to differentiate between these formation channels.
Other gravitational-wave observables may also offer clues about binary formation.
The observation of orbital eccentricity~\citep{Samsing2014,Samsing2017,Samsing2018,Romero-Shaw2019,Lower2018} or hierarchical merger candidates~\cite{Fishbach:2017dwv,2019PhRvD.100j4015C,Doctor2019,Kimball} could also provide strong circumstantial evidence for the role of dynamical mergers in the LIGO--Virgo catalogs.}

\begin{figure*}
    \centering
    \includegraphics[width = 0.7\textwidth]{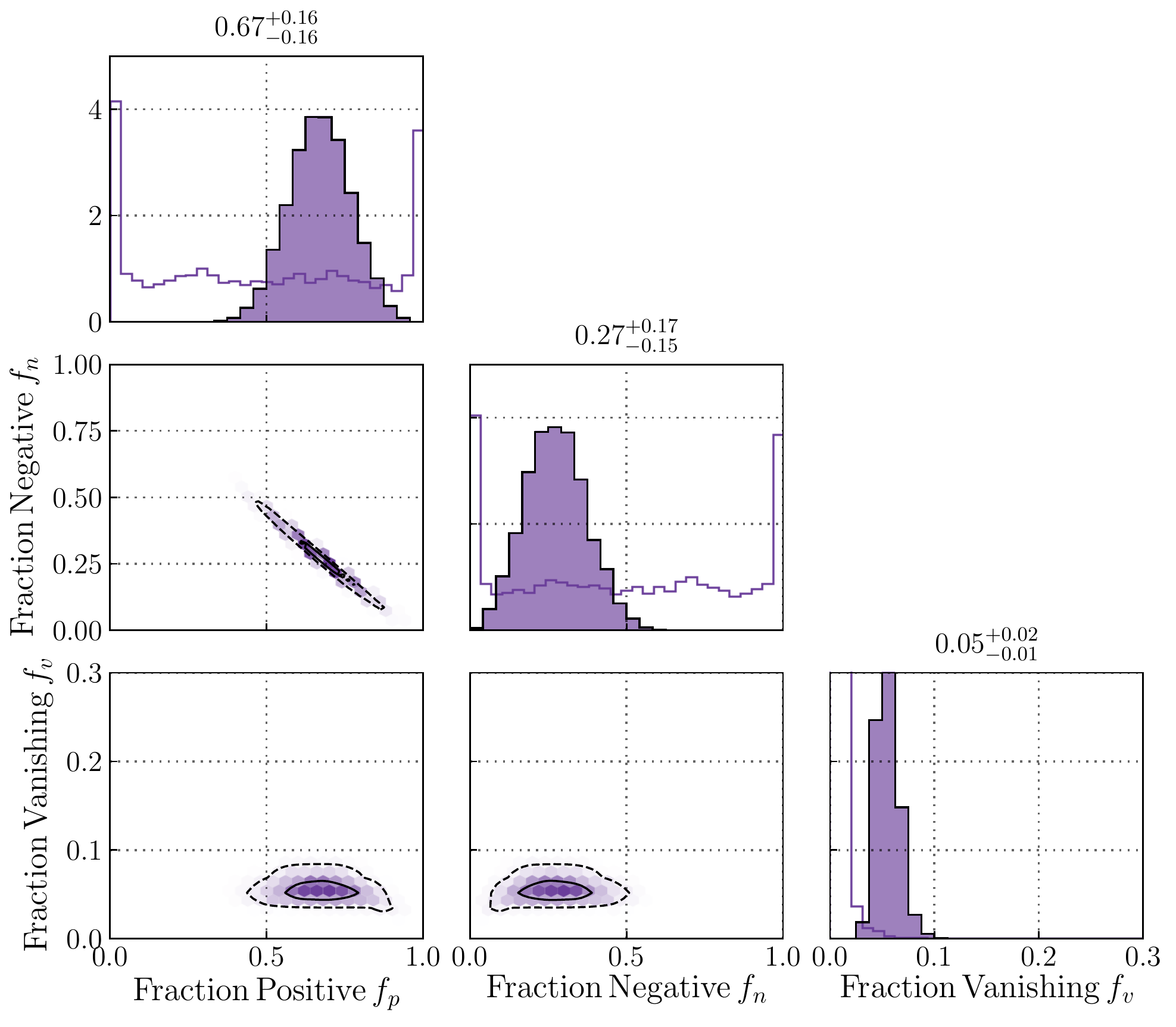}
    \caption{
    Posterior distribution for the fraction of \bbhevnt{} with positive or negative $\chi_\text{eff}$ (corresponding to alignment or anti-alignment of the BH spin with the orbital angular momentum; see Eq.~\ref{eq:chi_eff}).
    We also include the fraction of events that is consistent with vanishingly small spins $|\chi_\text{eff}|<0.01$.
    \replaced{We confidently infer that a nonzero fraction of events are spinning with (at least one) tilt less than $90^\circ$ (positive $\chi_\text{eff}$). A smaller fraction of events has (at least one) spin tilt $> 90^\circ$ (negative $\chi_\text{eff}$).} {We confidently infer that a nonzero fraction of events have positive $\chi_\text{eff}$, 
which requires at least one component to have spin tilt less than $90^\circ$. 
A smaller fraction of events have negative $\chi_\text{eff}$, which requires 
at least one component with spin tilt $> 90^\circ$.}
    A nonzero fraction of events have vanishingly small spins.
    The \textit{prior} distributions on the parameters are marked with the non-filled histograms.
    }
    \label{fig:fraction_negative}
\end{figure*}

Using the posteriors for $\mu_\mathrm{eff}$ and $\sigma_\mathrm{eff}$ from Fig.~\ref{fig:chi_eff_cornerPlot}, in Fig.~\ref{fig:fraction_negative} we show posteriors for the implied fractions of \bbhsys{} with negative ($\chi_\mathrm{eff}<-0.01$) and positive \chieff{} ($\chi_\mathrm{eff}>0.01$).
Motivated by recent work suggesting that BHs are born with natal spins as small as $\chi\approx 10^{-2}$~\citep{Qin2018,Fuller2019a,Fuller2019b,Bavera2019}, as well as the tendency of vanishingly small spins to confound efforts to distinguish between positive and negative $\chi_\mathrm{eff}$~\citep{BFarrSpin,O2pop}, we include a third bin containing vanishingly small spins between $-0.01\leq\chi_\mathrm{eff}\leq0.01$.
At 90\% credibility, we find that fractions \result{$f_p = \fractionChiEffPositive{}^{+\fractionChiEffPositiveErrorHigh{}}_{-\fractionChiEffPositiveErrorLow{}}$}, \result{$f_n = \fractionChiEffNegative{}^{+\fractionChiEffNegativeErrorHigh{}}_{-\fractionChiEffNegativeErrorLow{}}$}, and \result{$f_v = \fractionChiEffVanishing{}^{+\fractionChiEffVanishingErrorHigh{}}_{-\fractionChiEffVanishingErrorLow{}}$} of \bbhsys{} have positive, negative, and vanishing $\chi_\mathrm{eff}$, respectively.
All three posterior distributions are peaked away from zero.
In particular, $f_n > \fractionChiEffNegativeExtremeLowerLimit\%$ at 99\% credibility.
This result is in contrast to results obtained using GWTC-1 alone, which did not exhibit a confidently non-zero fraction of events with negative $\chi_\mathrm{eff}$~\citep{O2pop,Miller2020}.
Additionally, the relatively small fraction $f_v$ of binaries with vanishing spins may provide clues about how BHs gain angular momentum, given recent studies suggesting that most BHs are born slowly rotating~\citep{Fuller2019b}. While we define the vanishing bin to be $-0.01\leq\chi_\mathrm{eff}\leq0.01$, the exact choice of width for the vanishing bin does not strongly affect the values of $f_p$ and $f_n$, relative to one another.
\added{
We obtain nearly identical results, albeit with a slightly weaker lower bound on $f_n$, when we additionally allow the minimum effective inspiral spin $\chi_\text{eff}^\text{min}$ to vary as described above; in this case \result{$f_p = \fractionChiEffPositiveVarMin{}^{+\fractionChiEffPositiveErrorHighVarMin{}}_{-\fractionChiEffPositiveErrorLowVarMin{}}$} and \result{$f_n = \fractionChiEffNegativeVarMin{}^{+\fractionChiEffNegativeErrorHighVarMin{}}_{-\fractionChiEffNegativeErrorLowVarMin{}}$}.
}

As mentioned above, dynamical formation in dense clusters is not the only astrophysical explanation of negative \chieff{}.
If stellar progenitors experience both strong natal kicks in supernovae \textit{and} inefficient spin realignment, $\lesssim10\%$ of \bbhsys{} formed through isolated binary evolution may have $\chi_\mathrm{eff}<0$~\citep{Rodriguez2016,2017PhRvL.119a1101O,Stevenson,2018PhRvD..97d3014W}, although these results depend on the poorly understood physics of natal kicks and binary interaction via torques and mass transfer.
Moreover, we have so far neglected the possibility of other formation channels that may operate in both the field and dynamical regimes.
Isolated hierarchical triples, for example, may produce binary mergers with preferentially in-plane component spins~\citep{rodriguez_triple_2018,Antonini2018,Liu2019,2020MNRAS.493.3920F}.
Mergers in the disks of active galactic nuclei, meanwhile, yield component spins that are preferentially parallel or anti-parallel to a binary's orbital angular momentum~\citep{mckernan_constraining_2018,Yang2018,mckernan_monte-carlo_2019}.

With these qualifications in mind, if we interpret negative $\chi_\mathrm{eff}$ as indicative of dynamical formation in stellar clusters, then our constraints on $f_n$ can be used to infer the fraction of dynamically assembled binaries.
We assume that dynamical assembly in dense stellar environments yields a $\chi_\mathrm{eff}$ distribution that is symmetric about zero, while isolated binary evolution produces only \textit{positive} $\chi_\mathrm{eff}$.
Among the binaries with non-negligible spin (excluding those in the ``vanishing'' category above), the fractions $f_d$ and $f_i$ of binaries arising from dynamical and isolated channels are
\begin{align}
    \label{eq:fraction_d_i}
    f_d = & \frac{2 f_n}{f_p + f_n} , \\
    f_i = & \frac{f_p - f_n}{f_p + f_n} .
\end{align}
We find $\fractionChiEffDynamicalLow{}\leq f_d \leq \fractionChiEffDynamicalHigh{}$ at 90\% credibility, suggesting that both the field and the dynamical cluster scenarios contribute to the BBH mergers observed in GWTC-2.
Because the relative values of $f_n$ and $f_p$ are not sensitive to the width of the vanishing $\chi_\mathrm{eff}$ bin, this conclusion does not depend strongly on the definition of vanishing spin.

At present, we are unable to include a systematic investigation of waveform error in our analysis of \antialigned{} spin and orbital precession.
However, preliminary studies suggest that waveform error is unlikely to significantly affect this and other results in this paper.
\added{As described in~\citet{O3acatalog}, there is good agreement regarding the parameters of the GWTC-2 events when inferred with different waveforms.
There is a caveat: our studies do not include eccentric waveforms.
\cite{GW190521_formation,gayathri} suggest that GW190521 may have been eccentric and \cite{confusing} point out that eccentricity can be confused with precession for high-mass events. Currently the only event likely to be affected is GW190521~\citep{confusing}.
It is not clear how our results would change if we accounted for eccentricity, but we note that eccentricity can be a signature of dynamical assembly~\citep{Samsing2014,Samsing2017,Samsing2018,Romero-Shaw2019,Lower2018,2019MNRAS.486.4781F,2019ApJ...871...91Z}.
We also note that \cite{2020ApJ...904L..26F,NitzCapano} find that GW190521 may be an intermediate mass ratio inspiral, which could potentially alter our conclusions if true.
}
For future work, it would be worthwhile to estimate the systematic error using different waveform approximants.

\begin{figure*}
    \centering
    \includegraphics[width=\textwidth]{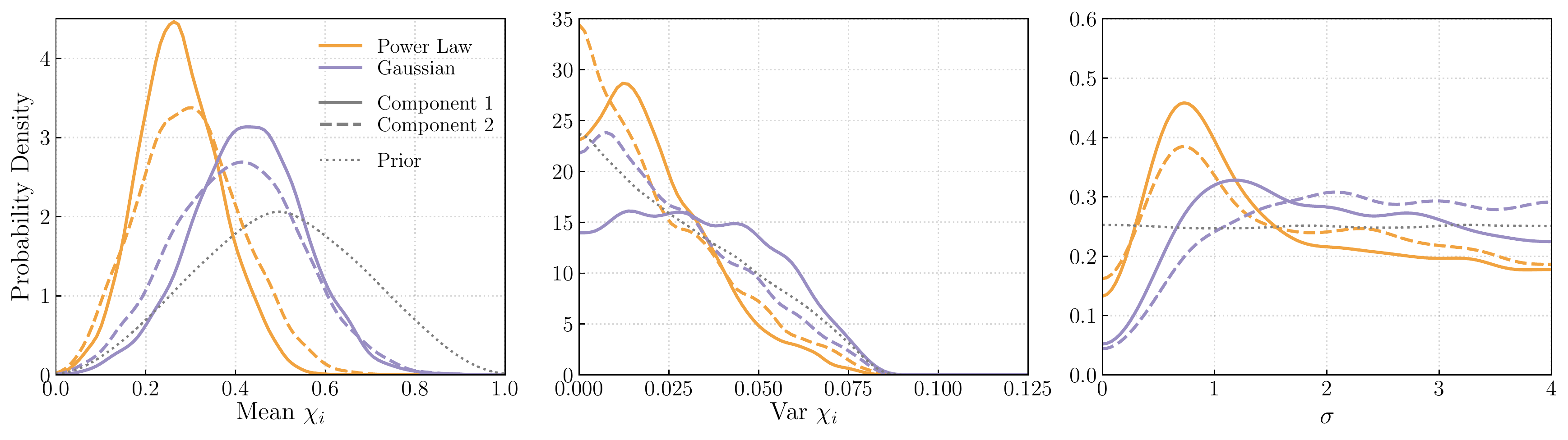}
    \caption{
    Posterior distribution for \ModelH{} spin hyper-parameters.
    Each subplot corresponds to a different parameter controlling the shape of the spin distribution (see Appendix~\ref{sec:modelH} for details).
    Each color corresponds to a different subpopulation (power-law or Gaussian) or component (primary mass or secondary mass) of the binary.
    The power-law subpopulation is slightly better measured than the Gaussian component, as a large number of detections are assigned to it.
    The results hint at a potential correlation between mass and spin, but the large measurement errors mean the spin distributions are consistent between the two subpopulations.
    }
    \label{fig:corner_spin_H}
\end{figure*}

\textbf{No strong evidence for variation of the spin distribution with mass.}
BHs born in hierarchical mergers inherit the orbital angular momenta of their progenitor systems, leading to significant spin magnitudes $\chi\approx0.7$ for nearly equal-mass systems~\citep{Pretorius:spin,2008ApJ...684..822B, Fishbach:2017dwv,Gerosa2018,Doctor2019,Rodriguez2018,Rodriguez2019,Kimball}.
If hierarchical mergers are present in GWTC-2, then one may expect correlations between the spins and masses of \bbhsys{}, with more massive hierarchical mergers also possessing larger spins.
We use the \ModelH{} model to explore possible trends in the BBH spin distribution with mass, allowing for a distinct low-mass and high-mass subpopulation (with primary mass distributions parameterized by a power-law and Gaussian, respectively; see Sec.~\ref{sec:modelH}), each with a distinct spin distribution.
The low-mass power-law has a weak preference for smaller spins, as compared to the high-mass Gaussian.
Both subpopulations disfavor perfectly aligned systems, though the low-mass subpopulation has more support for small misalignments.
In spite of these differences, the uncertainties on both of these subpopulations are broad enough that the two are fully consistent with each other, and we cannot confidently claim to detect a mass-dependence to the spin distribution at this stage.
This is demonstrated in Fig.~\ref{fig:corner_spin_H}, which shows the posteriors for the spin distribution hyper-parameters associated with each mass subpopulation.
These findings support the results of previous studies on GWTC-1, which could neither exclude nor confidently detect variation of the spin distribution with mass~\citep{2020ApJ...894..129S,2020arXiv200615047T}.

\subsection{Merger rate and redshift evolution}
In this subsection we use the \ppsn{} and \tapered{} mass models with the \textsc{Default} spin model, and infer the merger rate using the \textsc{Non-Evolving} redshift model and the \textsc{Power-law} evolution redshift model.

\textbf{We better constrain the binary black hole merger rate.}
Assuming a log-uniform prior, we find a BBH merger rate of $R_\text{BBH} = {\unit[\peakNoAugNoEvolutionrate]{Gpc^{-3}\,yr^{-1}}}$ using the \ppsn{} mass distribution and the assumption of a \textsc{non-evolving} merger rate density.
We find that estimates of the BBH merger rate are robust to our choice of mass model, with excellent agreement between the \ppsn{}, \tapered{}, and \multipeak{} models. 
The \truncated{} model yields a higher merger rate than the other models, but the results agree within statistical uncertainties: $R_\text{BBH} = {\unit[\truncatedNoAugNoEvolutionrate]{Gpc^{-3}\,yr^{-1}}}$.
\added{Our BBH merger rate estimate is consistent with the hypothesis that a significant fraction of the merger rate is due to dynamically assembled binaries in globular clusters~\citep{2000ApJ...528L..17P,2006ApJ...637..937O,2009ApJ...690.1370M,2011MNRAS.416..133D,Rodriguez:2015,2017MNRAS.469.4665P,2018PhRvL.121p1103F,2020arXiv200901861A}, young/open star clusters~\citep{2010MNRAS.402..371B,2014MNRAS.441.3703Z}, nuclear star clusters~\citep{2009MNRAS.395.2127O,2009ApJ...692..917M,2016ApJ...831..187A}, or active galactic nuclei discs~\citep{mckernan_constraining_2018,Tagawa_2020,Grobner}.}

\begin{table*}
    \centering
    \begin{tabular}{lcccc}
        \hline
        \multirow{2}{*}{\textbf{Mass model}} & \multicolumn{4}{c}{$\mathcal{R}\, [\mathrm{Gpc}^{-3}\mathrm{yr}^{-1}]$} \\
        \cline{2-5}
            & $4\,M_\odot \leq m_1 < 10\,M_\odot$ & $10\,M_\odot < m_1 < 20\,M_\odot$ & $20\,M_\odot \leq m_1 < 30\,M_\odot$ & $30\,M_\odot \leq m_1 < 40\,M_\odot$ \\
        \hline
        \hline
        \tapered{}   & \BPLNoAugNoEvolutionRateMassBandFourToTen & \BPLNoAugNoEvolutionRateMassBandTenToTwenty & \BPLNoAugNoEvolutionRateMassBandTwentyToThirty & \BPLNoAugNoEvolutionRateMassBandThirtyToForty \\
        \ppsn{}  & \peakNoAugNoEvolutionRateMassBandFourToTen  & \peakNoAugNoEvolutionRateMassBandTenToTwenty & \peakNoAugNoEvolutionRateMassBandTwentyToThirty & \peakNoAugNoEvolutionRateMassBandThirtyToForty\\
        \multipeak{} &  \multipeakNoAugNoEvolutionRateMassBandFourToTen & \multipeakNoAugNoEvolutionRateMassBandTenToTwenty & \multipeakNoAugNoEvolutionRateMassBandTwentyToThirty & \multipeakNoAugNoEvolutionRateMassBandThirtyToForty \\
        \hline
    \end{tabular}
    \caption{
        Merger rate estimate in different primary mass bins. These results assume a \textsc{non-evolving}  merger rate density, and exclude GW190814 from the analysis.
    }
    \label{tab:rates by mass}
\end{table*}

\added{In Table~\ref{tab:rates by mass}, we show the merger rate in different primary mass bins, as inferred with the \tapered{}, \ppsn{} and \multipeak{} models. Taking the range between the lowest 5\% and  highest 95\% estimate across these three models in each mass bin, we find the merger rate to be $\sim\unit[4\text{--}14]{Gpc^{-3}yr^{-1}}$ in the range of $10$--$20\ {M_\odot}$, $\sim\unit[1.3\text{--}5.3]{Gpc^{-3}yr^{-1}}$ in the range of $20$--$30\ {M_\odot}$, and $\sim\unit[1.3\text{--}5.2]{Gpc^{-3}yr^{-1}}$ in the range of $30$--$40\ {M_\odot}$.
  In Fig.~\ref{fig:deltam_mmin}, we showed that the primary mass spectrum turns over between $4$--$10\ {M_\odot}$.
  We estimate the merger rate in this range to be $\sim\unit[3\text{--}21]{Gpc^{-3}yr^{-1}}$.
}

Our estimate of the BBH merger rate include only systems with $m_1 \geq m_2 > 3 \ M_\odot$, which notably excludes \NAME{GW190814A}{}.
If we calculate the merger rate for all systems down to $m_2 \geq 2 \ M_\odot$ using our models, thereby including \NAME{GW190814A}{}, we infer a higher merger rate: $R_\text{BBH} = {\unit[\peakAllNoEvolutionrate]{Gpc^{-3}\,yr^{-1}}}$ for the \ppsn{} model.
The reason for this change is that including \NAME{GW190814A}{} increases the low-mass rate (see Fig.~\ref{fig:pm1_withlowmass}).
However, because our mass distribution models do not extrapolate well to $m_2 < 3 \ M_\odot$ (see Section~\ref{mass_results}), the fit with \NAME{GW190814A}{} likely overestimates the rate of systems with masses between $\sim 2.6\ M_\odot$ and $\sim 6 \ M_\odot$. 
Because of the uncertainty regarding the nature of \NAME{GW190814A}{} and the low significance of \NAME{GW190426A}{} (the other NSBH candidate in GWTC-2), we do not attempt to model the NSBH mass distribution, and do not calculate an NSBH merger rate. 
An estimate of the merger rate for \NAME{GW190814A}{}-like systems can be found in~\citet{GW190814}.

\textbf{We update the binary neutron star merger rate.}
We give an update to the BNS rate based on the two confident BNS detections in GWTC-2, GW170817 and \NAME{GW190425A}.
We estimate a rate $\mathcal{R}_\text{BNS}= \unit[\BNSrate]{Gpc^{-3}\,yr^{-1}}$ by Monte-Carlo sampling from a non-spinning distribution of systems, with component masses uniformly distributed between $1\ M_\odot$ and $2.5\ M_\odot$,\footnote{See~\cite{bns-mass,heavy_dns} for fits to the mass distribution of Galactic BNSs.} and uniformly distributed in comoving volume and detector-frame time with isotropic orientations and sky locations.
The detectability of each simulated system was approximated by requiring a network signal-to-noise ratio above 10 with signal-to-noise ratios above 5 in at least two detectors.
Because of the longer observing time and the lack of additional detections, we find a slightly smaller value for the BNS rate than previously reported: $R_{\rm BNS} = {\unit[\BNSrate]{Gpc^{-3}yr^{-1}}}$.
Assuming that there are $0.01$ Milky Way equivalent galaxies (MWEG) in $\unit[1]{Mpc^3}$~\citep{2008ApJ...675.1459K}, this implies a rate of $R_\text{BNS}=\result{\unit[32^{+49}_{-24}]{MWEG^{-1}Myr^{-1}}}$.

\begin{figure}
    \centering
    \includegraphics[width=0.48\textwidth]{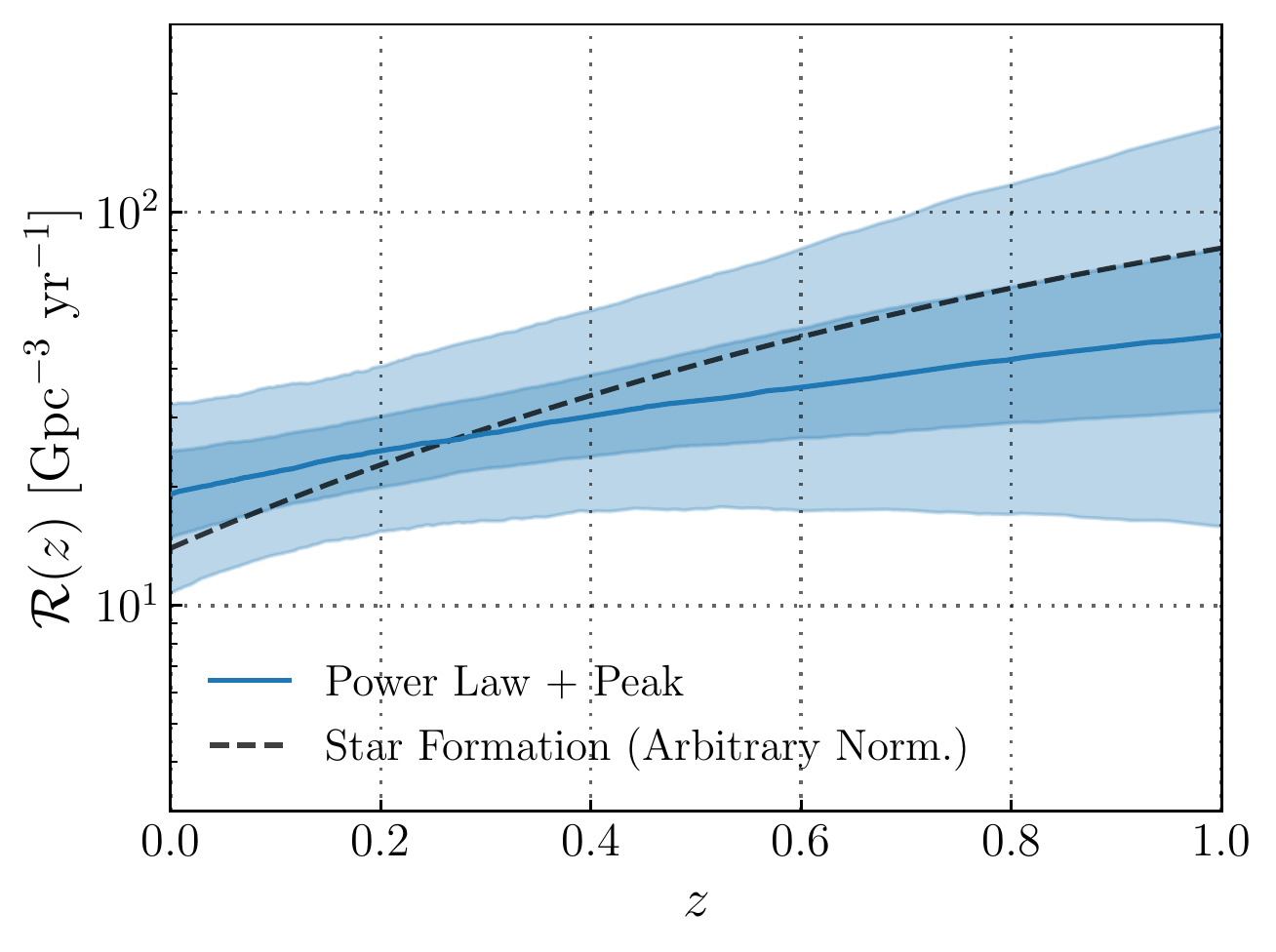}
    \caption{Merger rate density as a function of redshift, fit to the \textsc{Power-law Evolution} model.
    The solid curve shows the median rate density, while the dark (light) shaded region shows 50\% (90\%) credible intervals. The dashed curve shows the shape of the SFR.
    The data exhibit a mild preference for the merger rate to increase with redshift, but are consistent with a flat distribution as well as one that tracks the SFR.
    }
    \label{fig:ratevz}
\end{figure}

\begin{figure}
    \centering
    \includegraphics[width=0.48\textwidth]{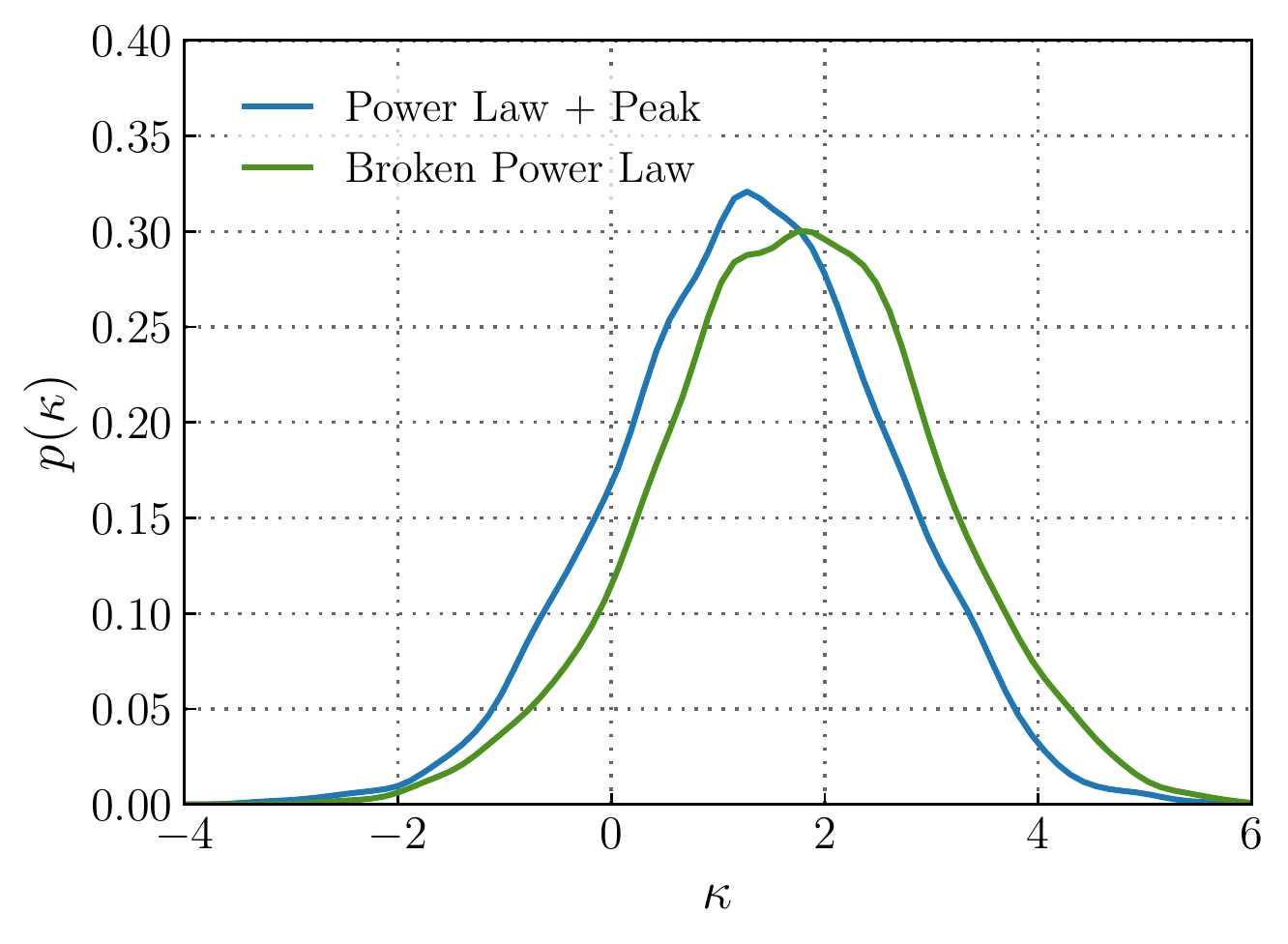}
    \caption{
    Posterior for the redshift evolution parameter $\kappa$ from the \textsc{Power-law Evolution} model, which assumes that rate density scales like $(1+z)^\kappa$.
    We assume the \ppsn{} and \tapered{} mass models, and take a flat prior on $\kappa$.
    }
    \label{fig:redshift-slope}
\end{figure}

\textbf{The BBH merger rate probably increases with redshift, but slower than the star-formation rate.}
Figure~\ref{fig:ratevz} shows the merger rate as a function of redshift using the \textsc{Power law} evolution model (see Appendix~\ref{Appendix:redshift} for additional details, and Fig.~\ref{fig:compare_cdfs_redshift} for a posterior predictive check).
When we allow the merger rate to evolve with redshift according to $(1+z)^\kappa$, we find that the $z = 0$ merger rate is $\mathcal{R}(z = 0) = \peakNoAugEvolutionrate \ \mathrm {Gpc}^{-3} \ \mathrm{yr}^{-1}$.
The posterior for the rate evolution parameter $\kappa$ is shown in Fig.~\ref{fig:redshift-slope}. 
Since GWTC-2 includes events with greater redshifts than the events in GWTC-1, we obtain a much tighter constraint on the evolution of the merger rate; compare our updated constraints of $\kappa = \peakNoAugEvolutionlamb$ (\ppsn{} model) and $\kappa = \BPLNoAugEvolutionlamb$ (\tapered{} model) to the GWTC-1 result of $\kappa = 8.4^{+9.6}_{-9.5}$.
We find that the merger rate is consistent with a non-evolving distribution ($\kappa = 0$), but is more likely to increase with increasing redshift, with $\kappa > 0$ at $\peakNoAugEvolutionLambdaPGrZero \%$ credibility (\ppsn{} model) or $\BPLNoAugEvolutionLambdaPGrZero \%$ (\tapered{} model).

Locally ($z\approx0$), the Madau--Dickinson star-formation rate~\citep{MadauDickinson} corresponds to $\kappa = 2.7$ in our \textsc{Power-Law Redshift} parameterization. We infer $\kappa < 2.7$ at \peakNoAugEvolutionMDPercentileInLambda\% credibility with the \ppsn{} mass model (\BPLNoAugEvolutionMDPercentileInLambda\% with \tapered{}).
Another way of comparing our inferred merger rate to the star-formation rate is by looking at the ratio between the rate at $z = 1$ and $z = 0$, $\mathcal{R}_\mathrm{BBH}(z = 1) / \mathcal{R}_\mathrm{BBH}(z = 0)$. For the star-formation rate $\mathcal{R}_\mathrm{SFR}(z = 1) / \mathcal{R}_\mathrm{SFR}(z = 0) \approx 6$, while for \bbhsys{}, we infer $\mathcal{R}_\mathrm{BBH}(z = 1) / \mathcal{R}_\mathrm{BBH}(z = 0) = \peakNoAugEvolutionRateAtRedshiftOneOverZero$(\ppsn{} model).
These results are consistent with most astrophysical formation channels, which predict a factor of $\sim 2$ increase between the merger rate at $z = 0$ and $z = 1$~\citep{2020arXiv200409533S,2013ApJ...779...72D,2019MNRAS.490.3740N,2019MNRAS.482..870E,2017MNRAS.472.2422M,2019PhRvD.100f4060B,Rodriguez:redshift}.

\section{Conclusions}\label{sec:conclusion}
The publication of the second LIGO--Virgo gravitational-wave transient catalog has increased the population of \bbhevnt{} by a factor of more than four.
The new catalog has highlighted the limitations of some early population models while yielding remarkable new signatures:

\begin{enumerate}
    \item We find that the BBH primary mass spectrum is not well-described as a simple power-law with an abrupt cut-off; there is a strong statistical preference for other models with non-trivial features such as a peak or a tapering.
    These features occur at $\approx\unit[37]{M_\odot}$, where one might expect pair instability supernovae (and pulsational pair-instability supernovae) to shape the mass distribution of BHs.
    \item At the opposite end of the spectrum, we observe a dearth of systems between NS and BH masses, suggesting that the BH mass spectrum likely turns over at $\sim \peakNoAugNoEvolutionTurnoverMass \ M_\odot$. 
      We constrain the minimum mass of BHs in BBH systems to be $\mmin\lesssim\peakNoAugNoEvolutionMminUpperNinety M_\odot$ at 90\% credibility.
      This is greater than the mass of BH candidates in Galactic binaries, e.g.,~\cite{Thompson}.
      \added{These results hold only when we restrict our analysis to events with both component masses above $3\,M_\odot$.}
    \item Meanwhile, we find that our models \added{fail to fit \NAME{GW190814A}{} together with the BBH systems with both components above $3\,M_\odot$.} \replaced{is an outlier in both secondary mass and mass ratio, indicating that it may belong to a distinct population compared to the other BBH systems.}{This may indicate that \NAME{GW190814A}{} belongs to a distinct population, or that there are additional features at the low mass end of the BBH mass spectrum that are missing in our models.} This is perhaps unsurprising as the combination of mass ratio, merger rate, and secondary mass inferred from this system pose a challenge to our current understanding of compact binary formation~\citep{GW190814,2020CmPhy...3...43A,Zevin:2020gma,Safarzadeh2020}.
    \item We detect clear evidence of spin-induced, general relativistic precession of the orbital plane.
    We determine that this signature is not due to a single precessing merger, but from the overall preference of the data for precessing waveforms.
    \item We observe that some fraction of the BHs in GWTC-2 are spinning with an orientation that is \antialigned{} with respect to the orbital angular momentum of the binary.
    If we plausibly assume that all binaries with \antialigned{} spins are assembled dynamically, this may imply that LIGO--Virgo events merge both dynamically and in the field. Based on the inferred mass and spin distributions, we find no clear evidence for or against hierarchical mergers in GWTC-2.
    \newline
    \item We compute the rate of compact binary mergers, finding $\mathcal{R}_\mathrm{BNS} = \BNSrate \ \mathrm{Gpc}^{-3} \ \mathrm{yr}^{-1}$ and $\mathcal{R}_\mathrm{BBH} = \peakNoAugNoEvolutionrate \ \mathrm{Gpc}^{-3} \ \mathrm{yr}^{-1}$.
    The data are consistent with both a merger rate that is constant in time and one that tracks the SFR in the local universe, though the data prefer a merger rate that is somewhere in between.
    We find that the merger rate at $z = 1$ differs from the merger rate at $z = 0$ by a factor of $\mathcal{R}_\mathrm{BBH}(z = 1) / \mathcal{R}_\mathrm{BBH}(z = 0) = \peakNoAugEvolutionRateAtRedshiftOneOverZero$, to be compared with the SFR, $\mathcal{R}_\mathrm{SFR}(z = 1) / \mathcal{R}_\mathrm{SFR}(z = 0) \sim 6$.
\end{enumerate}

While a clearer picture is emerging of the population properties of compact binaries, key questions remain.
How do we best characterize the deviations from power-law in the primary BH mass spectrum, and what is the physical origin of these new features?
What is the origin of BBH mergers in the high-mass gap: hierarchical mergers, stars producing remnants heavier than expected from pair instability supernovae theory, or something else?
What is the shape of the mass spectrum between NS and BH masses, and does the current dearth of systems between $\sim 3 \ M_\odot$ and $\sim 6 \ M_\odot$ represent an empty low-mass gap? If so, do systems like the secondary mass of \NAME{GW190814A}{} belong to the NS or BH side of the gap?
Is the observation of \antialigned{} spins indicative of dynamically assembled binaries?
As the sensitivity of LIGO, Virgo, and KAGRA improves, and as more gravitational-wave transients are detected, we expect to begin to answer these questions.
\added{As future observations subject our models to increasing scrutiny, it is inevitable that refinements will be required to fit newly resolved features.
  This cycle of refining models to account for new data will reveal new questions while providing an evolving understanding of the conclusions presented here.}

\acknowledgments
The authors gratefully acknowledge the support of the United States
National Science Foundation (NSF) for the construction and operation of the
LIGO Laboratory and Advanced LIGO as well as the Science and Technology Facilities Council (STFC) of the
United Kingdom, the Max-Planck-Society (MPS), and the State of
Niedersachsen/Germany for support of the construction of Advanced LIGO 
and construction and operation of the GEO600 detector. 
Additional support for Advanced LIGO was provided by the Australian Research Council.
The authors gratefully acknowledge the Italian Istituto Nazionale di Fisica Nucleare (INFN),  
the French Centre National de la Recherche Scientifique (CNRS) and
the Netherlands Organization for Scientific Research, 
for the construction and operation of the Virgo detector
and the creation and support  of the EGO consortium. 
The authors also gratefully acknowledge research support from these agencies as well as by 
the Council of Scientific and Industrial Research of India, 
the Department of Science and Technology, India,
the Science \& Engineering Research Board (SERB), India,
the Ministry of Human Resource Development, India,
the Spanish Agencia Estatal de Investigaci\'on,
the Vicepresid\`encia i Conselleria d'Innovaci\'o, Recerca i Turisme and the Conselleria d'Educaci\'o i Universitat del Govern de les Illes Balears,
the Conselleria d'Innovaci\'o, Universitats, Ci\`encia i Societat Digital de la Generalitat Valenciana and
the CERCA Programme Generalitat de Catalunya, Spain,
the National Science Centre of Poland,
the Swiss National Science Foundation (SNSF),
the Russian Foundation for Basic Research, 
the Russian Science Foundation,
the European Commission,
the European Regional Development Funds (ERDF),
the Royal Society, 
the Scottish Funding Council, 
the Scottish Universities Physics Alliance, 
the Hungarian Scientific Research Fund (OTKA),
the French Lyon Institute of Origins (LIO),
the Belgian Fonds de la Recherche Scientifique (FRS-FNRS), 
Actions de Recherche Concertées (ARC) and
Fonds Wetenschappelijk Onderzoek – Vlaanderen (FWO), Belgium,
the Paris \^{I}le-de-France Region, 
the National Research, Development and Innovation Office Hungary (NKFIH), 
the National Research Foundation of Korea,
Industry Canada and the Province of Ontario through the Ministry of Economic Development and Innovation, 
the Natural Science and Engineering Research Council Canada,
the Canadian Institute for Advanced Research,
the Brazilian Ministry of Science, Technology, Innovations, and Communications,
the International Center for Theoretical Physics South American Institute for Fundamental Research (ICTP-SAIFR), 
the Research Grants Council of Hong Kong,
the National Natural Science Foundation of China (NSFC),
the Leverhulme Trust, 
the Research Corporation, 
the Ministry of Science and Technology (MOST), Taiwan
and
the Kavli Foundation.
The authors gratefully acknowledge the support of the NSF, STFC, INFN and CNRS for provision of computational resources.

{\it We would like to thank all of the essential workers who put their health at risk during the COVID-19 pandemic, without whom we would not have been able to complete this work.}

This is LIGO document number LIGO-P2000077.

\bibliographystyle{apj}
\bibliography{o3a_rates}

\begin{thebibliography}{}
\expandafter\ifx\csname natexlab\endcsname\relax\def\natexlab#1{#1}\fi

\bibitem[{Aasi {et~al.}(2015)}]{aLIGO}
Aasi, J., {et~al.} 2015, CQGra, 32, 074001

\bibitem[{Abbott {et~al.}(2013)Abbott, Abbott, Abbott, {et~al.}}]{ObsProspects}
Abbott, B., Abbott, R., Abbott, T., {et~al.} 2013, Prospects for Observing and
  Localizing Gravitational-Wave Transients with Advanced LIGO, Advanced Virgo
  and KAGRA, arXiv:1304.0670

\bibitem[{Abbott {et~al.}(2016{\natexlab{a}})Abbott, Abbott, Abbott,
  {et~al.}}]{2016PhRvD..94f4035A}
---. 2016{\natexlab{a}}, PhRvD, 94, 064035

\bibitem[{Abbott {et~al.}(2016{\natexlab{b}})Abbott, Abbott, Abbott,
  {et~al.}}]{GW151226}
---. 2016{\natexlab{b}}, PhRvL, 116, 241103

\bibitem[{Abbott {et~al.}(2016{\natexlab{c}})Abbott, Abbott, Abbott,
  {et~al.}}]{GW150914PE}
---. 2016{\natexlab{c}}, PhRvL, 116, 241102

\bibitem[{Abbott {et~al.}(2019{\natexlab{a}})Abbott, Abbott, Abbott,
  {et~al.}}]{O2pop}
---. 2019{\natexlab{a}}, ApJL, 882, L24

\bibitem[{Abbott {et~al.}(2019{\natexlab{b}})Abbott, Abbott, Abbott,
  {et~al.}}]{GWTC1}
---. 2019{\natexlab{b}}, PhRvX, 9, 031040

\bibitem[{Abbott {et~al.}(2019{\natexlab{c}})Abbott, Abbott, Abbott,
  {et~al.}}]{Abbott:2019ebz}
---. 2019{\natexlab{c}}, arXiv:1912.11716

\bibitem[{Abbott {et~al.}(2020{\natexlab{a}})Abbott, Abbott, Abraham,
  {et~al.}}]{GW190412}
Abbott, R., Abbott, T., Abraham, S., {et~al.} 2020{\natexlab{a}},
  arXiv:2004.08342

\bibitem[{Abbott {et~al.}(2020{\natexlab{b}})Abbott, Abbott, Abraham,
  {et~al.}}]{GW190814}
---. 2020{\natexlab{b}}, ApJL, 896, L44

\bibitem[{Abbott {et~al.}(2020{\natexlab{c}})Abbott, Abbott, Abraham,
  {et~al.}}]{O3acatalog}
---. 2020{\natexlab{c}}, arXiv e-prints, arXiv:2010.14527

\bibitem[{Abbott {et~al.}(2020{\natexlab{d}})}]{P2000434}
Abbott, R., {et~al.} 2020{\natexlab{d}}

\bibitem[{Abbott {et~al.}(2020{\natexlab{e}})}]{GW190521}
---. 2020{\natexlab{e}}, PhRvL, 125, 101102

\bibitem[{Abbott {et~al.}(2020{\natexlab{f}})}]{Abbott:GW190521_implications}
---. 2020{\natexlab{f}}, ApJL, 900, L13

\bibitem[{Acernese {et~al.}(2015)}]{aVirgo}
Acernese, F., {et~al.} 2015, CQGra, 32, 024001

\bibitem[{Adams {et~al.}(2016)Adams, Buskulic, Germain, Guidi, Marion, Montani,
  Mours, Piergiovanni, \& Wang}]{MBTA}
Adams, T., Buskulic, D., Germain, V., {et~al.} 2016, Classical and Quantum
  Gravity, 33, 175012

\bibitem[{Allen(2005)}]{PyCBC3}
Allen, B. 2005, PhRvD, 71, "062001"

\bibitem[{Allen {et~al.}(2012)Allen, Anderson, Brady, Brown, \&
  Creighton}]{PyCBC2}
Allen, B., Anderson, W.~G., Brady, P.~R., Brown, D.~A., \& Creighton, J. D.~E.
  2012, PhRvD, 85, "122006"

\bibitem[{{Antonini} \& {Gieles}(2020)}]{2020arXiv200901861A}
{Antonini}, F., \& {Gieles}, M. 2020, arXiv e-prints, arXiv:2009.01861

\bibitem[{Antonini {et~al.}(2014)Antonini, Murray, \&
  Mikkola}]{2014ApJ...781...45A}
Antonini, F., Murray, N., \& Mikkola, S. 2014, ApJ, 781, 45

\bibitem[{Antonini \& Rasio(2016)}]{2016ApJ...831..187A}
Antonini, F., \& Rasio, F.~A. 2016, ApJ, 831, 187

\bibitem[{Antonini {et~al.}(2018)Antonini, Rodriguez, Petrovich, \&
  Fischer}]{Antonini2018}
Antonini, F., Rodriguez, C.~L., Petrovich, C., \& Fischer, C.~L. 2018, Mon.
  Not. R. Ast. Soc. Lett., 480, L58

\bibitem[{Antonini {et~al.}(2017)Antonini, Toonen, \&
  Hamers}]{2017ApJ...841...77A}
Antonini, F., Toonen, S., \& Hamers, A.~S. 2017, ApJ, 841, 77

\bibitem[{Apostolatos {et~al.}(1994)Apostolatos, Cutler, Sussman, \&
  Thorne}]{Apostolatos}
Apostolatos, T.~A., Cutler, C., Sussman, G.~J., \& Thorne, K.~S. 1994, PhRvD,
  49, 6274

\bibitem[{{Arca Sedda}(2020{\natexlab{a}})}]{2020ApJ...891...47A}
{Arca Sedda}, M. 2020{\natexlab{a}}, \apj, 891, 47

\bibitem[{{Arca Sedda}(2020{\natexlab{b}})}]{2020CmPhy...3...43A}
---. 2020{\natexlab{b}}, Communications Physics, 3, 43

\bibitem[{{Arca Sedda} {et~al.}(2020){Arca Sedda}, {Mapelli}, {Spera},
  {Benacquista}, \& {Giacobbo}}]{2020ApJ...894..133A}
{Arca Sedda}, M., {Mapelli}, M., {Spera}, M., {Benacquista}, M., \& {Giacobbo},
  N. 2020, \apj, 894, 133

\bibitem[{Ashton \& Thrane(2020)}]{Ashton}
Ashton, G., \& Thrane, E. 2020, Mon. Not. Roy. Astron. Soc., 498, 1905

\bibitem[{{Ashton} {et~al.}(2019){Ashton}, {H{\"u}bner}, {Lasky}, {Talbot},
  {Ackley}, {Biscoveanu}, {Chu}, {Divakarla}, {Easter}, {Goncharov}, {Hernandez
  Vivanco}, {Harms}, {Lower}, {Meadors}, {Melchor}, {Payne}, {Pitkin},
  {Powell}, {Sarin}, {Smith}, \& {Thrane}}]{bilby}
{Ashton}, G., {H{\"u}bner}, M., {Lasky}, P.~D., {et~al.} 2019, ApJS, 241, 27

\bibitem[{Babak {et~al.}(2017)Babak, Taracchini, \& Buonanno}]{Babak:2016tgq}
Babak, S., Taracchini, A., \& Buonanno, A. 2017, Phys. Rev. D, 95, 024010

\bibitem[{{Baibhav} {et~al.}(2019){Baibhav}, {Berti}, {Gerosa}, {Mapelli},
  {Giacobbo}, {Bouffanais}, \& {Di Carlo}}]{2019PhRvD.100f4060B}
{Baibhav}, V., {Berti}, E., {Gerosa}, D., {et~al.} 2019, PhRvD, 100, 064060

\bibitem[{Bailyn {et~al.}(1998)Bailyn, Jain, Coppi, \& Orosz}]{Bailyn1998}
Bailyn, C., Jain, R.~K., Coppi, P., \& Orosz, J.~A. 1998, ApJ, 499, 367

\bibitem[{{Banerjee} {et~al.}(2010){Banerjee}, {Baumgardt}, \&
  {Kroupa}}]{2010MNRAS.402..371B}
{Banerjee}, S., {Baumgardt}, H., \& {Kroupa}, P. 2010, \mnras, 402, 371

\bibitem[{{Barkat} {et~al.}(1967){Barkat}, {Rakavy}, \&
  {Sack}}]{1967PhRvL..18..379B}
{Barkat}, Z., {Rakavy}, G., \& {Sack}, N. 1967, PhRvL, 18, 379

\bibitem[{Bartos {et~al.}(2017)Bartos, Kocsis, Haiman, \&
  M{\'a}rka}]{2017ApJ...835..165B}
Bartos, I., Kocsis, B., Haiman, Z., \& M{\'a}rka, S. 2017, ApJ, 835, 165

\bibitem[{Bavera {et~al.}(2019)Bavera, Fragos, Qin, Zapartas, Neijssel, Mandel,
  Batta, Gaebel, Kimball, \& Stevenson}]{Bavera2019}
Bavera, S.~S., Fragos, T., Qin, Y., {et~al.} 2019, A\&A, 635, A97

\bibitem[{Belczynski {et~al.}(2002)Belczynski, Kalogera, \&
  Bulik}]{2002ApJ...572..407B}
Belczynski, K., Kalogera, V., \& Bulik, T. 2002, ApJ, 572, 407

\bibitem[{{Belczynski} {et~al.}(2016){Belczynski}, {Heger}, {Gladysz},
  {Ruiter}, {Woosley}, {Wiktorowicz}, {Chen}, {Bulik}, {O'Shaughnessy}, {Holz},
  {Fryer}, \& {Berti}}]{2016A&A...594A..97B}
{Belczynski}, K., {Heger}, A., {Gladysz}, W., {et~al.} 2016, \aap, 594, A97

\bibitem[{{Belczynski} {et~al.}(2020){Belczynski}, {Klencki}, {Fields},
  {Olejak}, {Berti}, {Meynet}, {Fryer}, {Holz}, {O'Shaughnessy}, {Brown}, \&
  et~al.}]{2020AnA...636A.104B}
{Belczynski}, K., {Klencki}, J., {Fields}, C.~E., {et~al.} 2020, \aap, 636,
  A104

\bibitem[{{Berti} \& {Volonteri}(2008)}]{2008ApJ...684..822B}
{Berti}, E., \& {Volonteri}, M. 2008, \apj, 684, 822

\bibitem[{Bethe \& Brown(1998)}]{1998ApJ...506..780B}
Bethe, H.~A., \& Brown, G.~E. 1998, ApJ, 506, 780

\bibitem[{Biscoveanu {et~al.}(2020)Biscoveanu, Isi, Vitale, \&
  Varma}]{Biscoveanu:2020are}
Biscoveanu, S., Isi, M., Vitale, S., \& Varma, V. 2020, arXiv:2007.09156

\bibitem[{Boh{\'e} {et~al.}(2016)Boh{\'e}, Hannam, Husa, Ohme, Puerrer, \&
  Schmidt}]{Bohe}
Boh{\'e}, A., Hannam, M., Husa, S., {et~al.} 2016

\bibitem[{Boh{\'e} {et~al.}(2017)}]{Bohe:2016gbl}
Boh{\'e}, A., {et~al.} 2017, PhRvD, 95, 044028

\bibitem[{{Calder{\'o}n Bustillo} {et~al.}(2020){Calder{\'o}n Bustillo},
  Sanchis-Gual, Torres-Forn{\'e}, \& Font}]{confusing}
{Calder{\'o}n Bustillo}, J., Sanchis-Gual, N., Torres-Forn{\'e}, A., \& Font,
  J.~A. 2020, arxiv/2009.01066

\bibitem[{Calderón~Bustillo {et~al.}(2017)Calderón~Bustillo, Laguna, \&
  Shoemaker}]{Calder_n_Bustillo_2017}
Calderón~Bustillo, J., Laguna, P., \& Shoemaker, D. 2017, Physical Review D,
  95, doi:10.1103/physrevd.95.104038

\bibitem[{Canton {et~al.}(2014)}]{PyCBC4}
Canton, T.~D., {et~al.} 2014, PhRvD, 90, 082004

\bibitem[{Carlo {et~al.}(2019{\natexlab{a}})}]{2019arXiv191101434D}
Carlo, U.~D., {et~al.} 2019{\natexlab{a}}, arxiv/1911.01434

\bibitem[{Carlo {et~al.}(2019{\natexlab{b}})}]{2019MNRAS.487.2947D}
---. 2019{\natexlab{b}}, MNRAS, 487, 2947

\bibitem[{Carpenter {et~al.}(2017)Carpenter, Gelman, Hoffman, Lee, Goodrich,
  Betancourt, Brubaker, Guo, Li, \& Riddell}]{stan}
Carpenter, B., Gelman, A., Hoffman, M.~D., {et~al.} 2017, Journal of
  Statistical Software, 76, doi:10.18637/jss.v076.i01

\bibitem[{Carr {et~al.}(2016)Carr, Kühnel, \& Sandstad}]{2016PhRvD..94h3504C}
Carr, B., Kühnel, F., \& Sandstad, M. 2016, PhRvD, 94, 083504

\bibitem[{Carr \& Hawking(1974)}]{1974MNRAS.168..399C}
Carr, B.~J., \& Hawking, S.~W. 1974, MNRAS, 168, 399

\bibitem[{Chatziioannou {et~al.}(2019)}]{2019PhRvD.100j4015C}
Chatziioannou, K., {et~al.} 2019, PhRvD, 100, 104015

\bibitem[{Chu(2017)}]{spiir}
Chu, Q. 2017, PhD thesis, The University of Western Australia

\bibitem[{Croon {et~al.}(2020)Croon, McDermott, \& Sakstein}]{Croon:2020oga}
Croon, D., McDermott, S.~D., \& Sakstein, J. 2020, arXiv:2007.07889

\bibitem[{Dai {et~al.}(2017)Dai, Venumadhav, \& Sigurdson}]{Dai:2016igl}
Dai, L., Venumadhav, T., \& Sigurdson, K. 2017, PhRvD, D95, 044011

\bibitem[{Damour(2001)}]{PhysRevD.64.124013}
Damour, T. 2001, PhRvD, 64, 124013

\bibitem[{de~Mink \& Mandel(2016)}]{2016MNRAS.460.3545D}
de~Mink, S.~E., \& Mandel, I. 2016, MNRAS, 460, 3545

\bibitem[{Doctor {et~al.}(2019)Doctor, Wysocki, O'Shaughnessy, Holz, \&
  Farr}]{Doctor2019}
Doctor, Z., Wysocki, D., O'Shaughnessy, R., Holz, D.~E., \& Farr, B. 2019, ApJ,
  893, 35

\bibitem[{{Dominik} {et~al.}(2013){Dominik}, {Belczynski}, {Fryer}, {Holz},
  {Berti}, {Bulik}, {Mand el}, \& {O'Shaughnessy}}]{2013ApJ...779...72D}
{Dominik}, M., {Belczynski}, K., {Fryer}, C., {et~al.} 2013, ApJ, 779, 72

\bibitem[{{Dominik} {et~al.}(2015){Dominik}, {Berti}, {O'Shaughnessy},
  {Mandel}, {Belczynski}, {Fryer}, {Holz}, {Bulik}, \&
  {Pannarale}}]{2015ApJ...806..263D}
{Dominik}, M., {Berti}, E., {O'Shaughnessy}, R., {et~al.} 2015, apj, 806, 263

\bibitem[{{Downing} {et~al.}(2011){Downing}, {Benacquista}, {Giersz}, \&
  {Spurzem}}]{2011MNRAS.416..133D}
{Downing}, J.~M.~B., {Benacquista}, M.~J., {Giersz}, M., \& {Spurzem}, R. 2011,
  \mnras, 416, 133

\bibitem[{{Eldridge} {et~al.}(2019){Eldridge}, {Stanway}, \&
  {Tang}}]{2019MNRAS.482..870E}
{Eldridge}, J.~J., {Stanway}, E.~R., \& {Tang}, P.~N. 2019, MNRAS, 482, 870

\bibitem[{{Eldridge} {et~al.}(2017){Eldridge}, {Stanway}, {Xiao}, {McClelland
  }, {Taylor}, {Ng}, {Greis}, \& {Bray}}]{2017PASA...34...58E}
{Eldridge}, J.~J., {Stanway}, E.~R., {Xiao}, L., {et~al.} 2017, \pasa, 34, e058

\bibitem[{Essick \& Landry(2020)}]{Essick:2020ghc}
Essick, R., \& Landry, P. 2020, arXiv:2007.01372

\bibitem[{Fairhurst {et~al.}(2019)Fairhurst, Green, Hannam, \&
  Hoy}]{Fairhurst:2019srr}
Fairhurst, S., Green, R., Hannam, M., \& Hoy, C. 2019, arXiv:1908.00555

\bibitem[{{Farmer} {et~al.}(2020){Farmer}, {Renzo}, {de Mink}, {Fishbach}, \&
  {Justham}}]{Farmer:2020}
{Farmer}, R., {Renzo}, M., {de Mink}, S., {Fishbach}, M., \& {Justham}, S.
  2020, arXiv e-prints, arXiv:2006.06678

\bibitem[{Farmer {et~al.}(2019)Farmer, Renzo, de~Mink, Marchant, \&
  Justham}]{Farmer_2019}
Farmer, R., Renzo, M., de~Mink, S.~E., Marchant, P., \& Justham, S. 2019, ApJ,
  887, 53

\bibitem[{{Farr} {et~al.}(2018){Farr}, {Holz}, \& {Farr}}]{BFarrSpin}
{Farr}, B., {Holz}, D.~E., \& {Farr}, W.~M. 2018, ApJL, 854, L9

\bibitem[{{Farr}(2019)}]{Farr:selection}
{Farr}, W.~M. 2019, Research Notes of the American Astronomical Society, 3, 66

\bibitem[{Farr {et~al.}(2011)Farr, Sravan, Cantrell, Kreidberg, Bailyn, Mandel,
  \& Kalogera}]{Farr2011}
Farr, W.~M., Sravan, N., Cantrell, A., {et~al.} 2011, ApJ, 741, 103

\bibitem[{Farr {et~al.}(2017)Farr, Stevenson, Coleman~Miller, Mandel, Farr, \&
  Vecchio}]{Farr:2017uvj}
Farr, W.~M., Stevenson, S., Coleman~Miller, M., {et~al.} 2017, Nature, 548, 426

\bibitem[{Farrell {et~al.}(2020)Farrell, Groh, Hirschi, Murphy, Kaiser,
  Ekstr\"om, Georgy, \& Meynet}]{Farrell:2020zju}
Farrell, E.~J., Groh, J.~H., Hirschi, R., {et~al.} 2020, arXiv:2009.06585

\bibitem[{Farrow {et~al.}(2019)Farrow, Zhu, \& Thrane}]{bns-mass}
Farrow, N., Zhu, X.-J., \& Thrane, E. 2019, ApJ, 876, 18

\bibitem[{{Fernandez} \& {Profumo}(2019)}]{2019JCAP...08..022F}
{Fernandez}, N., \& {Profumo}, S. 2019, \jcap, 2019, 022

\bibitem[{Fishbach {et~al.}(2020{\natexlab{a}})Fishbach, Essick, \&
  Holz}]{Fishbach:2020ryj}
Fishbach, M., Essick, R., \& Holz, D.~E. 2020{\natexlab{a}}, arXiv:2006.13178

\bibitem[{Fishbach {et~al.}(2020{\natexlab{b}})Fishbach, Farr, \&
  Holz}]{Fishbach:2019ckx}
Fishbach, M., Farr, W.~M., \& Holz, D.~E. 2020{\natexlab{b}}, ApJL, 891, L31

\bibitem[{Fishbach \& Holz(2017)}]{Fishbach2017}
Fishbach, M., \& Holz, D.~E. 2017, ApJL, 851, L25

\bibitem[{{Fishbach} \& {Holz}(2020{\natexlab{a}})}]{2020ApJ...904L..26F}
{Fishbach}, M., \& {Holz}, D.~E. 2020{\natexlab{a}}, \apjl, 904, L26

\bibitem[{{Fishbach} \& {Holz}(2020{\natexlab{b}})}]{Fishbach:picky}
---. 2020{\natexlab{b}}, ApJL, 891, L27

\bibitem[{Fishbach {et~al.}(2017)Fishbach, Holz, \& Farr}]{Fishbach:2017dwv}
Fishbach, M., Holz, D.~E., \& Farr, B. 2017, ApJL, 840, L24

\bibitem[{Fishbach {et~al.}(2018)Fishbach, Holz, \& Farr}]{Fishbach2018}
Fishbach, M., Holz, D.~E., \& Farr, W.~M. 2018, ApJL, 863, L41

\bibitem[{{Foreman-Mackey} {et~al.}(2013){Foreman-Mackey}, {Hogg}, {Lang}, \&
  {Goodman}}]{emcee}
{Foreman-Mackey}, D., {Hogg}, D.~W., {Lang}, D., \& {Goodman}, J. 2013, Publ.
  Astron. Soc. Pac, 125, 306

\bibitem[{{Fowler} \& {Hoyle}(1964)}]{1964ApJS....9..201F}
{Fowler}, W.~A., \& {Hoyle}, F. 1964, ApJS, 9, 201

\bibitem[{{Fragione} {et~al.}(2019){Fragione}, {Grishin}, {Leigh}, {Perets}, \&
  {Perna}}]{2019MNRAS.488...47F}
{Fragione}, G., {Grishin}, E., {Leigh}, N. W.~C., {Perets}, H.~B., \& {Perna},
  R. 2019, \mnras, 488, 47

\bibitem[{{Fragione} \& {Kocsis}(2018)}]{2018PhRvL.121p1103F}
{Fragione}, G., \& {Kocsis}, B. 2018, \prl, 121, 161103

\bibitem[{{Fragione} \& {Kocsis}(2019)}]{2019MNRAS.486.4781F}
---. 2019, \mnras, 486, 4781

\bibitem[{{Fragione} \& {Kocsis}(2020)}]{2020MNRAS.493.3920F}
---. 2020, \mnras, 493, 3920

\bibitem[{{Fuller} \& {Ma}(2019)}]{Fuller2019b}
{Fuller}, J., \& {Ma}, L. 2019, ApJL, 881, L1

\bibitem[{{Fuller} {et~al.}(2019){Fuller}, {Piro}, \& {Jermyn}}]{Fuller2019a}
{Fuller}, J., {Piro}, A.~L., \& {Jermyn}, A.~S. 2019, MNRAS, 485, 3661

\bibitem[{Galaudage {et~al.}(2020)Galaudage, Adamcewicz, Zhu, \&
  Thrane}]{heavy_dns}
Galaudage, S., Adamcewicz, C., Zhu, X.-J., \& Thrane, E. 2020, arxiv/2011.01495

\bibitem[{Galaudage {et~al.}(2019)Galaudage, Talbot, \& Thrane}]{Galaudage}
Galaudage, S., Talbot, C., \& Thrane, E. 2019, arxiv/1912.09708

\bibitem[{Gayathri {et~al.}(2020)Gayathri, Healy, Lange, O'Brien, Szczepanczyk,
  Bartos, Campanelli, Klimenko, Lousto, \& O'Shaughnessy}]{gayathri}
Gayathri, V., Healy, J., Lange, J., {et~al.} 2020, arxiv/2009.05461

\bibitem[{{Gerosa} \& {Berti}(2017)}]{2017PhRvD..95l4046G}
{Gerosa}, D., \& {Berti}, E. 2017, PhRvD, 95, 124046

\bibitem[{Gerosa {et~al.}(2018)Gerosa, Berti, O’Shaughnessy, Belczynski,
  Kesden, Wysocki, \& Gladysz}]{Gerosa2018}
Gerosa, D., Berti, E., O’Shaughnessy, R., {et~al.} 2018, PhRvD, 98, 084036

\bibitem[{Giacobbo {et~al.}(2017)Giacobbo, Mapelli, \&
  Spera}]{2018MNRAS.474.2959G}
Giacobbo, N., Mapelli, M., \& Spera, M. 2017, MNRAS, 474, 2959

\bibitem[{{Gr{\"o}bner} {et~al.}(2020){Gr{\"o}bner}, {Ishibashi}, {Tiwari},
  {Haney}, \& {Jetzer}}]{Grobner}
{Gr{\"o}bner}, M., {Ishibashi}, W., {Tiwari}, S., {Haney}, M., \& {Jetzer}, P.
  2020, \aap, 638, A119

\bibitem[{G{\"u}ltekin {et~al.}(2004)G{\"u}ltekin, Miller, \&
  Hamilton}]{2004ApJ...616..221G}
G{\"u}ltekin, K., Miller, M.~C., \& Hamilton, D.~P. 2004, ApJ, 616, 221

\bibitem[{Hanna {et~al.}(2020)}]{Hanna:2019ezx}
Hanna, C., {et~al.} 2020, PhRvD, 101, 022003

\bibitem[{Hannam {et~al.}(2014)}]{Hannam:2013oca}
Hannam, M., {et~al.} 2014, PhRvL, 113, 151101

\bibitem[{Hannuksela {et~al.}(2019)Hannuksela, Haris, Ng, Kumar, Mehta, Keitel,
  Li, \& Ajith}]{Hannuksela:2019kle}
Hannuksela, O.~A., Haris, K., Ng, K. K.~Y., {et~al.} 2019, ApJ, 874, L2

\bibitem[{Harry {et~al.}(2016)Harry, Privitera, Bohé, \&
  Buonanno}]{Harry_2016}
Harry, I., Privitera, S., Bohé, A., \& Buonanno, A. 2016, Physical Review D,
  94, doi:10.1103/physrevd.94.024012

\bibitem[{Heger {et~al.}(2003)Heger, Fryer, Woosley, Langer, \&
  Hartmann}]{Heger2003}
Heger, A., Fryer, C.~L., Woosley, S.~E., Langer, N., \& Hartmann, D.~H. 2003,
  ApJ, 591, 288

\bibitem[{Heger \& Woosley(2002)}]{Heger2002}
Heger, A., \& Woosley, S.~E. 2002, ApJ, 567, 532

\bibitem[{{Huang} {et~al.}(2020){Huang}, {Haster}, {Vitale}, {Zimmerman},
  {Roulet}, {Venumadhav}, {Zackay}, {Dai}, \& {Zaldarriaga}}]{Huang2020}
{Huang}, Y., {Haster}, C.-J., {Vitale}, S., {et~al.} 2020, arXiv e-prints,
  arXiv:2003.04513

\bibitem[{Husa {et~al.}(2016)}]{Husa:2015iqa}
Husa, S., {et~al.} 2016, PhRvD, 93, 044006

\bibitem[{Inayoshi {et~al.}(2017)Inayoshi, Hirai, Kinugawa, \&
  Hotokezaka}]{Inayoshi:2017mrs}
Inayoshi, K., Hirai, R., Kinugawa, T., \& Hotokezaka, K. 2017, Mon. Not. Roy.
  Astron. Soc., 468, 5020

\bibitem[{Jeffreys(1961)}]{Jeffreys1961}
Jeffreys, H. 1961, Theory of Probability, 3rd edn. (Oxford University Press)

\bibitem[{{Kalogera}(2000)}]{2000ApJ...541..319K}
{Kalogera}, V. 2000, apj, 541, 319

\bibitem[{Khan {et~al.}(2020)Khan, Ohme, Chatziioannou, \&
  Hannam}]{Khan2020_IMRPhenomPv3HM}
Khan, S., Ohme, F., Chatziioannou, K., \& Hannam, M. 2020, PhRvD, 101, 024056

\bibitem[{Khan {et~al.}(2016)}]{Khan:2015jqa}
Khan, S., {et~al.} 2016, PhRvD, 93, 044007

\bibitem[{Kidder(1995)}]{Kidder1995}
Kidder, L.~E. 1995, Phys. Rev. D, 52, 821

\bibitem[{Kimball {et~al.}(2020{\natexlab{a}})Kimball, Talbot, Berry, Carney,
  Zevin, Thrane, \& Kalogera}]{Kimball}
Kimball, C., Talbot, C., Berry, C.~P., {et~al.} 2020{\natexlab{a}}, ApJ, 900,
  177

\bibitem[{Kimball {et~al.}(2020{\natexlab{b}})Kimball, Talbot, Berry, Zevin,
  Thrane, {et~al.}}]{gwtc2_hierarchical}
---. 2020{\natexlab{b}}, arxiv/2011.05332

\bibitem[{Kimpson {et~al.}(2016)Kimpson, Spera, Mapelli, \&
  Ziosi}]{2016MNRAS.463.2443K}
Kimpson, T.~O., Spera, M., Mapelli, M., \& Ziosi, B.~M. 2016, MNRAS, 463, 2443

\bibitem[{Klimenko \& Mitselmakher(2004)}]{Klimenko:2004qh}
Klimenko, S., \& Mitselmakher, G. 2004, Class. Quant. Grav., 21, S1819

\bibitem[{Klimenko {et~al.}(2016)}]{Klimenko:2015ypf}
Klimenko, S., {et~al.} 2016, PhRvD, 93, 042004

\bibitem[{{Kopparapu} {et~al.}(2008){Kopparapu}, {Hanna}, {Kalogera},
  {O'Shaughnessy}, {Gonz{\'a}lez}, {Brady}, \&
  {Fairhurst}}]{2008ApJ...675.1459K}
{Kopparapu}, R.~K., {Hanna}, C., {Kalogera}, V., {et~al.} 2008, apj, 675, 1459

\bibitem[{Kovetz {et~al.}(2017)Kovetz, Cholis, Breysse, \&
  Kamionkowski}]{Kovetz}
Kovetz, E.~D., Cholis, I., Breysse, P.~C., \& Kamionkowski, M. 2017, PhRvD, 95,
  103010

\bibitem[{Kreidberg {et~al.}(2012)Kreidberg, Bailyn, Farr, \&
  Kalogera}]{Kreidberg:2012ud}
Kreidberg, L., Bailyn, C.~D., Farr, W.~M., \& Kalogera, V. 2012, ApJ, 757, 36

\bibitem[{Kulkarni {et~al.}(1993)Kulkarni, Hut, \&
  McMillan}]{1993Natur.364..421K}
Kulkarni, S.~R., Hut, P., \& McMillan, S. 1993, Nature, 364, 421

\bibitem[{{Kushnir} {et~al.}(2016){Kushnir}, {Zaldarriaga}, {Kollmeier}, \&
  {Waldman}}]{2016MNRAS.462..844K}
{Kushnir}, D., {Zaldarriaga}, M., {Kollmeier}, J.~A., \& {Waldman}, R. 2016,
  \mnras, 462, 844

\bibitem[{Lange {et~al.}(2018)Lange, O'Shaughnessy, \& Rizzo}]{Lange:2018pyp}
Lange, J., O'Shaughnessy, R., \& Rizzo, M. 2018, arxiv/1805.10457

\bibitem[{Li {et~al.}(2018)Li, Mao, Zhao, \& Lu}]{Li:2018prc}
Li, S.-S., Mao, S., Zhao, Y., \& Lu, Y. 2018, MNRAS, 476, 2220

\bibitem[{{LIGO Scientific Collaboration}(2018)}]{lalsuite}
{LIGO Scientific Collaboration}. 2018, {LIGO} {A}lgorithm {L}ibrary -
  {LALS}uite, Free Software (GPL), doi:10.7935/GT1W-FZ16

\bibitem[{LIGO-Virgo(2020)}]{P2000217}
LIGO-Virgo. 2020, {GWTC-2 Data Release: Sensitivity of Matched Filter Searches
  to Binary Black Hole Merger Populations},
  doi:https://dcc.ligo.org/LIGO-P2000217/public

\bibitem[{{Liu} \& {Bromm}(2020)}]{2020arXiv200911447L}
{Liu}, B., \& {Bromm}, V. 2020, arXiv e-prints, arXiv:2009.11447

\bibitem[{Liu {et~al.}(2019)Liu, Lai, \& Wang}]{Liu2019}
Liu, B., Lai, D., \& Wang, Y.-H. 2019, ApJ, 881, 41

\bibitem[{{Loredo}(2004)}]{Loredo2004}
{Loredo}, T.~J. 2004, in American Institute of Physics Conference Series, Vol.
  735, 195--206

\bibitem[{Lower {et~al.}(2018)Lower, Thrane, Lasky, \& Smith}]{Lower2018}
Lower, M.~E., Thrane, E., Lasky, P.~D., \& Smith, R. J.~E. 2018, PhRvD, 98,
  083028

\bibitem[{{Madau} \& {Dickinson}(2014)}]{MadauDickinson}
{Madau}, P., \& {Dickinson}, M. 2014, Annual Review of Astronomy and
  Astrophysics, 52, 415

\bibitem[{Madau \& Rees(2001)}]{Madau:2001sc}
Madau, P., \& Rees, M.~J. 2001, ApJL, 551, L27

\bibitem[{Magee {et~al.}(2019)}]{Magee:2019vmb}
Magee, R., {et~al.} 2019, ApJ, 878, L17

\bibitem[{Mandel \& de~Mink(2016)}]{2016MNRAS.458.2634M}
Mandel, I., \& de~Mink, S.~E. 2016, MNRAS, 458, 2634

\bibitem[{Mandel {et~al.}(2019)Mandel, Farr, \& Gair}]{Mandel2019}
Mandel, I., Farr, W.~M., \& Gair, J.~R. 2019, MNRAS, 486, 1086

\bibitem[{{Mandel} {et~al.}(2021){Mandel}, {M{\"u}ller}, {Riley}, {de Mink},
  {Vigna-G{\'o}mez}, \& {Chattopadhyay}}]{Mandel:2020cig}
{Mandel}, I., {M{\"u}ller}, B., {Riley}, J., {et~al.} 2021, \mnras, 500, 1380

\bibitem[{{Mandel} \& {O'Shaughnessy}(2010)}]{2010CQGra..27k4007M}
{Mandel}, I., \& {O'Shaughnessy}, R. 2010, CQGra, 27, 114007

\bibitem[{{Mapelli}(2016)}]{2016MNRAS.459.3432M}
{Mapelli}, M. 2016, MNRAS, 459, 3432

\bibitem[{{Mapelli} {et~al.}(2017){Mapelli}, {Giacobbo}, {Ripamonti}, \&
  {Spera}}]{2017MNRAS.472.2422M}
{Mapelli}, M., {Giacobbo}, N., {Ripamonti}, E., \& {Spera}, M. 2017, MNRAS,
  472, 2422

\bibitem[{Mapelli {et~al.}(2020)Mapelli, Spera, Montanari, Limongi, Chieffi,
  Giacobbo, Bressan, \& Bouffanais}]{2020ApJ...888...76M}
Mapelli, M., Spera, M., Montanari, E., {et~al.} 2020, ApJ, 888, 76

\bibitem[{Marchant {et~al.}(2016)Marchant, Langer, Podsiadlowski, Tauris, \&
  Moriya}]{2016A&A...588A..50M}
Marchant, P., Langer, N., Podsiadlowski, P., Tauris, T.~M., \& Moriya, T.~J.
  2016, A\&A, 588, A50

\bibitem[{Marchant \& Moriya(2020)}]{Marchant:2020haw}
Marchant, P., \& Moriya, T. 2020, arXiv:2007.06220

\bibitem[{McKernan {et~al.}(2012)McKernan, Ford, Lyra, \&
  Perets}]{2012MNRAS.425..460M}
McKernan, B., Ford, K. E.~S., Lyra, W., \& Perets, H.~B. 2012, MNRAS, 425, 460

\bibitem[{McKernan {et~al.}(2020)McKernan, Ford, O'Shaughnessy, \&
  Wysocki}]{mckernan_monte-carlo_2019}
McKernan, B., Ford, K. E.~S., O'Shaughnessy, R., \& Wysocki, D. 2020, MNRAS,
  494, 1203

\bibitem[{McKernan {et~al.}(2018)McKernan, Saavik~Ford, Bellovary, Leigh,
  Haiman, Kocsis, Lyra, Mac~Low, Metzger, O’Dowd, Endlich, \&
  Rosen}]{mckernan_constraining_2018}
McKernan, B., Saavik~Ford, K.~E., Bellovary, J., {et~al.} 2018, ApJ, 866, 66

\bibitem[{Messick {et~al.}(2017)}]{Messick:2016aqy}
Messick, C., {et~al.} 2017, PhRvD, 95, 042001

\bibitem[{Miller \& Hamilton(2002{\natexlab{a}})}]{Miller2002}
Miller, C.~M., \& Hamilton, D.~P. 2002{\natexlab{a}}, ApJ, 576, 894

\bibitem[{Miller \& Hamilton(2002{\natexlab{b}})}]{2002MNRAS.330..232C}
Miller, M.~C., \& Hamilton, D.~P. 2002{\natexlab{b}}, MNRAS, 330, 232

\bibitem[{{Miller} \& {Lauburg}(2009)}]{2009ApJ...692..917M}
{Miller}, M.~C., \& {Lauburg}, V.~M. 2009, \apj, 692, 917

\bibitem[{{Miller} {et~al.}(2020){Miller}, {Callister}, \& {Farr}}]{Miller2020}
{Miller}, S., {Callister}, T.~A., \& {Farr}, W. 2020, ApJ, 895, 128

\bibitem[{{Moody} \& {Sigurdsson}(2009)}]{2009ApJ...690.1370M}
{Moody}, K., \& {Sigurdsson}, S. 2009, \apj, 690, 1370

\bibitem[{Most {et~al.}(2020)Most, Papenfort, Weih, \& Rezzolla}]{Most:2020bba}
Most, E.~R., Papenfort, L.~J., Weih, L.~R., \& Rezzolla, L. 2020,
  arXiv:2006.14601

\bibitem[{{Natarajan}(2020)}]{2020arXiv200909156N}
{Natarajan}, P. 2020, arXiv e-prints, arXiv:2009.09156

\bibitem[{{Neijssel} {et~al.}(2019){Neijssel}, {Vigna-G{\'o}mez}, {Stevenson},
  {Barrett}, {Gaebel}, {Broekgaarden}, {de Mink}, {Sz{\'e}csi}, {Vinciguerra},
  \& {Mandel}}]{2019MNRAS.490.3740N}
{Neijssel}, C.~J., {Vigna-G{\'o}mez}, A., {Stevenson}, S., {et~al.} 2019,
  MNRAS, 490, 3740

\bibitem[{Ng {et~al.}(2018{\natexlab{a}})Ng, Vitale, Zimmerman, Chatziioannou,
  Gerosa, \& Haster}]{Ng2018}
Ng, K.~K., Vitale, S., Zimmerman, A., {et~al.} 2018{\natexlab{a}}, PhRvD, 98,
  083007

\bibitem[{Ng {et~al.}(2018{\natexlab{b}})Ng, Wong, Broadhurst, \&
  Li}]{Ng:2017yiu}
Ng, K. K.~Y., Wong, K. W.~K., Broadhurst, T., \& Li, T. G.~F.
  2018{\natexlab{b}}, PhRvD, 97, 023012

\bibitem[{Nitz {et~al.}(2019{\natexlab{a}})}]{PyCBC1}
Nitz, A., {et~al.} 2019{\natexlab{a}}, "gwastro/pycbc: Pycbc release v1.15.0”

\bibitem[{Nitz {et~al.}(2019{\natexlab{b}})Nitz, Capano, Nielsen, Reyes, White,
  Brown, \& Krishnan}]{Nitz:2018imz}
Nitz, A.~H., Capano, C., Nielsen, A.~B., {et~al.} 2019{\natexlab{b}}, ApJ, 872,
  195

\bibitem[{Nitz \& Capano(2020)}]{NitzCapano}
Nitz, A.~H., \& Capano, C.~D. 2020, arxiv/2010.12558

\bibitem[{Nitz {et~al.}(2017)Nitz, Dent, Canton, Fairhurst, \& Brown"}]{PyCBC6}
Nitz, A.~H., Dent, T., Canton, T.~D., Fairhurst, S., \& Brown", D.~A. 2017,
  ApJ, 849, 118

\bibitem[{Nitz {et~al.}(2020)Nitz, Dent, Davies, Kumar, Capano, Harry, Mozzon,
  Nuttall, Lundgren, \& Tápai}]{Nitz:2019hdf}
Nitz, A.~H., Dent, T., Davies, G.~S., {et~al.} 2020, ApJ, 891, 123

\bibitem[{{Oguri}(2018)}]{2018MNRAS.480.3842O}
{Oguri}, M. 2018, MNRAS, 3842

\bibitem[{{O'Leary} {et~al.}(2009){O'Leary}, {Kocsis}, \&
  {Loeb}}]{2009MNRAS.395.2127O}
{O'Leary}, R.~M., {Kocsis}, B., \& {Loeb}, A. 2009, \mnras, 395, 2127

\bibitem[{{O'Leary} {et~al.}(2006){O'Leary}, {Rasio}, {Fregeau}, {Ivanova}, \&
  {O'Shaughnessy}}]{2006ApJ...637..937O}
{O'Leary}, R.~M., {Rasio}, F.~A., {Fregeau}, J.~M., {Ivanova}, N., \&
  {O'Shaughnessy}, R. 2006, \apj, 637, 937

\bibitem[{{Olejak} {et~al.}(2020){Olejak}, {Fishbach}, {Belczynski}, {Holz},
  {Lasota}, {Miller}, \& {Bulik}}]{2020arXiv200411866O}
{Olejak}, A., {Fishbach}, M., {Belczynski}, K., {et~al.} 2020, arXiv e-prints,
  arXiv:2004.11866

\bibitem[{{O'Shaughnessy} {et~al.}(2017){O'Shaughnessy}, {Gerosa}, \&
  {Wysocki}}]{2017PhRvL.119a1101O}
{O'Shaughnessy}, R., {Gerosa}, D., \& {Wysocki}, D. 2017, PhRvL, 119, 011101

\bibitem[{Ossokine {et~al.}(2020)}]{Ossokine:2020kjp}
Ossokine, S., {et~al.} 2020, arXiv:2004.09442

\bibitem[{{\"O}zel {et~al.}(2011){\"O}zel, Psaltis, Narayan, \&
  McClintock}]{Ozel2011}
{\"O}zel, F., Psaltis, D., Narayan, R., \& McClintock, J.~E. 2011, ApJ, 725,
  1918

\bibitem[{Pan {et~al.}(2014)Pan, Buonanno, Taracchini, Kidder, Mrou'e,
  Pfeiffer, Scheel, \& Szil'agyi}]{Pan:2013rra}
Pan, Y., Buonanno, A., Taracchini, A., {et~al.} 2014, Phys. Rev. D, 89, 084006

\bibitem[{{Park} {et~al.}(2017){Park}, {Kim}, {Lee}, {Bae}, \&
  {Belczynski}}]{2017MNRAS.469.4665P}
{Park}, D., {Kim}, C., {Lee}, H.~M., {Bae}, Y.-B., \& {Belczynski}, K. 2017,
  \mnras, 469, 4665

\bibitem[{{Portegies Zwart} \& {McMillan}(2000)}]{2000ApJ...528L..17P}
{Portegies Zwart}, S.~F., \& {McMillan}, S. L.~W. 2000, \apjl, 528, L17

\bibitem[{Portegies~Zwart \& Yungelson(1998)}]{1998A&A...332..173P}
Portegies~Zwart, S.~F., \& Yungelson, L.~R. 1998, A\&A, 332, 173

\bibitem[{Pratten \& Vecchio(2020)}]{Pratten:2020ruz}
Pratten, G., \& Vecchio, A. 2020, arXiv:2008.00509

\bibitem[{Pretorius(2005)}]{Pretorius:spin}
Pretorius, F. 2005, PhRvL, 95, 121101

\bibitem[{Qin {et~al.}(2018)Qin, Fragos, Meynet, Andrews, Sørensen, \&
  Song}]{Qin2018}
Qin, Y., Fragos, T., Meynet, G., {et~al.} 2018, A\&A, 616, A28

\bibitem[{{Rice} \& {Zhang}(2020)}]{2020arXiv200911326R}
{Rice}, J.~R., \& {Zhang}, B. 2020, arXiv e-prints, arXiv:2009.11326

\bibitem[{Riddell {et~al.}(2018)Riddell, Hartikainen, Lee, {Riddell-Stan},
  {et~al.}}]{pystan}
Riddell, A., Hartikainen, A., Lee, D., {Riddell-Stan}, {et~al.} 2018,
  Stan-Dev/Pystan: V2.18.0.0, doi:10.5281/ZENODO.1456206

\bibitem[{Rodriguez {et~al.}(2018)Rodriguez, Amaro-Seoane, Chatterjee, \&
  Rasio}]{Rodriguez2018}
Rodriguez, C.~L., Amaro-Seoane, P., Chatterjee, S., \& Rasio, F.~A. 2018,
  PhRvL, 120, 151101

\bibitem[{Rodriguez \& Antonini(2018)}]{rodriguez_triple_2018}
Rodriguez, C.~L., \& Antonini, F. 2018, ApJ, 863, 7

\bibitem[{Rodriguez \& Loeb(2018)}]{Rodriguez:redshift}
Rodriguez, C.~L., \& Loeb, A. 2018, ApJ, 866, L5

\bibitem[{{Rodriguez} {et~al.}(2015){Rodriguez}, {Morscher}, {Pattabiraman},
  {Chatterjee}, {Haster}, \& {Rasio}}]{Rodriguez:2015}
{Rodriguez}, C.~L., {Morscher}, M., {Pattabiraman}, B., {et~al.} 2015, \prl,
  115, 051101

\bibitem[{Rodriguez {et~al.}(2019)Rodriguez, Zevin, Amaro-Seoane, Chatterjee,
  Kremer, Rasio, \& Ye}]{Rodriguez2019}
Rodriguez, C.~L., Zevin, M., Amaro-Seoane, P., {et~al.} 2019, PhRvD, 100,
  043027

\bibitem[{Rodriguez {et~al.}(2016)Rodriguez, Zevin, Pankow, Kalogera, \&
  Rasio}]{Rodriguez2016}
Rodriguez, C.~L., Zevin, M., Pankow, C., Kalogera, V., \& Rasio, F.~A. 2016,
  ApJL, 832, L2

\bibitem[{Romero-Shaw {et~al.}(2019)Romero-Shaw, Lasky, \&
  Thrane}]{Romero-Shaw2019}
Romero-Shaw, I.~M., Lasky, P.~D., \& Thrane, E. 2019, MNRAS, 490, 5210

\bibitem[{Romero-Shaw {et~al.}(2020{\natexlab{a}})Romero-Shaw, Lasky, Thrane,
  \& Bustillo}]{GW190521_formation}
Romero-Shaw, I.~M., Lasky, P.~D., Thrane, E., \& Bustillo, J.~C.
  2020{\natexlab{a}}, ApJL, 903, L5

\bibitem[{Romero-Shaw {et~al.}(2020{\natexlab{b}})}]{bilby_gwtc1}
Romero-Shaw, I.~M., {et~al.} 2020{\natexlab{b}}, arxiv/2006.00714

\bibitem[{{Roulet} {et~al.}(2020){Roulet}, {Venumadhav}, {Zackay}, {Dai}, \&
  {Zaldarriaga}}]{Roulet:2020}
{Roulet}, J., {Venumadhav}, T., {Zackay}, B., {Dai}, L., \& {Zaldarriaga}, M.
  2020, arXiv e-prints, arXiv:2008.07014

\bibitem[{{Roulet} \& {Zaldarriaga}(2019)}]{Roulet:2019}
{Roulet}, J., \& {Zaldarriaga}, M. 2019, \mnras, 484, 4216

\bibitem[{{Roupas} \& {Kazanas}(2019)}]{2019A&A...632L...8R}
{Roupas}, Z., \& {Kazanas}, D. 2019, \aap, 632, L8

\bibitem[{Sachdev {et~al.}(2019)}]{Sachdev:2019vvd}
Sachdev, S., {et~al.} 2019, arXiv:1901.08580

\bibitem[{Safarzadeh(2020)}]{Safarzadeh2020}
Safarzadeh, M. 2020, Nature Astronomy, 4, 735

\bibitem[{{Safarzadeh} {et~al.}(2020){Safarzadeh}, {Farr}, \&
  {Ramirez-Ruiz}}]{2020ApJ...894..129S}
{Safarzadeh}, M., {Farr}, W.~M., \& {Ramirez-Ruiz}, E. 2020, \apj, 894, 129

\bibitem[{{Safarzadeh} \& {Haiman}(2020)}]{2020arXiv200909320S}
{Safarzadeh}, M., \& {Haiman}, Z. 2020, arXiv e-prints, arXiv:2009.09320

\bibitem[{Samsing(2018)}]{Samsing2018}
Samsing, J. 2018, PhRvD, 97, 103014

\bibitem[{Samsing {et~al.}(2014)Samsing, MacLeod, \&
  Ramirez-Ruiz}]{Samsing2014}
Samsing, J., MacLeod, M., \& Ramirez-Ruiz, E. 2014, ApJ, 784, 71

\bibitem[{Samsing \& Ramirez-Ruiz(2017)}]{Samsing2017}
Samsing, J., \& Ramirez-Ruiz, E. 2017, ApJL, 840, L14

\bibitem[{{Santoliquido} {et~al.}(2020){Santoliquido}, {Mapelli}, {Bouffanais},
  {Giacobbo}, {Di Carlo}, {Rastello}, {Artale}, \&
  {Ballone}}]{2020arXiv200409533S}
{Santoliquido}, F., {Mapelli}, M., {Bouffanais}, Y., {et~al.} 2020, arXiv
  e-prints, arXiv:2004.09533

\bibitem[{Schmidt {et~al.}(2012)Schmidt, Hannam, \& Husa}]{Schmidt2012}
Schmidt, P., Hannam, M., \& Husa, S. 2012, PhRvD, 86, 104063

\bibitem[{Schmidt {et~al.}(2015)Schmidt, Ohme, \& Hannam}]{Schmidt2015}
Schmidt, P., Ohme, F., \& Hannam, M. 2015, PhRvD, 91, 024043

\bibitem[{Sigurdsson \& Hernquist(1993)}]{1993Natur.364..423S}
Sigurdsson, S., \& Hernquist, L. 1993, Nature, 364, 423

\bibitem[{{Smith} {et~al.}(2018){Smith}, {Jauzac}, {Veitch}, {Farr}, {Massey},
  \& {Richard}}]{2018MNRAS.475.3823S}
{Smith}, G.~P., {Jauzac}, M., {Veitch}, J., {et~al.} 2018, MNRAS, 475, 3823

\bibitem[{{Speagle}(2020)}]{dynesty}
{Speagle}, J.~S. 2020, MNRAS, 493, 3132

\bibitem[{Spera \& Mapelli(2017)}]{2017MNRAS.470.4739S}
Spera, M., \& Mapelli, M. 2017, MNRAS, 470, 4739

\bibitem[{Stevenson {et~al.}(2017)Stevenson, Berry, \& Mandel}]{Stevenson}
Stevenson, S., Berry, C. P.~L., \& Mandel, I. 2017, MNRAS, 471, 2801

\bibitem[{Stone {et~al.}(2017)Stone, Metzger, \& Haiman}]{2017MNRAS.464..946S}
Stone, N.~C., Metzger, B.~D., \& Haiman, Z. 2017, MNRAS, 464, 946

\bibitem[{{Sun} {et~al.}(2020){Sun}, {Goetz}, {Kissel}, {Betzwieser}, {Karki},
  {Viets}, {Wade}, {Bhattacharjee}, {Bossilkov}, {Covas}, \&
  et~al.}]{2020CQGra..37v5008S}
{Sun}, L., {Goetz}, E., {Kissel}, J.~S., {et~al.} 2020, Classical and Quantum
  Gravity, 37, 225008

\bibitem[{Tagawa {et~al.}(2020)Tagawa, Haiman, \& Kocsis}]{Tagawa_2020}
Tagawa, H., Haiman, Z., \& Kocsis, B. 2020, The Astrophysical Journal, 898, 25

\bibitem[{Talbot {et~al.}(2019)Talbot, Smith, Thrane, \& Poole}]{gwpopulation}
Talbot, C., Smith, R., Thrane, E., \& Poole, G.~B. 2019, PhRvD, 100, 043030

\bibitem[{Talbot \& Thrane(2017)}]{Talbot2017}
Talbot, C., \& Thrane, E. 2017, PhRvD, 96, 023012

\bibitem[{Talbot \& Thrane(2018)}]{Talbot2018}
---. 2018, ApJ, 856, 173

\bibitem[{Tan {et~al.}(2020)Tan, Noronha-Hostler, \& Yunes}]{Tan:2020ics}
Tan, H., Noronha-Hostler, J., \& Yunes, N. 2020, arXiv:2006.16296

\bibitem[{{Tanikawa} {et~al.}(2020){Tanikawa}, {Susa}, {Yoshida}, {Trani}, \&
  {Kinugawa}}]{2020arXiv200801890T}
{Tanikawa}, A., {Susa}, H., {Yoshida}, T., {Trani}, A.~A., \& {Kinugawa}, T.
  2020, arXiv e-prints, arXiv:2008.01890

\bibitem[{Taracchini {et~al.}(2014)}]{Taracchini:2013rva}
Taracchini, A., {et~al.} 2014, Phys. Rev. D, 89, 061502

\bibitem[{Tews {et~al.}(2020)Tews, Pang, Dietrich, Coughlin, Antier, Bulla,
  Heinzel, \& Issa}]{Tews:2020ylw}
Tews, I., Pang, P.~T., Dietrich, T., {et~al.} 2020, arXiv:2007.06057

\bibitem[{Thompson {et~al.}(2019)}]{Thompson}
Thompson, T.~A., {et~al.} 2019, Science, 366

\bibitem[{Thrane \& Talbot(2019)}]{Thrane2019}
Thrane, E., \& Talbot, C. 2019, PASA, 36, E010

\bibitem[{{Tiwari}(2018)}]{2018CQGra..35n5009T}
{Tiwari}, V. 2018, Classical and Quantum Gravity, 35, 145009

\bibitem[{{Tiwari}(2020)}]{2020arXiv200615047T}
---. 2020, arXiv e-prints, arXiv:2006.15047

\bibitem[{Tiwari {et~al.}(2018)Tiwari, Fairhurst, \& Hannam}]{Tiwari:2018qch}
Tiwari, V., Fairhurst, S., \& Hannam, M. 2018, ApJ, 868, 140

\bibitem[{Trovato {et~al.}(2020)}]{Trovato:2019liz}
Trovato, A., {et~al.} 2020, PoS, Asterics2019, 082

\bibitem[{Usman {et~al.}(2016)}]{PyCBC5}
Usman, S.~A., {et~al.} 2016, CQGra, 33, 215004

\bibitem[{{van Son} {et~al.}(2020){van Son}, {de Mink}, {Broekgaarden},
  {Renzo}, {Justham}, {Laplace}, {Moran-Fraile}, {Hendriks}, \&
  {Farmer}}]{2020arXiv200405187V}
{van Son}, L.~A.~C., {de Mink}, S.~E., {Broekgaarden}, F.~S., {et~al.} 2020,
  arXiv e-prints, arXiv:2004.05187

\bibitem[{Varma {et~al.}(2019)Varma, Field, Scheel, Blackman, Gerosa, Stein,
  Kidder, \& Pfeiffer}]{Varma2019_NRSur7dq4}
Varma, V., Field, S.~E., Scheel, M.~A., {et~al.} 2019, Phys. Rev. Research, 1,
  033015

\bibitem[{{Veitch} {et~al.}(2015){Veitch}, {Raymond}, {Farr}, {Farr}, {Graff},
  {Vitale}, {Aylott}, {Blackburn}, {Christensen}, {Coughlin}, {Del Pozzo},
  {Feroz}, {Gair}, {Haster}, {Kalogera}, {Littenberg}, {Mandel},
  {O'Shaughnessy}, {Pitkin}, {Rodriguez}, {R{\"o}ver}, {Sidery}, {Smith}, {Van
  Der Sluys}, {Vecchio}, {Vousden}, \& {Wade}}]{lalinference}
{Veitch}, J., {Raymond}, V., {Farr}, B., {et~al.} 2015, PhRvD, 91, 042003

\bibitem[{Venumadhav {et~al.}(2019)Venumadhav, Zackay, Roulet, Dai, \&
  Zaldarriaga}]{Venumadhav:2019tad}
Venumadhav, T., Zackay, B., Roulet, J., Dai, L., \& Zaldarriaga, M. 2019,
  PhRvD, 100, 023011

\bibitem[{Venumadhav {et~al.}(2020)Venumadhav, Zackay, Roulet, Dai, \&
  Zaldarriaga}]{Venumadhav:2019lyq}
---. 2020, PhRvD, 101, 083030

\bibitem[{{Vigna-G{\'o}mez} {et~al.}(2020){Vigna-G{\'o}mez}, {Toonen},
  {Ramirez-Ruiz}, {Leigh}, {Riley}, \& {Haster}}]{2020arXiv201013669V}
{Vigna-G{\'o}mez}, A., {Toonen}, S., {Ramirez-Ruiz}, E., {et~al.} 2020, arXiv
  e-prints, arXiv:2010.13669

\bibitem[{Vitale(2020)}]{Vitale:2020aaz}
Vitale, S. 2020, arXiv:2007.05579

\bibitem[{Vitale {et~al.}(2017{\natexlab{a}})Vitale, Gerosa, Haster,
  Chatziioannou, \& Zimmerman}]{Vitale2017}
Vitale, S., Gerosa, D., Haster, C.-J., Chatziioannou, K., \& Zimmerman, A.
  2017{\natexlab{a}}, PhRvL, 119, 251103

\bibitem[{Vitale {et~al.}(2017{\natexlab{b}})Vitale, Lynch, Sturani, \&
  Graff}]{Vitale:2015tea}
Vitale, S., Lynch, R., Sturani, R., \& Graff, P. 2017{\natexlab{b}}, Class.
  Quant. Grav., 34, 03LT01

\bibitem[{Woosley(2017)}]{2017ApJ...836..244W}
Woosley, S.~E. 2017, ApJ, 836, 244

\bibitem[{Woosley \& Heger(2015)}]{Woosley2015}
Woosley, S.~E., \& Heger, A. 2015, in Very Massive Stars Local Universe
  (Springer, Cham), 199--225

\bibitem[{Woosley {et~al.}(2002)Woosley, Heger, \&
  Weaver}]{2002RvMP...74.1015W}
Woosley, S.~E., Heger, A., \& Weaver, T.~A. 2002, Rev. of Mod. Phys., 74, 1015

\bibitem[{Wysocki(2020)}]{Wysocki:VTcalibration}
Wysocki, D. 2020

\bibitem[{Wysocki {et~al.}(2019{\natexlab{a}})Wysocki, Lange, \&
  O’Shaughnessy}]{Wysocki2019}
Wysocki, D., Lange, J., \& O’Shaughnessy, R. 2019{\natexlab{a}}, PhRvD, 100,
  043012

\bibitem[{Wysocki \& O'Shaughnessy(2017)}]{RITpop}
Wysocki, D., \& O'Shaughnessy, R. 2017, Bayesian Parametric Population Models,
  bayesian-parametric-population-models.readthedocs.io

\bibitem[{Wysocki {et~al.}(2019{\natexlab{b}})Wysocki, O’Shaughnessy, Lange,
  \& Fang}]{Wysocki_2019}
Wysocki, D., O’Shaughnessy, R., Lange, J., \& Fang, Y.-L.~L.
  2019{\natexlab{b}}, PhRvD, 99

\bibitem[{Wysocki {et~al.}(2018)}]{2018PhRvD..97d3014W}
Wysocki, D., {et~al.} 2018, PhRvD, 97, 043014

\bibitem[{{Yang} {et~al.}(2019){Yang}, {Bartos}, {Gayathri}, {Ford}, {Haiman},
  {Klimenko}, {Kocsis}, {M{\'a}rka}, {M{\'a}rka}, {McKernan}, \&
  {O'Shaughnessy}}]{Yang2018}
{Yang}, Y., {Bartos}, I., {Gayathri}, V., {et~al.} 2019, PhRvL, 123, 181101

\bibitem[{Zackay {et~al.}(2020)Zackay, Dai, Venumadhav, Roulet, \&
  Zaldarriaga}]{Zackay:2019btq}
Zackay, B., Dai, L., Venumadhav, T., Roulet, J., \& Zaldarriaga, M. 2020,
  PhRvD, 101, 083030

\bibitem[{Zackay {et~al.}(2019)Zackay, Venumadhav, Dai, Roulet, \&
  Zaldarriaga}]{Zackay:2019tzo}
Zackay, B., Venumadhav, T., Dai, L., Roulet, J., \& Zaldarriaga, M. 2019,
  PhRvD, 100, 023007

\bibitem[{{Zaldarriaga} {et~al.}(2018){Zaldarriaga}, {Kushnir}, \&
  {Kollmeier}}]{2018MNRAS.473.4174Z}
{Zaldarriaga}, M., {Kushnir}, D., \& {Kollmeier}, J.~A. 2018, \mnras, 473, 4174

\bibitem[{Zevin {et~al.}(2017)Zevin, Pankow, Rodriguez, Sampson, Chase,
  Kalogera, \& Rasio}]{Zevin2017}
Zevin, M., Pankow, C., Rodriguez, C.~L., {et~al.} 2017, ApJ, 846, 82

\bibitem[{{Zevin} {et~al.}(2019){Zevin}, {Samsing}, {Rodriguez}, {Haster}, \&
  {Ramirez-Ruiz}}]{2019ApJ...871...91Z}
{Zevin}, M., {Samsing}, J., {Rodriguez}, C., {Haster}, C.-J., \&
  {Ramirez-Ruiz}, E. 2019, \apj, 871, 91

\bibitem[{Zevin {et~al.}(2020)Zevin, Spera, Berry, \& Kalogera}]{Zevin:2020gma}
Zevin, M., Spera, M., Berry, C.~P., \& Kalogera, V. 2020, arXiv:2006.14573

\bibitem[{Zhang \& Li(2020)}]{Zhang:2020zsc}
Zhang, N.-B., \& Li, B.-A. 2020, arXiv:2007.02513

\bibitem[{{Ziegler} \& {Freese}(2020)}]{2020arXiv201000254Z}
{Ziegler}, J., \& {Freese}, K. 2020, arXiv e-prints, arXiv:2010.00254

\bibitem[{{Ziosi} {et~al.}(2014){Ziosi}, {Mapelli}, {Branchesi}, \&
  {Tormen}}]{2014MNRAS.441.3703Z}
{Ziosi}, B.~M., {Mapelli}, M., {Branchesi}, M., \& {Tormen}, G. 2014, \mnras,
  441, 3703

\end{thebibliography}

\appendix

\section{Estimating the detection fraction}\label{Appendix:xi}
A key ingredient in Eqs.~\eqref{eq:generic-likelihood} and \eqref{eq:generic-likelihood-marginalized} is the detection fraction $\xi(\Lambda)$, the fraction of systems within some prior volume (redshift $z < 2.3$) that we expect to successfully detect.
The detection fraction quantifies selection biases, and so it is critical to accurately characterize.
For a population described by hyper-parameters $\Lambda$, the detection fraction is
\begin{equation}
    \label{eq:selection term}
    \xi(\Lambda) = \int P_\mathrm{det}(\theta) \pi(\theta|\Lambda) d\theta.
\end{equation}
Here, $P_\mathrm{det}(\theta)$ is the detection probability: the probability that an event with parameters $\theta$ is detectable.
The detection probability depends primarily on the masses and redshift of a system, and, to a lesser degree, on the spins.

We calculate $\xi(\Lambda)$ using injections.
We simulate compact binary signals from a reference population and record which ones are successfully detected by the \texttt{PyCBC} and \texttt{GstLAL} search pipelines; see~\cite{O3acatalog}. 
Following \cite{2018CQGra..35n5009T,Farr:selection,Vitale:2020aaz,Loredo2004}, the point estimate for Eq.~\ref{eq:selection term} is calculated using a Monte Carlo integral over found injections:
\begin{equation}
    \hat\xi(\Lambda) =  \frac{1}{N_\mathrm{inj}} \sum_{j=1}^{ N_\mathrm{found}} \frac{\pi(\theta_j \mid \Lambda)}{p_\mathrm{draw}(\theta_j)} ,
\end{equation}
where $N_\mathrm{inj}$ is the total number of injections, $N_\mathrm{found}$ are the injections that are successfully detected, and $p_\mathrm{draw}$ is the probability distribution from which the injections are drawn; see~\cite{P2000217} for additional details.
When sampling the population likelihood, we marginalize over the uncertainty in $\hat\xi(\Lambda)$ following \cite{Farr:selection}, and ensure that the effective number of found injections remaining after population re-weighting is sufficiently high ($N_\mathrm{eff} > 4 N_\mathrm{det}$).

For the O3a observing period, we use the injection campaign described in \citet{O3acatalog} and characterize the found injections as those recovered with a $\text{FAR}$ below our threshold of $1~\mathrm{yr}^{-1}$ in either \texttt{PyCBC} or \texttt{GstLAL}.
For the O1 and O2 observing period, we supplement the O3a pipeline injections with mock injections drawn from the same distribution $p_\mathrm{draw}$ above.
For the mock injections, we calculate $P_\mathrm{det}(m_1, m_2, z, \chi_{1,z}, \chi_{2,z})$ according to the semi-analytic approximation described in~\citet{O2pop}, based on a single-detector signal-to-noise ratio threshold $\rho = 8$ and the \textsc{Advanced LIGO Early-High Noise PSD}~\citep{ObsProspects}.
We combine O1, O2 and O3 injection sets ensuring a constant rate of injections across the total observing time, yielding $N_\mathrm{inj} \approx 7.7\times10^7$ injections for O3a and $N_\mathrm{inj} \approx 7.1\times10^7$ for O1 and O2. 
To control computational costs, not all of the injections are performed in real data. 
Before injecting, the expected network signal-to-noise ratio of the injections is computed, and the hopeless injections with signal-to-noise ratio $< 6$  are removed.

Due to the finite number of injections, we approximate Eq.~\eqref{eq:selection term} with a fixed spin distribution instead of the distribution implied by $\Lambda$.
When combining the \truncated{}, \ppsn{}, \tapered{} and \multipeak{} mass models together with the \textsc{Default} spin distribution, we assume that the aligned spin components $\chi_{1,z}, \chi_{2,z}$ are independently drawn from a uniform distribution $U(-0.5, 0.5)$. 
By making this approximation, we are in effect ignoring selection effects due to spin. 
Nevertheless, we expect this approximation to have a negligible impact on the inferred spin distribution compared to the statistical uncertainties.
For aligned spin components in the range $(-0.5, 0.5)$, the detection probability varies by no more than a factor of 2~\citep{Ng2018}.
Furthermore, our main conclusions regarding the spin distribution inferred from the \textsc{Default} model are supported by the \textsc{Gaussian} model, which requires no approximations for spin selection effects.
The \ModelH{} model calculates Eq.~\eqref{eq:selection term} by calibrating a semi-analytic approximation to the list of found injections~\citep{Wysocki:VTcalibration}.

\added{
  The transfer function between the observed strain and astrophysical strain is subject to a systematic calibration uncertainty.
  We neglect this calibration uncertainty in our estimates of the search sensitivity above.
  For the O3a observing run, the amplitude uncertainty was $\lesssim 3\%$~\citep{2020CQGra..37v5008S}, which leads to a $\lesssim 10\%$ systematic uncertainty in the sensitive spacetime volume and the inferred merger rate. This systematic uncertainty is subdominant to our uncertainties from Poisson counting error.
}

\section{Details of mass population models}\label{details}
In this section we provide details about the population models described above in Section~\ref{models}; see also Fig.~\ref{fig:thumbs}.
Each subsection includes a table with a summary of the parameters for that model and the prior distribution used for each parameter.
The prior distributions are indicated using abbreviations: for example, $\mathrm{U}(0,1)$ translates to uniform on the interval $(0,1)$ and LU($10^{-6},10^5$) translates to log-uniform on the interval $10^{-6},10^5$.

\subsection{\truncated{} mass model}\label{truncated}
This model is equivalent to ``Model~B'' in \cite{O2pop}.
The primary mass distribution for this model follows a power-law with spectral index $\alpha$, and with a sharp cut-off at the lower end $\mmin$ and the upper end of the distribution $\mmax$:
\begin{align}
    \pi(m_1 | \alpha, \mmin, \mmax) \propto 
    \begin{cases}
    m_1^{-\alpha} & \mmin < m_1 < \mmax \\
    0 & \text{otherwise},
    \end{cases}
\end{align}
Meanwhile, the mass ratio $q \equiv m_2/m_1$ follows a power-law distribution with spectral index $\beta_q$
\begin{align}
    \pi(q | \beta_q, \mmin, m_1) \propto 
    \begin{cases}
    q^{\beta_q} & \mmin < m_2 < m_1 \\
    0 & \text{otherwise} .
    \end{cases}
\end{align}
The hyper-parameters for this model are summarized in Table~\ref{tab:parameters_truncated}.

\begin{table}[t]
    \centering
    \begin{tabular}{ c  p{11cm} p{2mm} p{3cm} }
        \hline
        {\bf Parameter} & \textbf{Description} &  & \textbf{Prior} \\\hline\hline
        $\alpha$ & Spectral index for the power-law of the primary mass distribution. &  & U($-4$, $12$) \\
        $\beta_q$ & Spectral index for the power-law of the mass ratio distribution. &  & U($-4$, $12$) \\
        $\mmin$ & Minimum mass of the power-law component of the primary mass distribution. &  & U($2\, M_{\odot}$, $10\, M_{\odot}$)\\
        $\mmax$ &  Maximum mass of the power-law component of the primary mass distribution. &  & U($30\, M_{\odot}$, $100\, M_{\odot}$)\\
        \hline
    \end{tabular}
    \caption{
    Summary of \truncated{} parameters.
    }
  \label{tab:parameters_truncated}
\end{table}

\subsection{\ppsn{} mass model}\label{ppsn}
This is equivalent to ``Model~C'' from ~\cite{O2pop}.
It is motivated by the idea that the mass loss undergone by pulsational pair-instability supernovae could lead to a pile-up of \bbhevnt{} before the pair-instability gap~\citep{Talbot2018}. 
The primary mass distribution is an extension of \truncated{} with the addition of tapering at the lower mass end of the distribution and a Gaussian component:
\begin{align}
    \pi(m_1 |& \lambda_\text{peak}, \alpha, \mmin, \delta_m, \mmax, \mu_m, \sigma_m) =
    \bigg[
    (1-\lambda_\text{peak})\mathfrak{P}(m_1|-\alpha, \mmax) +
    \lambda_\text{peak} G(m_1|\mu_m,\sigma_m)
    \bigg] S(m_1|\mmin, \delta_m) .
\end{align}
Here, $\mathfrak{P}(m_1|-\alpha,\mmax)$ is a normalized power-law distribution with spectral index $-\alpha$ and high-mass cut-off $\mmax$.
Meanwhile, $G(m_1|\mu_m, \sigma_m)$ is a normalized Gaussian distribution with mean $\mu_m$ and width $\sigma_m$.
The parameter $\lambda_\text{peak}$ is a mixing fraction determining the relative prevalence of mergers in $\mathfrak{P}$ and $G$.
Finally, $S(m_1, \mmin, \delta_m)$ is a smoothing function, which rises from 0 to 1 over the interval $(\mmin, \mmin+\delta_m)$:
\begin{equation}
\label{eq:smoothing}
S(m \mid \mmin, \delta_m) = \begin{cases}
    0 & \left(m< \mmin\right) \\
    \left[f(m - \mmin, \delta_m) + 1\right]^{-1} & \left(\mmin \leq m < \mmin+\delta_m\right) \\
    1 & \left(m\geq \mmin + \delta_m\right)
\end{cases}
\end{equation}
with
\begin{equation}
    f(m', \delta_m) = \exp \left(\frac{\delta_m}{m'} + \frac{\delta_m }{m' - \delta_m}\right).
\end{equation}
The conditional mass ratio distribution in this model also includes the smoothing term:
\begin{align}
\label{eq:pq_smoothing}
\pi(q \mid \beta, m_1, \mmin, \delta_m) \propto q^{\beta_q} S(q m_1 \mid \mmin, \delta_m).
\end{align}
The hyper-parameters for this model are summarized in Table~\ref{tab:parameters_ppsn_peak}.

\begin{figure*}[h]
    \centering
    \includegraphics[width = \textwidth]{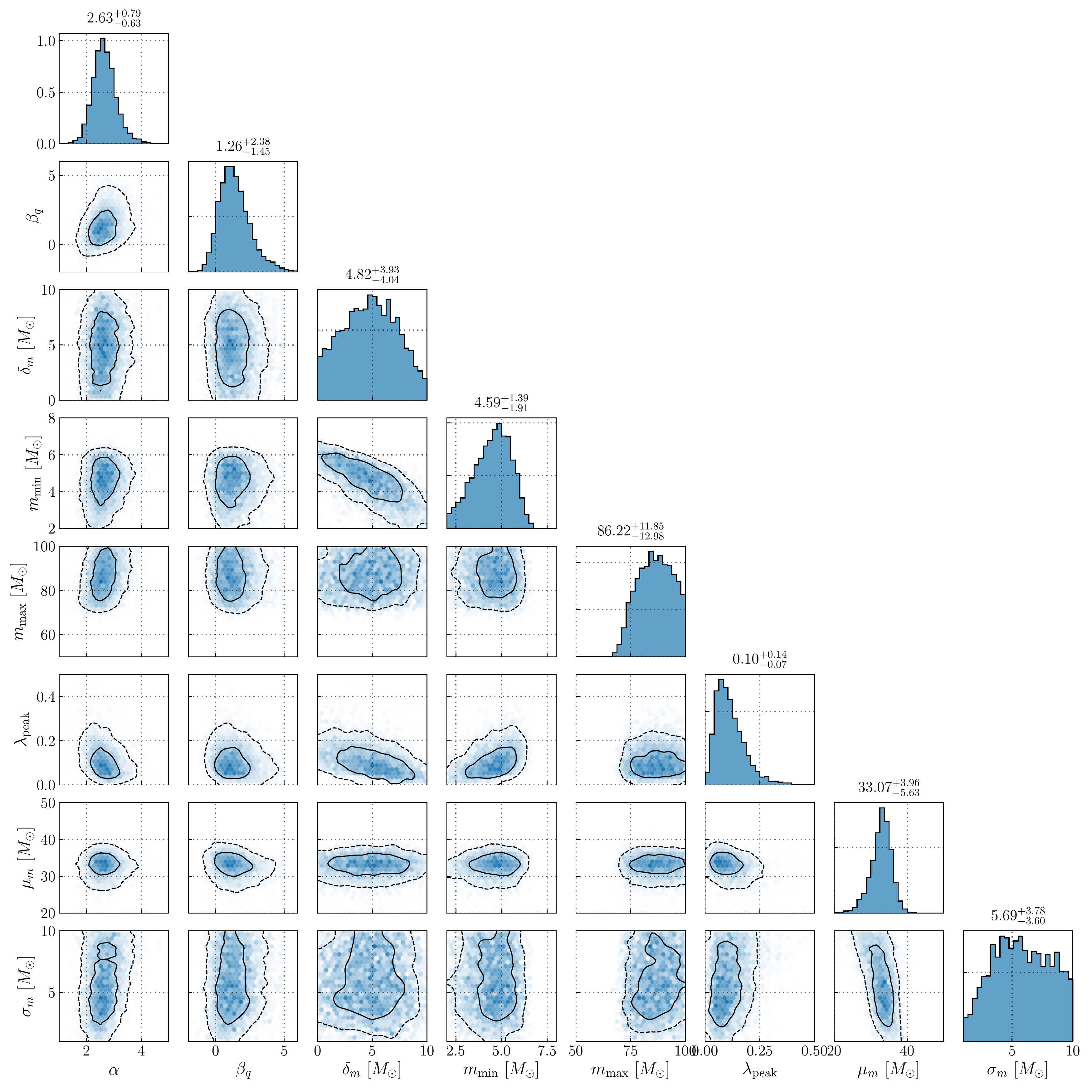}
    \caption{
    Posterior distribution for mass hyper-parameters for \ppsn{}.
    The fit excludes \NAME{GW190814A}{}. The contours represent $50\%$ and $90\%$ credible bounds.
    }
    \label{fig:PPSN_corner}
\end{figure*}

In Fig.~\ref{fig:PPSN_corner}, we provide a corner plot representation of the posterior distribution for the \ppsn{} hyper-parameters.
The $(\mu_m, \lambda_\text{peak})$ panel describes the Gaussian component: $\mu_m$ is the center of the Gaussian while $\lambda_\text{peak}$ is the fraction of mergers taking place in the Gaussian (as opposed to the power-law distribution).
Judging from this panel, it appears at first that $\lambda_\text{peak}$ peaks close to $0$ (corresponding to no Gaussian peak).
However, if we zoom in as in Fig.~\ref{fig:zoomed_in_lambda}, we see that the posterior for $\lambda_\text{peak}$ is peaked clearly away from zero at $\sim 0.02$.
This is consistent with the large Bayes factor indicating preference for \ppsn{} over \truncated{}.

\begin{table}[t]
    \centering
    \begin{tabular}{ c p{11cm} p{2mm} p{3cm} }
        \hline
        {\bf Parameter} & \textbf{Description} &  & \textbf{Prior} \\\hline\hline
        $\alpha$ & Spectral index for the power-law of the primary mass distribution. &  & U($-4$, $12$)\\
        $\beta_q$ & Spectral index for the power-law of the mass ratio distribution. &  & U($-4$, $12$)\\
        $\mmin$ & Minimum mass of the power-law component of the primary mass distribution. &  & U($2\, M_{\odot}$, $10\, M_{\odot}$)\\
        $\mmax$ &  Maximum mass of the power-law component of the primary mass distribution. &  & U($30\, M_{\odot}$, $100\, M_{\odot}$)\\
        $\lambda_\text{peak}$ & Fraction of \bbhsys{} in the Gaussian component. &  & U(0, 1) \\
        $\mu_{m}$ & Mean of the Gaussian component in the primary mass distribution.  &  & U($20\, M_{\odot}$, $50\, M_{\odot}$) \\
        $\sigma_{m}$ & Width of the Gaussian component in the primary mass distribution.  &  & U($1\, M_{\odot}$, $10\, M_{\odot}$)\\
        $\delta_{m}$ & Range of mass tapering at the lower end of the mass distribution.  &  & U($0\, M_{\odot}$, $10\, M_{\odot}$)\\
        \hline
    \end{tabular}
    \caption{The parameters that describe the BBH mass distribution for Model \ppsn{}.}
  \label{tab:parameters_ppsn_peak}
\end{table}

\subsection{\tapered{} mass model}\label{tapered}
\begin{figure*}[h]
    \centering
    \includegraphics[width = 0.95\textwidth]{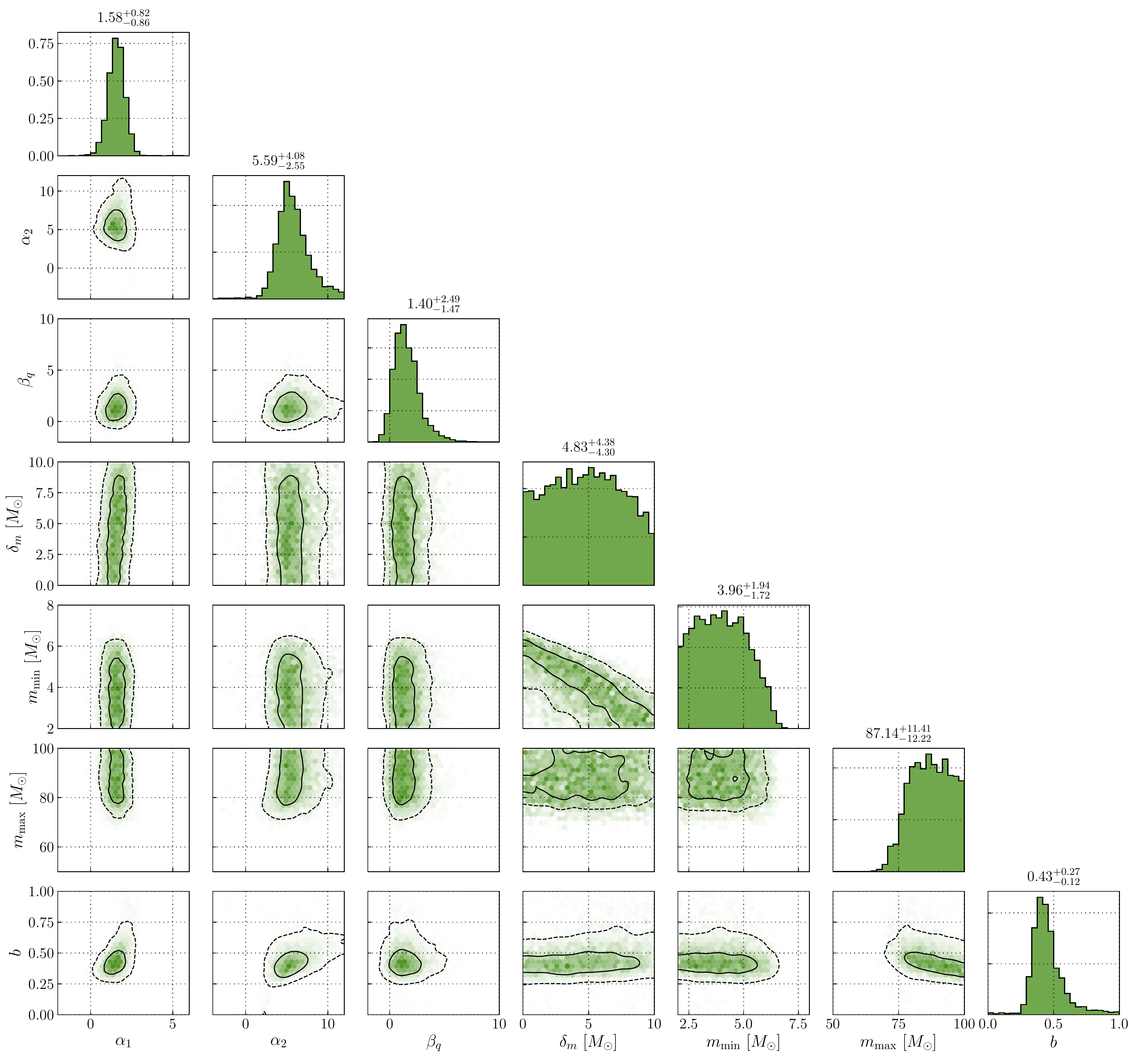}
    \caption{
    Posterior distribution for mass hyper-parameters for \tapered{}.
    The fit excludes \NAME{GW190814A}{}. The contours represent $50\%$, $90\%$ credible bounds.
    }
    \label{fig:BPL_corner}
\end{figure*}

This model is an extension of \truncated{}.
The primary mass distribution consists of a broken power-law.
This is motivated by the potential tapering of the primary mass distribution at high masses. 
Also, the model employs a smoothing function to prevent a sharp cut-off at low masses.
\begin{align}
    \pi(m_1 | \alpha_1, \alpha_2, \mmin, \mmax) \propto
    \begin{cases}
        m_1^{-\alpha_1} S(m_1|\mmin,\delta_m) & \mmin < m_1 < m_\text{break} \\
        m_1^{-\alpha_2} S(m_1|\mmin,\delta_m) & m_\text{break} < m_1 < \mmax \\
        0 & \text{otherwise} .
    \end{cases}
\end{align}
Here,
\begin{align}
    m_\text{break} = \mmin +
    b(\mmax-\mmin) ,
\end{align}
is the mass where there is a break in the spectral index and $b$ is the fraction of the way between $\mmin$ and $m_\text{max}$ at which the primary mass distribution undergoes a break.
Meanwhile, $S(m_1, \mmin, \delta_m)$ is a smoothing function as in Eq.~\eqref{eq:smoothing}. The conditional mass ratio distribution is the same as in the \ppsn{} model; see Eq.~\ref{eq:pq_smoothing}.
The hyper-parameters for this model are summarized in Table~\ref{tab:parameters_tapered}. 
In Fig.~\ref{fig:BPL_corner} we provide a corner plot for \tapered{}.
In the limit of no low-mass smoothing ($\delta_m =0$), and in the limit of a second power-law with a steep slope that mimics a sharp cutoff ($m_\mathrm{break} = \mmax$), this model reduces to \truncated{}.
Above, we noted that the \tapered{} model prefers a break in the primary mass spectrum near $40\ M_\odot$.
On the other hand, if we believe that the feature represented by $m_\mathrm{break}$ should be closer to a sharp cutoff, then the cut-off must occur at higher masses approaching the maximum mass of \truncated{} at $\mmax = 74.6^{+ 15.4}_{- 8.6} \ M_\odot$.
This can be seen by the correlation between $b$ and $\alpha_2$ in Fig.~\ref{fig:BPL_corner}.

\begin{table}[t]
    \centering
    \begin{tabular}{ c p{11cm} p{2mm} p{3cm} }
        \hline
        {\bf Parameter} & \textbf{Description} &  & \textbf{Prior} \\\hline\hline
        $\alpha_1$ & Power-law slope of the primary mass distribution for masses below $m_\mathrm{break}$. &  & U($-4$, $12$)\\
        $\alpha_2$ & Power-law slope for the primary mass distribution for masses above $m_\mathrm{break}$. &  & U($-4$, $12$) \\
        $\beta_q$ & Spectral index for the power-law of the mass ratio distribution. &  & U($-4$, $12$)\\
        $\mmin$ & Minimum mass of the power-law component of the primary mass distribution. &  & U($2\, M_{\odot}$, $10\, M_{\odot}$) \\
        $m_\mathrm{max}$ & Maximum mass of the primary mass distribution. &  & U($30\, M_{\odot}$, $100\, M_{\odot}$) \\
        $b$ & The fraction of the way between $m_\text{min}$ and $m_\text{max}$ at which the primary mass distribution breaks, e.g. a break fraction of 0.4 between $\mmin=5$ and $m_\mathrm{max}=85$ means the break occurs at $m_1=32$. &  & U(0, 1)\\
        $\delta_{m}$ & Range of mass tapering on the lower end of the mass distribution.  &  & U($0\, M_{\odot}$, $10\, M_{\odot}$)\\\hline
    \end{tabular}
    \caption{
    Summary of \tapered{} parameters.
    }
  \label{tab:parameters_tapered}
\end{table}

\subsection{\multipeak{} mass model}\label{Multi-Peak}
This model in an extension of \ppsn{}, where there is an additional Gaussian component at the upper end of the mass distribution motivated by a possible subpopulation of objects in the upper mass gap:

\begin{align}
    \pi(m_1 |& \lambda, \alpha, \mmin, \delta_m, \mmax, \mu_m, \sigma_m) = \nonumber \\
    & \bigg[
    (1-\lambda)\mathfrak{P}(m_1|-\alpha, \mmax) +
    \lambda \lambda_1 G(m_1|\mu_{m,1},\sigma_{m,1}) +
    \lambda (1-\lambda_1) G(m_1|\mu_{m,2},\sigma_{m,2})
    \bigg] S(m_1|\mmin, \delta_m) .
\end{align}

\begin{figure*}[h]
    \centering
    \includegraphics[width = \textwidth]{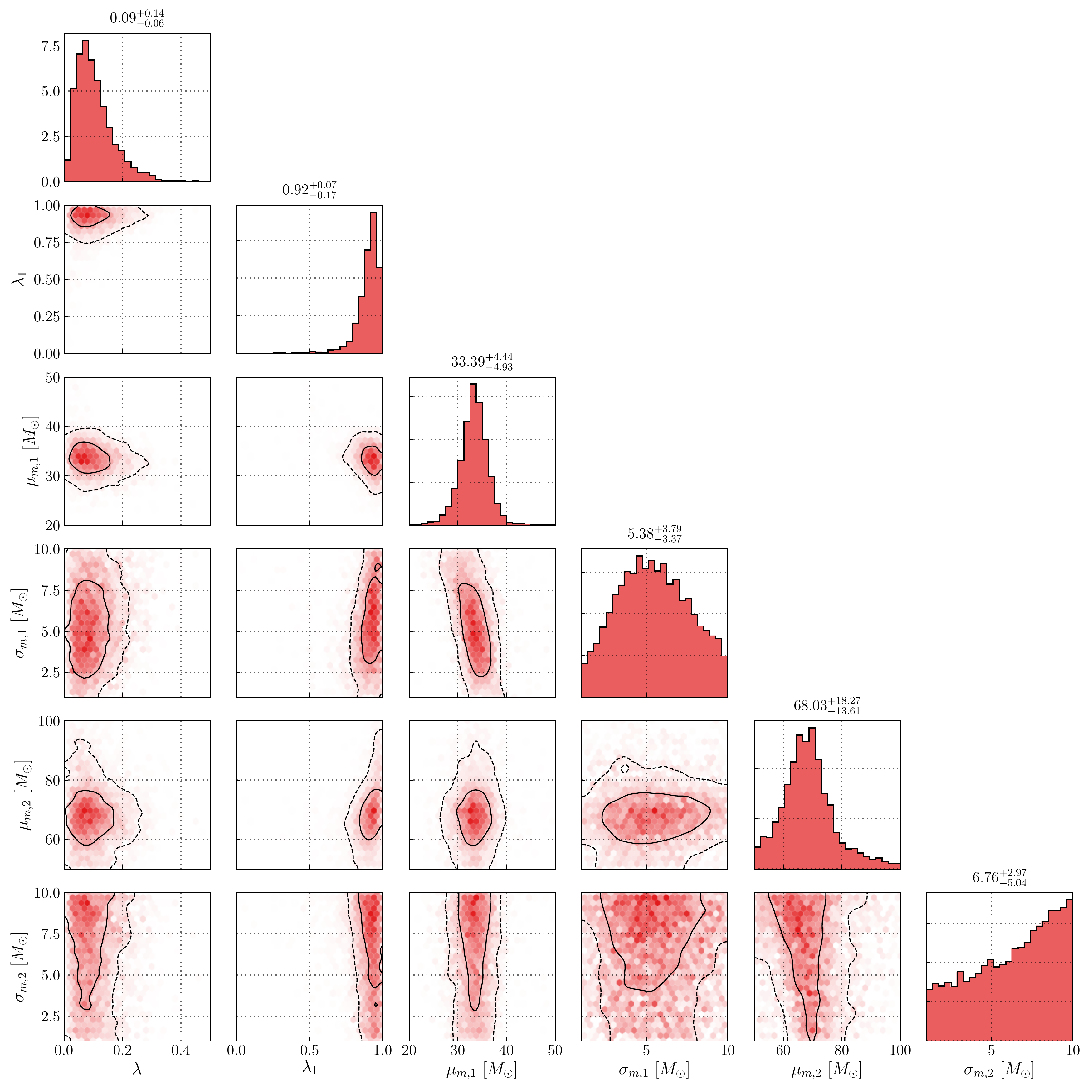}
    \caption{
    Posterior distribution for mass hyper-parameters for \multipeak{}.
    The fit excludes \NAME{GW190814A}{}. The contours represent $50\%$ and $90\%$ credible bounds.
    }
    \label{fig:Multipeak_corner}
\end{figure*}

Here, the parameters $\lambda$ and $\lambda_1$ correspond to the fraction of binaries in any Gaussian component and the fraction of binaries in the lower-mass Gaussian of the Gaussian components, respectively. 
The distribution $G(m_1|\mu_{m,1},\sigma_{m,1})$ is a normalized Gaussian distribution for the lower-mass peak with mean $\mu_{m,1}$ and width $\sigma_{m,2}$ and $G(m_1|\mu_{m,2},\sigma_{m,1})$ is a normalized Gaussian distribution for the upper-mass peak with mean $\mu_{m,2}$ and width $\sigma_{m,2}$.
The hyper-parameters for this model are summarized in Table~\ref{tab:parameters_multi_peak}.
\added{In Fig.~\ref{fig:Multipeak_corner} we provide a corner plot for \multipeak{} for parameters corresponding to the two Gaussian peaks. The mean of the upper-mass peak $\mu_{m,2}= \multipeakNoAugNoEvolutionmppTwo\ M_\odot$ is located at approximately twice the mean of the lower-mass peak $\mu_{m,1}= \multipeakNoAugNoEvolutionmppOne\ M_\odot$. The remaining parameters are: $\alpha= \multipeakNoAugNoEvolutionalpha$, $\beta_q= \multipeakNoAugNoEvolutionbeta$, $\mmin=\multipeakNoAugNoEvolutionmmin \ M_\odot$ and $\mmax=\multipeakNoAugNoEvolutionmmax\ M_\odot$.} 
In Fig.~\ref{fig:compare_cdfs}, we provide a posterior predictive check for all of the mass models used in this analysis.

\begin{table}[t]
    \centering
    \begin{tabular}{ c p{11cm} p{2mm} p{3cm} }
        \hline
        {\bf Parameter} & \textbf{Description} &  & \textbf{Prior} \\\hline\hline
        $\alpha$ & Spectral index for the power-law of the primary mass distribution. &  & U($-4$, $12$) \\
        $\beta_q$ & Spectral index for the power-law of the mass ratio distribution. &  & U($-4$, $12$)\\
        $\mmin$ & Minimum mass of the power-law component of the primary mass distribution. & & U($2\, M_{\odot}$, $10\, M_{\odot}$)\\
        $\mmax$ &  Maximum mass of the power-law component of the primary mass distribution. & & U($30\, M_{\odot}$, $100\, M_{\odot}$)\\
        $\lambda$ & Fraction of \bbhsys{} in the Gaussian components. &  & U(0, 1) \\
        $\lambda_1$ & Fraction of \bbhsys{} iin the Gaussian components belonging to the lower-mass component. &  & U(0, 1) \\
        $\mu_{m,1}$ & Mean of the lower-mass Gaussian component in the primary mass distribution.  &  & U($20\, M_{\odot}$, $50\, M_{\odot}$) \\
        $\sigma_{m,1}$ & Width of the lower-mass Gaussian component  in the primary mass distribution.  &  & U($1\, M_{\odot}$, $10\, M_{\odot}$)\\
        $\mu_{m,2}$ & Mean of the upper-mass Gaussian component in the primary mass distribution.  &  & U($50\, M_{\odot}$, $100\, M_{\odot}$) \\
        $\sigma_{m,2}$ & Width of the upper-mass Gaussian component  in the primary mass distribution. &  & U($1\, M_{\odot}$, $10\, M_{\odot}$) \\
        $\delta_{m}$ & Range of mass tapering on the lower end of the mass distribution. &  & U($0\, M_{\odot}$, $10\, M_{\odot}$) \\
        \hline
    \end{tabular}
    \caption{
    Parameters for the BBH mass distribution for Model \multipeak{}.
    }
  \label{tab:parameters_multi_peak}
\end{table}

\section{Mass Model Checking}
\label{Appendix:massmodelchecking}

\begin{figure*}
    \centering
    \includegraphics[width=0.9\textwidth]{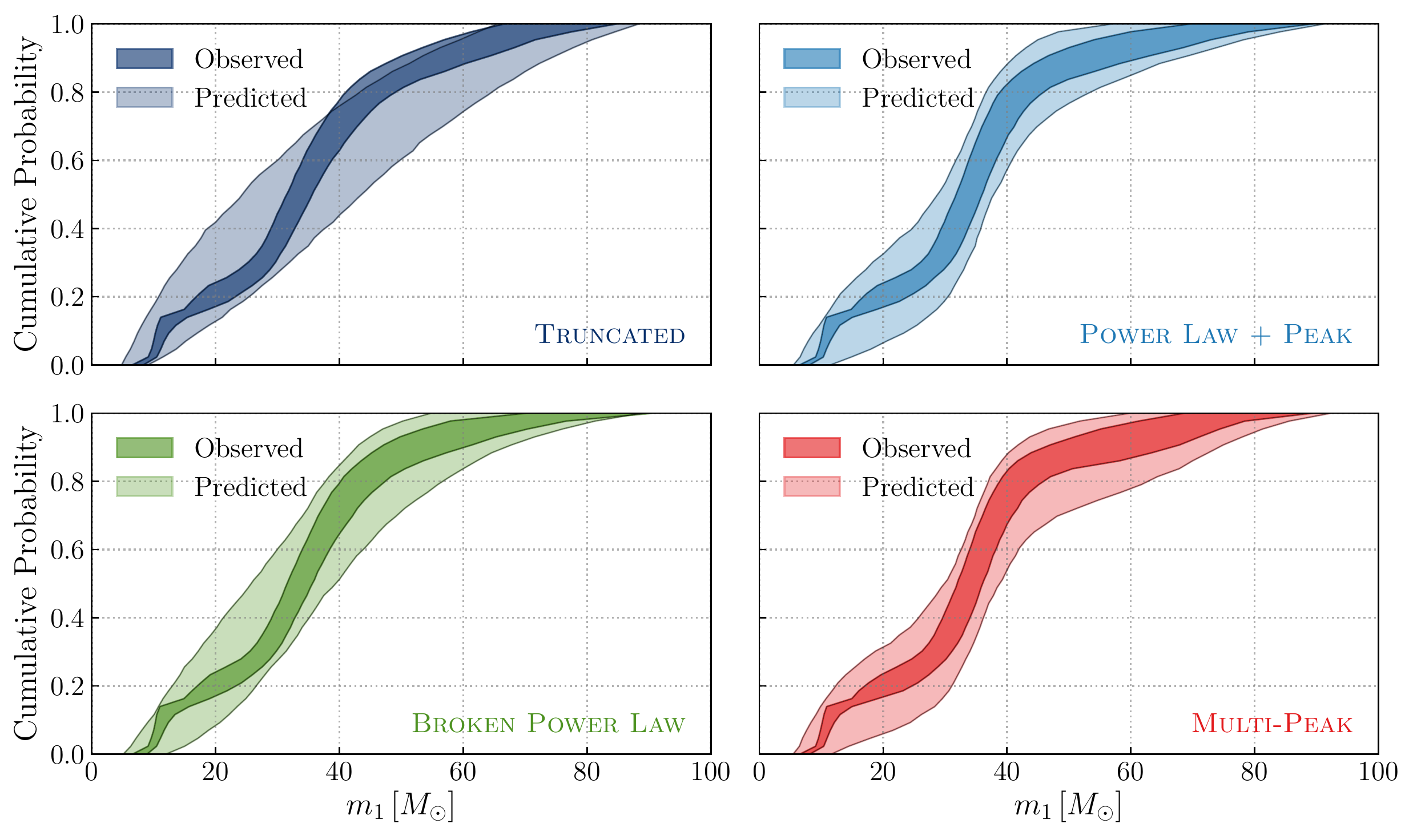}
    \caption{A posterior predictive check: the cumulative density function (CDF) of the observed primary mass distribution for the \truncated{}, \ppsn{}, \tapered{} and \multipeak{} models. 
    The observed event distribution is shown in the darker colors.
    The thickness of the bands indicates the 90\% credibility range. 
    The \ppsn{}, \tapered{} and \multipeak{} models are a better fit than the \truncated{} model; the dark band overlaps entirely with the light colored band.
    This is due to the \ppsn{}, \tapered{} and \multipeak{} models having more flexibility to fit the relative excess of binaries in the $\unit[30]{M_\odot}$--$\unit[40]{M_\odot}$ region compared to the $\gtrsim 40 M_\odot$ region.
    \NAME{GW190814A}{} is excluded from this analysis.
    \label{fig:compare_cdfs}}
\end{figure*}

Section~\ref{mass_results} describes the inferred mass distribution obtained with the \truncated{}, \tapered{}, \ppsn{}, and \multipeak{} models, and compares the different models by calculating their Bayes factors.
Here we assess the goodness-of-fit of the models using posterior predictive checks, comparing predicted and empirical catalogs of  observed $m_1$ distributions in Fig.~\ref{fig:compare_cdfs}.
The light colored bands show the cumulative distribution of $m_1$ as predicted by the model, while the darker bands show the empirical distribution based on the actual events observed in GWTC-2.
The bands represent a family of curves, where each curve corresponds to a different draw from the population hyper-posterior.
Each draw from the hyper-posterior updates both the predicted distribution (in the lighter color) and the empirical distribution (in the darker color), as the individual event posteriors are updated according to the inferred population distribution~\citep{Fishbach:2019ckx,Galaudage,Miller2020}.
If the model is a good fit to the data, the dark colored bands should overlap with the light colored bands.
Figure~\ref{fig:compare_cdfs} shows the relatively poor fit for the \truncated{} model, which cannot capture the excess of events at $\sim30$--$40 \ M_\odot$ compared to $\gtrsim 40 \ M_\odot$.
The remaining panels show the improved fits with the \ppsn{}, \tapered{} and \multipeak{} models.
These results are consistent with the Bayes factors in Table~\ref{tab:BF}, which conclude that the \truncated{} model is disfavored by a Bayes factor of \result{10--80} relative to the other models.

\subsection{On GW190412}
Other than \NAME{GW190814A}{}, we find that \NAME{GW190412A}{}~\citep{GW190412}, when analyzed with a population informed prior, remains the only system for which we can confidently bound the mass ratio away from unity, yielding $q < \MassRatioAprNinetyNinePercentile$ at 99\% credibility (using the \ppsn{} mass model).
All other events, when analyzed with a population-informed prior, are consistent with $q = 1$ at 99\% credibility.
Repeating the analysis in \citet{GW190412}, we perform a leave-one-out analysis without  \NAME{GW190412A}{} and find $\beta_q = \peakNoAprNoEvolutionbeta$ ($\beta_q = \BPLNoAprNoEvolutionbeta$) for the \ppsn{} (\tapered{}) model. The $\beta_q$ posterior inferred with the inclusion of GW190412 ($\beta_q = \peakNoAugNoEvolutionbeta$ for the \ppsn{} model; $\beta_q = \BPLNoAugNoEvolutionbeta$ for the \tapered{} model) has moderate ($\sim 50\%$) overlap with the leave-one-out $\beta_q$ posterior, indicating that, consistent with the conclusion in~\citet{GW190412},  \NAME{GW190412A}{} likely belongs to the low mass ratio tail of the distribution rather than a distinct subpopulation of asymmetric systems.

\subsection{On GW190521}
\label{GW190521}
\begin{figure*}
    \centering
    \includegraphics[width = 0.55\textwidth]{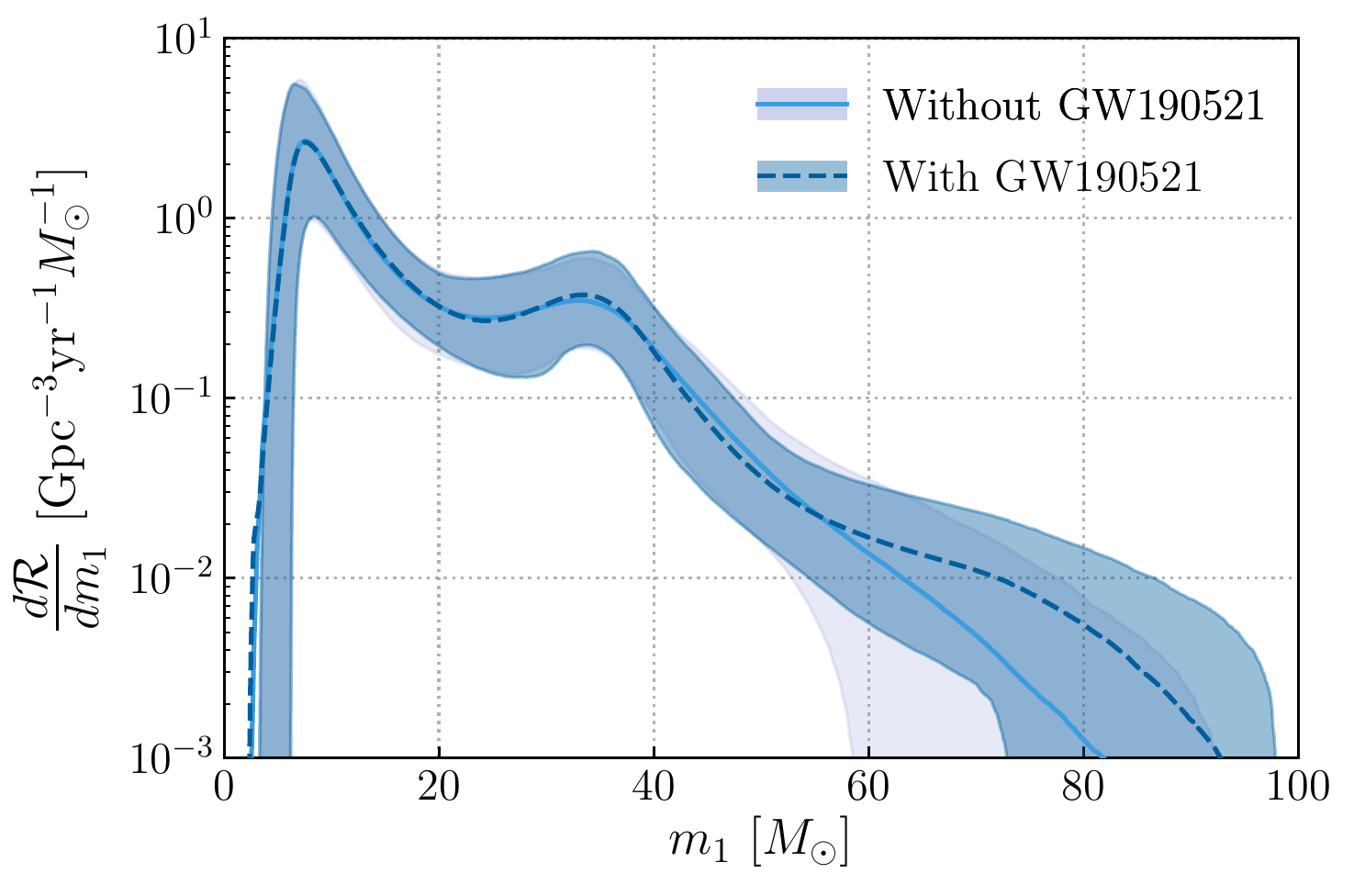}
    \caption{
    A comparison of the primary BH mass distribution for the population with and without \NAME{GW190521A}.
    The data are fit using the \ppsn{} model; the \tapered{} model produces similar results.
    The solid curves are the posterior predictive distributions while the shaded regions show the 90\% credible interval.
    The inclusion/exclusion of \NAME{GW190521A}{} does not have a significant effect on the fit.
    \NAME{GW190814A}{} is excluded from this analysis.
    }
    \label{fig:excluding_190521}
\end{figure*}

\begin{figure}
    \centering
    \includegraphics[width = 0.5\textwidth]{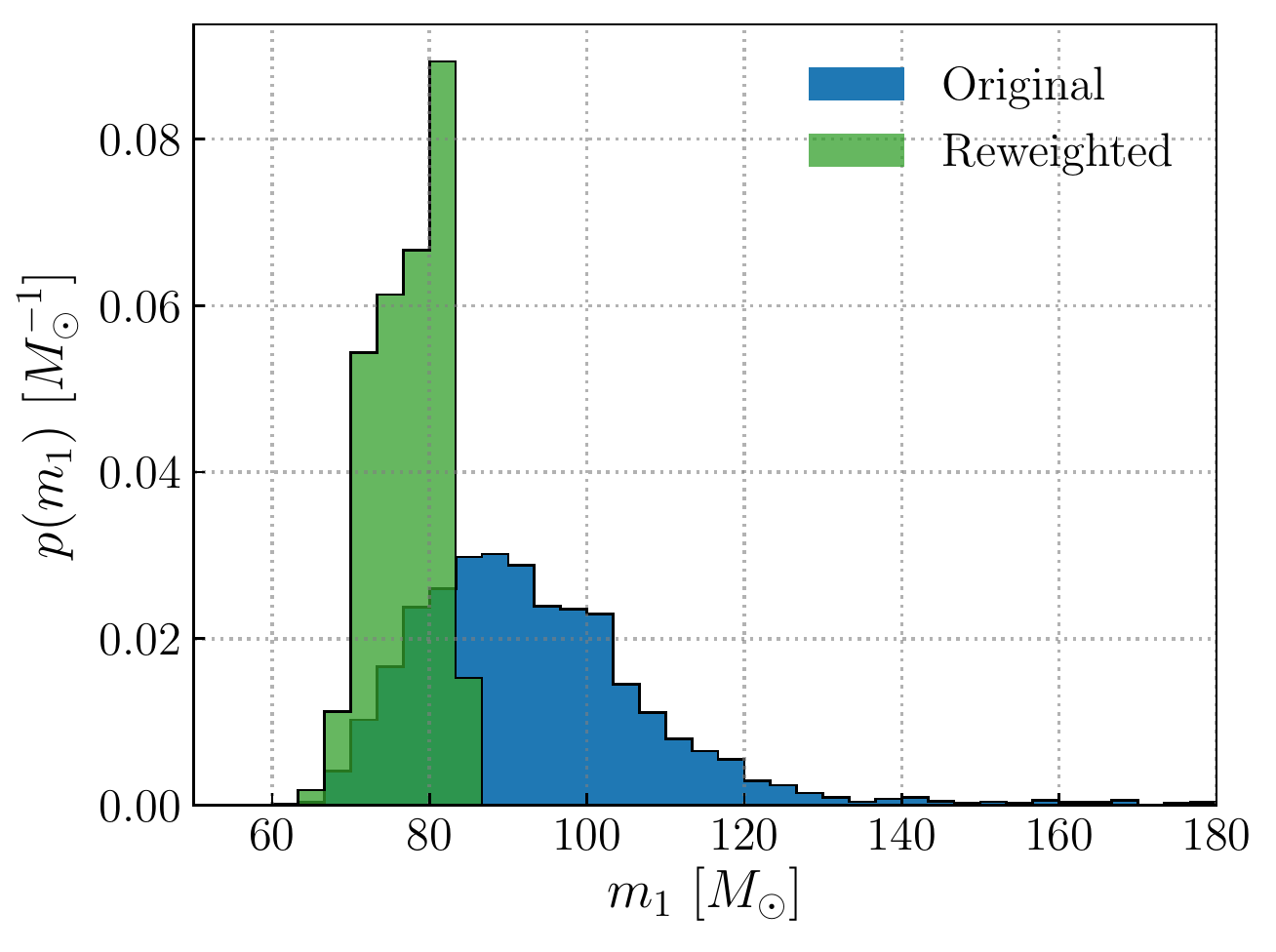}
    \caption{
    The posterior probability density for the primary mass of \NAME{GW190521A}{} using the original default prior (blue; flat in redshifted masses) and a reweighted version (green) obtained by using the \ppsn{} mass model. Results using the \tapered{} mass model are similar.
    The reweighted version shifts the posterior support to lower masses, with $m_1 < \PrimaryMassMayNinetyNinePercentile \ M_\odot$ (99\% credibility).
    }
    \label{fig:GW190521}
\end{figure}

As discussed in Section~\ref{mass_results}, the most massive event, \NAME{GW190521A}{}~\citep{GW190521}, is an outlier with respect to the \truncated{} model (see Fig.~\ref{fig:mmax_compare}), but fits well within the mass distributions inferred from the other models.
In Fig.~\ref{fig:excluding_190521}, we show the effect of \NAME{GW190521A}{} on the primary mass distribution.
This event shifts the best-fit mass distribution, but this shift is within the statistical uncertainties.
Thus, we find no evidence that \NAME{GW190521A}{} is an outlier within the framework of the \ppsn{} and \tapered{} mass models. This finding is supported by the posterior predictive check in subsection~\ref{sec:cwb}.
In Fig.~\ref{fig:GW190521}, we show how the primary mass posterior distribution for \NAME{GW190521A}{} changes when we use the \ppsn{} model to inform our prior.
While the population-informed posterior on the primary mass prefers smaller masses~\citep{Fishbach:2019ckx}, the conclusion that the primary mass of GW190521 is above $\PrimaryMassMayOnePercentile \ M_\odot$ (99\% credibility) is robust to the choice of prior, consistent with the claims in~\citet{GW190521}.

\subsection{On GW190814}\label{Appendix:GW190814}
\begin{figure*}
    \subfloat[The model-averaged astrophysical mass distribution $p(m_1, m_2)$.]{
    \includegraphics[width=0.49\textwidth]{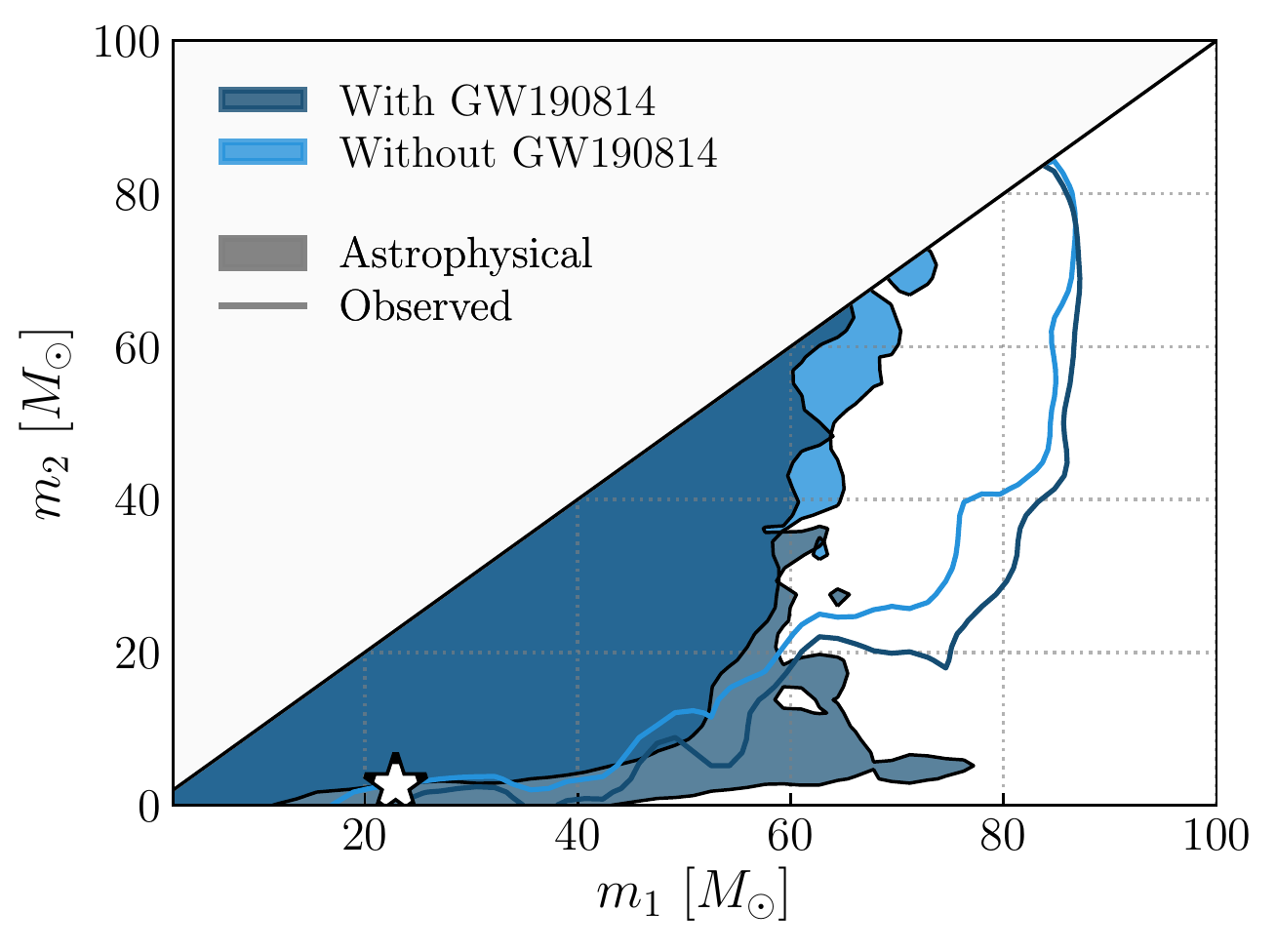}
    \label{fig:ppd_pm1m2_190814}
    }
    \subfloat[Observed minimum secondary mass distribution.]{
    \includegraphics[width=0.49\textwidth]{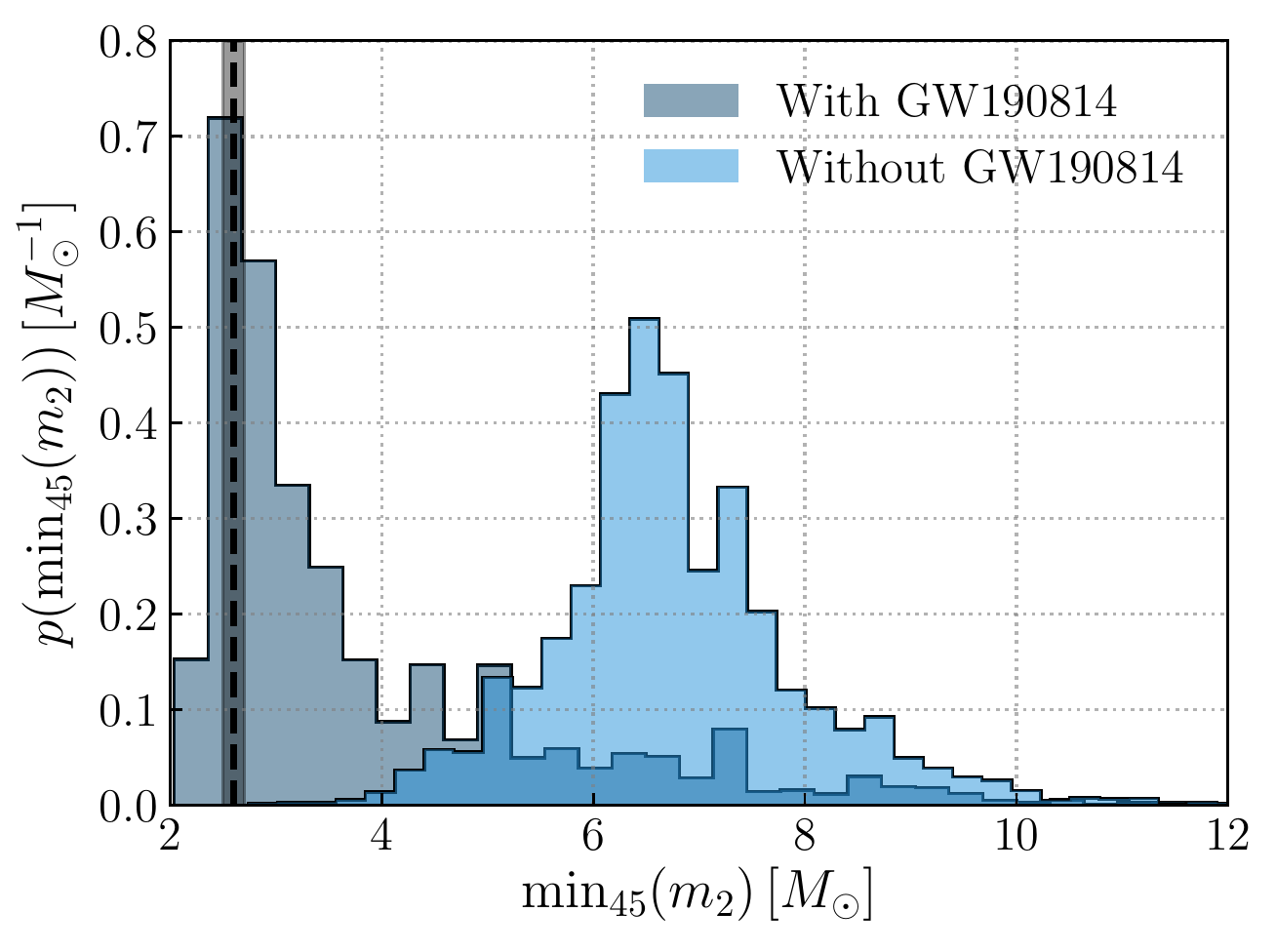}
    \label{fig:minm2_190814compare}
    }
    \caption{
    Left: the \appd{} for primary and secondary mass with and without  \NAME{GW190814A}{}.
    Shown here are the 99\% credible intervals.
    Dark blue is with \NAME{GW190814A}{} and light blue is without.
    The shaded regions show the \astrophysical, while the colored contours show the \observed{} distribution (as it appears in the catalog due to selection effects).
    The median values of the posterior of \NAME{GW190814A}{} is marked with a star. 
    Right: Distribution of the minimum secondary mass detected out of 45 detections, predicted from the fit to the \ppsn{} model to the BBH population excluding \NAME{GW190814A}{} (light blue) and the BBH population including \NAME{GW190814A}{} (dark blue).
    The dashed line and shaded region (gray) denote the median and 90\% symmetric credible interval on the secondary mass of \NAME{GW190814A}{}. 
    This distribution is qualitatively similar to the distribution predicted from the fit to the \tapered{} model.
    }
\end{figure*}

On the other hand, we see clear indication that \NAME{GW190814A}{} is an outlier with respect to the BBH population within the framework of the \ppsn{} and \tapered{} mass models, as discussed in Section~\ref{mass_results}.
As an additional posterior predictive check, following the analysis described in~\citet{Fishbach:2019ckx} and \citet{GW190412}, we use the \oppd{}, inferred without \NAME{GW190814A}{}, to construct a distribution for the minimum $m_2$ detected in a sample of 45 events.
When using both the \ppsn{} and \tapered{} models, we find that the observation of a system with a secondary mass equal to or smaller than the that of \NAME{GW190814A}{} ($2.59_{-0.09}^{+0.08} \ M_\odot$) is highly improbable, with probability {$< \peakNoAugNoEvolutionMTwoMinPercentBelow \%$} for both \ppsn{} and \tapered{}; see Fig.~\ref{fig:minm2_190814compare} for the distribution of the minimum observed secondary mass in a sample of 45 events predicted by the \ppsn{} model.
The distribution for \tapered{} is qualitatively similar.
The mass ratio of \NAME{GW190814A}{} is also somewhat unusual according to this posterior predictive check; see Fig.~\ref{fig:ppd_pm1m2_190814}.
Observing an event with the mass ratio of \NAME{GW190814A}{} or smaller, based on the fit to the other 44 \bbhevnt{}, has probability {$<\peakNoAugNoEvolutionMassRatioMinPercentBelow \%$} in both the \ppsn{} and \tapered{} models. These posterior predictive checks suggest that \NAME{GW190814A}{} is not a typical BBH, and support the conclusion that there may be a dearth of systems between $\sim 2.6\ M_\odot$ and $\sim 6 \ M_\odot$. Future observations will reveal the precise shape of the mass distribution at low masses and extreme mass ratios, and better determine the nature of \NAME{GW190814A}{}.

\subsection{Mass and distance checks with a burst analysis}
\label{sec:cwb}
\added{The earlier posterior predictive checks in this section compared simulated sets of BBH masses to the observed set of catalog events. As a complimentary posterior predictive check, we can simulate the gravitational-wave signals from these predicted events, run it through our search pipelines, and compare the synthetic data, as detected by the pipeline, to the observed data. As a proof of principle, we carry out this posterior predictive check with the \textsc{Coherent WaveBurst} (\textsc{cWB}) pipeline~\citep{Klimenko:2004qh,Klimenko:2015ypf}, which is designed to detect unmodeled gravitational-wave transients.}
\deleted{The \textsc{Coherent WaveBurst} (\textsc{cWB}) pipeline~\citep{Klimenko:2004qh,Klimenko:2015ypf} is designed to detect unmodeled gravitational-wave transients also known as bursts.}
The \textsc{cWB} analysis resulted in the detection of 22 BBH events in GWTC-2.
We investigate whether the set of \textsc{cWB} observations is consistent with the model predictions. 
\added{We focus on assessing possible outliers at high masses and high redshifts, where \textsc{cWB} is especially sensitive to BBH signals.}
In particular, we examine whether \NAME{GW190521A}{} \added{,which was recovered with higher significance by the \textsc{cWB} saerch than the templated searches,} is an outlier in the context of our mass and redshift models. 
Following \cite{Klimenko:2015ypf}, we calculate the expected distribution of the central frequency $f$ \added{(which depends on the redshifted mass of the BBH)} and coherent signal-to-noise ratio $\rho$ \added{(which, for a given redshifted mass, depends on the distance of the BBH)} for two different population models: \ppsn{} and \tapered{}, using the \textsc{Non-Evolving} redshift model.
We then compare the empirical distribution of $(f, \rho)$, as recovered by the \textsc{cWB} pipeline to the distribution predicted by the population model. 
We quantify the comparison by calculating a $p$-value for each event $i$, which measures how unusual its observed $(f_i,\rho_i)$ is, given the distribution of predicted $(f, \rho)$.
The central frequency is a proxy for the redshifted mass while the signal-to-noise ratio is a proxy for the distance.

\begin{figure}
    \centering
    \includegraphics[width = 0.65\textwidth]{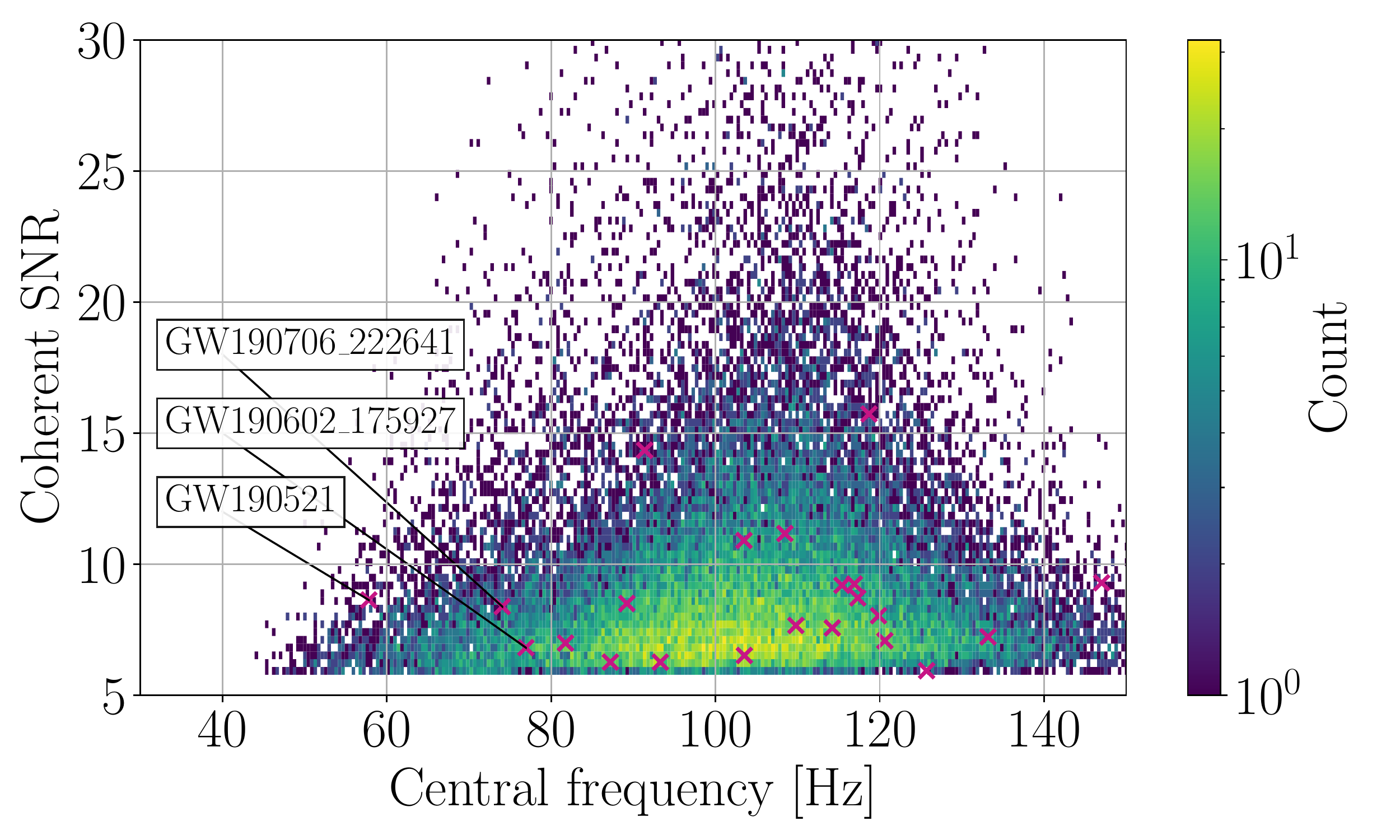}
    \caption{The coherent signal-to-noise ratio and central frequency for 22 BBH events detected by one of our detection pipelines, \textsc{cWB}, in O1, O2 and O3a (violet dots) compared to simulated BBH events from the \ppsn{} model detected by \textsc{cWB}. \added{The consistency between the distributions of simulated and observed triggers shows that the \ppsn{} mass model coupled with the \textsc{non-evolving} redshift distribution is a good fit to the data. Results for the \tapered{} model are similar. }
    }
    \label{fig:cwb}
\end{figure}

To compute the predicted distribution of $(f, \rho)$, we inject simulated waveforms drawn from the \ppsn{} and \tapered{} distributions into the O1, O2 and O3a data and compile injections recovered by \textsc{cWB} with a $\mathrm{FAR} < \unit[1]{yr^{-1}}$. 
The central frequencies and coherent signal-to-noise ratios of the recovered injections generated according to \ppsn{} are plotted in Fig.~\ref{fig:cwb}. 
The results for \tapered{} are similar.
The locations of the 22 detections on this plane are visually consistent with the model predictions, indicating that the model is a reasonably good fit to the data.
The event with the lowest $p$-value (least consistent with predictions) is \NAME{GW190521A}, with $p$-values of {0.053} and {0.077} for the \tapered{} model and the \ppsn{} model respectively.
These $p$-values indicate that \NAME{GW190521A}{} is a moderately unusual detection, but it is consistent with the population models.

\section{Details of spin population models}
\label{sec:spindetails}

\subsection{\textsc{Default} spin model}\label{sec:default}
This model was introduced in~\cite{O2pop}.
Following~\cite{Wysocki2019}, the dimensionless spin magnitude distribution is taken to be a Beta distribution,
\begin{align}
    \pi(\chi_{1,2} | \alpha_\chi, \beta_\chi) = \text{Beta}(\alpha_\chi, \beta_\chi) ,
\end{align}
where $\alpha_\chi$ and $\beta_\chi$ are the standard shape parameters that determine the distribution's mean and variance.
The Beta distribution is convenient because it is bounded on (0,1).
The distributions for $\chi_1$ and $\chi_2$ are assumed to be the same.
Following~\cite{Talbot2017}, we define $z=\cos \theta_{1,2}$ as the cosine of the tilt angle between component spin and a binary's orbital angular momentum, and assume that $z$ is distributed as a mixture of two populations:
\begin{align}
    \pi(z | \zeta, \sigma_t) = \zeta \, G_t(z|\sigma_t) + (1-\zeta) \mathfrak{I}(z) .
\end{align}
Here, $\mathfrak{I}(z)$ is an isotropic distribution, while $G_t(z|\sigma_t)$ is a truncated Gaussian, peaking at $z=0$ (perfect alignment) with width $\sigma_t$.
The mixing parameter $\zeta$ controls the relative fraction of mergers drawn from the isotropic distribution and Gaussian subpopulations.
The isotropic subpopulation is intended to accommodate dynamically assembled binaries, while $G_t$ is a model for field mergers.
The hyper-parameters for this model and their priors are summarized in Table~\ref{tab:parameters_default}. Additional constraints to the priors on $\mu_\chi$ and $\sigma^2_\chi$ are applied by setting $\alpha_\chi$, $\beta_\chi > 1$.  

\begin{figure*}
    \centering
    \includegraphics[width = 0.75\textwidth]{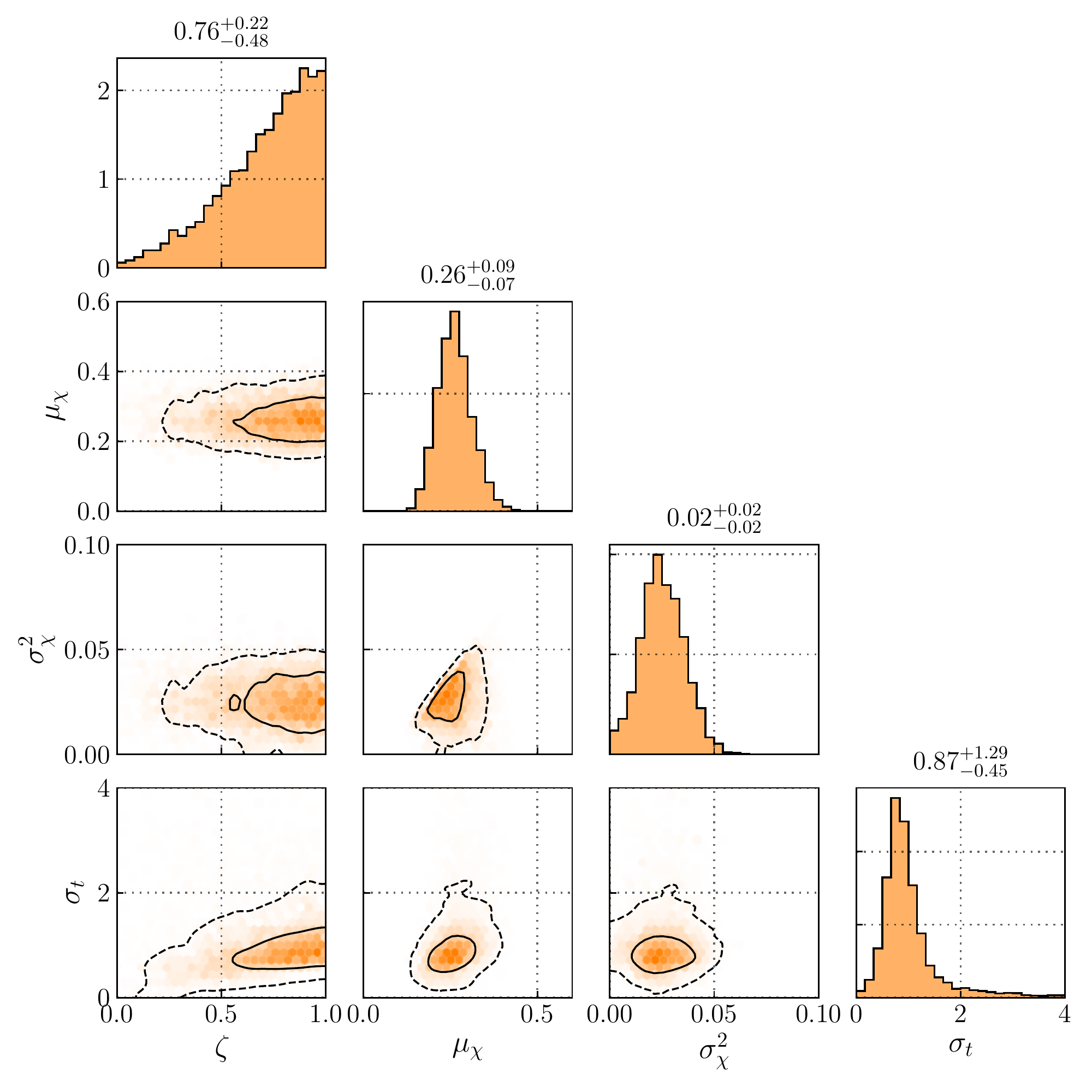}
    \caption{
    Posterior distribution for spin hyper-parameters for \textsc{Default}, assuming the \ppsn{} mass model and \textsc{Non-Evolving} redshift distribution.
    The fit excludes \NAME{GW190814A}{}.
    The contours represent $50\%$, $90\%$ credible bounds. A \emph{perfectly} aligned spin distribution ($\sigma_t = 0$, $\zeta = 1$) is ruled out at $> 99\%$ credibility, consistent with the results of the \textsc{Gaussian} model, but the data disfavor a purely isotropic distribution ($\zeta = 0$ or $\sigma_t \gtrsim 2$).
    }
    \label{fig:default_spin}
\end{figure*}

In Fig.~\ref{fig:default_spin} we provide a corner plot for the \textsc{Default} spin model.
This model prefers modest spin magnitudes.
It favors the hypothesis that binaries are preferentially aligned ($\zeta \rightarrow 1$ and $\sigma_t \lesssim 2$), albeit with potentially large misalignment angles ($\sigma_t>0$).
The case of \textit{perfect} alignment, which would correspond to $\zeta=1$ and $\sigma=0$, is disfavored, lying outside the 99\% credible bound on $\zeta$ and $\sigma_t$.
Within the main text, Fig.~\ref{fig:default-spin-ppds} shows the implied distributions of component spin magnitudes and tilt angles.
The implied distributions of the \chip{} ($\chi_\mathrm{p}$) and the \chieff{} ($\chi_\mathrm{eff}$) are shown in Figs.~\ref{fig:chiP-corner_combined} and~\ref{fig:chi_eff_cornerPlot}, respectively; these distributions are in good agreement with the results obtained using the \textsc{Gaussian} model described below. In particular, both models predict the existence of systems with \antialigned{} spins (negative $\chi_\mathrm{eff}$), and in-plane spin components (non-zero $\chi_\mathrm{p}$).

\begin{table}[t]
    \centering
    \begin{tabular}{ c p{11cm} p{2mm} p{3cm} }
        \hline
        {\bf Parameter} & \textbf{Description} &  & \textbf{Prior} \\\hline\hline
        $\mu_\chi$ & Mean of the Beta distribution of spin magnitudes. &  & U(0,1) \\
        $\sigma^2_\chi$ & Variance of the Beta distribution of spin magnitudes. &  & U(0,0.25) \\
        $\zeta$ & Mixing fraction of mergers from truncated Gaussian distribution. &  & U(0,1)\\
        $\sigma_t$ &  Width of truncated Gaussian, determining typical spin misalignment. &  & U(0.01,4) \\
        \hline
    \end{tabular}
    \caption{
    Summary of \textsc{Default} spin parameters.
    }
  \label{tab:parameters_default}
\end{table}

\subsection{\textsc{Gaussian} spin model}\label{sec:Gaussian}
The \textsc{Gaussian} spin model offers an alternative description of BBH spins.
It is convenient to measure the distribution of the \chieff{} ($\chi_\mathrm{eff}$) and the \chip{} ($\chi_\mathrm{p}$), which are better constrained than individual component spin magnitudes or tilts.
We parameterize the distributions of $\chi_\mathrm{eff}$ and $\chi_\mathrm{p}$, using a bivariate Gaussian:
    \begin{equation}
    \pi(\chi_\mathrm{eff},\chi_\mathrm{p}|\mu_\mathrm{eff},\sigma_\mathrm{eff},\mu_p,\sigma_p,\rho) \propto G(\chi_\mathrm{eff},\chi_\mathrm{p}|\pmb\mu,\pmb\Sigma) .
    \label{eq:gaussian-spin}
    \end{equation}
The distribution has a mean $\pmb\mu = (\mu_\mathrm{eff},\mu_p)$ and a covariance matrix
    \begin{equation}
    \pmb \Sigma = \begin{pmatrix}
    \sigma^2_\mathrm{eff} & \rho \sigma_\mathrm{eff}\sigma_p \\
    \rho\sigma_\mathrm{eff}\sigma_p & \sigma^2_p
    \end{pmatrix}.
    \label{eq:gaussian-cov}
    \end{equation}
The population parameters appearing in Eqs.~\eqref{eq:gaussian-spin} and \eqref{eq:gaussian-cov} and their associated priors are summarized in Table~\ref{tab:parameters_gaussian}.
We truncate and normalize Eq.~\eqref{eq:gaussian-spin} based on the allowed regions of the \chieff{}: $\chi_\mathrm{eff} \in (-1,1)$ and $\chi_\mathrm{p} \in (0,1)$.
The results from the \textsc{Gaussian} model are obtained assuming a \truncated{} mass model with $\alpha = -2.2$, $\beta_q = 1.3$, $m_\mathrm{min}=5\,M_\odot$, and $m_\mathrm{max}=75\,M_\odot$, consistent with the median values obtained when fitting the \textsc{Truncated} model to GWTC-2.
We additionally assume a comoving merger rate density that grows as $(1+z)^{2.7}$.
Although the \truncated{} model is disfavored relative to the more complex mass models discussed above, it is sufficient for purposes of constructing an informed mass ratio distribution, the primary confounding factor in efforts to measure $\chi_\mathrm{eff}$ and $\chi_\mathrm{p}$~\citep{Ng2018}.

\begin{table}[t]
    \centering
    \begin{tabular}{ c p{11cm} p{2mm} p{3cm} }
        \hline
        {\bf Parameter} & \textbf{Description} &  & \textbf{Prior} \\\hline\hline
        $\mu_\mathrm{eff}$ & Mean of the $\chi_\mathrm{eff}$ distribution. &  & U($-1$, $1$) \\
        $\sigma_\mathrm{eff}$ & Standard deviation of the $\chi_\mathrm{eff}$ distribution. &  & U(0.01,1) \\
        $\mu_p$ & Mean of the $\chi_\mathrm{p}$ distribution. &  & U($0.01$, $1$) \\
        $\sigma_p$ &  Standard deviation of the $\chi_\mathrm{p}$ distribution. &  & U($0.01$, $1$) \\
        $\rho$ & Degree of correlation between $\chi_\mathrm{eff}$ and $\chi_\mathrm{p}$. &  & U($-0.75$, $0.75$) \\
        \hline
    \end{tabular}
    \caption{
    Summary of \textsc{Gaussian} spin parameters.
    }
  \label{tab:parameters_gaussian}
\end{table}

\begin{figure*}
    \centering
    \includegraphics[width = 0.8\textwidth]{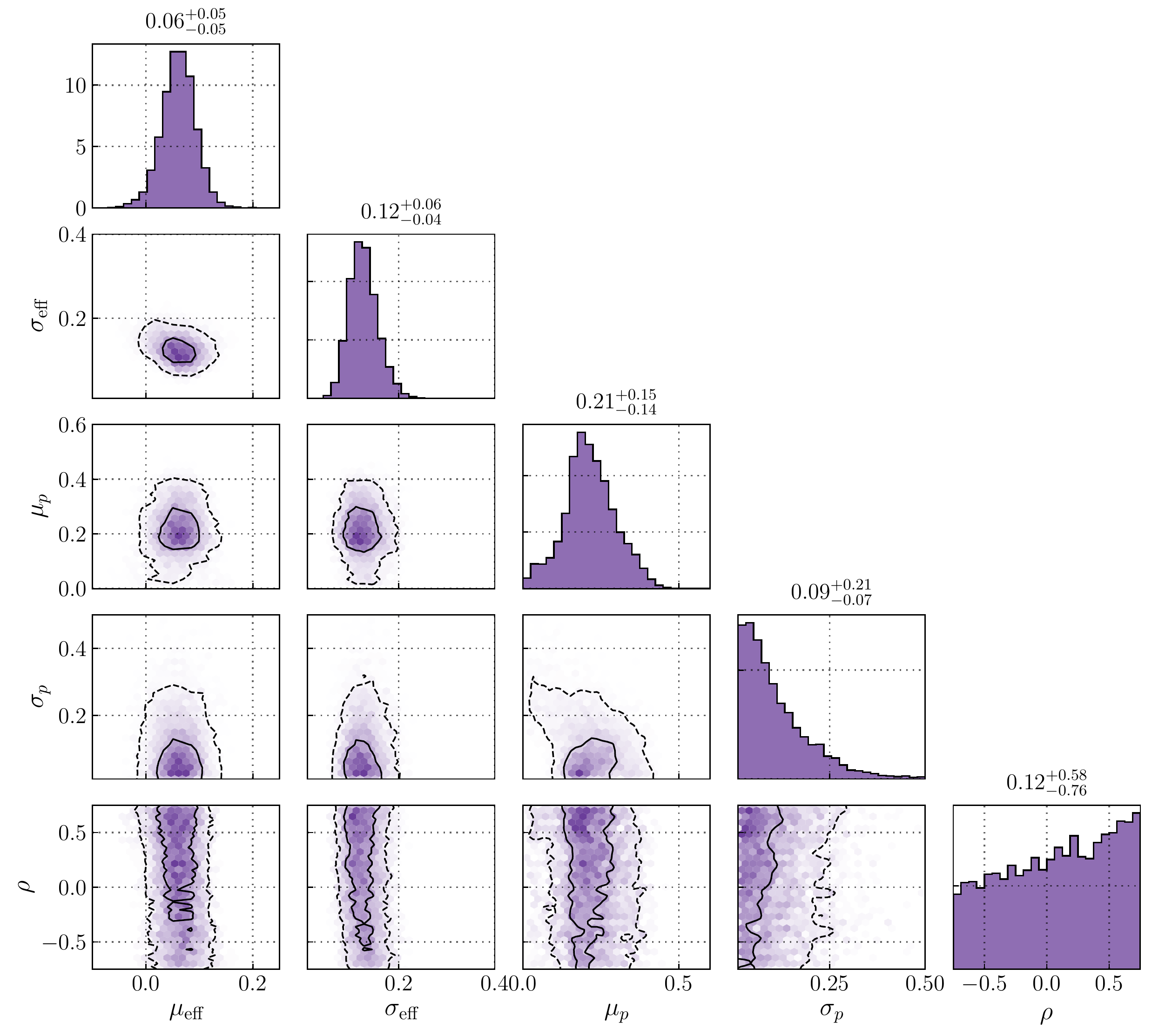}
    \caption{
    \added{
    Posterior distribution for spin hyper-parameters under the \textsc{Gaussian} model.
    The fit again excludes \NAME{GW190814A}{}, and contours represent $50\%$ and $90\%$ credible bounds.
    Consistent with the results of the \textsc{Default} model, which results out a perfectly aligned spin distribution at $99\%$ credibility, here we find that a vanishing $\chi_{\rm p}$ distribution ($\mu_p = \sigma_p = 0$) is disfavored.}
    }
    \label{fig:gaussian_corner_full}
\end{figure*}

\begin{figure}
    \centering
    \includegraphics[width=0.48\textwidth]{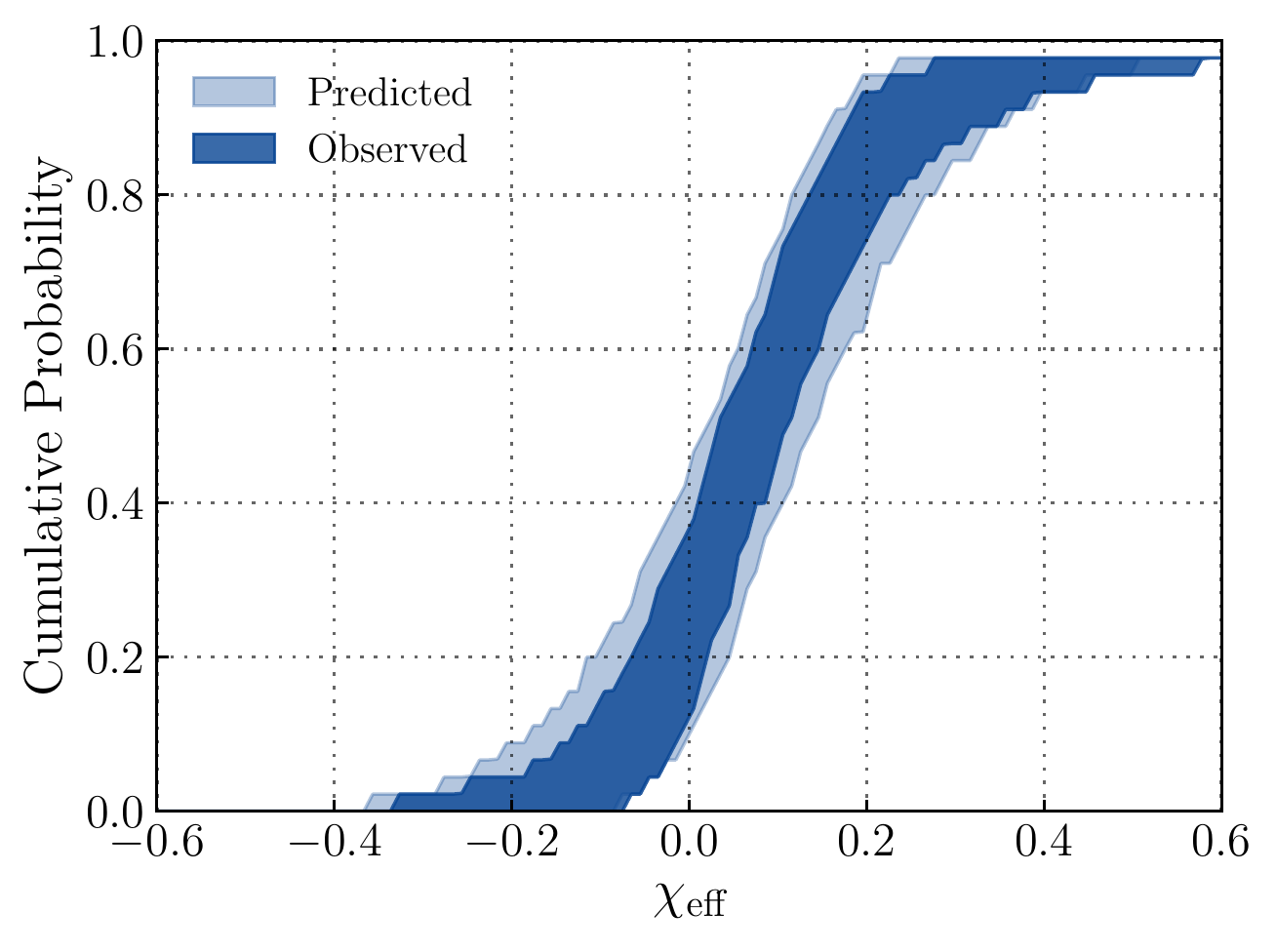} 
    \hspace{5mm}
    \includegraphics[width=0.48\textwidth]{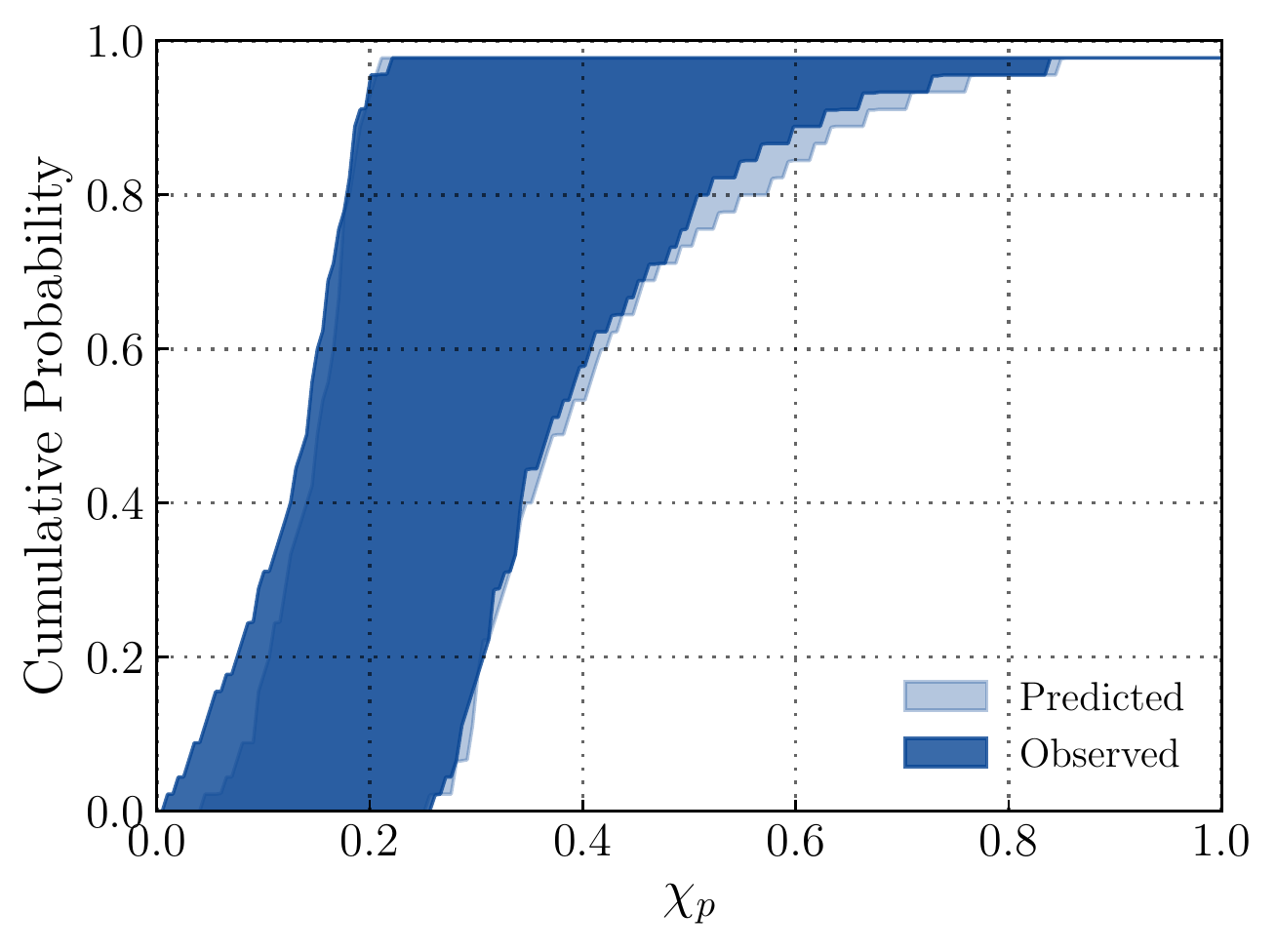}
    \caption{
    Population predictive checks for the \chieff{} $\chi_\mathrm{eff}$ (\textit{left}) and the \chip{} $\chi_\mathrm{p}$ (\textit{right}) of BBH mergers using the \textsc{Gaussian} spin model.
    The light shaded regions show the central 90\% credible bounds on the posterior predictive distributions. According to the model, we expect the observed distributions on $\chi_\mathrm{eff}$ and $\chi_\mathrm{p}$ to lie within the light shaded region 90\% of the time. The dark shaded regions show the 90\% credible bounds on the observed distributions in GWTC-2, found using the population-informed posteriors of the confident BBH events in GWTC-2. The overlap between the dark and light regions shows that the model passes the posterior predictive check.
    The results for the \textsc{Default} model are similar, indicating that both models are a good fit to the data.
    }
    \label{fig:gaussian-spin-ppc}
\end{figure}

\deleted{
The marginal posterior distributions on $\mu_p$ and $\sigma_p$ is shown in Fig.~\ref{fig:chiP-corner_combined}, while the marginal posterior on $\mu_\mathrm{eff}$ and $\sigma_\mathrm{eff}$ is shown in Fig.~\ref{fig:chi_eff_cornerPlot}.
}
\added{The full posterior on parameters of the \textsc{Gaussian} model is shown in Fig.~\ref{fig:gaussian_corner_full}.}
We find no correlation between the parameters of the $\chi_\mathrm{eff}$ and $\chi_\mathrm{p}$ distributions, nor do we obtain any information regarding the degree of correlation $\rho$ between the \chieff{} and the \chip{}.
As discussed in Sec.~\ref{sec:results-spin}, analysis with the \textsc{Gaussian} spin model
\deleted{suggests the BBH population exhibits non-vanishing $\chi_\mathrm{p}$, such that $\mu_p = \sigma_p = 0$ is ruled out.}
\added{is consistent with the identification of spin-orbit misalignment using the \textsc{Default} spin model; with $\mu_p = \sigma_p = 0$ disfavored.
For the \textsc{Default} spin model, we verified that the signature of spin-orbit misalignment was not a spurious prior artifact, finding that our posterior lies safely outside the artificial exclusion region in Fig.~\ref{fig:chiP-corner_combined}. A similar exclusion region also exists for the \textsc{Gaussain} model around $\mu_p = \sigma_p = 0$, but our estimate of its exact size is subject to sampling uncertainties driven by the relatively small number of prior samples close to $\chi_\mathrm{p} = 0$ for each event.} 

With the \textsc{Gaussian} model, we also find evidence that at least some BHs have \antialigned{} spins, with $\theta>90^\circ$, such that $\chi_\mathrm{eff}<0$.
To further evaluate the robustness of our \textsc{Gaussian} model fits, in Fig.~\ref{fig:gaussian-spin-ppc} we show posterior predictive comparisons between predicted and empirical catalogs of $\chi_\mathrm{eff}$ and $\chi_\mathrm{p}$ measurements.
The light blue bands mark 90\% credible bounds on the predicted cumulative distribution of observed \chieff{} values, given our posterior on the \textsc{Gaussian} model parameters.
The dark shaded regions, meanwhile, show 90\% credible bounds on the true distribution observed within GWTC-2, achieved by reweighting single event $\chi_\mathrm{eff}$ and $\chi_\mathrm{p}$ by repeated random draws from the \textsc{Gaussian} hyper-posterior.
\added{
A similar predictive comparison between observation and an alternative \textit{strictly positive} model, in which the effective inspiral spin distribution is truncated on the interval $0\leq \chi_{\rm eff} \leq 1$, reveals possible tension; when asserting that all effective inspiral spins are positive, the resulting population model underpredicts the number of observations with $\chi_{\rm eff} < 0.1$ approximately 75\% of the time.
}

\begin{figure}
    \centering
    \includegraphics[width=0.48\textwidth]{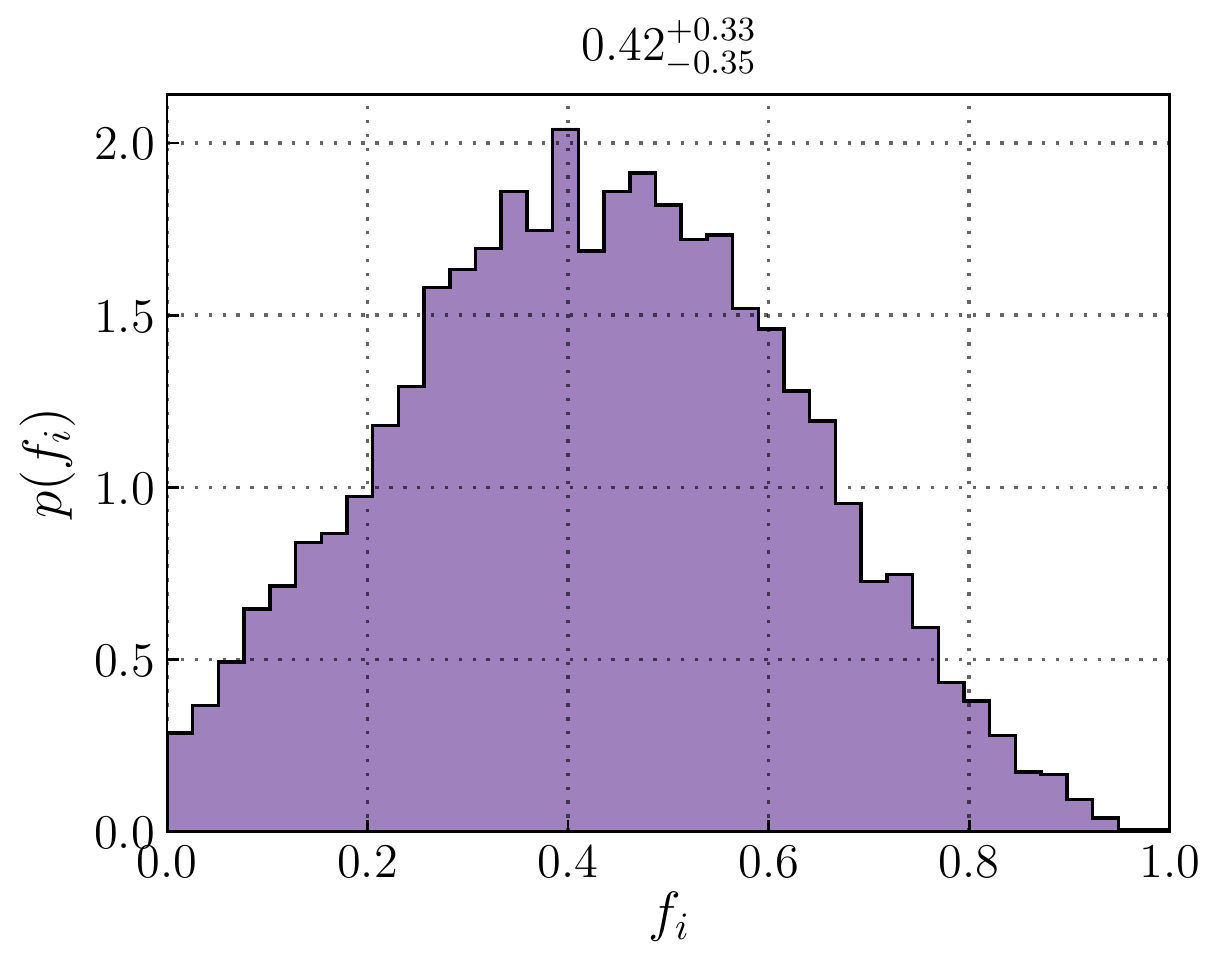}
    \hspace{5mm}
    \includegraphics[width=0.48\textwidth]{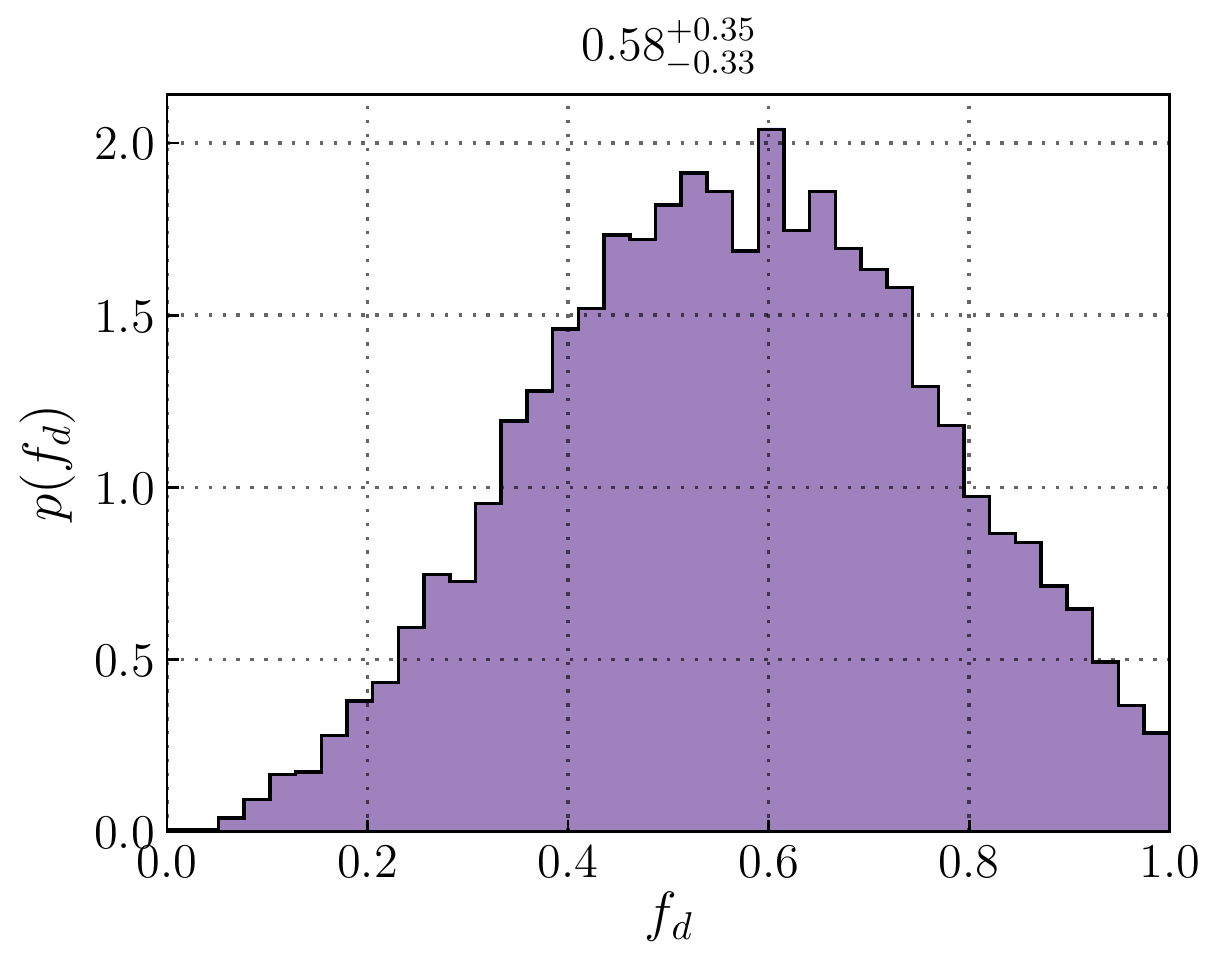}
    \caption{
    Posterior distributions for $f_i$ (\textit{left}) and $f_d$ (\textit{right}) of BBH mergers as defined Eq.~\ref{eq:fraction_d_i}.
    Assuming all binaries with $\chi_\mathrm{eff}<0$ are dynamically formed, $f_d$ corresponds to the fraction of dynamically assembled BBH systems. $f_i$ corresponds to the fraction of BBH systems formed through the isolated channel.
    }
    \label{fig:fraction_d_i_posterior}
\end{figure}

\added{As described in the main text, we can leverage the assumption that BBH with negative values of $\chi_\mathrm{eff}$ are formed dynamically to infer the fraction of binaries formed via dynamical ($f_d$) and isolated ($f_i$) channels; see Eq.~\eqref{eq:fraction_d_i}.
In Fig.~\ref{fig:fraction_d_i_posterior} we show the corresponding posterior distributions for $f_i$ and $f_d$.
}

\begin{figure}
    \centering
    \includegraphics[width=0.5\textwidth]{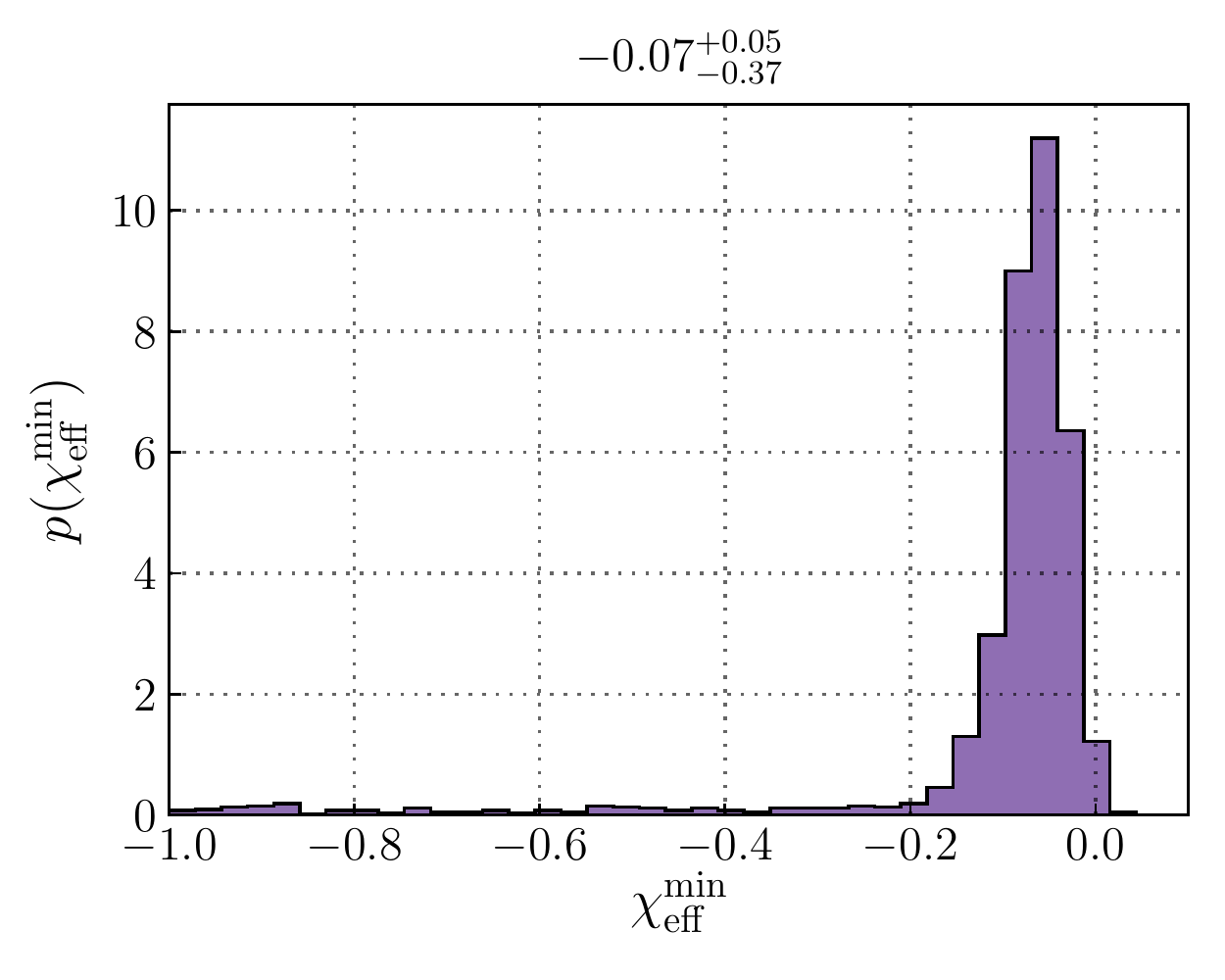}
    \caption{
    Posterior distribution for $\chi_\mathrm{eff}^\mathrm{min}$, below which we truncate the \textsc{Gaussian} $\chi_\mathrm{eff}$ distribution.
    While the results shown in the main text presume $\chi_\mathrm{eff}^\mathrm{min}=-1$, in Sec.~\ref{sec:Gaussian} we elevate $\chi_\mathrm{eff}^\mathrm{min}$ to a free hyper-parameter to be determined by the data; the resulting marginalized posterior distribution is shown here.
    In this case, we exclude $\chi_\mathrm{eff}^\mathrm{min}\geq 0$ at \percentMinChiLessThanZero{} credibility.
    This finding affirms that the signatures of \antialigned{} BH spins are present in our BBH catalog, and not a bias due to our choice of parameterized spin model.
    }
    \label{fig:chiMin_posterior}
\end{figure}

There are several sources of possible bias that might influence our \textsc{Gaussian} model conclusions.
One possible source of bias is the mass model presumed for the \textsc{Gaussian} spin analysis.
As noted above, measurements of a binary's $\chi_\mathrm{eff}$ and mass ratio $q$ are generally anti-correlated~\citep{Ng2018}.
Therefore, our particular choice of $\beta_q=1.3$ could conceivably affect conclusions regarding the $\chi_\mathrm{eff}$ distribution.
We have directly verified that the results in Fig.~\ref{fig:chiMin_posterior} remain robust under different fiducial choices of $\beta_q$ between $-1.5$ and $2$.

\begin{figure}
    \centering
    \includegraphics[width=0.48\textwidth]{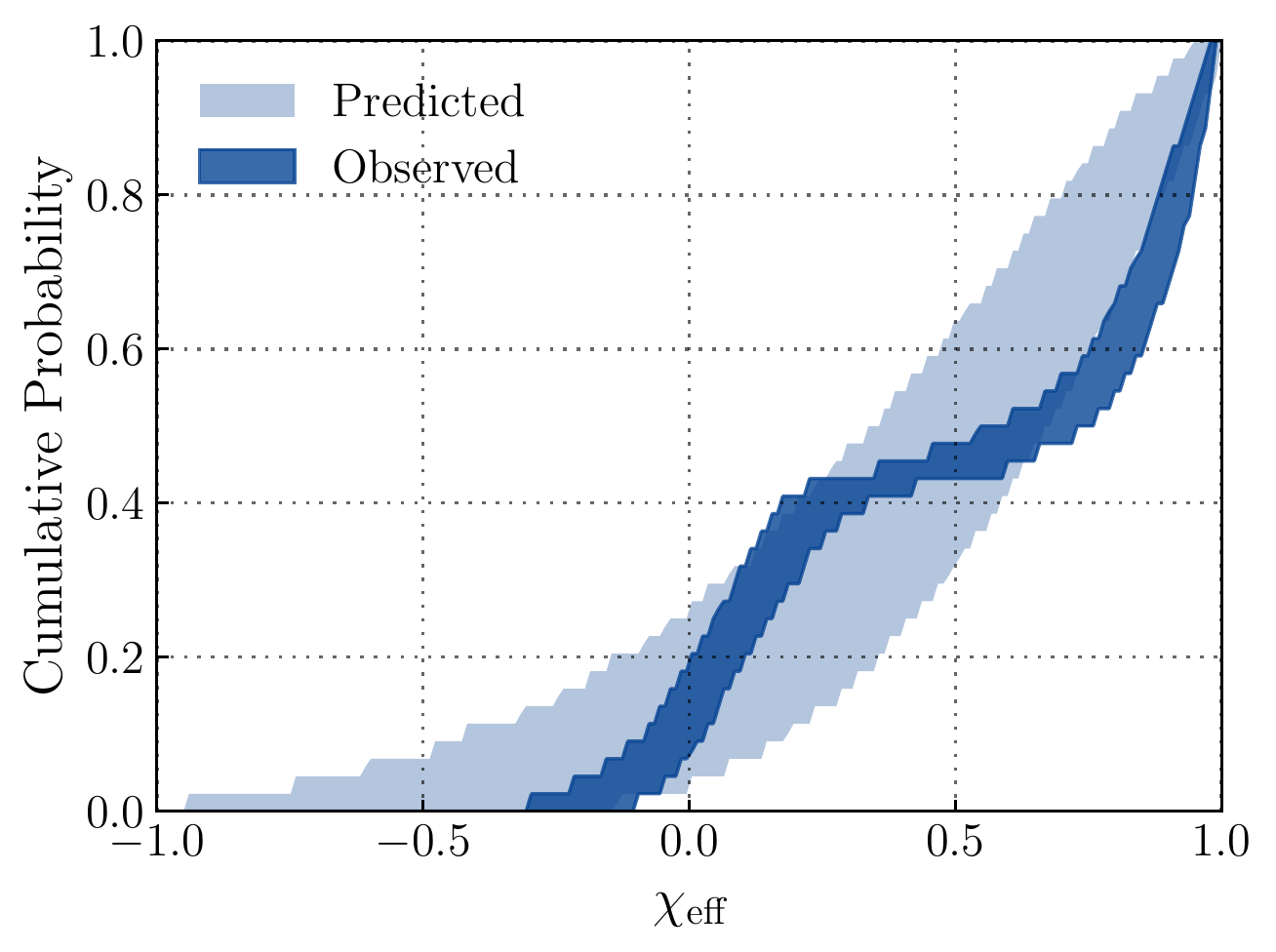}
    \caption{
	Example of a failed posterior predictive check for the \chieff{} $\chi_\mathrm{eff}$ distribution. We fit the \textsc{Gaussian} spin model to a mock catalog drawn from the strongly bimodal $\chi_{\rm eff}$ distributions predicted in the presence of tidal torques~\citep{2016MNRAS.462..844K, 2018MNRAS.473.4174Z,Bavera2019,2020AnA...636A.104B}.
	The dark shaded region shows the central 90\% credible bounds on the mock observed cumulative $\chi_{\rm eff}$ distribution, while the dark region corresponds to that predicted by the model.
	In this case, our \textsc{Gaussian} model is an extremely poor representation of the underlying $\chi_{\rm eff}$ distribution, and so significant tension is seen between the observed and predicted distributions, with the model overpredicting the fraction of observations with $\chi_{\rm eff} < 0$ as well the fraction of observations with $0.3 \lesssim \chi_{\rm eff} \lesssim 0.8$.
    }
    \label{fig:tidal_ppc}
\end{figure}

Another source of bias may be the Gaussian functional form we impose on the $\chi_\mathrm{eff}$ and $\chi_\mathrm{p}$ distributions, enforcing a unimodal distribution with smooth tails.
As discussed in Sec.~\ref{sec:results-spin}, though, the \textsc{Default} spin model yields near-identical $\chi_\mathrm{eff}$ and $\chi_\mathrm{p}$ distributions, despite its different parametrization and different physical assumptions.
As an additional check, we repeat the \textsc{Gaussian} spin analysis on our data, truncating the $\chi_\mathrm{eff}$ distribution not on $(-1,1)$, but on $(\chi_\mathrm{eff}^\mathrm{min},1)$, where $\chi_\mathrm{eff}^\mathrm{min}$ is inferred from the data.
Figure~\ref{fig:chiMin_posterior} shows the marginal posterior for $\chi_\mathrm{eff}^\mathrm{min}$.
We find that $\chi_\mathrm{eff}^\mathrm{min}$ is constrained to be negative at \result{\percentMinChiLessThanZero} credibility, confirming that support for negative \chieff{} is a feature of the data and not simply an artifact of the \textsc{Gaussian} model.
\added{
In contrast, when we repeat the measurement of $\chi_\mathrm{eff}^\mathrm{min}$ using simulated catalogs drawn from (\textit{i}) a Gaussian population truncated to strictly positive values, and (\textit{ii}) a pair of delta functions at $\chi_\mathrm{eff} = 0$ and $0.1$, we correctly observe consistency with $\chi_\mathrm{eff}^\mathrm{min} = 0$.
Examining $\chi_\mathrm{eff}^\mathrm{min}$ is therefore a useful safeguard even when the true population is poorly fit by a Gaussian.
}
\added{However, if the $\chi_{\rm eff}$ distribution deviates too strongly from a Gaussian functional form, then it may remain possible to spuriously conclude that  $\chi_\mathrm{eff}^\mathrm{min} < 0$.
In the case of significant tidal torques, for example, it is predicted that the $\chi_{\rm eff}$ distribution is strongly bimodal, with effective inspiral spins clustered near $\chi_{\rm eff} = 0$ or $1$, with no support at $\chi_{\rm eff} < 0$~\citep{2016MNRAS.462..844K, 2018MNRAS.473.4174Z,Bavera2019,2020AnA...636A.104B}.
To illustrate how results can be biased from model misspecification, we analyze mock observations drawn from a similar bimodal distribution. Using this intentionally misspecified model, we incorrectly conclude that $\chi_\mathrm{eff}^\mathrm{min} < 0$.
However, in cases of such extreme model mismatch, we expect that our data would fail the predictive check of Fig.~\ref{fig:gaussian-spin-ppc}. Indeed, Fig.~\ref{fig:tidal_ppc} shows the result of such a predictive check in the case of the bimodal, tidally-torqued $\chi_{\rm eff}$ distribution.
When fitting this population with a Gaussian model, we find that the model overpredicts the fraction of mock observations with negative $\chi_\mathrm{eff}$ as well as the range $0.3 \lesssim \chi_\mathrm{eff} \lesssim 0.8$, showing a clear deviation in the predicted and observed cumulative $\chi_{\rm eff}$ distributions.
}

\subsection{\ModelH{} model}\label{modelH}
\label{sec:modelH}
This model is an extension of the \truncated{} mass model with an additional Gaussian component.
It is similar to the \ppsn{} model, but there are several differences.
First, the high-mass subpopulation in \ModelH{} is described by a Gaussian in both $m_1$ and $m_2$ (up to the $m_1 \geq m_2$ truncation) while \ppsn{} only models $m_1$ as a Gaussian, and assumes that all \bbhsys{} are described by a power-law distribution in mass ratio $q$.
Most importantly, as its name suggests, \ModelH{} allows each subpopulation to have its own independent spin distribution, each of which follows the \textsc{Default} model, with $\zeta = 1$.
This allows us to probe whether the spin distribution varies with mass.
The parameters for \ModelH{} are summarized in Table~\ref{tab:parameters_multispin}.

\begin{table*}[h]
    \centering
    \begin{tabular}{ c p{11cm} p{2mm} p{20mm} }
        \hline
        {\bf Parameter} & \textbf{Description} &  & \textbf{Prior} \\\hline\hline
        $\mathcal{R}_{\mathrm{pl}}$ & Local merger rate for the low-mass power-law subpopulation. & & U($0, 5000$) \\
        $\mathcal{R}_{\mathrm{g}}$ & Local merger rate for the high-mass Gaussian subpopulation. & & U($0, 5000$) \\
        $\alpha_m$ & Power-law slope of the primary mass distribution for the low-mass subpopulation. & & U($-4, 12$) \\
        $\beta_q$ & Power-law slope of the mass ratio distribution for the low-mass subpopulation & & U($-4, 10$) \\
        $m_{\mathrm{min}}$ & Minimum mass of the primary mass distribution for the low-mass subpopulation. & & U($2, 10$) \\
        $m_{\mathrm{max}}$ & Maximum mass of the primary mass distribution for the low-mass subpopulation. & & U($30, 100$) \\
        $\mu_{m_1}$ & Centroid of the primary mass distribution for the high-mass subpopulation & & U($20,50$) \\
        $\sigma_{m_1}$ & Width of the primary mass distribution for the high-mass subpopulation & & U($0.4,10$) \\
        $\mu_{m_2}$ & Centroid of the secondary mass distribution for the high-mass subpopulation & & U($20,50$) \\
        $\sigma_{m_2}$ & Width of the secondary mass distribution for the high-mass subpopulation & & U($0.4,10$) \\
        $\mathrm{Mean} \chi_{1,\mathrm{pl}}$ & Mean of the beta distribution of primary spin magnitudes for the low-mass subpopulation. & & U($0,1$) \\
        $\mathrm{Var} \chi_{1,\mathrm{pl}}$ & Variance of the beta distribution of primary spin magnitudes for the low-mass subpopulation. &  & U($0,0.25$) \\
        $\sigma_{1,\mathrm{pl}}$ & Width of the truncated Gaussian distribution of $\cos$(primary spin tilt angle) for the low-mass subpopulation. &  & U($0,4$) \\
        $\mathrm{Mean} \chi_{2,\mathrm{pl}}$ & Mean of the beta distribution of secondary spin magnitudes for the low-mass subpopulation. & & U($0,1$) \\
        $\mathrm{Var} \chi_{2,\mathrm{pl}}$ & Variance of the beta distribution of secondary spin magnitudes for the low-mass subpopulation. &  & U($0,0.25$) \\
        $\sigma_{2,\mathrm{pl}}$ & Width of the truncated Gaussian distribution of $\cos$(secondary spin tilt angle) for the low-mass subpopulation. &  & U($0,4$) \\
        $\mathrm{Mean} \chi_{1,\mathrm{g}}$ & Mean of the beta distribution of primary spin magnitudes for the high-mass subpopulation. & & U($0,1$) \\
        $\mathrm{Var} \chi_{1,\mathrm{g}}$ & Variance of the beta distribution of primary spin magnitudes for the high-mass subpopulation. &  & U($0,0.25$) \\
        $\sigma_{1,\mathrm{g}}$ & Width of the truncated Gaussian distribution of $\cos$(primary spin tilt angle) for the high-mass subpopulation. &  & U($0,4$) \\
        $\mathrm{Mean} \chi_{2,\mathrm{g}}$ & Mean of the beta distribution of secondary spin magnitudes for the high-mass subpopulation. & & U($0,1$) \\
        $\mathrm{Var} \chi_{2,\mathrm{g}}$ & Variance of the beta distribution of secondary spin magnitudes for the high-mass subpopulation. &  & U($0,0.25$) \\
        $\sigma_{2,\mathrm{g}}$ & Width of the truncated Gaussian distribution of $\cos$(secondary spin tilt angle) for the high-mass subpopulation. &  & U($0,4$) \\
        \hline
    \end{tabular}
    \caption{
    Summary of \ModelH{} parameters.
    }
  \label{tab:parameters_multispin}
\end{table*}

\section{Redshift Evolution Models} \label{Appendix:redshift}
\begin{figure*}
    \centering
    \includegraphics[width = 0.5\textwidth]{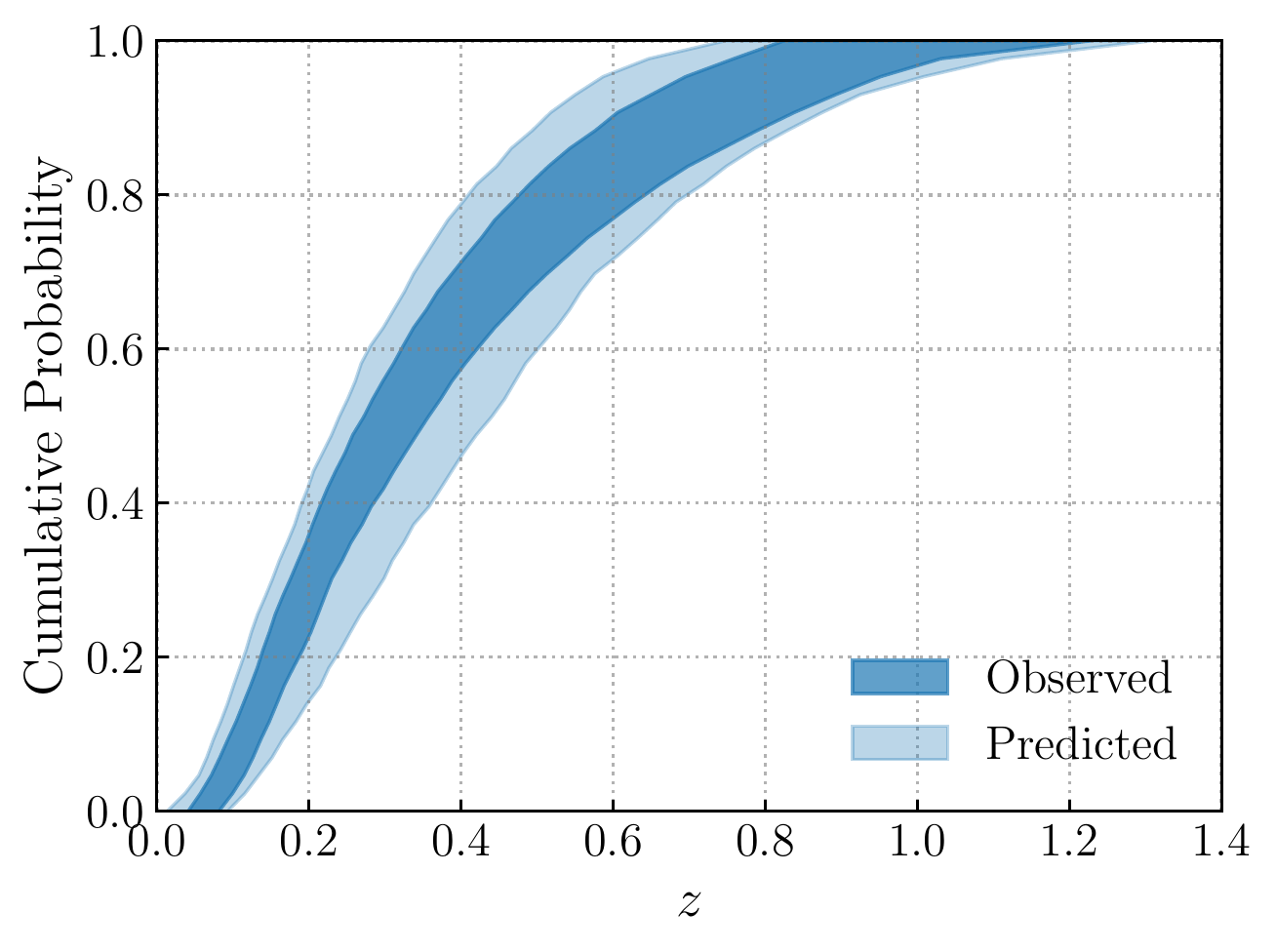}
    \caption{
    A posterior predictive check: the cumulative density function (CDF) for the \textsc{Power Law Evolution} model.
    The model is a good fit to the data.
    \NAME{GW190814A}{} is excluded from this analysis.
    }
    \label{fig:compare_cdfs_redshift}
\end{figure*}
The \textsc{power-law} redshift evolution model parameterizes the merger rate density as
\begin{equation}
    \mathcal{R}(z) = \mathcal{R}_0 (1+z)^\kappa,
\end{equation}
where $\mathcal{R}_0$ denotes the merger rate density at $z = 0$. 
This implies that the redshift distribution is
\begin{equation}
    \frac{dN}{dz} =  \mathcal{C} \frac{dV_c}{dz} (1 + z)^{\kappa-1},
\end{equation}
where $dV_c/dz$ is the differential comoving volume, and $\mathcal{C}$ is related to $\mathcal{R}_0$ by 
\begin{equation}
   \mathcal{R}_0 = \mathcal{C} \frac{dV_c}{dz} \left[{\int_0^{z_\mathrm{max}} \frac{dV_c}{dz} (1 + z)^{\kappa-1}}\right]^{-1}.
\end{equation}
We take $z_\mathrm{max} = 2.3$ in the analysis, as this is a conservative upper bound on the redshift at which we could detect BBH systems during O3a within the mass range considered here.
When fitting this model, we employ a uniform prior on $\kappa$ centered at $\kappa = 0$.
The value $\kappa = 0$ corresponds to no evolution; i.e., a merger rate that is uniform-in-comoving volume and source-frame time. We take a sufficiently wide prior so that the likelihood is entirely within the prior range, $\kappa \in (-6,6)$.

\section{Gravitational-wave lensing} \label{other}

It has been suggested that gravitational-wave lensing could bias the estimate of binary masses~\citep{Dai:2016igl,Ng:2017yiu,Li:2018prc,2018MNRAS.480.3842O,Hannuksela:2019kle}, which could lead to a biased population inference.
However, based on the predictions on the number of expected gravitational-wave sources and the distribution of galaxy lenses in the Universe, \citet{Li:2018prc,2018MNRAS.480.3842O} predict that only around one in a thousand observed events are lensed, although this estimate can vary depending on the redshift evolution of the merger-rate density.
The lensing rate is expected to be similarly rare for galaxy clusters~\citep{2018MNRAS.475.3823S}.
As the expected lensing rate is low compared to the number of events in GWTC-2, we assume that all events in our sample are unlensed.

\end{document}